\RequirePackage{fix-cm}
\RequirePackage{rotating}
\documentclass[smallextended]{svjour3}       
\smartqed  
\usepackage{multirow}
\usepackage{graphicx}
\usepackage{textcomp}
\usepackage{xcolor}
\usepackage{subfigure}
\usepackage{colortbl}
\usepackage{pifont}
\usepackage{multirow}
\usepackage{enumitem}
\usepackage{amsmath, amsfonts}
\usepackage{listings}
\usepackage{framed}
\usepackage{float}
\usepackage{fancyhdr}
\usepackage{url}
\usepackage{pifont}
\usepackage[noend]{algpseudocode}
\usepackage{algorithm}
\usepackage{xspace}
\usepackage{paralist}
\usepackage{booktabs}
\usepackage{pifont}
\usepackage{tcolorbox}
\usepackage{todonotes}

\usepackage[switch]{lineno}

\usepackage{wrapfig}
\usepackage{hyperref}

\usepackage{rotating}

\usepackage{soul, color,xcolor}     
 \definecolor{myColor}{rgb}{0,0,255}   
\makeatletter
\newcommand*{\revision}{\@ifnextchar\bgroup{\revision@}{\color{myColor}}}
\newcommand*{\revision@}[1]{{\textcolor{myColor}{#1}}}
\makeatother

\begin{document}

\title{Boosting Source Code Learning with Text-Oriented Data Augmentation: An Empirical Study 
}


\author{Zeming Dong$^{1}$,  Qiang Hu*$^{2}$, Yuejun Guo$^{3}$, Zhenya Zhang$^{4}$, Maxime Cordy$^{1}$, Mike Papadakis$^{1}$, Yves Le Traon$^{1}$ and Jianjun Zhao$^{4}$ \\
	\normalsize $^{1}$University of Luxembourg, Luxembourg\\
        \normalsize $^{2}$Tianjin University, China\\
	\normalsize $^{3}$Luxembourg Institute of Science and Technology, Luxembourg\\
        \normalsize $^{4}$Kyushu University, Japan\\ 
}

\authorrunning{Dong and Hu et.al} 

\institute{*Corresponding author 
              \normalsize 
}


\maketitle

\begin{abstract}
Recent studies have demonstrated remarkable advancements in \emph{source code learning}, which applies \emph{deep neural networks (DNNs)} to tackle various software engineering tasks. Similar to other DNN-based domains, source code learning also requires massive high-quality training data to achieve the success of these applications. Data augmentation, a technique used to produce additional training data, is widely adopted in other domains (e.g. \emph{computer vision}). However, the existing practice of data augmentation in source code learning is limited to simple syntax-preserved methods, such as code refactoring. In this paper, considering that source code can also be represented as text data, we take an early step to investigate the effectiveness of data augmentation methods originally designed for natural language texts in the context of source code learning. To this end, we focus on code classification tasks and conduct a comprehensive empirical study across four critical code problems and four DNN architectures to assess the effectiveness of 25 data augmentation methods. Our results reveal specific data augmentation methods that yield more accurate and robust models for source code learning. Additionally, we discover that the data augmentation methods remain beneficial even when they slightly break source code syntax. 


\end{abstract}

\section{Introduction}
\label{sec: intro}

In recent years, \emph{source code learning} that applies \emph{machine learning (ML)} in the domain of big code (\textbf{ML4Code})~\cite{allamanis2018survey} has gained significant attention. ML4Code leverages the power of ML, especially \emph{deep learning (DL)}, to extract patterns from large code corpora. The remarkable performance of ML4Code in various downstream code tasks, such as \emph{program repair}~\cite{goues2019automated,dinella2020hoppity}, \emph{clone detection}~\cite{svajlenko2014towards,wang2020detecting}, \emph{code generation}~\cite{hu2018deep,wan2018improving}, \emph{bug detection}~\cite{hu2019re,zhou2019devign}, and \emph{problem classification}~\cite{puri2021codenet,dong2023mixcode}, demonstrate its immense potential in facilitating software developers in daily activity.

Well-designed model architectures and high-quality training data are essential factors in producing \emph{programming language~(PL)} models with outstanding performance. In practice, the model architecture which can be informed by ideas from the \emph{natural language processing (NLP)} field~\cite{allamanis2018survey,10.1145/2666356.2594321,hindle2016naturalness} has been extensively studied by researchers and many models can achieve state-of-the-art performance, e.g., \emph{GraphCodeBERT}~\cite{guo2020graphcodebert} and \emph{CodeBERT}~\cite{feng2020codebert}. However, there is no shortcut to preparing high-quality training data. The challenge comes from the expensive human efforts required for collecting, cleaning and annotating data. For example, annotating only four libraries of code can take 600 man-hours~\cite{zhou2019devign}. As a result, it remains a challenging problem to prepare sufficient training data for PL models.

To address the issue of data scarcity, integrating data augmentation into training can be a promising solution. Essentially, data augmentation generates new training data by modifying existing labeled data under the premise that these new data preserve the original semantics. For example, when training an image classification model, instead of using the original training data only, a common practice is to utilize image transformation techniques~\cite{cifar10c2019dan} to produce more diverse images. Despite the remarkable attention data augmentation has already gained in other fields, such as \emph{computer vision (CV)} and NLP, its application in source code learning has received limited attention and its potential has not been fully exploited. Most existing studies in ML4Code still stick to the design of more powerful model architectures (e.g., GraphCodeBERT~\cite{guo2020graphcodebert}), more generic code representation techniques (e.g., \emph{token-based}~\cite{white2015toward} or \emph{tree-based} representation~\cite{alon2019code2vec}), or learning strategies (e.g., using contrastive learning for code tasks~\cite{bui2021self}). Only a few works tend to study the problem of automatically enriching the code-related training data, such as recent works~\cite{allamanis2021self,yu2022data} in which a series of code refactoring methods are introduced for data augmentation. However, as reported in their works, the performance of those methods is very limited, and more effective data augmentation methods in source code learning are still in demand. In this work, to bridge this gap, we investigate new data augmentation methods to improve the performance of model training in source code learning.

\smallskip
\noindent\textbf{Our Contributions:}
In source code learning, raw code is often converted into machine-readable data format by representing it as a sequence of text tokens due to its analogy to texts~\cite{allamanis2018survey}, where each token is represented by an integer and then fed into the code model. Inspired by this token-based representation, we empirically study the problem of whether existing data augmentation approaches in NLP (that handles text data) are effective in improving the training quality in source code learning. Concretely, we first survey and categorize existing text-oriented data augmentation methods in the literature, and we find seven data augmentation methods that are applicable to the source code. Then, we adapt these methods to train code models and investigate their effectiveness in improving the accuracy and robustness of those models. Overall, our study considers two mainstream programming languages (Java and Python), two low-resource programming languages (Ruby and Go), four crucial downstream classification tasks (problem classification, bug detection, authorship attribution, and clone detection), and four DNN model architectures including two pre-trained PL models (CodeBERT and GraphCodeBERT). 

This work builds on and extends our previous publication presented at an earlier conference~\cite{dong2023Boosting}. Specifically, the following contributions in the present paper are novel compared to our earlier conference publication:
\begin{compactitem} 
\item We explore a new and important problem (see Section~\ref{sec:section4}) to investigate whether text-oriented data augmentation methods can improve the robustness of source code models.

\item We add new experiments (see Section~\ref{sec:section7_2}) to discuss and delve deeper into the reasons why Mixup-based data augmentation techniques produce code models with superior performance compared to other techniques.

\item We conduct statistical tests (see Sections~\ref{sec:RQ1},~\ref{sec:RQ2},~\ref{sec:RQ3},~\ref{sec:RQ4}) using the \emph{Wilcoxon signed-rank test}~\cite{woolson2007wilcoxon} and apply the \emph{Bonferroni correction}~\cite{armstrong2014use,dong2024effectiveness_jss} to adjust \emph{p}-values for multiple comparisons, assessing the significance of each data augmentation method across all code models.  

\item We expand the scale of the experiment (see Sections~\ref{sec:RQ1},~\ref{sec:RQ2},~\ref{sec:RQ3}) by incorporating the \emph{Python800} dataset and \emph{CodRep1} dataset, thereby increasing the number of trained models from earlier 1,080 to 4,560.

\item We propose a new research question (see Section~\ref{sec:RQ4})) to explore whether text-oriented data augmentation methods are effective for low-resource programming languages.

\end{compactitem}

\begin{figure}[!tb]
	\centering
	\includegraphics[width=1.0\linewidth]{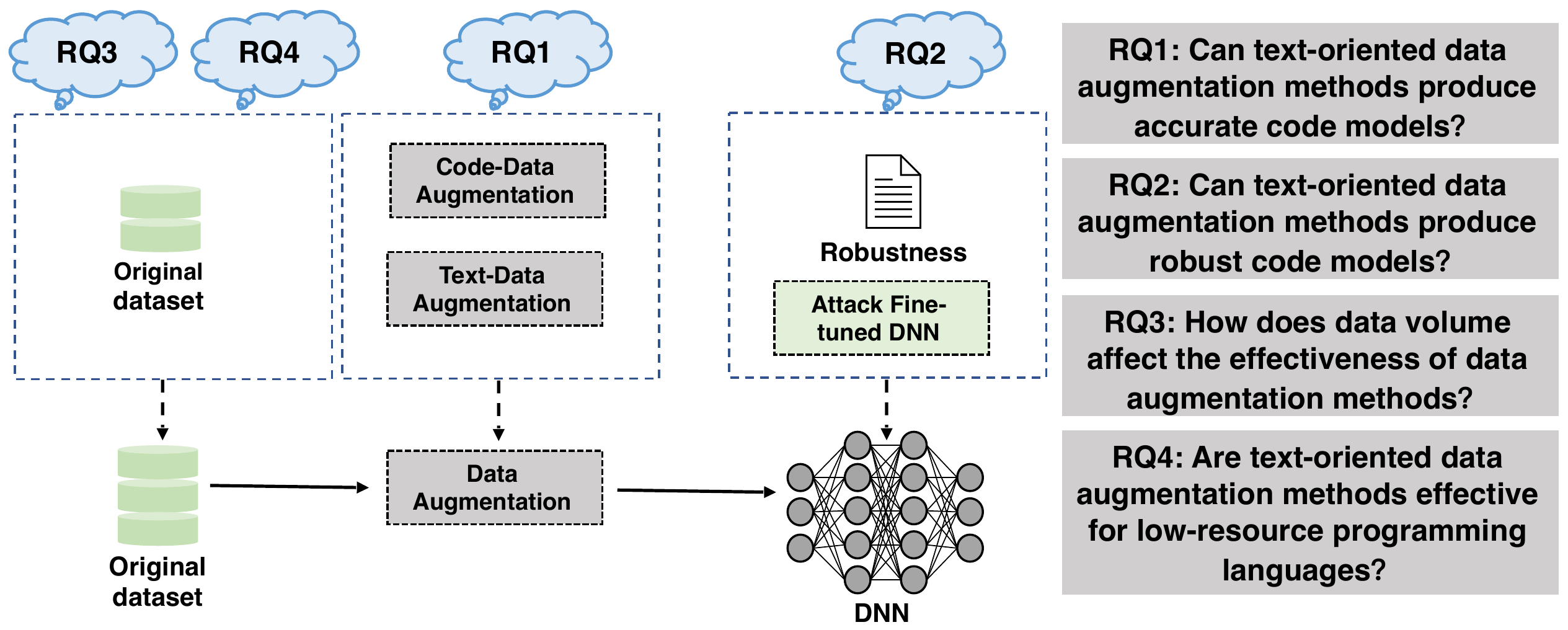}
	\caption{Overview of our empirical study}
	\label{fig:RQ_design}
\end{figure}

\noindent As shown in Fig.~\ref{fig:RQ_design}, we design our large-scale study to answer four research questions: 

\smallskip
\noindent\textbf{RQ1: Can text-oriented data augmentation methods produce accurate code models?} The results show that data augmentation methods that linearly mix feature vectors in code embedding, e.g., \emph{SenMixup}, can enhance the accuracy by up to 8.74\%, compared to the training without using data augmentation. Remarkably, the methods adapted from NLP are more effective than the code-specific data augmentation technique, namely, code refactoring. 

\smallskip
\noindent\textbf{RQ2: Can text-oriented data augmentation methods produce robust code models?} 
The results show that using text-oriented data augmentation brings limited robustness improvement. Specifically, \emph{Random Deletion (RD)}, a method from NLP,  performs the best and can reduce the attack success rate by 5.67\%. 

\smallskip
\noindent\textbf{RQ3: How does data volume affect the effectiveness of data augmentation methods?} The results demonstrate that when training data is scarce, incorporating data augmentation can help to improve both accuracy and robustness. For example, using \emph{SenMixup} can improve the accuracy of CodeBERT by up to 12.92\%, and \emph{Random Insertion (RI)} can enhance the robustness by up to 38.47\%.

\smallskip
\noindent\textbf{RQ4: Are text-oriented data augmentation methods effective for low-resource programming languages?}
The results suggest that text-oriented program transformation methods, especially \emph{RD}, \emph{Random Swap (RS)}, and \emph{SenMixup}, can help improve the robustness of DNN models for low-resource programming languages. For example, \emph{RD} enhances the robustness of CodeBERT by reducing ASR by 3.63\%.

Via the large-scale experiments, we found that data augmentation methods from NLP outperform the existing simple code refactoring data augmentation methods in most cases, for example, in clone detection-BigCloneBench, \emph{Back-translation (BT)} outperforms code refactoring method in 3 out of 4 cases. Even though some data augmentation methods (e.g., \emph{RS}) can produce training data that slightly breaks the syntax of the source code, they are still useful in improving the quality of training in source code learning. Besides, when training data are scarce, data augmentation is especially important since it can significantly improve the quality of trained code models. 

To summarize, our main contributions are:

\begin{compactitem}
\item This is the first study that adapts data augmentation methods from NLP to source code learning and empirically evaluates the effectiveness of training code models.

\item Our study yields several insights to help developers select optimal data augmentation methods for training code models. A key finding is that the methods that slightly break the syntax rules of the source code can still be beneficial in model training.

\item We make all our implementations, datasets, and models publicly available\footnote{\url{https://github.com/zemingd/PT4Code}\label{site}} to support and advance future research in data augmentation for source code learning.
\end{compactitem}

\section{Background}
\label{sec:background}
\subsection{Similarities between programming languages and natural languages}
\label{sec:background_representation}
Programming languages and natural languages exhibit similarities in terms of data representation and the dependency on syntactic and semantic rules. For instance, source code can be processed into a sequence of tokens just as text data in natural languages. In NLP, a sentence can be broken down into tokens such as words and punctuation marks. Similarly, in programming languages, a piece of code can be tokenized into separators, operators, reserved words, constants, and identifiers. For example, the source code ``\verb|def func(a, b)|'' is processed to a number of tokens as ``[\verb|`def', `func', `(', `a', `b', `)'|]''. 

The similarities make the adaptation of NLP techniques to code learning both intuitive and effective. For instance, representation techniques such as \emph{word2vec} ~\cite{church2017word2vec} in NLP have inspired \emph{code2vec} ~\cite{alon2019code2vec} for source code learning. word2vec is a technique that captures relationships between different words and then transforms them into numerical vectors and code2vec represents code snippets as single fixed-length vectors, capturing semantic relationships within code. In addition, model architectures originally designed for NLP tasks have been successfully adapted for code learning. BERT~\cite{devlin2018bert}, a transformer-based model~\cite{vaswani2017attention} that is designed for language modeling with a bidirectional training strategy, has led to the development of CodeBERT~\cite{feng2020codebert}, which is tailed for source code. BERT can be extended to handle multi-modality. CodeBERT, a bimodal pre-trained model extended from BERT, incorporates both natural language and source code for its training. Similarly, the Llama~\cite{touvron2023llama}, developed by Meta AI, has been leveraged to develop CodeLlama~\cite{roziere2023code} for source code learning. CodeLlama is a specialized variant of Llama 2, adapted by training on code-specific datasets and increasing data sampling from these datasets over extended training periods.

\subsection{Data augmentation in source code learning}
\label{sec:codeRefactor}
Despite the great advantages of DNN, there are two main bottlenecks that prevent DNNs from achieving high performance, 
1) the lack of high-quality labeled training data and 
2) the different data distribution between training data and testing data. 
One simple solution to these two problems is to increase the size and diversity of training data. Data augmentation is proposed to automatically produce additional synthetic training data by modifying existing data without further human effort~\cite{shorten2019survey}. Generally, data augmentation involves a family of well-designed data transformation methods. For instance, in image processing, commonly used data augmentation methods include re-scaling, zooming, random rotating, padding, and adding noise.

Recently, software engineering researchers also considered data augmentation in source code learning~\cite{allamanis2021self,wang2022bridging}, and the proposed methods are known as \emph{code refactoring}. 
In general, code refactoring, originally used for code simplification, involves a family of techniques that rewrite the syntactic structure of source code while keeping the semantic information~\cite{kaur2016analysis}. 
Commonly used code refactoring techniques include \emph{local variable renaming}, \emph{duplication}, \emph{dead store}, etc. For instance, \emph{local variable renaming} is a method that changes the names of a code element, including symbols, files, directories, packages, and modules. Technically, this method modifies the source code slightly but does not change the semantic behavior of the program.

However, existing studies~\cite{yu2022data,bielik2020adversarial} have shown that these simple strategies have limited advantages in improving the performance of code models. 
In this study, inspired by the analogy of source code to texts (as mentioned in Section~\ref{sec:background_representation}), we empirically investigate whether data augmentation methods from NLP (that handles text data) can effectively enrich the diversity of training data for source code learning. 

\section{Adapting Data Augmentation Methods for Source Code Learning}
\label{sec: secction3}

\begin{figure}[!tb]
	\centering
	\includegraphics[width=1.0\linewidth]{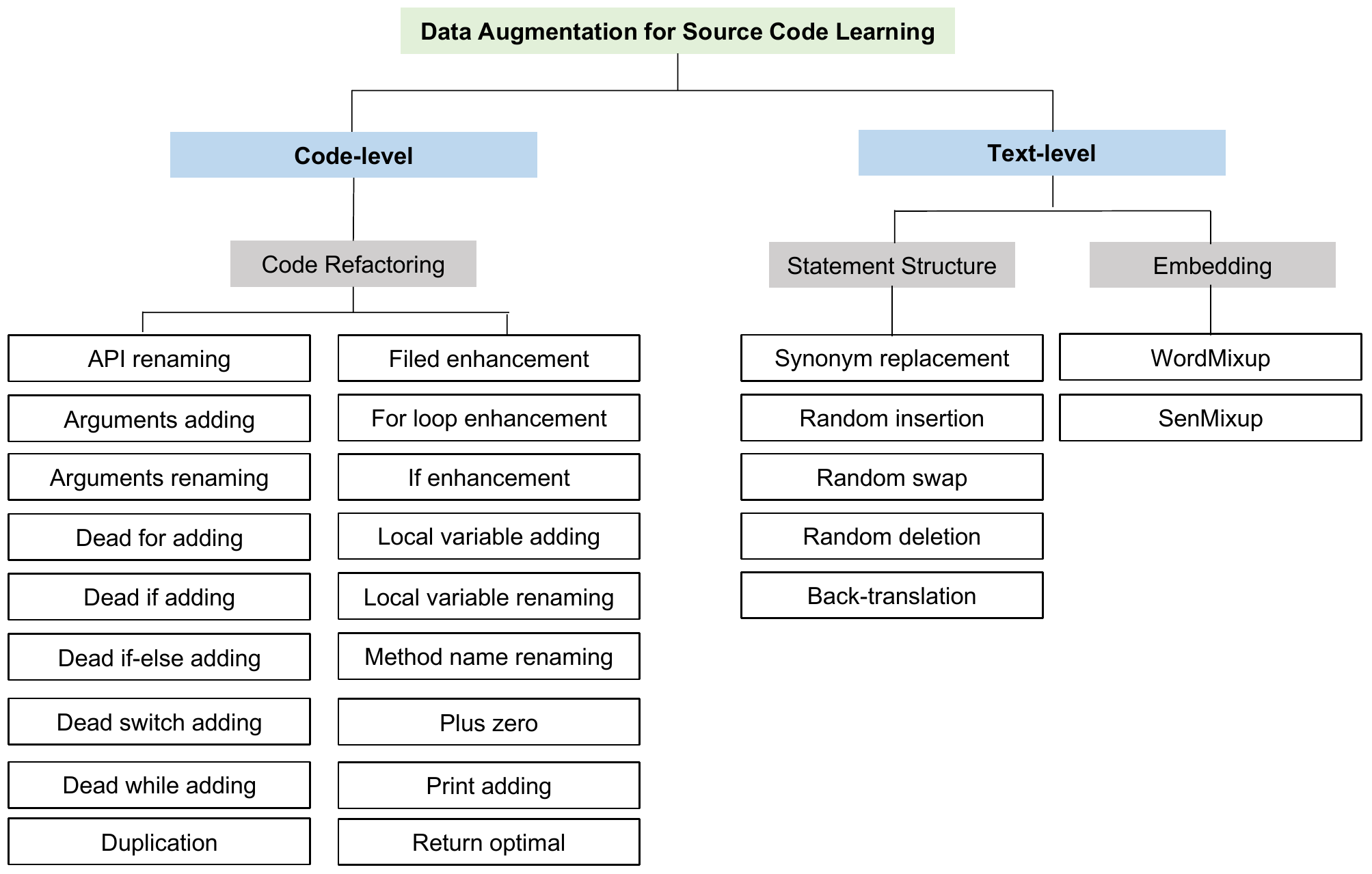}
	\caption{Data augmentation methods}
	\label{fig:DA_survey}
\end{figure}

In our study, we explore the adaptation of data augmentation methods from NLP to source code learning. This is motivated by 1) the similarities between natural language and programming language; 2) For instance, both can be processed as sequential tokens and follow defined syntactic and semantic rules. Concretely, we study seven augmentation methods shown in Fig.~\ref{fig:DA_survey}. These methods can be grouped into three categories:

\begin{compactitem}[$\bullet$]
    \item \textbf{Paraphrasing} can express the same information as the original form and has been commonly used in NLP~\cite{feng2021survey}. In this paper, we select the \emph{Back-translation}~\cite{xie2020unsupervised} method. 
    \item \textbf{Noising-based methods} slightly add noise to the original data but keep their semantic information~\cite{wei-zou-2019-eda}. Four types of noise injection methods are employed in this study, namely, \emph{Synonym Replacement}, \emph{Random Insertion}, \emph{Random Swap}, and \emph{Random Deletion}. 
    \item \textbf{Sampling-based methods} generate new synthetic data by linearly mixing the latent embeddings instead of directly operating on the raw text data. Unlike other methods, sampling-based methods are task-specific and require both data and label formats~\cite{feng2021survey}. 
    In this paper, we select two advanced sampling-based data augmentation methods used in NLP, namely \emph{WordMixup} and \emph{SenMixup}~\cite{guo2019augmenting}. 
\end{compactitem}

In the following sections, we elaborate on these seven data augmentation methods and particularly highlight the adaptations we have made to handle source code data.

\begin{figure}[!tb]
\centering
\subfigure[BT program transformation]{
\begin{minipage}[t]{0.5\linewidth}
\centering
\label{fig:subfig_a}
\includegraphics[width=1.0\linewidth]{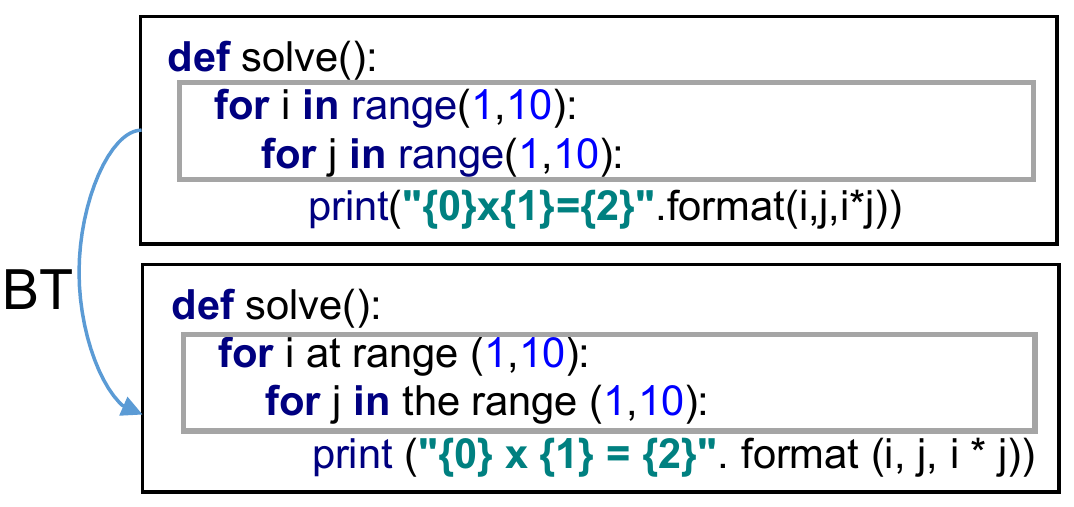}
\end{minipage}%
}%
\subfigure[SR program transformation]{
\begin{minipage}[t]{0.5\linewidth}
\centering
\label{fig:subfig_b}
\includegraphics[width=1.0\linewidth]{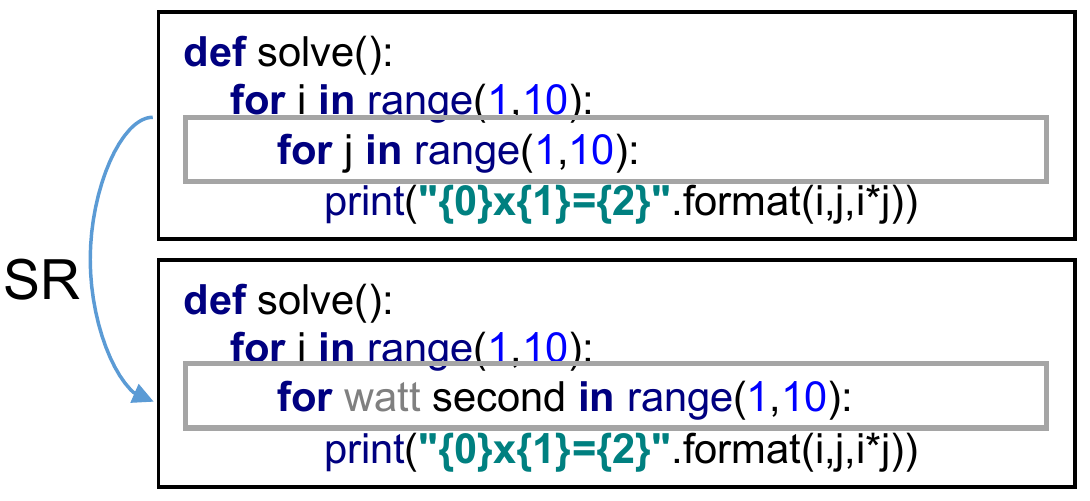}
\end{minipage}%
}%

\centering
\subfigure[RI program transformation]{
\begin{minipage}[t]{0.5\linewidth}
\centering
\label{fig:subfig_c}
\includegraphics[width=1.0\linewidth]{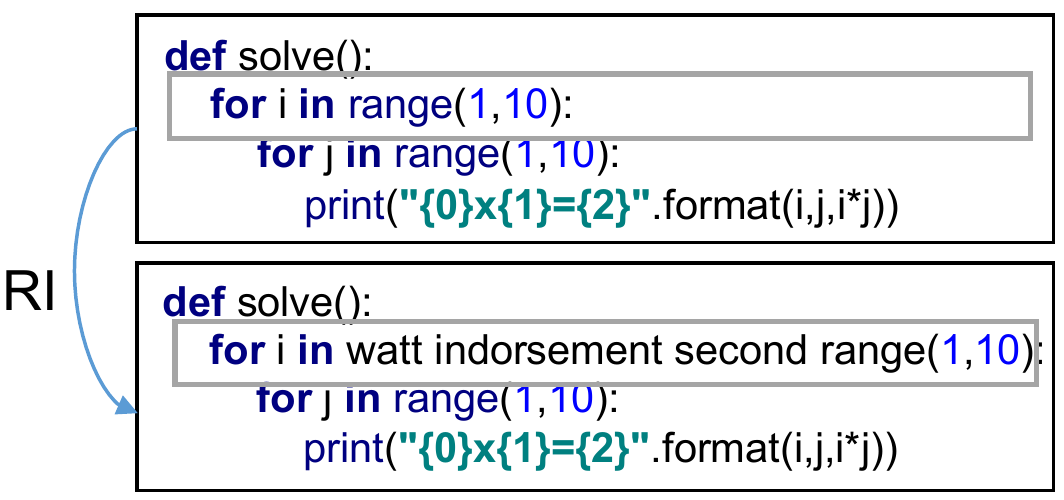}
\end{minipage}%
}%
\subfigure[RS program transformation]{
\begin{minipage}[t]{0.5\linewidth}
\centering
\label{fig:subfig_d}
\includegraphics[width=1.0\linewidth]{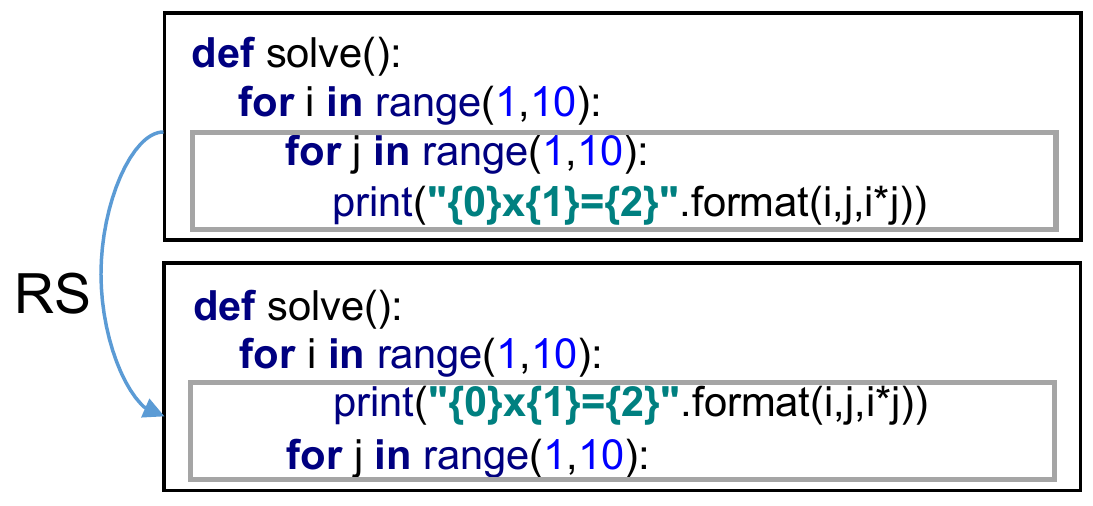}
\end{minipage}%
}%

\centering
\subfigure[RD program transformation]{
\begin{minipage}[t]{0.5\linewidth}
\centering
\label{fig:subfig_e}
\includegraphics[width=1.0\linewidth]{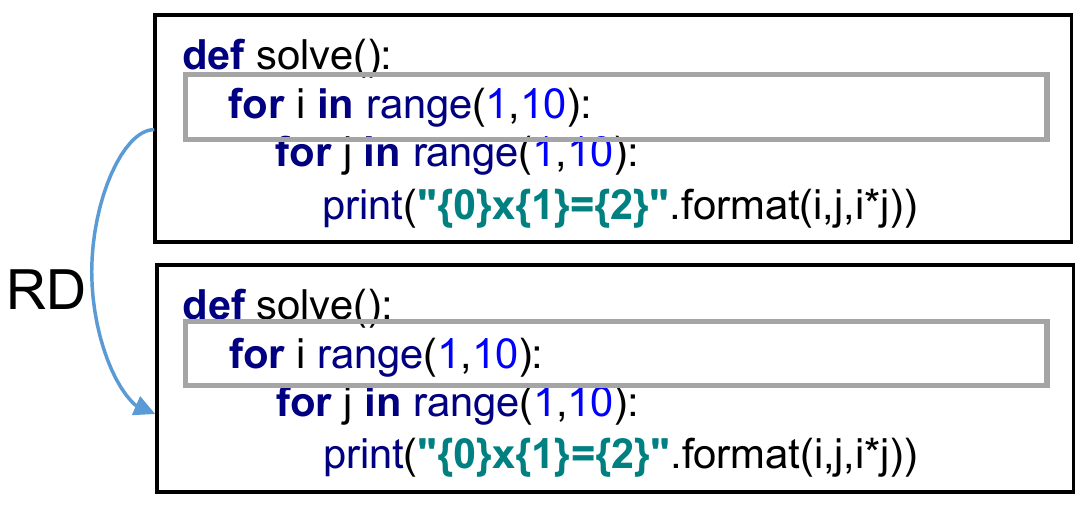}
\end{minipage}%
}%
\caption{Examples of data augmentation methods from NLP to source code learning, with a code snippet from Python800-p00000-s024467653.py (For each sub-figure, the upper part shows the code without data augmentation, and the lower part shows the code after applying data augmentation.) }
\label{fig:Section3_DAs}
\end{figure}

\subsection{Paraphrasing}
\label{subsec:da4nlp_BT}
\noindent\textbf{Back-translation (BT)} This method translates the original text into another language and then translates it back to the original one to generate additional data. In source code learning, we implement \emph{BT} by applying the English-French translation model bidirectionally for each statement in a program. A program $P$ with $k$ lines is represented as $P=\left\{S_{1},S_{2},\ldots,S_{k}\right\}$, where $S$ represents the tokenization of the statement and $S_{k}$ specifically is a statement $S$ at the $k$th line in a program. In our experiment, \emph{BT} is used as one program transformation method to produce the transformed code data $P_{bt}=\left\{S_{1}^{bt}, S_{2}^{bt},\ldots, S_{k}^{bt}\right\}$ from its original code data $P$. Specifically, we apply the English-French translation model (in both directions)~\cite{xie2020unsupervised,yu2018qanet,fabbri2020improving} to perform back-translation on each statement $S_{i}\left(S_{i}\in P\right)$ and obtain its paraphrases $S_{i}^{bt}\left(S_{i}^{bt}\in P_{bt}\right)$, where $S_{i}^{bt}$ is the new transformed statement generated by the bi-directional neural machine translation model. For example, in Fig.~\ref{fig:subfig_a}, after \emph{BT}, we replace the statement ``\verb|in range|'' in the original code with ``\verb|at range|'' and ``\verb|in the range|'' respectively.


\subsection{Noising-based methods}
\label{subsec:da4nlp_EDA}
\noindent
\textbf{Synonym Replacement (SR)} In NLP, this method randomly selects $n$ words from a sentence and then replaces the selected words with one of its randomly chosen synonyms. Different from \emph{BT}, to further enrich the diversity, \emph{SR} usually refrains from substituting strings that are semantically similar to the original text data. Specifically, in source code learning, we first randomly select $n$ statements from a program. Then, each of the $n$ words is replaced with one of its synonyms that is selected at random. In Fig.~\ref{fig:subfig_b}, we randomly select one statement from a program and then replace it with another string. This string is generated by selecting certain words from the statement and replacing them with randomly chosen synonyms. In this example, we first randomly chose the statement  ``\verb|for i in range(1,10)|'' and then replace the selected word ``\verb|j|'' with ``\verb|watt second|''. 
 
\smallskip
\noindent
\textbf{Random Insertion (RI)} \emph{RI} randomly inserts a random synonym of a
random word into a sentence to generate augmented text data. Different from \emph{RI} used in text data, we first select a random synonym of a random word in the chosen statement, then randomly insert this selected synonym into a random position of this statement. Generally, this process is repeated $n$ times. In Fig.~\ref{fig:subfig_c}, we randomly insert the string that is generated from synonyms in a random position of the selected statement from the original code, i.e., ``\verb|for i in range(1,10)|''. 

\smallskip
\noindent
\textbf{Random Swap (RS)} \emph{RS} randomly chooses two words in a sentence and then swaps their positions. Although the semantics of text data is, in general, sensitive to the order of words, within a limited level of word swapping, the text after \emph{RS} is often still understandable to humans. Therefore, \emph{RS} can be used to produce augmented text data. In source code learning, we randomly select two statements of a program and swap their positions, and this process is usually repeated $n$ times. In Fig.~\ref{fig:subfig_d}, we randomly select two statement ``\verb|for j in range(1,10)|'' and ``\verb|print("{0}x{1}={2}".format(i,j,i*j))|'', and swap their positions. 

\smallskip
\noindent
\textbf{Random Deletion (RD)} \emph{RD} randomly removes some words in a sentence or some sentences in a document, with a probability $p$, to generate augmented text data. In source code learning, we randomly delete words in a randomly chosen statement. In Fig.~\ref{fig:subfig_e}, we delete words in a statement with probability $p=0.01$. As a result, the operator ``\verb|in|'' is removed from the statement ``\verb|for i in range(1,10);|'' after \emph{RD}. 

\begin{tcolorbox}
\textbf{Note}: Although \textit{SR}, \textit{RI}, \textit{RS}, \textit{RD}, and \textit{BT} may slightly break the syntax and change the label of code, especially in the bug detection task, we found that these code augmentation methods are still useful in boosting the training of source code models.
\end{tcolorbox}

\begin{figure}[!tb]
	\centering
	\includegraphics[width=1.0\linewidth]{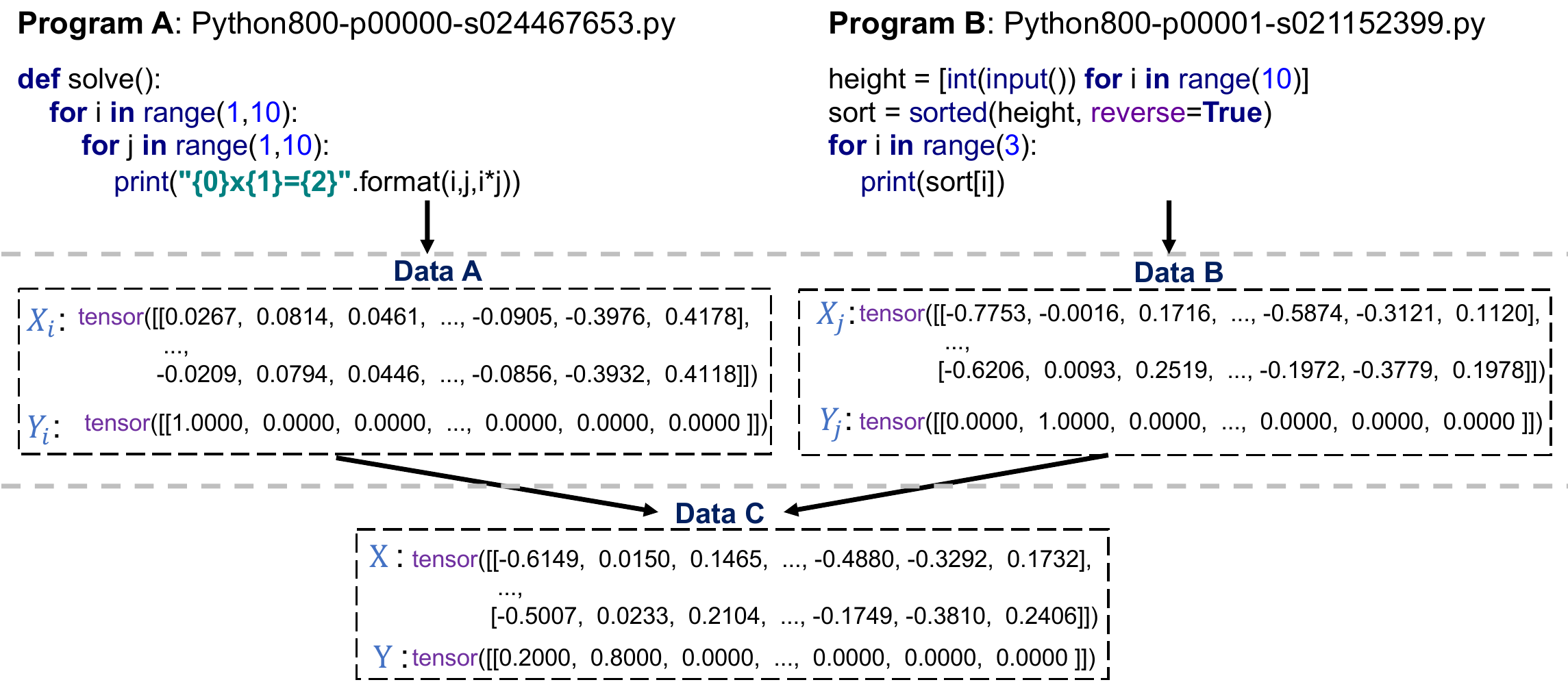}
	\caption{An example of linear interpolation of two programs}
	\label{fig:Section3_Mixup}
\end{figure}

\subsection{Sampling-based methods}
\label{subsec:da4nlp_Mixup}
We introduce two data augmentation methods based on \emph{Mixup}~\cite{zhang2017mixup}, a popular data augmentation approach in CV. In that context, \emph{Mixup} synthesizes new image data and their labels by linearly mixing the image features and the labels of two selected images. By taking two different data points and generating a new data point as a linear combination of the original data points, \emph{Mixup} effectively smooths the data distribution in the feature space, which aids in mitigating the sharpness of decision boundaries between different classes, making them less susceptible to overfitting~\cite{Masanari-why}. Furthermore, it has inspired the development of many data augmentation methods in other fields, including \emph{WordMixup} and \emph{SenMixup} in NLP, introduced as follows.

\smallskip
\noindent
\textbf{WordMixup and SenMixup.} 
Originally, \emph{WordMixup} interpolates the samples in the word embedding space, and \emph{SenMixup} interpolates the hidden states of sentence representations~\cite{guo2019augmenting}. We slightly modify \emph{WordMixup} and \emph{SenMixup} to adapt to source code learning. As shown in Eq.~(\ref{eqn:mixup}), there are two variants of \emph{Mixup} in our study. The first one, denoted as \emph{WordMixup}, interpolates samples in the embedding space of statement representation, and the second one, denoted as \emph{SenMixup}, conducts the interpolation after a linear transformation and before it is passed to a standard classifier that generates the predictive distribution over different labels. Given two pairs $\left(x^{i},y^{i}\right)$ and $\left(x^{j},y^{j}\right)$, where $x^{i}$ and $x^{j}$ represent the code data, and $y^{i}$ and $y^{j}$ are their corresponding labels, 
the interpolated new data  are obtained via \emph{WordMixup} and \emph{SenMixup}, as follows:
\begin{equation}\label{eqn:mixup}
\begin{aligned}
    &x_{\mathit{WordMix}}^{ij} = \lambda x^{i} + (1 - \lambda) x^{j} & \qquad &
     x_{\mathit{SenMix}}^{ij} = \lambda f(x^{i}) + (1 - \lambda) f\left(x^{j}\right) \\
    & y_{\mathit{WordMix}}^{ij} = \lambda y^{i} + \left(1 - \lambda\right) y^{j} & \qquad &
    y_{\mathit{SenMix}}^{ij} = \lambda y^{i} + \left(1 - \lambda\right) y^{j}
\end{aligned}
\end{equation}
\noindent Here \emph{SenMixup} follows a similar workflow with~\cite{guo2019augmenting}, and $f\left(\cdot\right)$ denotes a linear transformation method that is able to ensure that the input and the output have the same dimension. Moreover, $x_{\mathit{WordMix}}^{ij}$ and $x_{\mathit{SenMix}}^{ij}$ represent the new synthetic training data obtained by \emph{WordMixup} and \emph{SenMixup} respectively, and $y_{\mathit{WordMix}}^{ij}$ and $y_{\mathit{SenMix}}^{ij}$ are their labels.  The parameter $\lambda$ denotes the \emph{Mixup} ratio, and according to the work~\cite{zhang2017mixup}, it is sampled from a \emph{Beta} distribution with a shape parameter $\alpha\left(\lambda\sim\textit{Beta}\left(\alpha,\alpha\right)\right)$. 

To better understand how the linear interpolation method works, we use an example to show the details of linearly mixing two different programs (Program A and B from Python800 with label 0 and label 1, respectively) as depicted in Fig.~\ref{fig:Section3_Mixup}. First, we map a pair of programs (Data A and Data B) into the vector space through CodeBERT~\cite{feng2020codebert} and transform their labels into one-hot vectors with 800 classes. Next, we linearly mix the code vectors and label vectors, respectively, of Data A and Data B as the augmented training data (Data C) that could be used to train the model. 

\section{Study Design}
\label{sec:section4} 

We design four research questions to assess the effectiveness of data augmentation methods in Section~\ref{sec: secction3} in source code classification. The first three questions focus on widely used programming languages (i.e., Java and Python), while the last one evaluates the performance of data augmentation methods for low-resource programming languages, specifically Go and Ruby. The details are as follows:


\smallskip
\noindent\textbf{RQ1: Can text-oriented data augmentation methods produce accurate code models?} 

\noindent
\underline{\textit{Motivation}}. Accuracy is a basic metric that calculates the \% of correctly classified data over the entire test data~\cite{li2022data,feng2021survey}. Here, accuracy defined in Eq.(~\ref{eqn:accuracy}), refers to the accuracy of models on the original test data. 
\begin{equation}\label{eqn:accuracy}
\begin{aligned}
    Accuracy = \frac{\# \text{correctly\_classified\_data}}{\# \text{entire\_data}} \%
\end{aligned}
\end{equation}
These data generally follow the same data distribution as the training data. Therefore, we first assess whether data augmentation methods from NLP can improve the accuracy of code models. These methods are compared to training without data augmentation and with code refactoring (as introduced in Section~\ref{sec:codeRefactor}). 

\noindent
\underline{\textit{Setup}}. In this RQ, we first prepare the original training data, randomly initialize code models, and then train these models using different data augmentation methods as listed in Fig.~\ref{fig:DA_survey}. 
Specifically, we compare these methods in terms of three metrics, namely, 1) the convergence speed of the model, 2) the final accuracy of the model with fixed training epochs, and 3) the statistical test using the \emph{Wilcoxon signed-rank test}. 

\smallskip
\noindent\textbf{RQ2: Can text-oriented data augmentation methods produce robust code models?} 

\begin{figure}[!tb]
	\centering
	\includegraphics[width=1.0\linewidth]{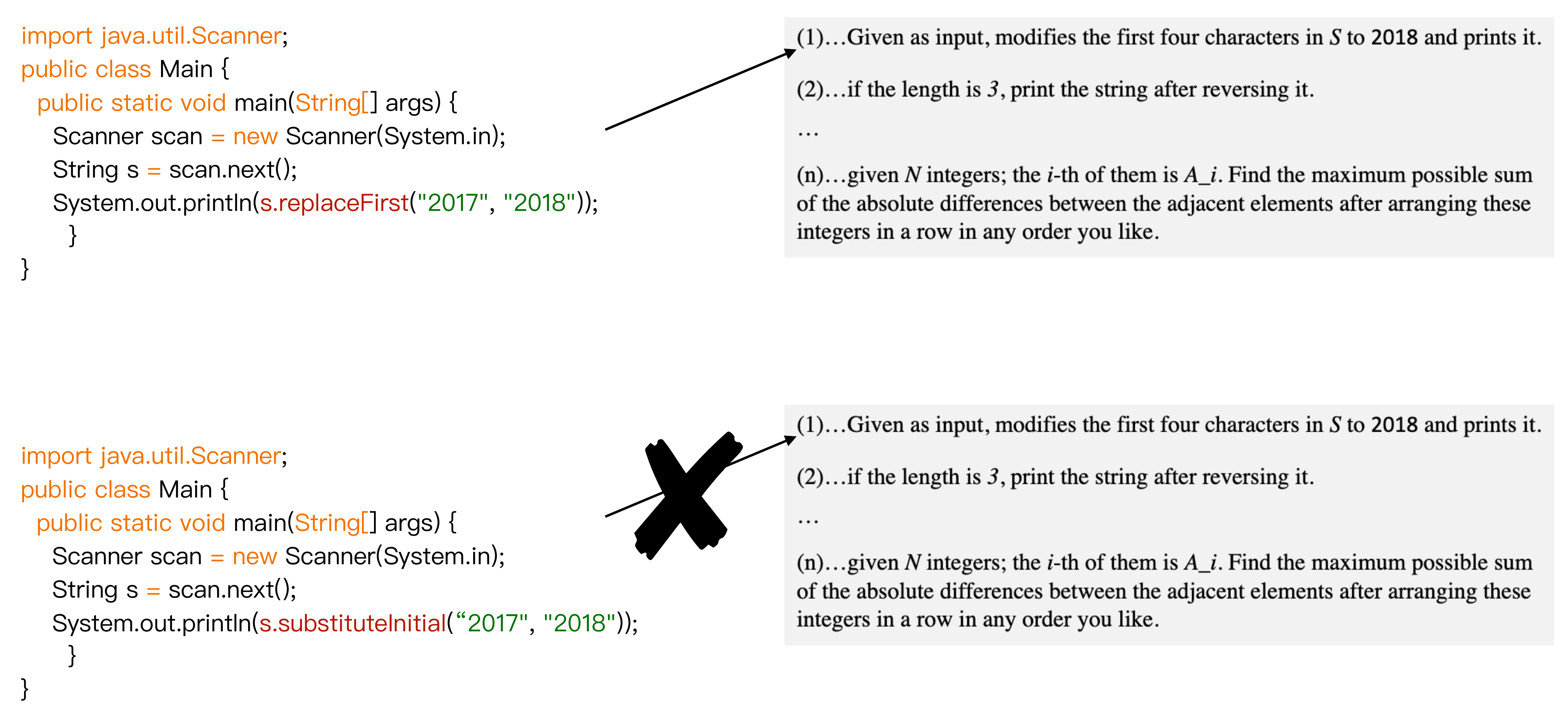}
	\caption{Example of a successful PL model attack.}
	\label{fig:robustness_motivation}
\end{figure}

\noindent
\underline{\textit{Motivation}}. Robustness reflects the generalization ability of DNN models. Given a multi-classification task, the PL model correctly identifies the first text description on the right of Fig.~\ref{fig:robustness_motivation}. However, when a perturbation renames the method from ``\verb|replaceFirst|'' to ``\verb|substituteInitial|'' in the source code, the model makes an incorrect decision.
Therefore, it is necessary to evaluate the robustness of well-trained PL models. 
The common way to measure the robustness of models is to use adversarial attack methods to attack models and check to what extent the model can defend against these attacks. Since robustness is another important metric that evaluates the generalization ability to handle unseen data of the trained model~\cite{bielik2020adversarial}, we obtain multiple trained code models and study their adversarial robustness in this RQ. According to the literature~\cite{rebuffi2021data}, data augmentation can help improve the adversarial robustness of DNN models in other fields (e.g., CV). Thus, we explore whether this conclusion also holds for code models. 

\noindent
\underline{\textit{Setup}}. Concretely, we follow the existing work~\cite{yang2022natural} and employ two state-of-the-art adversarial attacks, namely, \emph{naturalness aware attack (ALERT)}~\cite{yang2022natural} and \emph{Metropolis-Hastings modifier (MHM)} algorithm~\cite{zhang2020generating}, to evaluate the robustness of two pre-trained PL models. Both attacks replace the names of local variables in the program and force the model to produce wrong predictions accordingly. MHM applies the Metropolis-Hastings sampling methodology to get the name of the replacement variable, and ALERT uses the masked language prediction function of pre-trained models to search for substitutes. 
\begin{equation}\label{eqn:robustness}
\begin{aligned}
    ASR = \frac{\# \text{successfully\_generated\_adversarial\_data}}{\# \text{correctly\_classified\_test\_data}} \% 
\end{aligned}
\end{equation} 

Attack success rate (ASR), defined in Eq.(~\ref{eqn:robustness}), is used as the measurement of robustness, which calculates the \% of successfully created adversarial examples. Only correctly predicted test samples are utilized to create adversarial examples while undertaking adversarial attacks. A higher ASR indicates that an attack method has strong performance; in turn, the robustness of the victim model is low. Furthermore, as Bagoftoen and Seqoftoken cannot generate adversarial examples, we only evaluate the robustness of CodeBERT and GraphCodeBERT.

\smallskip
\noindent\textbf{RQ3: How does data volume affect the effectiveness of data augmentation methods?}

\noindent
\underline{\textit{Motivation}}. Given the original goal of data augmentation to address the issue of limited labeled data~\cite{li2022data,feng2021survey}, it is essential to explore the effectiveness of data augmentation methods in a practical scenario, namely, the case when there is no sufficient training data. 

\noindent
\underline{\textit{Setup}}. To this end, we reduce the size of the training set and repeat the evaluation as in RQ1 and RQ2, and we then check whether these data augmentation methods are still useful.

\smallskip
\noindent\textbf{RQ4: Are text-oriented data augmentation methods effective for low-resource programming languages?} 

\noindent
\underline{\textit{Motivation}}. Data augmentation not only addresses the limitation of insufficient training data but also considers the challenges involving low-resource languages~\cite{xia-etal-2019-generalized}. Therefore, we investigate another critical scenario where code data augmentation is necessary and conduct experiments on low-resource programming languages.

\noindent
\underline{\textit{Setup}}. We examine the statistics of the \emph{Project CodeNet} Dataset~\cite{puri2021codenet}, a large-scale, high-quality curated dataset that is widely used in the ML4Code research community. The dataset comprises 13,916,868 submissions across 55 different languages. For our experimental evaluation, we select Ruby and Go as our target programming languages because Ruby submissions constitute less than 2\% and Go submissions constitute less than 1\% of the total dataset. Lastly, we evaluate the effectiveness of text-oriented data augmentation methods for Ruby and Go in terms of accuracy and robustness.

\section{Experimental Setup}
\label{sec:section5}
Our study considers four programming languages, four crucial downstream tasks, four DNN model architectures, and two pre-trained PL models. Table~\ref{tab:data_model} shows the details of datasets and models used in the experiments.

\subsection{Baseline data augmentation}

In our experiments, we apply the data augmentation method uniformly to each code sample in the training dataset. We also maintain an equal number of training samples per epoch to ensure a fair comparison in terms of the training cost. Specifically, the number of augmented data samples matches the number of original training samples used in the standard training process.

Moving to methods, we totally collect 18 code refactoring methods from the existing literature~\cite{allamanis2021self,pour2021search,wei2022cocofuzzing} as baselines, such as \emph{local variable renaming}, \emph{if loop enhance}, and \emph{argument adding} in Fig.~\ref{fig:DA_survey} (please visit our project site\footref{site} for more details). In our experiment, for each code data, we randomly select one of these 18 code refactoring methods and apply it to the original code data to generate augmented code data\footnote{For simplicity, we refer to this method as \textit{Refactor} in this paper.}. 

Besides, to address the potential bias that augmented training data may become sparse and noisy due to random selection, we divide code refactoring methods into four groups, which are \emph{Rename Operator}, \emph{Dead Operator}, \emph{Inside Operator}, and \emph{Outside Operator}. We comprehensively examine both the robustness and accuracy performance of each code refactoring method and select the most representative method from each group. Specifically, we choose the method that demonstrates optimal robustness and accuracy within each group based on the findings from the prior work~\cite{dong2023mixcode}. Table~\ref{tab:refactor_class}  provides detailed information about the code refactoring groups. The \emph{Method name renaming} from the Rename Operator group, \emph{Dead if adding} from the Dead Operator group, \emph{If enhancement} from the Inside Operator group, and \emph{Filed enhancement} from the Outside Operator group are used for source code model training.

\begin{table*}[]
\centering
\caption{Description of code refactoring groups. The representative refactoring method from each group is highlighted in gray.}
\label{tab:refactor_class}
\resizebox{1.0\textwidth}{!}{
\begin{tabular}{clll}
\toprule
\textbf{No.} & \textbf{Group name} & \textbf{Refactoring method} & \textbf{Functionality} \\ \hline
1 & Rename Operator & \begin{tabular}[c]{@{}l@{}} API renaming\\ Local variable renaming \\ Arguments renaming \\  \cellcolor[HTML]{C0C0C0}{Method name renaming} \end{tabular} & Rename variables including API, argument, and method. \\ \hline

2 & Dead Operator &\begin{tabular}[c]{@{}l@{}} Dead for adding \\  \cellcolor[HTML]{C0C0C0}{Dead if adding} \\ Dead if else adding \\ Dead switch adding \\ Dead while adding \end{tabular} & Add unreachable loops and conditional statements.
 \\ \hline

3 & Inside Operator & \begin{tabular}[c]{@{}l@{}} Arguments adding\\   \cellcolor[HTML]{C0C0C0}{If enhancement} \\ For loop enhancement\\ Plus zero   \end{tabular}  & 
\begin{tabular}[c]{@{}l@{}} Transformations occur within the specific statement, \\ such as function definitions and numerical assignments.  \end{tabular} \\ \hline

4 & Outside Operator & \begin{tabular}[c]{@{}l@{}} Local variable adding \\ Duplication \\ \cellcolor[HTML]{C0C0C0}{Filed enhancement}\\ Print adding \\ Return optimal \end{tabular} & \begin{tabular}[c]{@{}l@{}} Insert new statements following the specific rule, \\ such as adding print lines and changing return content. \end{tabular} \\ \bottomrule

\end{tabular}
}
\end{table*}

\subsection{Tasks and datasets}
Code classification serves to estimate programs’ functionality automatically, which is crucial for software reuse~\cite{puri2021codenet}. Therefore, we first focus on code classification. Table~\ref{tab:data_model} presents the dataset and model details. 

\noindent\textbf{Dataset split.} For all considered datasets, we directly follow the settings provided by the official projects~\cite{hu2019re,puri2021codenet,yang2022natural} to split the data into training, validation, and test sets.
\begin{table}[!tb]
\centering
\caption{Details of tasks, datasets, and DNN models. \textbf{\#Training}: number of training data.  \textbf{\#Test}: number of test data.}
\label{tab:data_model}
\resizebox{1\columnwidth}{!}{
\begin{tabular}{llllll}
\hline
\multicolumn{6}{c}{\textbf{Most common programming languages}} \\ \hline
\textbf{Dataset} & \textbf{Language} & \textbf{Task} & \textbf{\#Training} & \textbf{\#Test}  \\ \hline
Java250 & Java & Problem classification & 48,000 & 15,000 \\
Python800 & Python & Problem classification & 153,600 & 48,000 &  \\
CodRep1 & Java & Bug detection & 6,944 & 772 &  \\
Refactory & Python & Bug detection & 3,380 & 423 &  \\
GCJ & Python & Authorship attribution & 528 & 132 &  \\
BigCloneBench & Java & Clone detection & 90,102 & 4,000 &  \\ \hline
\multicolumn{6}{c}{\textbf{Low-resource programming languages}} \\ \hline
Ruby300 & Ruby & Problem classification & 48,000 & 6,000 \\
Go250 & Go & Problem classification & 20,000 & 2,500 & \\ \hline
\end{tabular}
}
\vspace{-3mm}
\end{table}

\begin{compactitem}[$\bullet$]
\item \textbf{Problem classification} is a typical source code learning task that classifies the target functions of source code. Given a series of problems with detailed descriptions and their corresponding candidate source code, the trained model will identify the problem that the code is trying to solve. In this task, we first introduce the datasets, which consist of code written in Java and Python. 
Additionally, as discussed in Section~\ref{sec:section4}, it is crucial to address the challenges of data augmentation in low-resource programming languages. Therefore, we also explore the performance of data augmentation methods in problem classification for Ruby and Go. 
Two recently released datasets, Java250 and Python800~\cite{puri2021codenet}, are used in our empirical study for this task. Java250 is built for Java program classification, which has 250 classification problems, including 300 Java programs for each problem. Python800 is a dataset that is specially used for Python program classification tasks, containing 800 different problems with 300 solutions written by the Python program for each. Similarly, Ruby300 is designed for Ruby, featuring 300 problems with 200 Ruby programs to solve each problem. Go250 is used for Go, with 250 problems and 100 Go programs to solve each problem. In the end, we release our dataset Ruby300 and Go250~\footnote{\url{https://drive.google.com/drive/folders/1MZfXLonaYtXSlD51yIhsMb9fS3aoAd72?usp=drive_link}\label{dataset}}.

\item  \textbf{Bug detection} is to determine whether a piece of code contains bugs. Generally, detecting bugs may be thought of as a binary classification problem. It is challenging to prepare a dataset for bug detection since it necessitates a pair of codes with and without bugs, where the process is ultimately identified by human programmers. The common method to gather such pairs is to automatically crawl versions of code before and after commits from GitHub. However, human effort is required to check if the commit is fixing a bug or causing a new bug. Refactory~\cite{hu2019re} and CodRep1~\cite{zhong2015empirical,chen2018codrep}, two open datasets designed for bug repair, are used in our study. Specifically, Refactory includes 2,242 correct and 1,783 buggy Python programs which are written by real-world undergraduate students to finish 5 programming assignments. CodRep1 is a program repair dataset for Java, which includes 3,858 program pairs (buggy program and its fixed version) that are from real bug fixes. 

\item  \textbf{Authorship attribution} task involves identifying the writer of a given code fragment by inferring the characteristics of programmers from their published source code, which is crucial for granting credit for a programmer's contribution and is also helpful for detecting plagiarism. We adopt the dataset from Google Code Jam (GCJ) provided by Yang~\emph{et al.}~\cite{yang2022natural}. 

\item \textbf{Clone detection} focuses on checking whether two codes are semantically identical or not, which helps prevent bug propagation and makes software maintenance easier. We use the broadly recognized clone detection benchmark dataset BigCloneBench~\cite{svajlenko2014towards}, written by Java. Note that Mixup and its variants are not suitable for this task since its input is code pairs.
\end{compactitem}

\subsection{Models}

There are two paradigms for code learning, 1) using task-specific PL models and 2) using pre-trained PL models. 

\noindent\textbf{Code learning with task-specific PL models.} This is a simple type of code learning where a code model is initialized randomly for a specific task and is trained using a task-related dataset from scratch. Generally, the trained models are lighter than the models using pre-trained PL models (e.g., 103 MB for BagofToken models vs.  487 MB for GraphCodeBERT models, as reported in our experiment) and can be deployed in machines with low computation resources. 

\noindent\textbf{Code learning with pre-trained PL models.} Different from task-specific models, pre-trained models are trained on a broad set of unlabeled data and can be used for a wide range of downstream tasks with minimal fine-tuning. Due to its large input volume, pre-trained models usually have better accuracy and higher generalization ability~\cite{hu2022codes}. First, pre-trained PL embedding models are trained using multi-language datasets, e.g., Java, C++, and Python. Then, given a dataset that targets a specific downstream task, such as code clone detection, we fine-tune the pre-trained model accordingly and produce the final model.

We prepare four types of DNN models for each dataset, including pre-trained PL models and models trained from scratch. Two types of models that need to be trained from scratch are studied, FNN (BagofToken)~\cite{puri2021codenet} and CNN (SeqofToken)~\cite{puri2021codenet}. FNN (BagofToken) only contains dense layers. CNN (SeqofToken) consists of both dense layers and convolutional layers. Besides, following the existing work~\cite{yang2022natural}, two well-known pre-trained models, CodeBERT and GraphCodeBERT, are also considered in our study. CodeBERT is a bimodal model trained by using data from multiple programming languages, such as C, C++, Java, and natural languages. It follows the same spirit as BERT~\cite{kenton2019bert} and treats programs as sequences during pre-training. To consider the semantic-level structure of programs, GraphCodeBERT adds data-flow information to the training data that can produce a more precise code representation. Notably, for pre-trained models, we fine-tune all the layers (including the encoder and decoder) in the models for downstream tasks.

\subsection{Implementation} 

We build this project on the top of the~\emph{ALERT}~\cite{yang2022natural} and \emph{Project\_CodeNet}~\cite{puri2021codenet}. The implementation of all methods is done using Python, making this study extensible with more techniques in the future. Moreover, we provide a code refactoring generator, including 18 different code refactoring methods that support Java, Python, Ruby, and Go. The models, including BagOfToken and SeqOfToken, are built using TensorFlow2.3 and Keras2.4.3. CodeBERT and GraphCodeBERT are built using PyTorch1.6.0. We set the training epoch to 50 for the above four models. For the Mixup ratio that is set in augmenting training data, $\alpha=0.1$ is our default setting. To lessen the impact of randomness, we train each model five times and report the average results with standard deviation. We conduct all experiments on a server with 4 GPUs of NVIDIA RTX A6000. 
 
\section{Evaluation Results}
\label{sec:section6}

\subsection{RQ1: Can text-oriented data augmentation methods produce accurate code models?}
\label{sec:RQ1}

\smallskip
\noindent
\textbf{Accuracy analysis.} First, we check the final accuracy of each trained model to explore if these data augmentation methods can improve the performance of models compared to models without data augmentation and with code refactoring, respectively. 

Table~\ref{tab:ACC_Bag_Seq} presents the test accuracy of BagofToken and SeqofToken models on original test data. First, for BagofToken, \emph{SenMixup} achieves the best performance in four (out of five) datasets. More specifically, the results show that it outperforms \emph{No Aug} by up to 4.09\% (2.27\% on average) and \emph{Refactor} by up to 6.63\% (3.82\% on average). Surprisingly, in most cases (four out of five), the code refactoring method cannot improve the accuracy of models compared to \emph{No Aug}. Only in the clone detection task, it brings a maximum 0.31\% accuracy improvement. For SeqofToken models, again, \emph{SenMixup} outperforms \emph{No Aug} in four (out of five) datasets with accuracy improvements by up to 8.74\% and 3.61\% on average, and \emph{Refactor} with accuracy improvements by up to 4.53\% and 2.52\% on average. The results recommend that when using traditional DNN models (e.g., FNN, CNN) to solve code classification tasks, \emph{SenMixup} is a better choice for training data augmentation. 
\begin{table}[h]
\caption{Effectiveness of data augmentation methods w.r.t. test accuracy $\uparrow$ (average $\pm$ standard deviation, \%) on original test data. \textbf{No Aug}: without data augmentation. The best results are highlighted in gray. The tested DNN models are BagofToken and SeqofToken. Tasks include \textbf{Problem Classification} (Java250, Python800), \textbf{Bug detection} (CodRep1, Refactory), \textbf{Authorship attribution} (GCJ), and \textbf{Clone detection} (BigCloneBench).}
\label{tab:ACC_Bag_Seq}
\resizebox{\columnwidth}{!}{
\begin{tabular}{llcccccc}
\cline{1-8}
\textbf{Model} & \textbf{DA method} & \textbf{Java250} & \textbf{Python800} & \textbf{Refactory} & \textbf{CodRep1} &  \textbf{GCJ} & \textbf{BigCloneBench} \\ \cline{1-8}
 & No Aug (Baseline) & {71.24 ± 0.04} & 67.21 ± 0.03 & {85.15 ± 0.12} & {56.47 ± 0.14} & {27.82 ± 0.14} & 85.23 ± 0.34\\
 &WordMixup & {73.12 ± 0.01} & {68.41 ± 0.09} & \cellcolor[HTML]{C0C0C0}{86.46 ± 0.28} & {57.87 ± 0.32} & {28.43 ± 0.25} & -\\
 &SenMixup & \cellcolor[HTML]{C0C0C0}{75.33 ± 0.02} & \cellcolor[HTML]{C0C0C0}{70.35 ± 0.07} & {85.42 ± 0.31} & \cellcolor[HTML]{C0C0C0}{58.89 ± 0.21} & \cellcolor[HTML]{C0C0C0}{29.23 ± 0.22} & -\\
 & Refactor & {68.95 ± 0.03} & {63.72 ± 0.08} & {85.03 ± 0.14} & {56.01 ± 0.14} & {26.43 ± 0.12} & 85.54 ± 0.42\\

 & Rename Operator & 75.15 ± 0.21 & 64.36 ± 0.13 & 86.03 ± 0.23 & 57.12 ± 0.13  &26.51 ± 0.21  & 85.89 ± 0.34 \\ 

 & Dead Operator  & 72.23 ± 0.12 & 63.53 ± 0.31 & 86.15 ± 0.27 & 56.34 ± 0.18 &  26.34 ± 0.19& 85.68 ± 0.31 \\ 

 & Inside Operator  & 72.86 ± 0.23 & 64.56 ± 0.18 & 86.49 ± 0.18 & 56.45 ± 0.24 & 26.89 ± 0.13 & 85.59 ± 0.24\\ 

 & Outside Operator  & 72.91 ± 0.14 & 64.55 ± 0.25 & 86.13 ± 0.15 & 56.59 ± 0.19 & 26.78 ± 0.24 & 85.61 ± 0.23\\ 
 
 & SR & {70.06 ± 0.05} & {66.74 ± 0.06} & {81.91 ± 0.23} & {56.23 ± 0.15} & {27.23 ± 0.21} & 85.45 ± 0.39\\
 & RI & {71.63 ± 0.03} & {66.62 ± 0.06} & {85.81 ± 0.12} & {56.67 ± 0.14} & {27.83 ± 0.33} & 86.28 ± 0.51\\
 & RS & {71.77 ± 0.02} & {67.42 ± 0.05} & {86.33 ± 0.11} & {57.34 ± 0.16} & {28.12 ± 0.24} & 86.48 ± 0.31\\
 & RD & {71.13 ± 0.05} & {67.17 ± 0.04} & {85.68 ± 0.12} & {56.37 ± 0.13} & {28.39 ± 0.26}&  \cellcolor[HTML]{C0C0C0}86.87 ± 0.44\\
\multirow{-13}{*}{BagofToken} & BT & {70.86 ± 0.02} & {67.36 ± 0.06} & {73.96 ± 0.16} & {55.23 ± 0.22} & {25.98 ± 0.21} & 85.65 ± 0.36 \\ \cline{1-8}
& No Aug (Baseline) & {86.61 ± 0.05} & {82.61 ± 0.35} & {85.55 ± 0.13} & {58.21 ± 0.12} & {38.67 ± 0.12} & 90.69 ± 0.25 \\ 
 &WordMixup & {94.42 ± 0.02} & {84.71 ± 0.29} & {87.51 ± 0.22} & {59.87 ± 0.22} & {38.98 ± 0.36} &-\\ 
 &SenMixup & \cellcolor[HTML]{C0C0C0}{95.35 ± 0.12} &\cellcolor[HTML]{C0C0C0}{85.11 ± 0.32} & \cellcolor[HTML]{C0C0C0}{90.02 ± 0.27} & {59.89 ± 0.24} & \cellcolor[HTML]{C0C0C0}{39.34 ± 0.31} &-\\ 
 & Refactor & {93.19 ± 0.28} & {83.38 ± 0.24} & {85.49 ± 0.16} & {57.03 ± 0.11} & {38.04 ± 0.24} & 91.14 ± 0.39\\ 

 & Rename Operator & 94.35 ± 0.19  & 84.56 ± 0.23 & 85.61 ± 0.22 & 58.56 ± 0.24 & 38.56 ± 0.22 & 92.09 ± 0.38\\ 

 & Dead Operator  & 94.01 ± 0.24 & 84.13 ± 0.21 & 86.13 ± 0.13 & 59.01 ± 0.31 & 37.78 ± 0.26 & 91.11 ± 0.31\\ 

 & Inside Operator  & 93.21 ± 0.14 & 83.58 ± 0.14 & 85.78 ± 0.25 & 58.11 ± 0.18 & 38.26 ± 0.31 & 91.31 ± 0.29\\ 

 & Outside Operator  & 93.35 ± 0.25 & 83.61 ± 0.19 & 86.98 ± 0.15 & 56.88 ± 0.14 & 38.01 ± 0.29 & 91.02 ± 0.31 \\ 
 
 & SR & {93.33 ± 0.35} & {84.25 ± 0.21} & {81.64 ± 0.13} & {57.34 ± 0.12} & {38.11 ± 0.21} & 91.23 ± 0.41\\ 
 & RI & {94.49 ± 0.16} & {83.53 ± 0.29} & {87.11 ± 0.22} & {59.33 ± 0.26} & {38.54 ± 0.33} & 91.09 ± 0.36\\ 
 & RS & {93.47 ± 0.22} & {85.02 ± 0.37} & {81.81 ± 0.24} & \cellcolor[HTML]{C0C0C0}{60.15 ± 0.15} & {38.87 ± 0.23} & \cellcolor[HTML]{C0C0C0}92.67 ± 0.31\\ 
 & RD & {94.25 ± 0.31} & {84.29 ± 0.23} & {84.64 ± 0.13} & {57.44 ± 0.32} & {38.75 ± 0.21} & 92.34 ± 0.29\\ 
\multirow{-13}{*}{SeqofToken} & BT & {93.81 ± 0.25} & {82.98 ± 0.31} & {81.38 ± 0.32} & {56.11 ± 0.23} & {36.56 ± 0.22} & 90.86 ± 0.54\\ \cline{1-8} 
\end{tabular}}
\end{table}

Table~\ref{tab:ACC_pre-trained} presents the results of CodeBERT and GraphCodeBERT. First of all, we observe that compared to the results of the above two types of models, \emph{SenMixup} is not sufficient to improve the accuracy of pre-trained PL models (only a maximum of 0.77\% accuracy improvement). By contrast, data augmentation methods that slightly break the syntax of source code, such as \emph{RS}, are more effective and have a clear improvement (by up to 2.25\%) compared to \emph{No Aug}, and (by up to 3.01\%) compared to \emph{Refactor}. In GraphCodeBERT, \emph{RS} achieves the best accuracy improvement in four (out of six) datasets. 

Moreover, compared to \emph{Refactor}, although code refactoring methods such as \emph{Rename Operator}, \emph{Dead Operator}, \emph{Inside Operator}, and \emph{Outside Operator} demonstrate higher accuracy improvements in both task-specific PL models and pre-trained PL models, their performance is inferior to text-oriented data augmentation methods, particularly \emph{SenMixup}, \emph{RS}, and \emph{RD}.

\begin{table}[h]
\caption{Effectiveness of data augmentation methods w.r.t. test accuracy $\uparrow$ (average $\pm$ standard deviation, \%) on original test data. \textbf{No Aug}: without data augmentation. The best results are highlighted in gray. The tested DNN models are CodeBERT and GraphCodeBERT. Tasks include \textbf{Problem Classification} (Java250, Python800), \textbf{Bug detection} (CodRep1, Refactory), \textbf{Authorship attribution} (GCJ), and \textbf{Clone detection} (BigCloneBench).}
\label{tab:ACC_pre-trained}
\resizebox{\columnwidth}{!}{
\begin{tabular}{clcccccc}
\cline{1-8}
\textbf{Model} & \textbf{DA method} & \textbf{Java250} & \textbf{Python800} & \textbf{Refactory} & \textbf{CodRep1} & \textbf{GCJ} & \textbf{BigCloneBench}\\ \cline{1-8} 
 & No Aug (Baseline) & {96.39 ± 0.03} & {96.07 ± 0.04} & {96.22 ± 0.11} & {69.13 ± 0.14} & {90.98 ± 0.14} & 96.89 ± 0.19\\ 
 &WordMixup & {96.31 ± 0.04} & {96.23 ± 0.03} & {96.16 ± 0.12} & {69.02 ± 0.16} & {91.34 ± 0.24} &-\\
 &SenMixup & {96.56 ± 0.02} & \cellcolor[HTML]{C0C0C0}{96.36 ± 0.02} & \cellcolor[HTML]{C0C0C0}{96.99 ± 0.24} & {69.26 ± 0.25} & {92.16 ± 0.29} &-\\ 
 & Refactor & {96.42 ± 0.05} & {96.11 ± 0.02} & {95.94 ± 0.12} & {70.18 ± 0.13} & {91.73 ± 0.17} & 96.95 ± 0.29\\  
  & Rename Operator & 96.45 ± 0.09 & 96.15 ± 0.03 & 96.23 ± 0.13 & 70.34 ± 0.12 & 91.89 ± 0.21 & 97.13 ± 0.24 \\ 

 & Dead Operator  & 96.44 ± 0.11 & 96.20 ± 0.07 & 96.14 ± 0.14 & 70.23 ± 0.14 & 92.65 ± 0.16 & 97.21 ± 0.21\\ 

 & Inside Operator  & 96.31 ± 0.08 & 96.14 ± 0.04 & 96.04 ± 0.18 & 70.09 ± 0.21 & 91.79 ± 0.18 & 96.97 ± 0.18 \\ 

 & Outside Operator  & 96.39 ± 0.05 & 96.01 ± 0.13 & 95.81 ± 0.15 & 70.11 ± 0.25 & 91.83 ± 0.21 & 97.02 ± 0.33\\ 
 & SR & {96.33 ± 0.02} & {96.01 ± 0.05} & {96.69 ± 0.14} & {69.24 ± 0.17} & {70.68 ± 0.08} & 96.98 ± 0.17\\ 
 & RI & {96.31 ± 0.06} & {95.96 ± 0.07} & {96.45 ± 0.12} & {70.14 ± 0.26} & {59.89 ± 0.34} & \cellcolor[HTML]{C0C0C0}97.31 ± 0.31\\ 
 & RS & {96.47 ± 0.04} & {96.17 ± 0.05} & {96.71 ± 0.13} & \cellcolor[HTML]{C0C0C0}{70.56 ± 0.21} & \cellcolor[HTML]{C0C0C0}{93.23 ± 0.09} & 96.99 ± 0.18\\ 
 & RD & \cellcolor[HTML]{C0C0C0}{96.58 ± 0.03} & {96.22 ± 0.02} & {96.45 ± 0.15} & {70.34 ± 0.12} & {76.99 ± 0.11} & 97.01 ± 0.37\\ 
\multirow{-13}{*}{CodeBERT} & BT & {96.21 ± 0.02} & {95.91 ± 0.03} & {94.33 ± 0.21} & {68.79 ± 0.33} & {82.71 ± 0.12} & 97.02 ± 0.19\\ \cline{1-8} 

 & No Aug (Baseline) & 96.47 ± 0.13 & 96.26 ± 0.07 & 96.82 ± 0.14 & 70.06 ± 0.15 & 93.98 ± 0.12 &  96.85 ± 0.15  \\ 
&WordMixup & 96.23 ± 0.05 & 96.09 ± 0.19 &  96.18 ± 0.24 & 70.05 ± 0.17 & 93.67 ± 0.23 & -   \\ 
&SenMixup & 96.52 ± 0.09  & 96.22 ± 0.14 & 96.46 ± 0.21 & \cellcolor[HTML]{C0C0C0}70.45 ± 0.23 & 94.51 ± 0.09 & -   \\ 
& Refactor  & 96.56 ± 0.12 & 96.28 ± 0.11 &  95.51 ± 0.26 & 70.11 ± 0.37 & 91.73 ± 0.14 &  97.08 ± 0.21  \\ 

  & Rename Operator & 96.58 ± 0.17 & 96.55 ± 0.13 & 95.96 ± 0.25 & 70.14 ± 0.31 & 92.35 ± 0.15 & 97.35 ± 0.23\\ 

 & Dead Operator  & 96.51 ± 0.13 & 96.52 ± 0.14 & 96.23 ± 0.17 & 70.08 ± 0.29 & 91.67 ± 0.14 & 97.13 ± 0.31\\ 

 & Inside Operator  & 96.56 ± 0.11 & 96.31 ± 0.17 & 95.67 ± 0.21 & 70.12 ± 0.34 & 91.97 ± 0.13 & 97.09 ± 0.15\\ 

 & Outside Operator  & 96.52 ± 0.14 & 96.29 ± 0.08 & 95.49 ± 0.23 & 70.09 ± 0.28 & 91.84 ± 0.17 & 97.31 ± 0.29 \\ 
 
& SR  & 96.54 ± 0.11 & 96.06 ± 0.21 & 95.54 ± 0.28  & 69.86 ± 0.29  & 89.47 ± 0.19 & 97.09 ± 0.16   \\   
& RI &96.58 ± 0.04  & 96.32 ± 0.24 &  94.56 ± 0.31  & 70.22 ± 0.19 & 80.45 ± 0.23 & 97.32 ± 0.36  \\ 
& RS & 96.49 ± 0.02 & \cellcolor[HTML]{C0C0C0}96.92 ± 0.16 &  \cellcolor[HTML]{C0C0C0}97.91 ± 0.19  & 70.34 ± 0.21 & \cellcolor[HTML]{C0C0C0}94.74 ± 0.12 & \cellcolor[HTML]{C0C0C0}97.52 ± 0.11   \\ 
& RD & \cellcolor[HTML]{C0C0C0}96.69 ± 0.21 & 96.24 ± 0.13 &  96.51 ± 0.18  & 70.42 ± 0.33 & 91.73 ± 0.21 & 97.18 ± 0.28    \\
\multirow{-13}{*}{GraphCodeBERT} & BT & 96.55 ± 0.18 & 96.31 ± 0.09 & 95.98 ± 0.32 & 69.59 ± 0.36 & 80.45 ± 0.24 & 97.12 ± 0.18\\ \cline{1-8}
\end{tabular}}
\vspace{-3mm}
\end{table}

\begin{figure}[h]
\centering
\subfigure[Java250-SeqofToken]{
\begin{minipage}[t]{0.45\linewidth}
\centering
\includegraphics[width=1.0\linewidth]{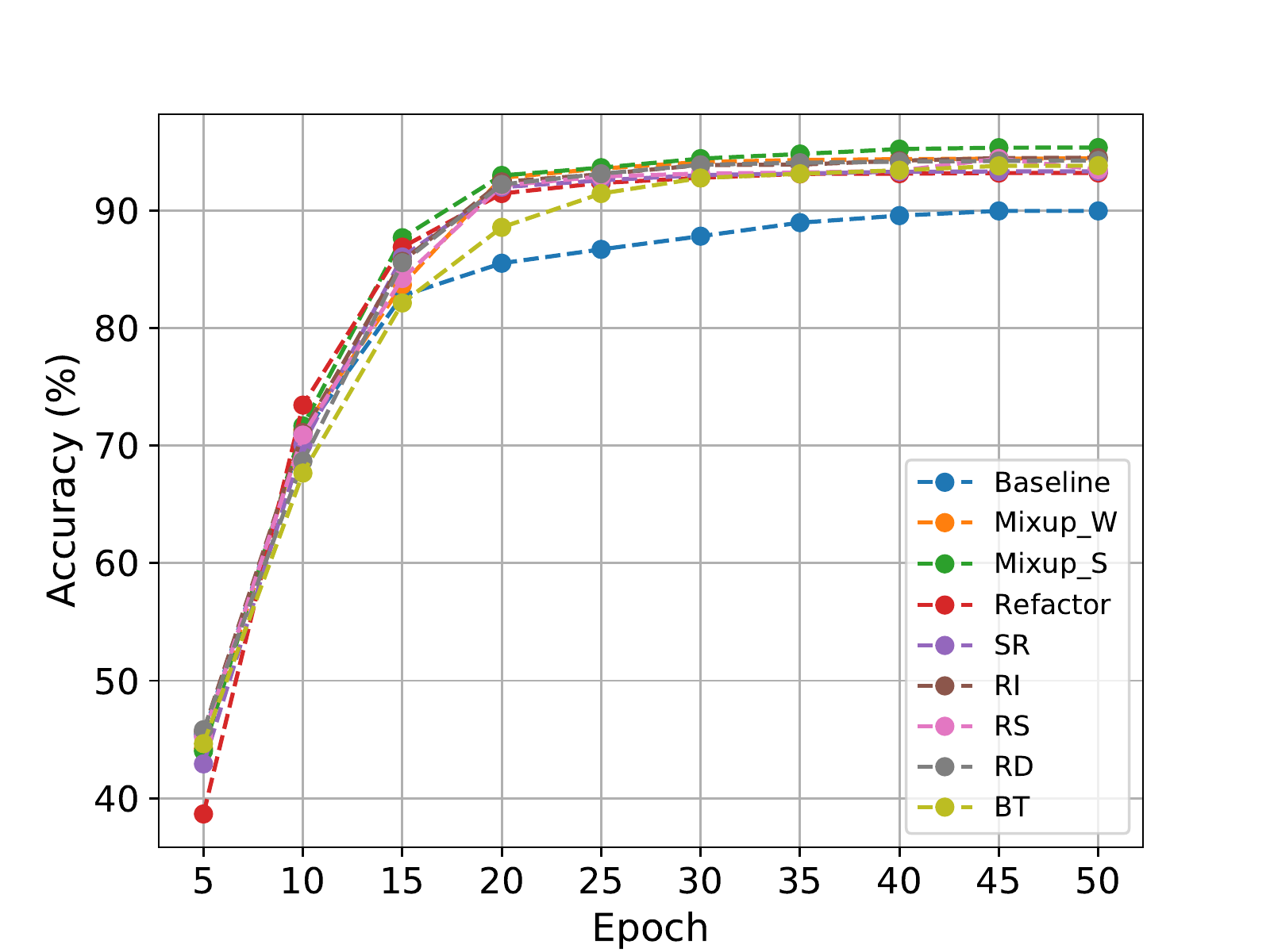}
\end{minipage}%
}%
\subfigure[Java250-GraphCodeBERT]{
\begin{minipage}[t]{0.45\linewidth}
\centering
\includegraphics[width=1.0\linewidth]{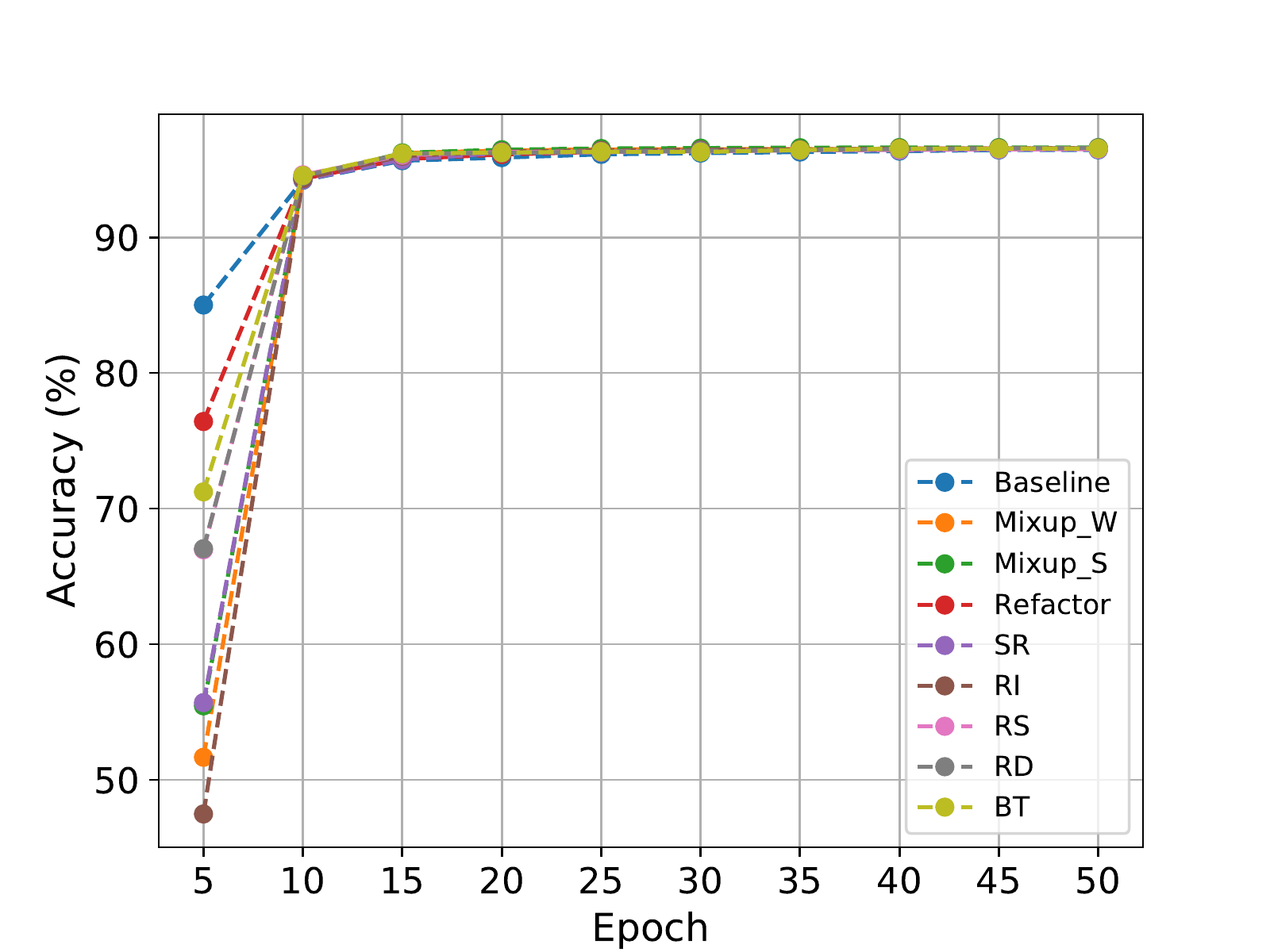}
\end{minipage}%
}%

\centering
\subfigure[Python800-SeqofToken]{
\begin{minipage}[t]{0.45\linewidth}
\centering
\includegraphics[width=1.0\linewidth]{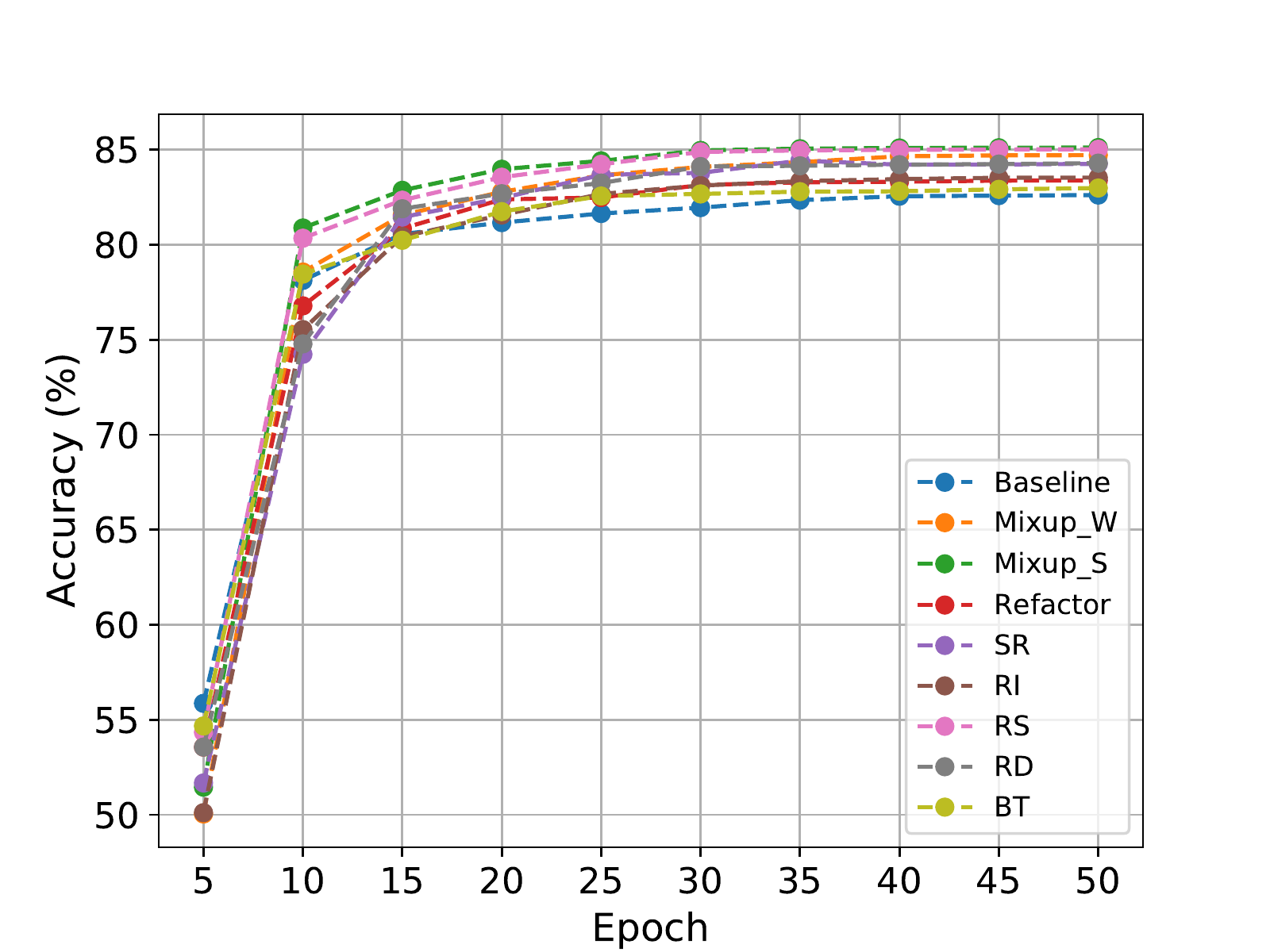}
\end{minipage}%
}%
\subfigure[Python800-GraphCodeBERT]{
\begin{minipage}[t]{0.45\linewidth}
\centering
\includegraphics[width=1.0\linewidth]{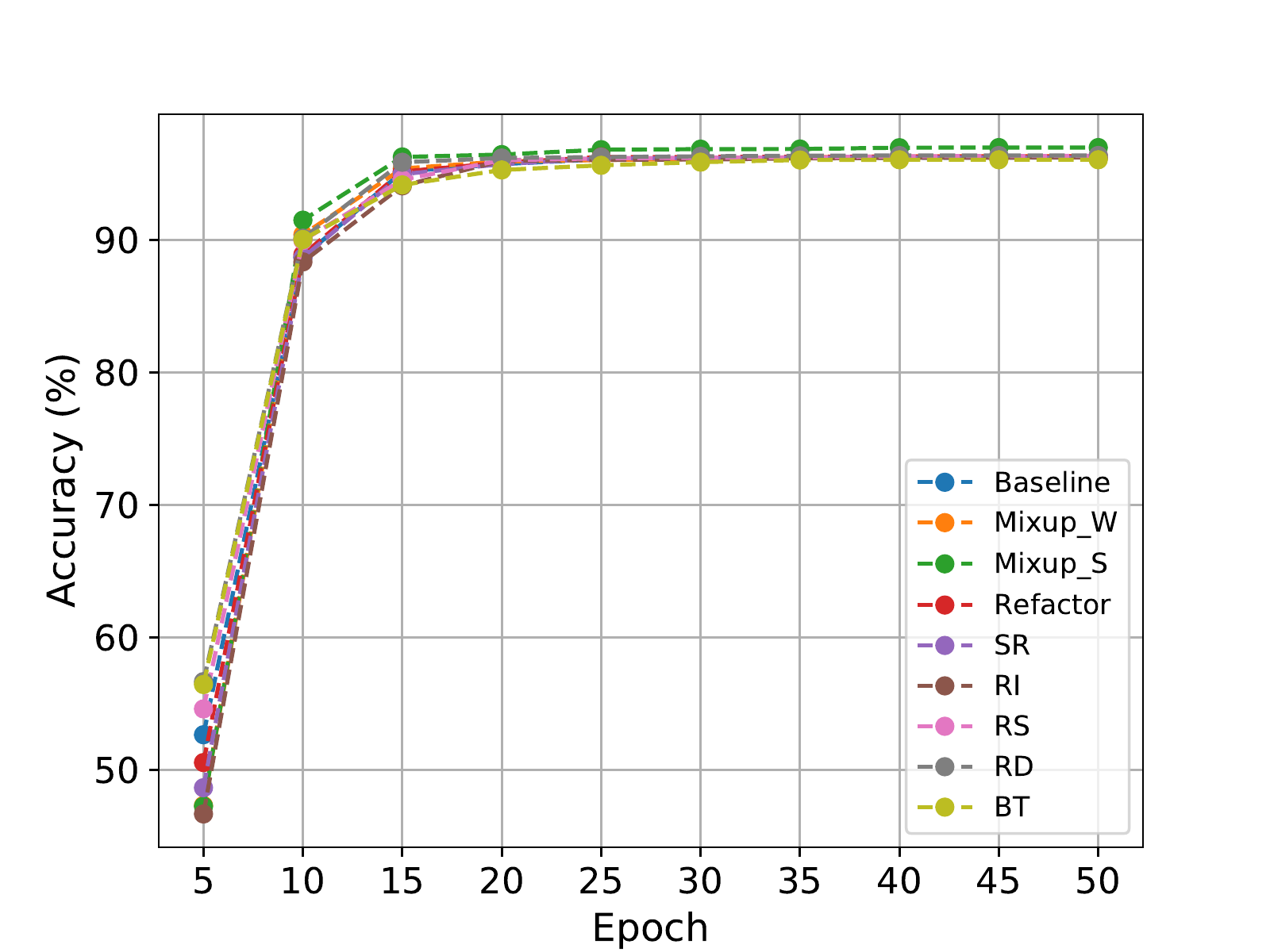}
\end{minipage}%
}%

\centering
\subfigure[Refactory-SeqofToken]{
\begin{minipage}[t]{0.45\linewidth}
\centering
\includegraphics[width=1.0\linewidth]{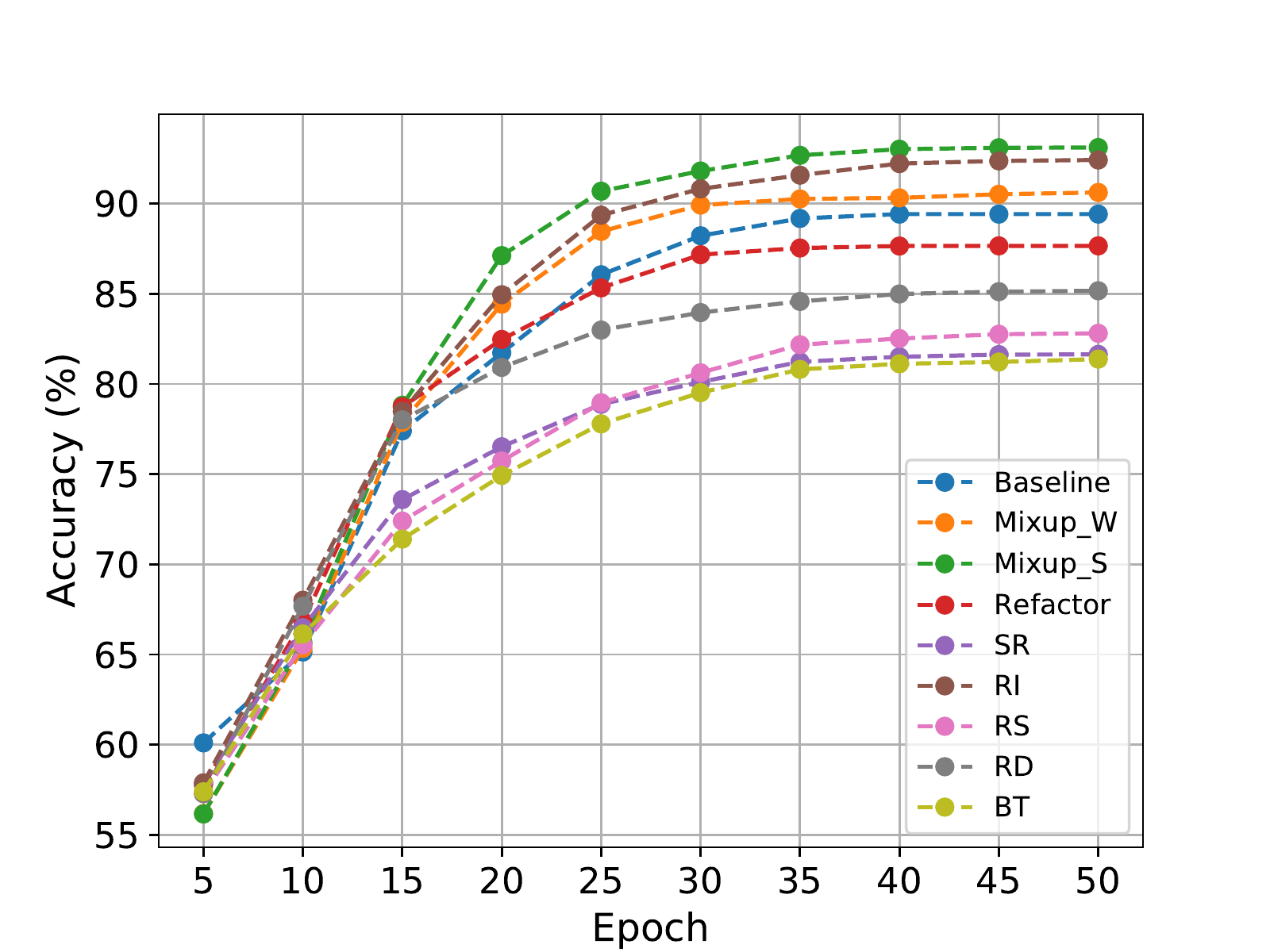}
\end{minipage}%
}%
\subfigure[Refactory-GraphCodeBERT]{
\begin{minipage}[t]{0.45\linewidth}
\centering
\includegraphics[width=1.0\linewidth]{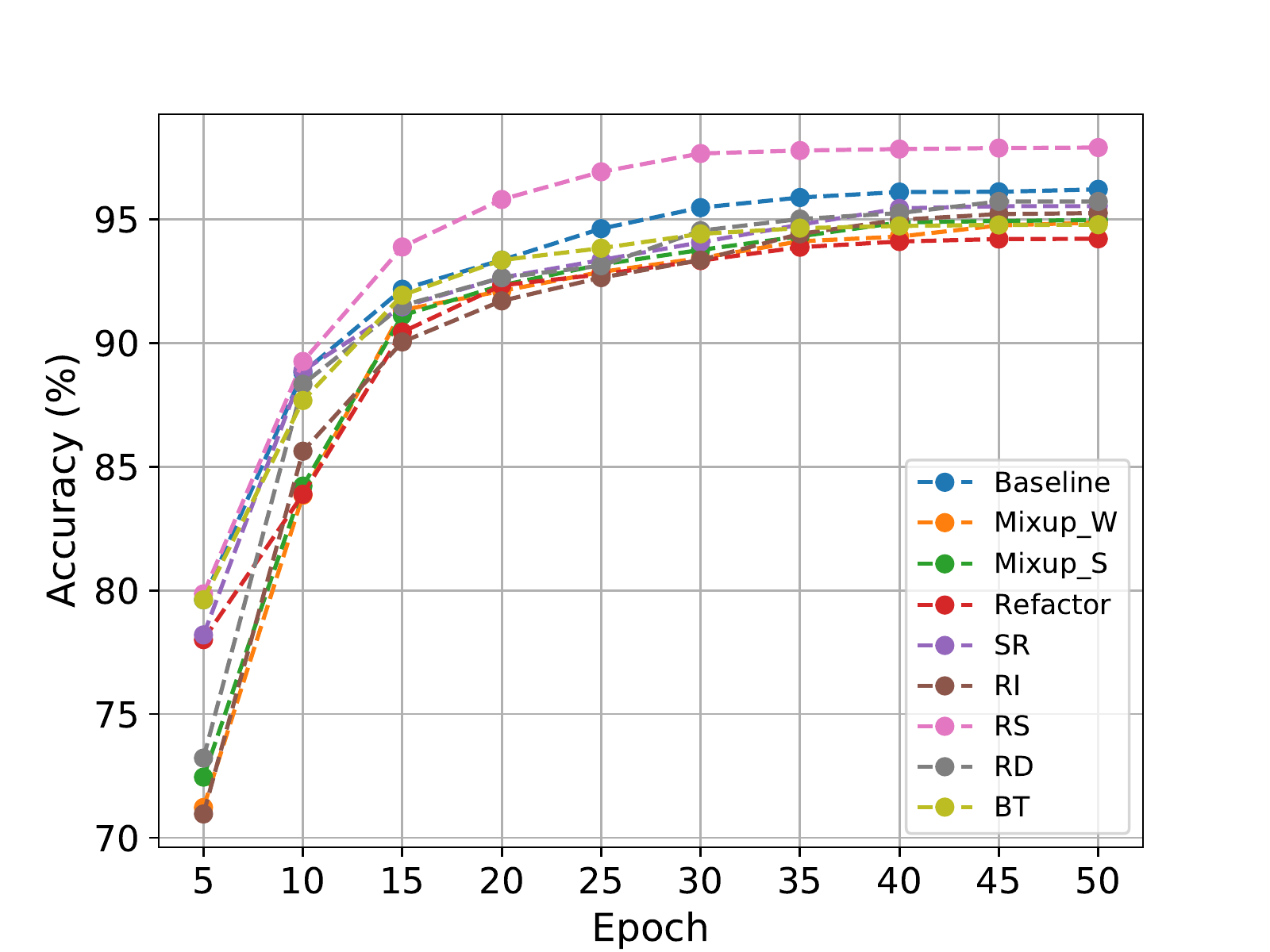}
\end{minipage}%
}%
\caption{Training log of models in Java250, Python800, and Refactory.}
\label{fig: gcj}
\end{figure}

\begin{figure}[htbp]
\centering
\subfigure[CodRep1-SeqofToken]{
\begin{minipage}[t]{0.45\linewidth}
\centering
\includegraphics[width=1.0\linewidth]{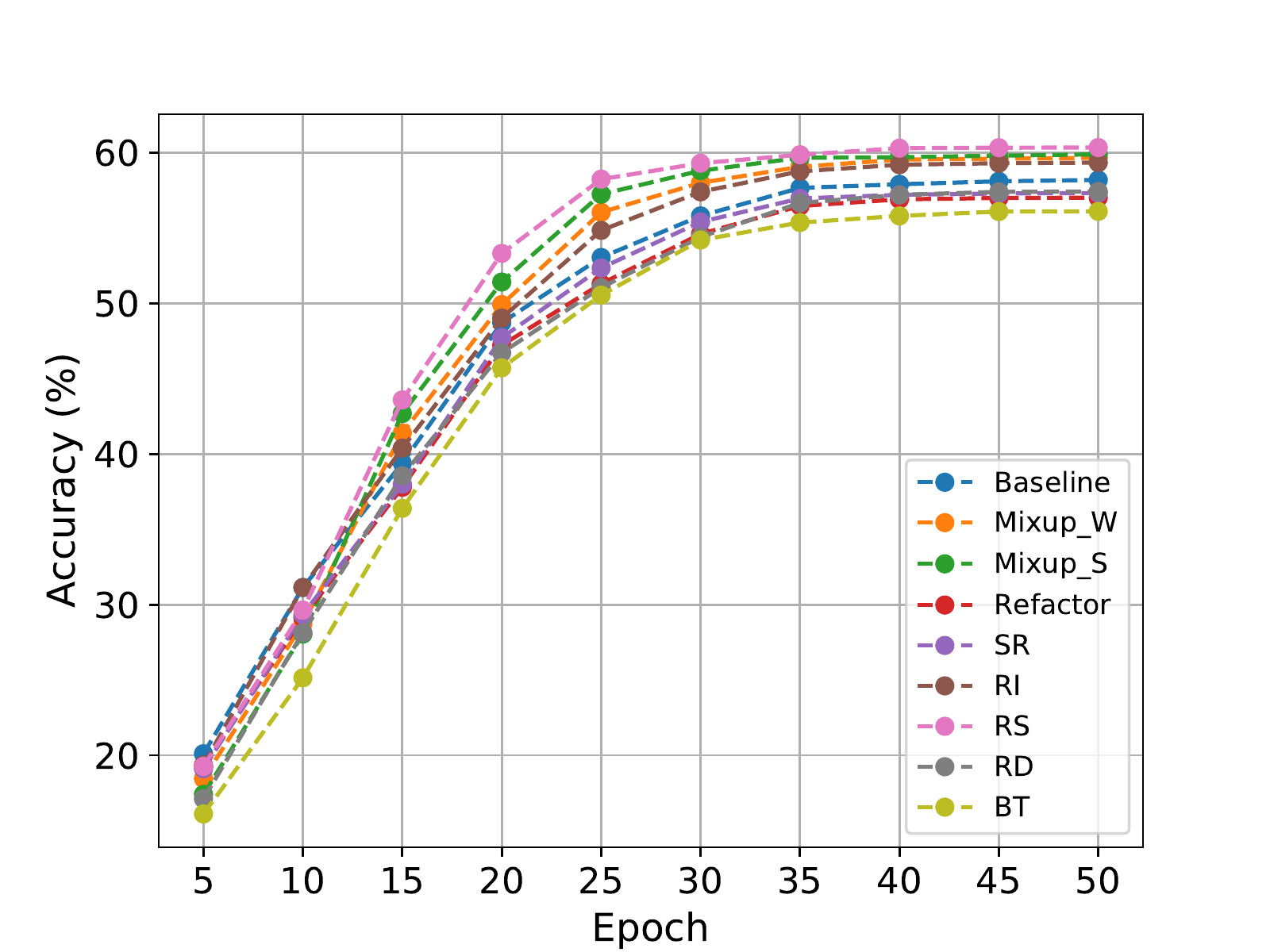}
\end{minipage}%
}%
\subfigure[CodRep1-GraphCodeBERT]{
\begin{minipage}[t]{0.45\linewidth}
\centering
\includegraphics[width=1.0\linewidth]{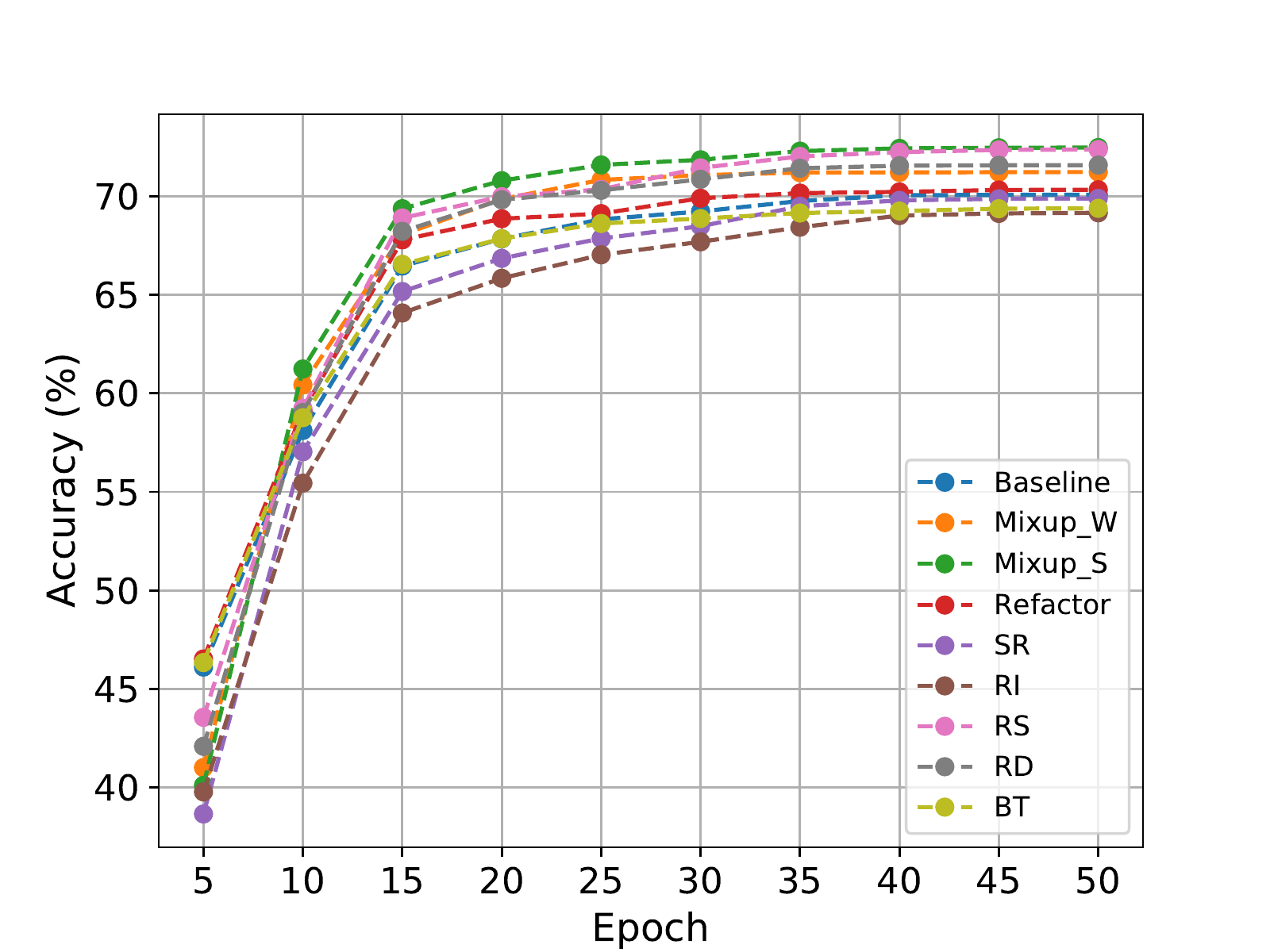}
\end{minipage}%
}%

\centering
\subfigure[GCJ-SeqofToken]{
\begin{minipage}[t]{0.45\linewidth}
\centering
\includegraphics[width=1.0\linewidth]{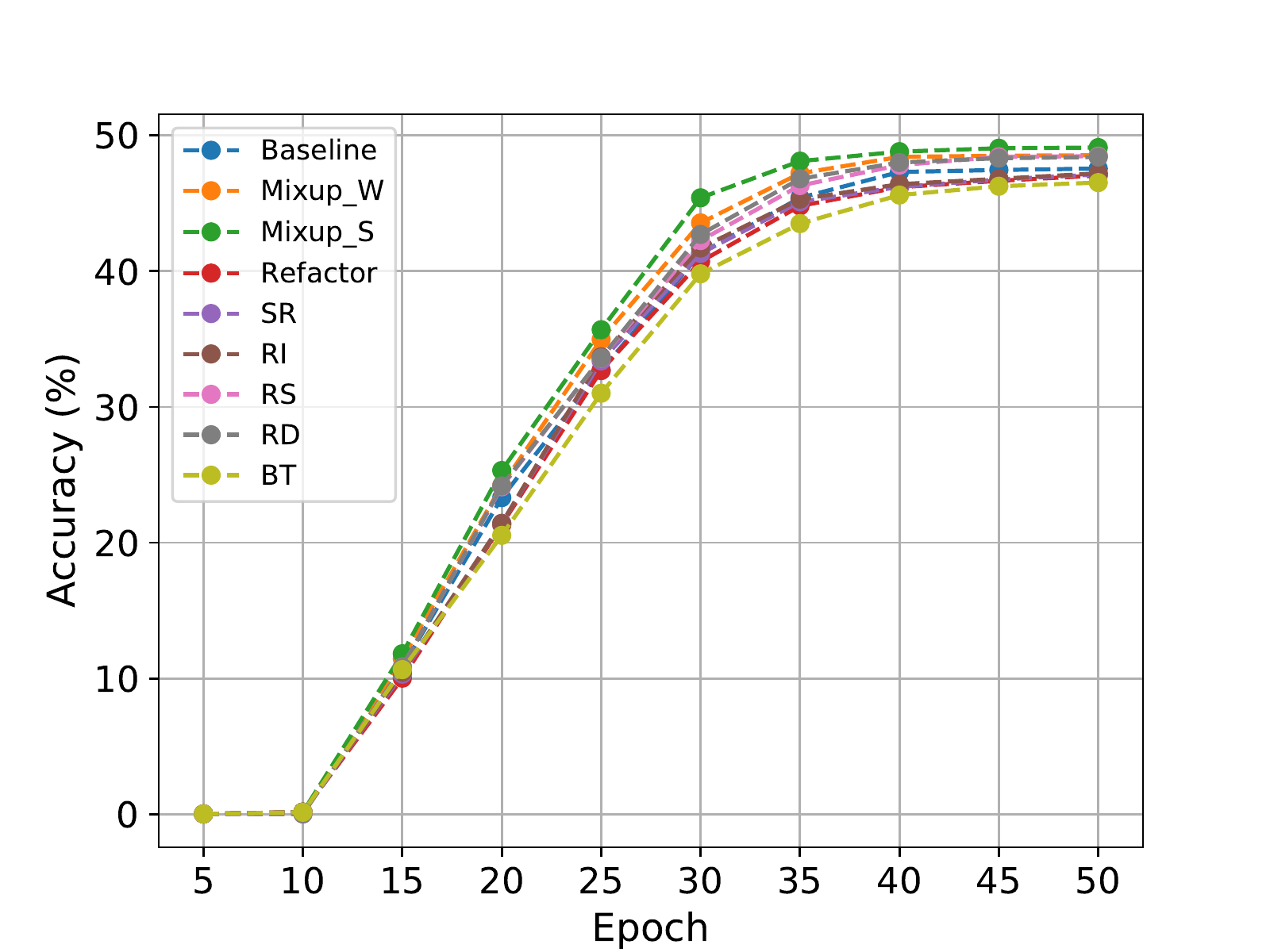}
\end{minipage}%
}%
\centering
\centering
\subfigure[GCJ-GraphCodeBERT]{
\begin{minipage}[t]{0.45\linewidth}
\centering
\includegraphics[width=1.0\linewidth]{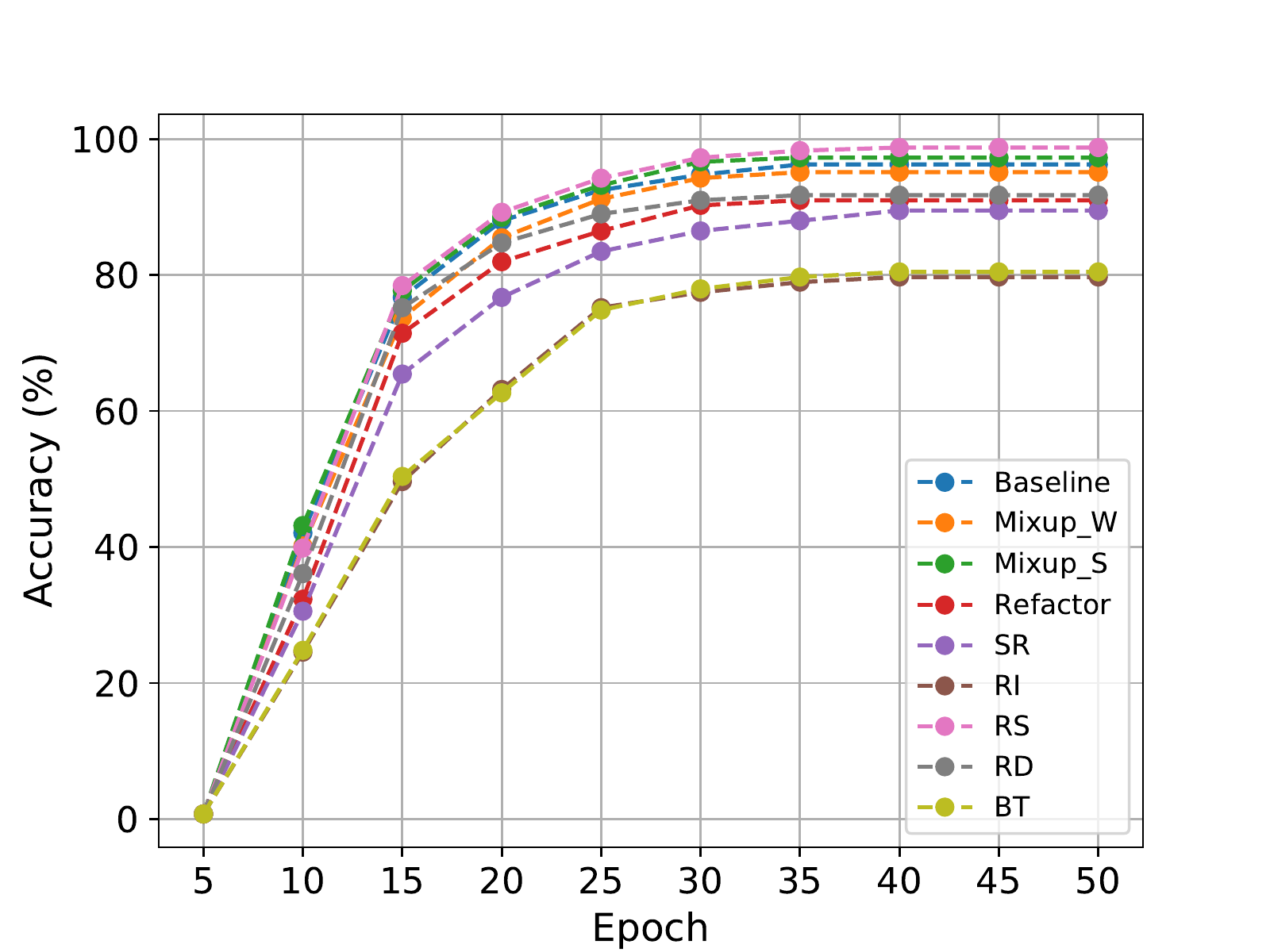}
\end{minipage}%
}%
\\
\centering
\subfigure[BigCloneBench-SeqofToken]{
\begin{minipage}[t]{0.45\linewidth}
\centering
\includegraphics[width=1.0\linewidth]{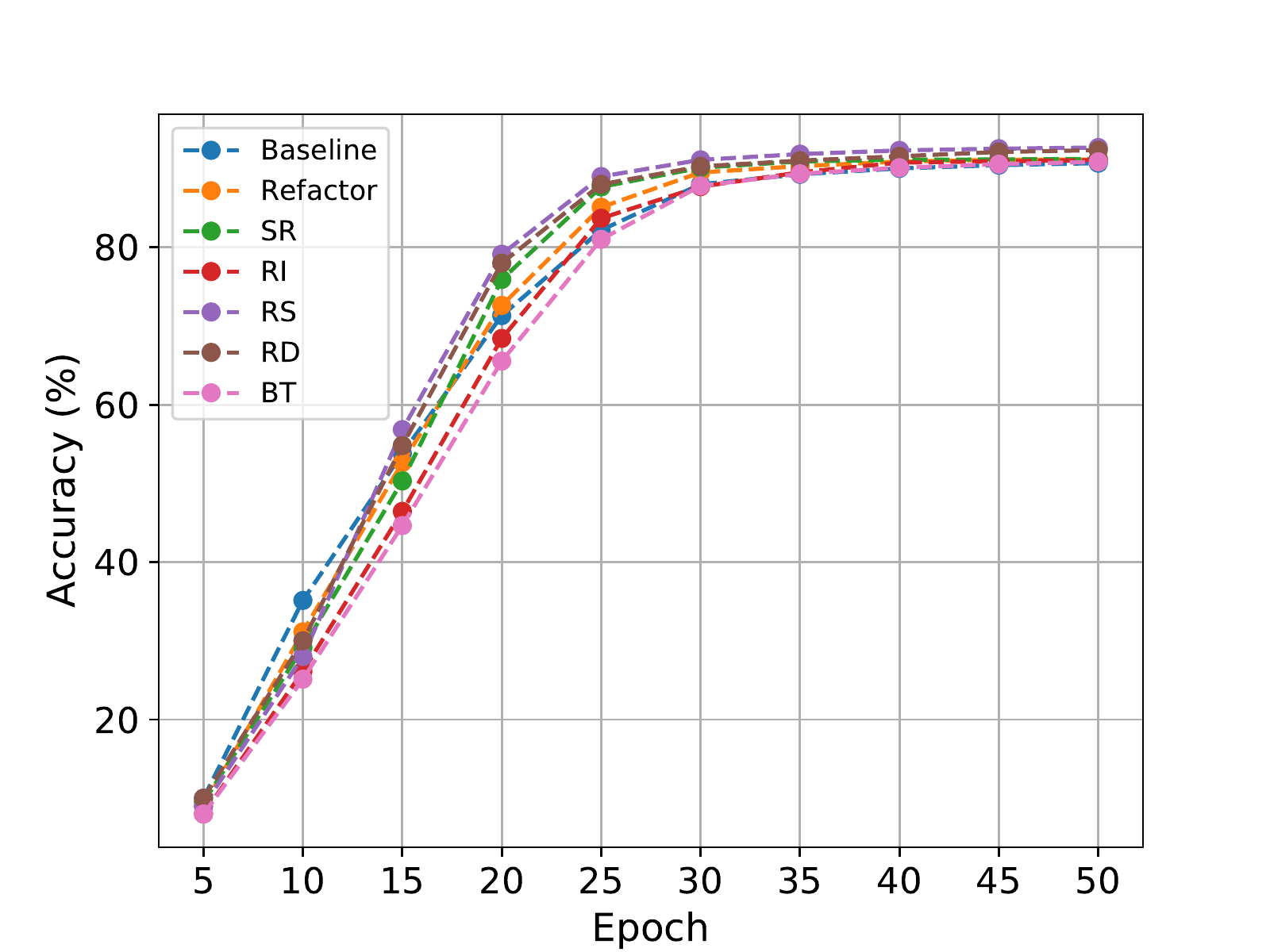}
\end{minipage}%
}%
\centering
\subfigure[BigCloneBench-GraphCodeBERT]{
\begin{minipage}[t]{0.45\linewidth}
\centering
\includegraphics[width=1.0\linewidth]{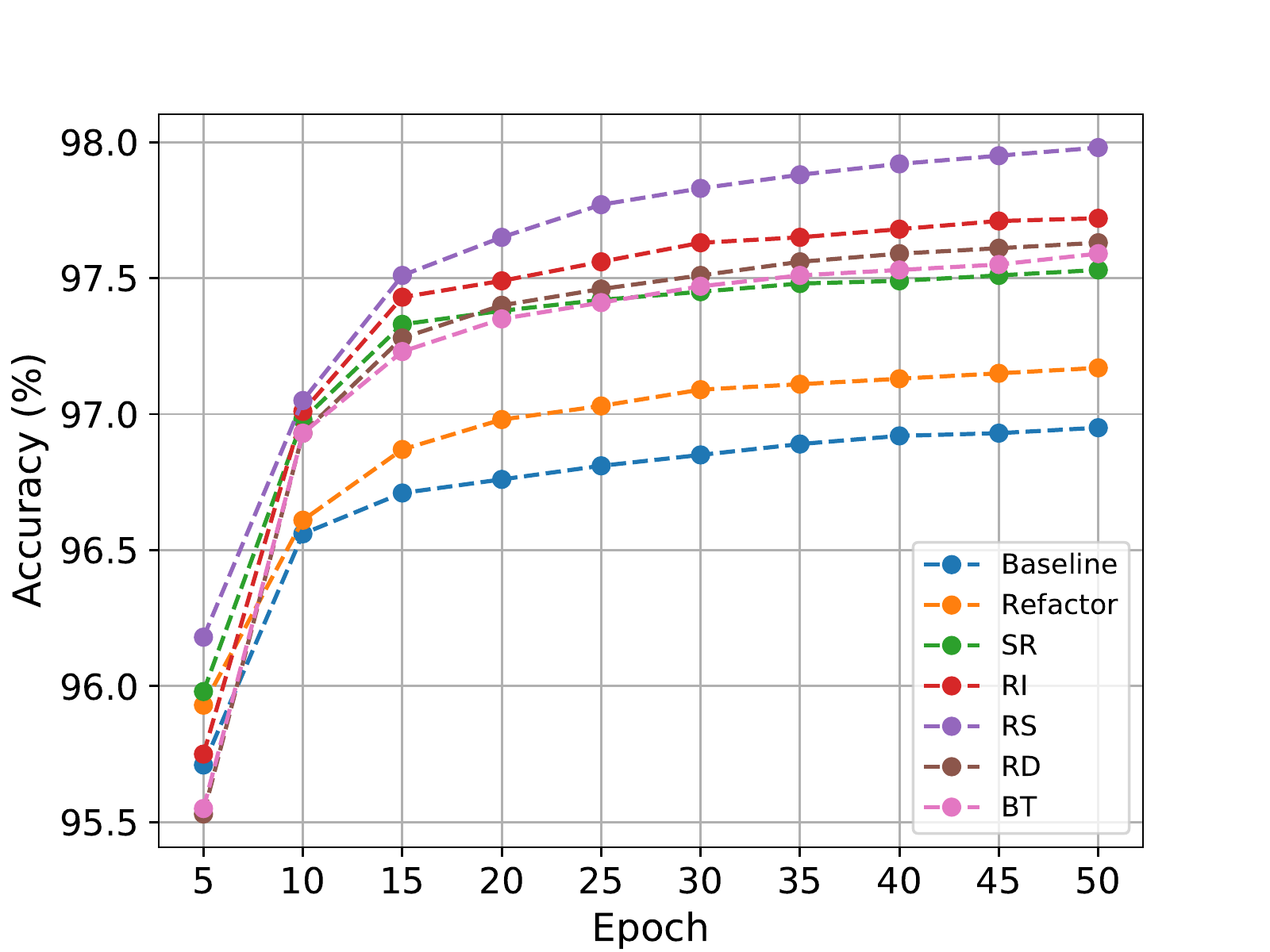}
\end{minipage}%
}%
\centering
\caption{Training log of models in CodRep1, GCJ, and BigCloneBench.}
\label{fig: log_clone}
\end{figure}

\smallskip
\noindent
\textbf{Convergence speed analysis.} We check the convergence speed of models using different data augmentation methods. Following existing works~\cite{dong2023mixcode,dong2024effectiveness}, we select \emph{Refactor} as the representative method from the code refactoring family in our visualization experiment. Fig.~\ref{fig: gcj} and Fig.~\ref{fig: log_clone} depict the training logs of SeqofToken and GraphCodeBERT models. From the results, we find that 1) models have similar convergence speed regardless of the used data augmentation methods, e.g., for Java250-SeqofToken, after 20 epochs, models have no significant improvements in terms of accuracy (except~\emph{BT}). This finding reflects that the execution (computation budget) cost of model training is similar regardless of the used data augmentation method. 2) Similar to findings that come from analyzing the final accuracy of models, there are two methods that have clearly better performance than others, \emph{SenMixup} and \emph{RS}. This indicates that, with limited computation budgets, these two methods are also recommended for practical use.

\smallskip
\noindent
\textbf{Statistical analysis.} We conduct statistical tests using the \emph{Wilcoxon signed-rank test}~\cite{woolson2007wilcoxon} and also adapt the \emph{Bonferroni correction}~\cite{armstrong2014use,dong2024effectiveness_jss} to adjust the \emph{p}-values for multiple comparisons. Even though  \emph{SenMixup} and \emph{RS} achieve relatively better results than other methods, our statistical analysis (more detailed results can be found in Appendix~\ref{sec:appendix}) demonstrates that their advantage is not significant. Specifically, as shown in Table~\ref{table:statistical_testing_acc_codebert_graphcodebert} in Appendix~\ref{sec:appendix}, compared to other data augmentation methods, most p-values of \emph{SenMixup} and \emph{RS} exceed 0.05. Furthermore, the results also suggest that the text-oriented data augmentation methods cannot significantly outperform code refactoring-based methods and training without data augmentation (with \emph{p}-values $>$ 0.05). Notably, \emph{Refactor} demonstrates the worst performance among light models including BagOfToken and SeqofToken, as indicated in Table~\ref{table:statistical_testing_acc_bag_seq} in Appendix~\ref{sec:appendix}. This finding indicates that proposing a more powerful data augmentation method for code learning is a promising direction.



\begin{tcolorbox}
\textbf{Answer to RQ1}: Data augmentation methods that linearly mix code embeddings (e.g., \emph{SenMixup}) are effective in enhancing model performance with an up to 8.74\% accuracy improvement compared to training without data augmentation. However, statistical analysis indicates that the advantage of existing text-oriented data augmentation methods is not significant. Exploring advanced code data augmentation methods still holds promising direction.



\end{tcolorbox}

\subsection{RQ2: Can text-oriented data augmentation methods produce robust code models?}
\label{sec:RQ2}

\smallskip
\noindent
\textbf{Robustness analysis.} Adversarial robustness reflects how models handle the data with noise, which is an important characteristic that should be evaluated by the models. Table~\ref{tab:Robustness_pre-trained} presents the ASR of ALERT and MHM attack on well-trained models. First, the results demonstrate that data augmentation can not always enhance the robustness of models. Compared to the \emph{No Aug}, only four out of eight models trained by using data augmentation have higher robustness. This phenomenon is consistent with the conclusion drawn by Yefet~\emph{et al.}~\cite{yefet2020adversarial} that simply increasing the training data is not sufficient for improving the robustness of code models. Also importantly, in some cases, although data augmentation can help train a more robust model, the robustness improvement is insignificant, e.g., the greatest improvement is 5.98\% (GraphCodeBERT-Refactor-BigCloneBench-ALERT).
\begin{table}[h]
\caption{ASR $\downarrow$ (\%) on test data. \textbf{No Aug}: without data augmentation. The best results are highlighted in gray. The victim DNN models are CodeBERT and GraphCodeBERT. Tasks include \textbf{Problem Classification} (Java250, Python800), \textbf{Bug detection} (CodRep1, Refactory), \textbf{Authorship attribution} (GCJ), and \textbf{Clone detection} (BigCloneBench).}
\label{tab:Robustness_pre-trained}
\resizebox{\columnwidth}{!}{
\begin{tabular}{clcccccccccccc}
\hline
 & \multicolumn{1}{c}{} & \multicolumn{2}{c}{\textbf{Java250}} & \multicolumn{2}{c}{\textbf{Python800}} & \multicolumn{2}{c}{\textbf{Refactory}} & \multicolumn{2}{c}{\textbf{CodRep1}} & \multicolumn{2}{c}{\textbf{GCJ}} & \multicolumn{2}{c}{\textbf{BigCloneBench}}\\
\multirow{-2}{*}{\textbf{Model}} & \multicolumn{1}{l}{\multirow{-2}{*}{\textbf{DA method}}} & \textbf{MHM} & \textbf{ALERT} & \textbf{\textbf{MHM}} & \textbf{ALERT} & \textbf{MHM} & \textbf{ALERT} & \textbf{MHM} & \textbf{ALERT} & \textbf{MHM} & \textbf{ALERT} & \textbf{MHM} & \textbf{ALERT} \\ \hline
 & No Aug (Baseline) & \multicolumn{1}{c}{{ 40.85}} & { 53.46} & \multicolumn{1}{c}{{ 42.63}} & { 58.32} & \multicolumn{1}{c}{{ 35.21}} & { 42.51} & \multicolumn{1}{c}{{ 42.18}} & { 55.32} & \multicolumn{1}{c}{{ 33.33}} & { 56.11}  & \multicolumn{1}{c}{{ 9.67}} & { 24.52}\\ 
 & WordMixup & \multicolumn{1}{c}{{ 40.68}} & { 53.85} & \multicolumn{1}{c}{{ 42.25}} & { 59.13} & \multicolumn{1}{c}{{ 37.54}} & { 48.98} & \multicolumn{1}{c}{{ 41.34}} & { 54.26} & \multicolumn{1}{c}{{ 32.15}} & { 53.34} & \multicolumn{1}{c}{{- }} & {- }\\ 
 & SenMixup & \multicolumn{1}{c}{\cellcolor[HTML]{C0C0C0}{ 38.13}} & { 51.19} & \multicolumn{1}{c}{{ 40.53}} & { 56.45} & \multicolumn{1}{c}{{ 32.24}} & { 41.13} & \multicolumn{1}{c}{{ 38.31}} & { 53.29} & \multicolumn{1}{c}{{ 29.27}} & { 52.56}& \multicolumn{1}{c}{{ -}} & {- } \\  
 & Refactor & \multicolumn{1}{c}{{ 38.34}} & { 52.23} & \multicolumn{1}{c}{{ 41.42}} & { 55.78} & \multicolumn{1}{c}{{ 38.89}} & { 48.78} & \multicolumn{1}{c}{{ 41.45}} & { 58.77} & \multicolumn{1}{c}{{ 35.54}} & { 52.89} & \multicolumn{1}{c}{{ 8.58}} & \cellcolor[HTML]{C0C0C0}{ 18.34 }\\ 
  & Rename Operator & \multicolumn{1}{c}{{ 40.13}} &   \multicolumn{1}{c}{{ 53.03}} & \multicolumn{1}{c}{{ 42.43}} & \multicolumn{1}{c}{{ 56.28}} & \multicolumn{1}{c}{{ 39.84}} & \multicolumn{1}{c}{{ 50.67}} & \multicolumn{1}{c}{{ 41.49}} & \multicolumn{1}{c}{{ 58.98}} & \multicolumn{1}{c}{{ 38.68}} & \multicolumn{1}{c}{{ 54.68}} & \multicolumn{1}{c}{{ 8.91}} & \multicolumn{1}{c}{{ 22.45}} \\ 

 & Dead Operator  & \multicolumn{1}{c}{{ 40.07}} &  \multicolumn{1}{c}{{ 52.96}} & \multicolumn{1}{c}{{ 42.52}}  & \multicolumn{1}{c}{{ 56.97}} & \multicolumn{1}{c}{{ 39.27}} & \multicolumn{1}{c}{{ 49.69}} & \multicolumn{1}{c}{{ 42.08}} & \multicolumn{1}{c}{{ 59.67}} & \multicolumn{1}{c}{{ 39.47}} & \multicolumn{1}{c}{{ 55.71}} & \multicolumn{1}{c}{{ 9.45}}  & \multicolumn{1}{c}{{ 24.78 }}\\ 

 & Inside Operator & \multicolumn{1}{c}{{ 39.76}} &  \multicolumn{1}{c}{{ 53.15}} & \multicolumn{1}{c}{{ 41.87}} & \multicolumn{1}{c}{{ 57.65}} & \multicolumn{1}{c}{{ 41.14}} & \multicolumn{1}{c}{{ 51.35}} & \multicolumn{1}{c}{{ 42.14}} & \multicolumn{1}{c}{{ 60.12}} & \multicolumn{1}{c}{{ 41.61}} & \multicolumn{1}{c}{{ 56.01}} & \multicolumn{1}{c}{{ 9.78}} & \multicolumn{1}{c}{{ 25.31}} \\ 

 & Outside Operator & \multicolumn{1}{c}{{ 39.81}} & \multicolumn{1}{c}{{ 53.19}} & \multicolumn{1}{c}{{ 41.92}} & \multicolumn{1}{c}{{ 58.02}} & \multicolumn{1}{c}{{ 42.56}} & \multicolumn{1}{c}{{ 52.75}} & \multicolumn{1}{c}{{ 42.11}} & \multicolumn{1}{c}{{ 61.82}} & \multicolumn{1}{c}{{ 40.97}} & \multicolumn{1}{c}{{ 54.56}} & \multicolumn{1}{c}{{ 9.61}} & \multicolumn{1}{c}{{ 26.81}} \\ 
 & SR & \multicolumn{1}{c}{{ 41.64}} & { 54.51} & \multicolumn{1}{c}{{ 43.83}} & { 60.82} & \multicolumn{1}{c}{{ 39.78}} & { 53.85} & \multicolumn{1}{c}{{ 43.76}} & { 56.98} & \multicolumn{1}{c}{{ 35.48}} & { 52.69} & \multicolumn{1}{c}{{12.49 }} & { 26.03}\\ 
 & RI & \multicolumn{1}{c}{{ 43.45}} & { 55.87} & \multicolumn{1}{c}{{ 45.55}} & { 61.21} & \multicolumn{1}{c}{{ 38.64}} & { 45.34} & \multicolumn{1}{c}{{ 40.01}} & { 54.44} & \multicolumn{1}{c}{{ 40.79}} & { 65.79} & \multicolumn{1}{c}{\cellcolor[HTML]{C0C0C0}{ 8.29}} & { 30.02} \\ 
 & RS & \multicolumn{1}{c}{{ 39.21}} & \cellcolor[HTML]{C0C0C0}{ 51.09} & \multicolumn{1}{c}{{ 40.98}} & { 59.22} & \multicolumn{1}{c}{{ 39.02}} & { 46.78} & \multicolumn{1}{c}{\cellcolor[HTML]{C0C0C0}{ 38.28}} & \cellcolor[HTML]{C0C0C0}{ 53.21} & \multicolumn{1}{c}{\cellcolor[HTML]{C0C0C0}{ 28.23}} & \cellcolor[HTML]{C0C0C0}{ 52.42} & \multicolumn{1}{c}{{ 13.53 }} & { 30.27 }\\ 
 & RD & \multicolumn{1}{c}{{ 38.79}} & { 52.11} & \multicolumn{1}{c}{\cellcolor[HTML]{C0C0C0}{ 39.69}} & \cellcolor[HTML]{C0C0C0}{ 55.66} & \multicolumn{1}{c}{\cellcolor[HTML]{C0C0C0}{ 29.54}} & \cellcolor[HTML]{C0C0C0}{ 40.91} & \multicolumn{1}{c}{{ 39.07}} & { 54.13} & \multicolumn{1}{c}{{ 35.64}} & { 55.45} & \multicolumn{1}{c}{{ 9.51}} & { 30.11 }\\ 
\multirow{-13}{*}{CodeBERT} & BT & \multicolumn{1}{c}{{ 41.98}} & { 53.19} & \multicolumn{1}{c}{{ 46.32}} & { 62.63} & \multicolumn{1}{c}{{ 63.63}} & { 69.69} & \multicolumn{1}{c}{{ 48.66}} & { 59.11} & \multicolumn{1}{c}{{ 40.37}} & { 55.96} & \multicolumn{1}{c}{{ 11.34}} & { 26.36}\\ \hline
 & No Aug (Baseline) & \multicolumn{1}{c}{{ 23.14}} & { 42.53} & \multicolumn{1}{c}{{ 30.56}} & { 46.73} & \multicolumn{1}{c}{{ 23.12}} & { 30.33} & \multicolumn{1}{c}{{ 30.23}} & { 41.34} & \multicolumn{1}{c}{{ 31.21}} & { 56.01} & \multicolumn{1}{c}{{ 4.11 }} & { 12.11}\\ 
 & WordMixup & \multicolumn{1}{c}{{ 23.12}} & { 42.44} & \multicolumn{1}{c}{{ 29.53}} & { 45.89} & \multicolumn{1}{c}{{ 22.19}} & { 28.32} & \multicolumn{1}{c}{{ 28.56}} & { 40.44} & \multicolumn{1}{c}{{ 31.67}} & { 56.45} & \multicolumn{1}{c}{{ -}} & {- }\\ 
 & SenMixup & \multicolumn{1}{c}{\cellcolor[HTML]{C0C0C0}{ 22.53}} & { 42.02} & \multicolumn{1}{c}{{ 27.45}} & { 44.67} & \multicolumn{1}{c}{{ 20.34}} & { 27.33} & \multicolumn{1}{c}{{ 26.22}} & { 38.62} & \multicolumn{1}{c}{{ 30.21}} & \cellcolor[HTML]{C0C0C0}{ 54.46} & \multicolumn{1}{c}{{- }} & {- }\\ 
 & Refactor & \multicolumn{1}{c}{{ 23.04}} & { 42.36} & \multicolumn{1}{c}{{ 28.35}} & { 44.45} & \multicolumn{1}{c}{{ 24.56}} & { 32.44} & \multicolumn{1}{c}{{ 29.45}} & { 39.56} & \multicolumn{1}{c}{{ 30.33}} & { 56.56} & \multicolumn{1}{c}{{ 6.02 }} & \cellcolor[HTML]{C0C0C0}{ 6.13 }\\ 
 
  & Rename Operator & \multicolumn{1}{c}{{ 23.11}} & \multicolumn{1}{c}{{ 42.46}} & \multicolumn{1}{c}{{ 29.35}} &  \multicolumn{1}{c}{{ 45.76}} &  \multicolumn{1}{c}{{ 27.51}} & \multicolumn{1}{c}{{ 34.45}} &\multicolumn{1}{c}{{ 30.15}} & \multicolumn{1}{c}{{ 41.23}} & \multicolumn{1}{c}{{ 30.89}} & \multicolumn{1}{c}{{ 56.98}} & \multicolumn{1}{c}{{ 6.81}} &  \multicolumn{1}{c}{{ 9.42}}\\ 

 & Dead Operator  & \multicolumn{1}{c}{{ 23.23}} &   \multicolumn{1}{c}{{ 42.39}} & \multicolumn{1}{c}{{ 30.23}} &  \multicolumn{1}{c}{{ 46.21}} &  \multicolumn{1}{c}{{ 28.96}}  & \multicolumn{1}{c}{{ 35.87}} & \multicolumn{1}{c}{{ 31.34}} & \multicolumn{1}{c}{{ 43.23}} & \multicolumn{1}{c}{{ 31.14}} & \multicolumn{1}{c}{{ 57.51}} & \multicolumn{1}{c}{{ 7.98}} &\multicolumn{1}{c}{{ 10.23}} \\ 

 & Inside Operator  & \multicolumn{1}{c}{{ 23.13}} & \multicolumn{1}{c}{{ 42.45}} & \multicolumn{1}{c}{{ 30.54}} &  \multicolumn{1}{c}{{ 46.56}} &  \multicolumn{1}{c}{{ 32.45}}  & \multicolumn{1}{c}{{ 37.76}} & \multicolumn{1}{c}{{ 29.97}} & \multicolumn{1}{c}{{ 41.02}} & \multicolumn{1}{c}{{ 32.65}} & \multicolumn{1}{c}{{ 59.93}} & \multicolumn{1}{c}{{ 8.11}} & \multicolumn{1}{c}{{ 10.81}}\\ 

 & Outside Operator  & \multicolumn{1}{c}{{ 23.09}} & \multicolumn{1}{c}{{ 42.37}} & \multicolumn{1}{c}{{ 29.98}} &  \multicolumn{1}{c}{{ 46.12}} &  \multicolumn{1}{c}{{ 31.86}}  &  \multicolumn{1}{c}{{ 36.91}} & \multicolumn{1}{c}{{ 32.13}} & \multicolumn{1}{c}{{ 43.69}} & \multicolumn{1}{c}{{ 31.42}} & \multicolumn{1}{c}{{ 59.12}} & \multicolumn{1}{c}{{ 7.34}} & \multicolumn{1}{c}{{ 9.93}}\\ 
 
 & SR & \multicolumn{1}{c}{{ 23.12}} & { 42.39} & \multicolumn{1}{c}{{ 32.57}} & { 48.97} & \multicolumn{1}{c}{{ 28.77}} & { 47.66} & \multicolumn{1}{c}{{ 31.34}} & { 42.52} & \multicolumn{1}{c}{{ 40.01}} & { 66.67} & \multicolumn{1}{c}{{ 6.15}} & {10.21 }\\ 
 & RI & \multicolumn{1}{c}{{ 22.56}} & { 42.41} & \multicolumn{1}{c}{{ 28.66}} & { 45.87} & \multicolumn{1}{c}{{ 23.24}} & { 33.12} & \multicolumn{1}{c}{{ 39.56}} & { 43.65} & \multicolumn{1}{c}{{ 43.52}} & { 65.74} & \multicolumn{1}{c}{\cellcolor[HTML]{C0C0C0}{ 2.18 }} & { 6.21}\\ 
 & RS & \multicolumn{1}{c}{{ 23.01}} & \cellcolor[HTML]{C0C0C0}{ 41.87} & \multicolumn{1}{c}{\cellcolor[HTML]{C0C0C0}{ 26.46}} & { 47.23} & \multicolumn{1}{c}{{ 21.45}} & { 29.65} & \multicolumn{1}{c}{\cellcolor[HTML]{C0C0C0}{ 26.16}} & \cellcolor[HTML]{C0C0C0}{ 38.32} & \multicolumn{1}{c}{\cellcolor[HTML]{C0C0C0}{ 30.16}} & { 54.76} & \multicolumn{1}{c}{{ 6.08}} & { 12.05}\\ 
 & RD & \multicolumn{1}{c}{{ 22.78}} & { 41.98} & \multicolumn{1}{c}{{ 33.97}} & \cellcolor[HTML]{C0C0C0}{ 43.66} & \multicolumn{1}{c}{\cellcolor[HTML]{C0C0C0}{ 20.21}} & \cellcolor[HTML]{C0C0C0}{ 27.31} & \multicolumn{1}{c}{{ 28.54}} & { 40.25} & \multicolumn{1}{c}{{ 31.71}} & { 60.98} & \multicolumn{1}{c}{{ 6.16}} & { 8.25}\\  
\multirow{-13}{*}{GraphCodeBERT} & BT & \multicolumn{1}{c}{{ 22.89}} & { 42.42} & \multicolumn{1}{c}{{ 26.86}} & { 45.32} & \multicolumn{1}{c}{{ 35.99}} & { 46.33} & \multicolumn{1}{c}{{ 33.56}} & { 42.76} & \multicolumn{1}{c}{{ 41.67}} & { 62.96}& \multicolumn{1}{c}{{ 8.21}} & {14.21}\\ \hline
\end{tabular}}
\end{table}
Then, we compare each data augmentation method. In CodeBERT, \emph{RS} performs the best and has a relatively better robustness improvement in five (out of 12) cases, and reduces ASR by up to 5.10\% under MHM attack and 3.69\% under ALERT attack compared to No Aug. Besides, the second best one, RD reduces the ASR by up to 5.67\% in MHM and 2.66\% in ALERT compared to \emph{No Aug}. In GraphCodeBERT, \emph{RS} still is the best choice for robustness improvement in five (out of 12) cases, which deduces ASR by up to 4.10\% in MHM and 3.02\% in ALERT compared to \emph{No Aug}. Interestingly, based on pre-trained PL models, compared to \emph{Refactor}, methods like \emph{RS} that could sightly break the syntax of programs can produce more accurate and robust code models to solve downstream tasks. This finding can inspire future research that, when proposing new data augmentation methods, it is unnecessary to follow the program constraint to design the method. Finally, moving to tasks, based on the results, we recommend using \emph{RS} to augment the training data when applying pre-trained PL models to source code learning because it has the best ability of robustness improvement in three (out of four) tasks.

Additionally, we find that code refactoring methods, especially \emph{Rename Operator} and \emph{Dead Operator}, achieve higher accuracy performance in DNN models compared to \emph{Refactor}. However, their robustness improvement is lower than that of \emph{Refactor}. This finding is consistent with phenomena observed in Dong~\emph{et al.}~\cite{dong2023mixcode}.

\smallskip
\noindent
\textbf{Statistical analysis.} First, when comparing text-oriented data augmentation methods, our findings indicate that no data augmentation method effectively improves robustness performance. This suggests that current text-oriented data augmentation methods may not be suitable for generating robust code models and there is a need to propose better data augmentation methods for code learning in the future. 
Interestingly, \emph{SenMixup} demonstrates significant robustness improvement compared to all code refactoring methods as shown in Table~\ref{table:statistical_testing_robust_codebert_graphcodebert} in Appendix~\ref{sec:appendix}.  Additionally, \emph{Refactor} demonstrates significant robustness improvement (with \emph{p}-values consistently $<$ 0.05) compared to model training using \emph{Rename Operator}, \emph{Dead Operator}, \emph{Inside Operator}, and \emph{Outside Operator}. This suggests that randomly combining different code refactoring methods can enhance the generalization of pre-trained PL models more effectively than adopting a specific code refactoring method.


\begin{tcolorbox}
\textbf{Answer to RQ2}: Although text-oriented data augmentation methods, e.g., \emph{RD}, can reduce the attack success rate by 5.67\%, our statistical tests find text-oriented data augmentation methods have limited benefits for enhancing the adversarial robustness of code models, i.e., no single method can consistently improve robustness across different datasets and models. Furthermore, randomly combining different refactoring methods is a better way to perform data augmentation than using a specific single refactoring method.



\end{tcolorbox}

\subsection{RQ3: How does data volume affect the effectiveness of data augmentation methods?}
\label{sec:RQ3}

Data augmentation is used to enrich the size and the diversity of training data, especially in the scenario -- lack of labeled training data. 
\begin{table}[h]
\caption{Effectiveness of data augmentation methods w.r.t. test accuracy $\uparrow$ (average $\pm$ standard deviation, \%) on original test data. \textbf{No Aug}: without data augmentation. The best results are highlighted in gray. The tested DNN model is CodeBERT. Particularly, the training data volume drops to 10\%, 5\%, 3\%, and 1\% of the original, respectively. Tasks include \textbf{Problem Classification} (Java250, Python800), \textbf{Bug detection} (CodRep1, Refactory), and \textbf{Clone detection} (BigCloneBench).}
\label{tab:ACC_CodeBERT_reduce}
\resizebox{\columnwidth}{!}{
\begin{tabular}{clccccc}
\hline
\textbf{Model} & \textbf{DA method} & \textbf{Java250} & \textbf{Python800} & \textbf{Refactory} & \textbf{CodRep1} & \textbf{BigCloneBench} \\ \hline
 & No Aug (Baseline) & { 92.28 ± 0.03} & { 92.66 ± 0.09} & { 82.51 ± 0.18} & { 57.12 ± 0.11} & 96.58 ± 0.15 \\ 
 & WordMixup & { 91.33 ± 0.04} & { 93.09 ± 0.02} & { 83.89 ± 0.13} & { 58.43 ± 0.21} & - \\ 
 & SenMixup & \cellcolor[HTML]{C0C0C0}{ 92.52 ± 0.07} & \cellcolor[HTML]{C0C0C0}{ 93.62 ± 0.04} & { 85.83 ± 0.12} & { 58.22 ± 0.12} & - \\ 
 & Refactor & { 92.36 ± 0.05} & { 92.48 ± 0.06} & { 85.58 ± 0.17} & { 56.21 ± 0.11} & 96.72 ± 0.33 \\ 
 & Rename Operator & { 92.38 ± 0.04} & { 92.52 ± 0.09} & { 85.89 ± 0.21} & { 56.82 ± 0.22} & 96.82 ± 0.14   \\ 

 & Dead Operator  & { 92.33 ± 0.05} & { 92.36 ± 0.11} & { 85.68 ± 0.15} & { 56.74 ± 0.18} & 96.69 ± 0.12 \\ 

 & Inside Operator  & { 92.36 ± 0.07} & { 92.41 ± 0.08} & { 85.97 ± 0.18} & { 56.32 ± 0.13} & 96.75 ± 0.19\\ 

 & Outside Operator  & { 92.39 ± 0.12} & { 92.43 ± 0.13} & { 85.74 ± 0.21} & { 56.41 ± 0.18} &  96.76 ± 0.16 \\ 
 & SR & { 92.09 ± 0.06} & { 92.28 ± 0.08} & { 81.81 ± 0.13} & { 55.67 ± 0.15} & 96.89 ± 0.27 \\ 
 & RI & { 91.91 ± 0.02} & { 92.12 ± 0.06} & { 79.67 ± 0.15} & { 57.68 ± 0.12} & 96.81 ± 0.19 \\ 
 & RS & { 92.39 ± 0.06} & { 92.69 ± 0.04} & { 85.82 ± 0.19} & \cellcolor[HTML]{C0C0C0}{ 58.44 ± 0.16} & \cellcolor[HTML]{C0C0C0}97.06 ± 0.18 \\ 
 & RD & { 92.41 ± 0.09} & { 92.55 ± 0.08} & { 84.87 ± 0.14} & { 57.87 ± 0.13} & 96.66 ± 0.21 \\ 
\multirow{-13}{*}{\begin{tabular}[c]{@{}c@{}}CodeBERT\\ (10\%)\end{tabular}} & BT & { 91.01 ± 0.02} & { 92.53 ± 0.07} & \cellcolor[HTML]{C0C0C0}{ 87.23 ± 0.16} & { 56.61 ± 0.14} & 96.93 ± 0.15 \\ \hline
 & No Aug (Baseline)& { 88.37 ± 0.13} & { 89.76 ± 0.12} & { 81.32 ± 0.26} & { 51.04 ± 0.22} & 79.24 ± 0.51 \\ 
 & WordMixup & { 88.52 ± 0.27} & \cellcolor[HTML]{C0C0C0}{ 89.89 ± 0.26} & { 87.43 ± 0.25} & { 52.21 ± 0.32} & - \\ 
 & SenMixup & { 88.14 ± 0.18} & { 89.64 ± 0.37} & { 88.89 ± 0.28} & \cellcolor[HTML]{C0C0C0}{ 53.33 ± 0.28} &- \\ 
 & Refactor & { 87.94 ± 0.15} & { 89.39 ± 0.15} & { 76.83 ± 0.35} & { 51.98 ± 0.37} & 78.99 ± 0.55 \\ 
  & Rename Operator & { 88.13 ± 0.22} & { 89.51 ± 0.24} & { 78.86 ± 0.34} & { 52.12 ± 0.25} &  79.45 ± 0.34  \\ 

 & Dead Operator & { 88.05 ± 0.26} & { 89.36 ± 0.21} & { 78.14 ± 0.28} & { 52.24 ± 0.35} & 79.13 ± 0.36  \\ 

 & Inside Operator & { 87.96 ± 0.16} & { 89.29 ± 0.31} & { 77.94 ± 0.21} & { 52.08 ± 0.31} & 79.07 ± 0.48  \\ 

 & Outside Operator& { 87.91 ± 0.14} & { 89.48 ± 0.27} & { 77.51 ± 0.19} & { 51.34 ± 0.25} & 79.19 ± 0.42 \\ 
 & SR & { 86.14 ± 0.12} & { 88.85 ± 0.28} & { 84.41 ± 0.24} & { 51.67 ± 0.23} & 78.58 ± 0.69 \\ 
 & RI & { 85.77 ± 0.14} & { 88.61 ± 0.37} & { 79.91 ± 0.43} & { 51.01 ± 0.22} & 79.02 ± 0.25 \\ 
 & RS & { 88.55 ± 0.16} & { 89.52 ± 0.29} & { 87.23 ± 0.27} & { 53.05 ± 0.23} & 80.02 ± 0.65 \\ 
 & RD & \cellcolor[HTML]{C0C0C0}{ 88.75 ± 0.15} & { 89.82 ± 0.48} & { 77.31 ± 0.16} & { 52.02 ± 0.35} & \cellcolor[HTML]{C0C0C0}80.06 ± 0.86 \\ \
\multirow{-13}{*}{\begin{tabular}[c]{@{}c@{}}CodeBERT\\  (5\%)\end{tabular}} & BT & { 85.18 ± 0.13} & { 89.69 ± 0.23} & \cellcolor[HTML]{C0C0C0}{ 90.07 ± 0.22} & { 53.11 ± 0.23} & 80.04 ± 0.45 \\ \hline
 & No Aug (Baseline) & { 78.15 ± 0.13} & { 82.39 ± 0.14} & { 75.89 ± 0.23} & { 44.12 ± 0.16} & 78.67 ± 0.44 \\  
 & WordMixup & { 78.89 ± 0.25} & { 82.57 ± 0.35} & { 87.33 ± 0.32} & { 48.09 ± 0.21} & - \\ 
 & SenMixup & \cellcolor[HTML]{C0C0C0}{ 79.19 ± 0.12} & \cellcolor[HTML]{C0C0C0}{ 82.84 ± 0.49} & \cellcolor[HTML]{C0C0C0}{ 88.81 ± 0.43} & \cellcolor[HTML]{C0C0C0}{ 49.88 ± 0.26} & - \\ 
 & Refactor & { 74.62 ± 0.15} & { 81.43 ± 0.39} & { 86.29 ± 0.36} & { 45.98 ± 0.22} & 79.53 ± 0.89 \\
& Rename Operator & { 76.86 ± 0.19} & { 81.95 ± 0.34} & { 88.14 ± 0.36} & { 46.45 ± 0.32} & 79.55 ± 0.56  \\ 

 & Dead Operator  & { 74.92 ± 0.13} & { 81.53 ± 0.31} & { 87.89 ± 0.41} & { 46.21 ± 0.29} &  79.61 ± 0.49 \\ 

 & Inside Operator  &{ 75.34 ± 0.21} & { 80.97 ± 0.29} & { 86.97 ± 0.32} & { 45.78 ± 0.28} &  79.58 ± 0.53 \\ 

 & Outside Operator  & { 75.51 ± 0.24} & { 81.67 ± 0.27} & { 86.51 ± 0.38} & { 45.67 ± 0.24} & 79.67 ± 0.47 \\ 
 & SR & { 75.03 ± 0.12} & { 80.54 ± 0.26} & { 84.41 ± 0.39} & { 44.67 ± 0.21} & 79.54 ± 0.67 \\ 
 & RI & { 73.05 ± 0.14} & { 79.65 ± 0.47} & { 87.71 ± 0.44} & { 45.97 ± 0.22} & 79.39 ± 0.39 \\ 
 & RS & { 77.59 ± 0.11} & { 82.01 ± 0.36} & { 81.81 ± 0.31} & { 44.41 ± 0.25} & \cellcolor[HTML]{C0C0C0}79.77 ± 0.22 \\ 
 & RD & { 78.18 ± 0.12} & { 82.42 ± 0.29} & { 76.12 ± 0.24} & { 44.19 ± 0.21} & 79.19 ± 0.39 \\ 
\multirow{-13}{*}{\begin{tabular}[c]{@{}c@{}}CodeBERT\\  (3\%)\end{tabular}} & BT & { 73.01 ± 0.16} & { 82.22 ± 0.34} & { 88.65 ± 0.33} & { 49.76 ± 0.33} & 79.67 ± 0.37 \\ \hline
 & No Aug (Baseline) & { 49.62 ± 0.14} & { 44.48 ± 0.38} & { 76.83 ± 0.21} & { 42.19 ± 0.06} & 72.02 ± 0.65 \\ 
 & WordMixup & { 53.21 ± 0.26} & { 54.65 ± 0.61} & { 87.11 ± 0.44} & { 46.56 ± 0.24} & - \\ 
 & SenMixup & \cellcolor[HTML]{C0C0C0}{ 55.26 ± 0.28} & \cellcolor[HTML]{C0C0C0}{ 56.17 ± 0.69} & { 88.12 ± 0.41} & \cellcolor[HTML]{C0C0C0}{ 48.89 ± 0.32} & - \\ 
 & Refactor & { 34.59 ± 0.26} & { 33.67 ± 0.56} & { 72.11 ± 0.53} & { 44.21 ± 0.31} &  72.03 ± 0.69\\
   & Rename Operator & { 37.31 ± 0.51} & { 37.43 ± 0.35} & { 77.84 ± 0.61} & { 45.56 ± 0.28} &  72.88 ± 0.58 \\ 

 & Dead Operator  & { 36.35 ± 0.42} & { 38.71 ± 0.44} & { 78.01 ± 0.42} & { 44.67 ± 0.31} &  72.56 ± 0.37 \\ 

 & Inside Operator  & { 35.86 ± 0.35} & { 36.68 ± 0.38} & { 75.69 ± 0.48} & { 43.13 ± 0.49} & 72.33 ± 0.38  \\ 

 & Outside Operator  & { 34.67 ± 0.32} & { 36.89 ± 0.39} & { 73.78 ± 0.34} & { 43.88  ± 0.38} & 72.18 ± 0.51 \\ 
 & SR & { 37.06 ± 0.35} & { 33.06 ± 0.57} & { 82.74 ± 0.43} & { 34.08 ± 0.43} & \cellcolor[HTML]{C0C0C0}73.43 ± 0.88 \\ 
 & RI & { 27.03 ± 0.49} & { 29.46 ± 0.42} & { 88.18 ± 0.54} & { 31.65 ± 0.55} & 72.67 ± 0.16 \\
 & RS & { 49.53 ± 0.38} & { 43.06 ± 0.48} & { 82.98 ± 0.52} & { 43.43 ± 0.58} & 72.51 ± 0.38 \\
 & RD & { 49.78 ± 0.33} & { 44.67 ± 0.33} & { 76.84 ± 0.44} & { 43.14 ± 0.64} & 72.47 ± 0.89 \\ 
\multirow{-13}{*}{\begin{tabular}[c]{@{}c@{}}CodeBERT\\ (1\%)\end{tabular}} & BT & { 36.68 ± 0.28} & { 44.75 ± 0.69} & \cellcolor[HTML]{C0C0C0}{ 88.42 ± 0.55} & { 48.64 ± 0.42} & 72.55 ± 0.59  \\ \hline
\end{tabular}}
\vspace{-3mm}
\end{table}
Therefore, it is necessary to explore whether data augmentation methods can be still effective in improving both the accuracy and robustness of DNN models when there is only a small size of training data. In this part, we keep only 10\%, 5\%, 3\%, and 1\% of training data (for the authorship attribution task, we keep 10\% and 50\% of training data since the size of original training data is relatively small, i.e., 528) and repeat the experiments conducted in Sections~\ref{sec:RQ1} and \ref{sec:RQ2}. We choose CodeBERT and GraphCodeBERT in our study.

\begin{table}[!tb]
\caption{Effectiveness of data augmentation methods w.r.t. test accuracy $\uparrow$ (average $\pm$ standard deviation, \%) on original test data. \textbf{No Aug}: without data augmentation. The best results are highlighted in gray. The tested DNN model is GraphCodeBERT. Particularly, the training data volume drops to 10\%, 5\%, 3\%, and 1\% of the original, respectively. Tasks include \textbf{Problem Classification} (Java250, Python800), \textbf{Bug detection} (CodRep1, Refactory), and \textbf{Clone detection} (BigCloneBench).}
\label{tab:ACC_GraphCodeBERT_reduce}
\resizebox{\columnwidth}{!}{
\begin{tabular}{clccccc}
\hline
\textbf{Model} & \textbf{DA method} & \textbf{Java250} & \textbf{Python800} & \textbf{Refactory} & \textbf{CodRep1} & \textbf{BigCloneBench} \\ \hline
 & No Aug (Baseline) & { 93.09 ± 0.12} & { 93.33 ± 0.08} & { 87.94 ± 0.12} & { 58.34 ± 0.19} & 96.31 ± 0.32 \\ 
 & WordMixup & { 93.25 ± 0.19} & { 93.39 ± 0.24} & { 88.02 ± 0.21} & { 57.86 ± 0.32} & - \\ 
 & SenMixup & \cellcolor[HTML]{C0C0C0}{ 93.41 ± 0.16} & { 93.44 ± 0.18} & { 88.25 ± 0.15} & { 59.67 ± 0.36} & - \\ 
 & Refactor & { 93.11 ± 0.24} & { 93.35 ± 0.21} & { 84.87 ± 0.34} & { 58.04 ± 0.24} & 96.35 ± 0.28 \\ 
  & Rename Operator & { 93.24 ± 0.24} & { 93.39 ± 0.31} & { 86.65 ± 0.24} & { 58.56 ± 0.37} & 96.55 ± 0.45  \\ 

 & Dead Operator  & { 93.13 ± 0.19} & { 93.36 ± 0.42} & { 85.34 ± 0.19} & { 59.13 ± 0.47}  &  96.34 ± 0.26 \\ 

 & Inside Operator  & { 92.68 ± 0.31} & { 93.29 ± 0.35} & { 84.96 ± 0.34} & { 58.89 ± 0.31} & 96.21 ± 0.67  \\ 

 & Outside Operator  & { 92.91 ± 0.26} & { 93.24 ± 0.38} &  { 85.56 ± 0.41} & { 58.33 ± 0.43} & 96.33 ± 0.51 \\ 
 & SR & { 92.79 ± 0.11} & { 92.98 ± 0.63} & { 81.14 ± 0.29} & { 57.78 ± 0.31} & 96.39 ± 0.43 \\ 
 & RI & { 92.51 ± 0.22} & { 92.32 ± 0.52} & { 78.87 ± 0.45} & { 56.34 ± 0.26} & \cellcolor[HTML]{C0C0C0}96.99 ± 0.33 \\ 
 & RS & { 92.91 ± 0.25} & { 93.41 ± 0.69} & { 88.34 ± 0.26} & { 59.33 ± 0.36} & 95.95 ± 0.52 \\ 
 & RD & { 92.84 ± 0.21} & { 92.93 ± 0.34} & { 88.75 ± 0.21} & \cellcolor[HTML]{C0C0C0}{ 59.84 ± 0.45} & 95.79 ± 0.25 \\ 
\multirow{-13}{*}{\begin{tabular}[c]{@{}c@{}}GraphCodeBERT\\ (10\%)\end{tabular}} & BT & { 92.86 ± 0.08} & \cellcolor[HTML]{C0C0C0}{ 93.45 ± 0.53} & \cellcolor[HTML]{C0C0C0}{ 89.18 ± 0.14} & { 59.35 ± 0.21} & 95.68 ± 0.21 \\ \hline
 & No Aug (Baseline) & { 91.03 ± 0.07} & { 90.87 ± 0.13} & { 77.31 ± 0.13} & { 55.53 ± 0.21} & 77.38 ± 0.29 \\ 
 & WordMixup & { 90.84 ± 0.14} & { 90.65 ± 0.21} & { 79.03 ± 0.34} & { 57.23 ± 0.37} &  -\\ 
 & SenMixup & \cellcolor[HTML]{C0C0C0}{ 91.22 ± 0.11} & { 90.92 ± 0.29} & { 80.12 ± 0.29} & \cellcolor[HTML]{C0C0C0}{ 58.94 ± 0.32} & - \\ 
 & Refactor & { 90.09 ± 0.18} & { 90.83 ± 0.19} & { 77.78 ± 0.27} & { 53.44 ± 0.39} & 77.46 ± 0.39 \\ 
    & Rename Operator & { 90.78 ± 0.25} & { 90.98 ± 0.31} & { 78.86 ± 0.33} & { 56.56 ± 0.46} & 77.53 ± 0.27  \\ 

 & Dead Operator  & { 90.31 ± 0.24} & { 90.84 ± 0.23} & { 77.79 ± 0.25} & { 56.31 ± 0.49} & 77.32 ± 0.31  \\ 

 & Inside Operator  & { 90.13 ± 0.16} & { 90.76 ± 0.22} & { 77.65 ± 0.31} & { 54.47 ± 0.37} & 76.89 ± 0.26  \\ 

 & Outside Operator  & { 90.22 ± 0.25} & { 90.88 ± 0.24} & { 78.35 ± 0.23} & { 53.83 ± 0.44} & 77.31 ± 0.38 \\ 
 & SR & { 90.26 ± 0.23} & { 90.53 ± 0.23} & { 76.31 ± 0.34} & { 56.32 ± 0.34} & 73.98 ± 0.23 \\
 & RI & { 89.41 ± 0.12} & { 90.88 ± 0.45} & { 78.23 ± 0.23} & { 55.34 ± 0.25} & 76.62 ± 0.15 \\ 
 & RS & { 90.91 ± 0.24} & \cellcolor[HTML]{C0C0C0}{ 91.08 ± 0.51} & { 80.21 ± 0.25} & { 58.76 ± 0.47} & 77.51 ± 0.42 \\ 
 & RD & { 90.84 ± 0.09} & { 91.03 ± 0.35} & { 77.98 ± 0.33} & { 58.54 ± 0.25} & 77.49 ± 0.32 \\ 
\multirow{-13}{*}{\begin{tabular}[c]{@{}c@{}}GraphCodeBERT\\ (5\%)\end{tabular}} & BT & { 91.04 ± 0.23} & { 90.89 ± 0.67} & \cellcolor[HTML]{C0C0C0}{ 81.85 ± 0.38} & { 55.66 ± 0.34} & \cellcolor[HTML]{C0C0C0}81.28 ± 0.26 \\ \hline
 & No Aug (Baseline) & { 84.61 ± 0.15} & { 84.56 ± 0.09} & { 78.96 ± 0.19} & { 44.35 ± 0.24} & 76.62 ± 0.36\\
 & WordMixup & { 84.64 ± 0.25} & { 84.97 ± 0.21} & { 79.01 ± 0.23} & { 46.12 ± 0.21} & - \\ 
 & SenMixup & { 85.26 ± 0.21} & { 85.39 ± 0.14} & { 79.45 ± 0.11} & \cellcolor[HTML]{C0C0C0}{ 46.78 ± 0.19} & - \\ 
 & Refactor & { 83.31 ± 0.09} & { 83.64 ± 0.23} & { 71.63 ± 0.31} & { 42.12 ± 0.13} & 75.65  ± 0.52 \\ 
    & Rename Operator & { 84.19 ± 0.14}  & { 84.36 ± 0.37} & { 73.64 ± 0.35} & { 44.56 ± 0.24} & 75.76 ± 0.56  \\ 

 & Dead Operator  & { 83.43 ± 0.12} & { 83.87 ± 0.33} & { 74.01 ± 0.28} & { 43.86 ± 0.19} & 75.74 ± 0.38 \\ 

 & Inside Operator  & { 83.67 ± 0.16} & { 83.25 ± 0.28} & { 72.16 ± 0.25} & { 42.56 ± 0.24} & 75.87 ± 0.45  \\ 

 & Outside Operator  & { 83.57 ± 0.18} & { 83.68 ± 0.36} & { 72.01 ± 0.34} & { 42.67 ± 0.31} & 75.05 ± 0.37 \\ 
 & SR & { 82.63 ± 0.27} & { 82.48 ± 0.62} & { 68.09 ± 0.24} & { 44.03 ± 0.35} & 70.98 ± 0.42 \\ 
 & RI & { 80.58 ± 0.24} & { 80.42 ± 0.57} & { 72.34 ± 0.35} & { 42.89 ± 0.34} & 74.28  ± 0.31 \\ 
 & RS & { 85.16 ± 0.13} & { 84.87 ± 0.63} & { 71.69 ± 0.24} & { 46.34 ± 0.45} & \cellcolor[HTML]{C0C0C0}76.92 ± 0.25  \\ 
 & RD & \cellcolor[HTML]{C0C0C0}{ 85.35 ± 0.11} & { 85.42 ± 0.46} & \cellcolor[HTML]{C0C0C0}{ 80.61 ± 0.11} & { 45.45 ± 0.47} & 76.77 ± 0.21 \\ 
\multirow{-13}{*}{\begin{tabular}[c]{@{}c@{}}GraphCodeBERT\\ (3\%)\end{tabular}} & BT & { 84.75 ± 0.04} & \cellcolor[HTML]{C0C0C0}{ 85.74 ± 0.74} & { 76.67 ± 0.36} & { 43.42 ± 0.56} & 74.35 ± 0.56 \\ \hline
 & No Aug (Baseline) & { 57.33 ± 0.14} & { 51.52 ± 0.06} & { 65.25 ± 0.13} & { 41.23 ± 0.17} & 72.51± 0.41 \\ 
 & WordMixup & \cellcolor[HTML]{C0C0C0}{ 57.86 ± 0.19} & { 51.14 ± 0.16} & { 68.13 ± 0.29} & { 44.87 ± 0.27} & - \\ 
 & SenMixup & { 57.13 ± 0.12} & \cellcolor[HTML]{C0C0C0}{ 52.04 ± 0.11} & \cellcolor[HTML]{C0C0C0}{ 69.21 ± 0.23} & \cellcolor[HTML]{C0C0C0}{ 45.67 ± 0.31} & - \\ 
 & Refactor & { 49.29 ± 0.25} & { 40.09 ± 0.13} & { 64.57 ± 0.36} & { 41.11 ± 0.12} & 69.48 ± 0.47 \\ 
 & Rename Operator & { 52.67 ± 0.25} & { 44.67 ± 0.37} & { 65.89 ± 0.41} & { 42.78 ± 0.31} & 70.64 ± 0.34  \\ 

 & Dead Operator  & { 51.45 ± 0.31} & { 43.89 ± 0.26} & { 63.78 ± 0.36} & { 43.07 ± 0.33} &  70.21 ± 0.38 \\ 

 & Inside Operator  & { 49.96 ± 0.35} & { 42.56 ± 0.46} & { 64.88 ± 0.31} & { 42.56 ± 0.43} &  69.86 ± 0.45 \\ 

 & Outside Operator  & { 50.76 ± 0.46} & { 42.98 ± 0.38} & { 64.67 ± 0.26} & { 41.78 ± 0.39} &  69.53 ± 0.51 \\ 
 & SR & { 38.49 ± 0.23} & { 34.56 ± 0.29} & { 47.05 ± 0.31} & { 36.87 ± 0.23} & 65.12 ± 0.26 \\ 
 & RI & { 32.79 ± 0.32} & { 35.99 ± 0.26} & { 44.21 ± 0.24} & { 38.22 ± 0.34} & 66.47 ± 0.32 \\ 
 & RS & { 56.46 ± 0.24} & { 51.26 ± 0.31} & { 66.45 ± 0.25} & { 45.09 ± 0.53} & \cellcolor[HTML]{C0C0C0}72.53 ± 0.26 \\ 
 & RD & { 57.59 ± 0.25} & { 50.67 ± 0.33} & { 65.76 ± 0.26} & { 39.14 ± 0.36} & 71.98 ± 0.55 \\ 
\multirow{-13}{*}{\begin{tabular}[c]{@{}c@{}}GraphCodeBERT\\ (1\%)\end{tabular}} & BT & { 57.21 ± 0.21} & { 51.03 ± 0.25} & { 55.32 ± 0.11} & { 44.64 ± 0.42} & 67.85 ± 0.58 \\ \hline
\end{tabular}}
\end{table}

\begin{table}[!tb]
\caption{Effectiveness of data augmentation methods w.r.t. test accuracy $\uparrow$ (average $\pm$ standard deviation, \%) and ASR $\downarrow$ (\%) on test data. \textbf{No Aug}: without data augmentation. The tested dataset is GCJ from task Authorship attribution. Particularly, the training data volume drops to 50\% and 10\% of the original, respectively.}
\label{tab:GCJ_reduce}
\resizebox{\columnwidth}{!}{
\begin{tabular}{clccccccccccc}
\hline
\textbf{Model} & \textbf{DA method} & \multicolumn{2}{c}{\textbf{GCJ}} & \textbf{Model} & \multicolumn{2}{c}{\textbf{GCJ}} & \textbf{Model} & \multicolumn{2}{c}{\textbf{GCJ}} & Model & \multicolumn{2}{c}{\textbf{GCJ}} \\ \hline
\multicolumn{13}{c}{\textbf{Test Accuracy}} \\ \hline
 & No Aug (Baseline) & \multicolumn{2}{c}{{86.47 ± 0.15}} & {} & \multicolumn{2}{c}{{40.69 ± 0.11}} & \multicolumn{1}{c}{{}} & \multicolumn{2}{c}{{86.47 ± 0.13}} & \multicolumn{1}{c}{{}} & \multicolumn{2}{c}{{17.29 ± 0.19}} \\ 
 & WordMixup & \multicolumn{2}{c}{{84.21 ± 0.23}} & {} & \multicolumn{2}{c}{{42.14 ± 2.01}} & \multicolumn{1}{c}{{}} & \multicolumn{2}{c}{{85.87 ± 0.31}} & \multicolumn{1}{c}{{}} & \multicolumn{2}{c}{{18.23 ± 0.24}} \\ 
 & SenMixup & \multicolumn{2}{c}{\cellcolor[HTML]{C0C0C0}{86.98 ± 0.31}} & {} & \multicolumn{2}{c}{\cellcolor[HTML]{C0C0C0}{44.51 ± 1.31}} & \multicolumn{1}{c}{{}} & \multicolumn{2}{c}{{86.92 ± 0.22}} & \multicolumn{1}{c}{{}} & \multicolumn{2}{c}{\cellcolor[HTML]{C0C0C0}{20.56 ± 0.12}} \\ 
 & Refactor & \multicolumn{2}{c}{{78.21 ± 0.17}} & {} & \multicolumn{2}{c}{{31.58 ± 0.21}} & \multicolumn{1}{c}{{}} & \multicolumn{2}{c}{{70.68 ± 0.27}} & \multicolumn{1}{c}{{}} & \multicolumn{2}{c}{{13.53 ± 0.28}} \\ 
    & Rename Operator & \multicolumn{2}{c}{{80.15 ± 0.21}} & \multicolumn{1}{c}{{}} & \multicolumn{2}{c}{{35.24 ± 0.15}} & &  \multicolumn{2}{c}{{74.45 ± 0.23}} &  & \multicolumn{2}{c}{{14.37 ± 0.35}}  \\ 

 & Dead Operator  & \multicolumn{2}{c}{{78.56 ± 0.16}} & \multicolumn{1}{c}{{}}&\multicolumn{2}{c}{{32.02 ± 0.11}} &  & \multicolumn{2}{c}{{73.14 ± 0.31}} &   &\multicolumn{2}{c}{{15.14 ± 0.31}}\\ 

 & Inside Operator  & \multicolumn{2}{c}{{78.76 ± 0.31}} & \multicolumn{1}{c}{{}}& \multicolumn{2}{c}{{32.68 ± 0.24}} &  &\multicolumn{2}{c}{{72.67 ± 0.28}}  &  &\multicolumn{2}{c}{{13.87 ± 0.29}}  \\ 

 & Outside Operator  & \multicolumn{2}{c}{{79.03 ± 0.18}} & \multicolumn{1}{c}{{}}& \multicolumn{2}{c}{{31.98 ± 0.31}} &  & \multicolumn{2}{c}{{71.59 ± 0.19}} & & \multicolumn{2}{c}{{13.49 ± 0.17}} \\ 
 
 & SR & \multicolumn{2}{c}{{57.89 ± 0.09}} & {} & \multicolumn{2}{c}{{7.52 ± 0.15}} & \multicolumn{1}{c}{{}} & \multicolumn{2}{c}{{52.88 ± 0.12}} & \multicolumn{1}{c}{{}} & \multicolumn{2}{c}{{9.77 ± 0.24}} \\ 
 & RI & \multicolumn{2}{c}{{22.56 ± 0.15}} & {} & \multicolumn{2}{c}{{3.76 ± 0.13}} & \multicolumn{1}{c}{{}} & \multicolumn{2}{c}{{23.56 ± 0.26}} & \multicolumn{1}{c}{{}} & \multicolumn{2}{c}{{3.01 ± 0.25}} \\
 & RS & \multicolumn{2}{c}{\cellcolor[HTML]{C0C0C0}{86.98 ± 0.19}} & {} & \multicolumn{2}{c}{{41.67 ± 0.17}} & \multicolumn{1}{c}{{}} & \multicolumn{2}{c}{\cellcolor[HTML]{C0C0C0}{88.22 ± 0.11}} & \multicolumn{1}{c}{{}} & \multicolumn{2}{c}{{18.54 ± 0.21}} \\ 
 & RD & \multicolumn{2}{c}{{55.64 ± 0.14}} & {} & \multicolumn{2}{c}{{12.78 ± 0.24}} & \multicolumn{1}{c}{{}} & \multicolumn{2}{c}{{82.23 ± 0.16}} & \multicolumn{1}{c}{{}} & \multicolumn{2}{c}{{15.04 ± 0.14}} \\ 
\multirow{-13}{*}{\begin{tabular}[c]{@{}c@{}}CodeBERT\\ (50\%)\end{tabular}} & BT & \multicolumn{2}{c}{{69.17 ± 0.06}} & \multirow{-13}{*}{{\begin{tabular}[c]{@{}c@{}}CodeBERT\\ (10\%)\end{tabular}}} & \multicolumn{2}{c}{{23.32 ± 0.27}} & \multicolumn{1}{c}{\multirow{-13}{*}{{\begin{tabular}[c]{@{}c@{}}GraphCodeBERT\\ (50\%)\end{tabular}}}} & \multicolumn{2}{c}{{60.15 ± 0.19}} & \multicolumn{1}{c}{\multirow{-13}{*}{{\begin{tabular}[c]{@{}c@{}}GraphCodeBERT\\ (10\%)\end{tabular}}}} & \multicolumn{2}{c}{{11.28 ± 0.27}} \\ \hline
\multicolumn{13}{c}{\textbf{Robustness}} \\ \hline
 &  & \textbf{{MHM}} & \textbf{ALERT} & {} & \textbf{MHM} & \textbf{ALERT} & & \textbf{MHM} & \textbf{ALERT} &  & \textbf{MHM} & \textbf{ALERT} \\ 
 & No Aug (Baseline)& \multicolumn{1}{c}{{28.07}} & \multicolumn{1}{c}{{52.63}} & {} & \multicolumn{1}{c}{{58.49}} & \multicolumn{1}{c}{{69.81}} & \multicolumn{1}{c}{} & \multicolumn{1}{c}{{44.83}} & \multicolumn{1}{c}{{68.11}} & \multicolumn{1}{c}{} & \multicolumn{1}{c}{{54.55}} & \multicolumn{1}{c}{{72.73}} \\ 
 & WordMixup & \multicolumn{1}{c}{{36.03}} & \multicolumn{1}{c}{{54.95}} & {} & \multicolumn{1}{c}{{52.98}} & \multicolumn{1}{c}{{63.56}} & \multicolumn{1}{c}{} & \multicolumn{1}{c}{{48.87}} & \multicolumn{1}{c}{{76.12}} & \multicolumn{1}{c}{} & \multicolumn{1}{c}{{45.78}} & \multicolumn{1}{c}{{52.34}} \\
 & SenMixup & \multicolumn{1}{c}{{27.87}} & \multicolumn{1}{c}{{49.55}} & {} & \multicolumn{1}{c}{{51.45}} & \multicolumn{1}{c}{{61.11}} & \multicolumn{1}{c}{} & \multicolumn{1}{c}{\cellcolor[HTML]{C0C0C0}{43.76}} & \multicolumn{1}{c}{\cellcolor[HTML]{C0C0C0}{67.56}} & \multicolumn{1}{c}{} & \multicolumn{1}{c}{{44.87}} & \multicolumn{1}{c}{{51.45}} \\ 
 & Refactor & \multicolumn{1}{c}{{33.01}} & \multicolumn{1}{c}{{60.19}} & {} & \multicolumn{1}{c}{{58.54}} & \multicolumn{1}{c}{{75.61}} & \multicolumn{1}{c}{} & \multicolumn{1}{c}{{51.58}} & \multicolumn{1}{c}{{71.58}} & \multicolumn{1}{c}{} & \multicolumn{1}{c}{\cellcolor[HTML]{C0C0C0}{44.44}} & \multicolumn{1}{c}{{66.67}} \\ 
     & Rename Operator & \multicolumn{1}{c}{{36.56}} & \multicolumn{1}{c}{{64.14}} & {} & \multicolumn{1}{c}{{62.14}} & \multicolumn{1}{c}{{77.65}} & \multicolumn{1}{c}{} & \multicolumn{1}{c}{{60.14}} & \multicolumn{1}{c}{{72.43}} & \multicolumn{1}{c}{} & \multicolumn{1}{c}{{48.68}} & \multicolumn{1}{c}{{67.46}} \\ 

 & Dead Operator & \multicolumn{1}{c}{{33.68}} & \multicolumn{1}{c}{{62.47}} & {} & \multicolumn{1}{c}{{63.36}} & \multicolumn{1}{c}{{70.24}} & \multicolumn{1}{c}{} & \multicolumn{1}{c}{{58.69}} & \multicolumn{1}{c}{{69.91}} & \multicolumn{1}{c}{} & \multicolumn{1}{c}{{46.45}} & \multicolumn{1}{c}{{65.59}} \\ 

 & Inside Operator  & \multicolumn{1}{c}{{38.27}} & \multicolumn{1}{c}{{63.72}} & {} & \multicolumn{1}{c}{{60.98}} & \multicolumn{1}{c}{{75.86}} & \multicolumn{1}{c}{} & \multicolumn{1}{c}{{59.11}} & \multicolumn{1}{c}{{68.37}} & \multicolumn{1}{c}{} & \multicolumn{1}{c}{{52.74}} & \multicolumn{1}{c}{{73.43}} \\  

 & Outside Operator & \multicolumn{1}{c}{{39.15}} & \multicolumn{1}{c}{{66.86}} & {} & \multicolumn{1}{c}{{61.39}} & \multicolumn{1}{c}{{76.45}} & \multicolumn{1}{c}{} & \multicolumn{1}{c}{{57.35}} & \multicolumn{1}{c}{{69.46}} & \multicolumn{1}{c}{} & \multicolumn{1}{c}{{51.53}} & \multicolumn{1}{c}{{73.72}} \\ 
 
 & SR & \multicolumn{1}{c}{{39.47}} & \multicolumn{1}{c}{{64.47}} & {} & \multicolumn{1}{c}{{70.01}} & \multicolumn{1}{c}{{70.01}} & \multicolumn{1}{c}{} & \multicolumn{1}{c}{{58.57}} & \multicolumn{1}{c}{{82.86}} & \multicolumn{1}{c}{} & \multicolumn{1}{c}{{69.23}} & \multicolumn{1}{c}{{84.62}} \\ 
 & RI & \multicolumn{1}{c}{{37.93}} & \multicolumn{1}{c}{\cellcolor[HTML]{C0C0C0}{48.28}} & {} & \multicolumn{1}{c}{\cellcolor[HTML]{C0C0C0}{20.02}} & \multicolumn{1}{c}{{60.01}} & \multicolumn{1}{c}{} & \multicolumn{1}{c}{{67.74}} & \multicolumn{1}{c}{{80.65}} & \multicolumn{1}{c}{} & \multicolumn{1}{c}{{50.01}} & \multicolumn{1}{c}{\cellcolor[HTML]{C0C0C0}{50.01}} \\
 & RS & \multicolumn{1}{c}{\cellcolor[HTML]{C0C0C0}{27.49}} & \multicolumn{1}{c}{{50.01}} & {} & \multicolumn{1}{c}{{60.38}} & \multicolumn{1}{c}{{71.69}} & \multicolumn{1}{c}{} & \multicolumn{1}{c}{{45.69}} & \multicolumn{1}{c}{{73.28}} & \multicolumn{1}{c}{} & \multicolumn{1}{c}{{54.54}} & \multicolumn{1}{c}{{72.72}} \\ 
 & RD & \multicolumn{1}{c}{{43.84}} & \multicolumn{1}{c}{{60.27}} & {} & \multicolumn{1}{c}{{41.18}} & \multicolumn{1}{c}{\cellcolor[HTML]{C0C0C0}{41.18}} & \multicolumn{1}{c}{} & \multicolumn{1}{c}{{50.01}} & \multicolumn{1}{c}{{77.27}} & \multicolumn{1}{c}{} & \multicolumn{1}{c}{{55.01}} & \multicolumn{1}{c}{{85.01}} \\ 
\multirow{-14}{*}{\begin{tabular}[c]{@{}c@{}}CodeBERT\\ (50\%)\end{tabular}} & BT & \multicolumn{1}{c}{{36.26}} & \multicolumn{1}{c}{{60.44}} & \multirow{-14}{*}{{\begin{tabular}[c]{@{}c@{}}CodeBERT\\ (10\%)\end{tabular}}} & \multicolumn{1}{c}{{61.29}} & \multicolumn{1}{c}{{67.74}} & \multicolumn{1}{c}{\multirow{-14}{*}{\begin{tabular}[c]{@{}c@{}}GraphCodeBERT\\ (50\%)\end{tabular}}} & \multicolumn{1}{c}{{47.51}} & \multicolumn{1}{c}{{75.01}} & \multicolumn{1}{c}{\multirow{-14}{*}{\begin{tabular}[c]{@{}c@{}}GraphCodeBERT\\ (10\%)\end{tabular}}} & \multicolumn{1}{c}{{60.01}} & \multicolumn{1}{c}{{86.67}} \\ \hline
\end{tabular}}
\vspace{-3mm}
\end{table}

\begin{table}[!tb]
\caption{Effectiveness of data augmentation methods w.r.t. ASR $\downarrow$ (\%) on test data. \textbf{No Aug}: without data augmentation. The best results are highlighted in gray. The victim DNN model is CodeBERT. Particularly, the training data volume drops to 10\%, 5\%, 3\%, and 1\% of the original, respectively. Tasks include \textbf{Problem Classification} (Java250, Python800), \textbf{Bug detection} (CodRep1, Refactory), and \textbf{Clone detection} (BigCloneBench)}
\label{tab:robust_CodeBERT_reduce}
\resizebox{\columnwidth}{!}{
\begin{tabular}{clcccccccccc}
\hline
 & & \multicolumn{2}{c}{\textbf{Java250}} & \multicolumn{2}{c}{\textbf{Python800}} & \multicolumn{2}{c}{\textbf{Refactory}} & \multicolumn{2}{c}{\textbf{CodRep1}} & \multicolumn{2}{c}{\textbf{BigCloneBench}} \\ 
\multirow{-2}{*}{\textbf{Model}} & \multicolumn{1}{l}{\multirow{-2}{*}{\textbf{DA method}}} & \textbf{MHM} & \textbf{ALERT} & \textbf{MHM} & \textbf{ALERT} & \textbf{MHM} & \textbf{ALERT} & \textbf{MHM} & \textbf{ALERT} & \textbf{MHM} & \textbf{ALERT} \\ \hline
 & No Aug (Baseline) & \multicolumn{1}{c}{{ 25.69}} & { 39.79} & \multicolumn{1}{c}{{ 38.63}} & { 53.63} & \multicolumn{1}{c}{{ 36.37}} & { 54.54} & \multicolumn{1}{c}{{ 44.56}} & { 59.65} & \multicolumn{1}{c}{15.56} & 26.33 \\ 
 & WordMixup & \multicolumn{1}{c}{{ 25.12}} & { 39.81} & \multicolumn{1}{c}{{ 39.12}} & { 55.42} & \multicolumn{1}{c}{{ 32.56}} & { 47.63} & \multicolumn{1}{c}{{ 41.22}} & { 60.43} & \multicolumn{1}{c}{-} & -\\ 
 & SenMixup & \multicolumn{1}{c}{{ 26.89}} & \cellcolor[HTML]{C0C0C0}{ 37.96} & \multicolumn{1}{c}{\cellcolor[HTML]{C0C0C0}{ 36.44}} & \cellcolor[HTML]{C0C0C0}{ 50.53} & \multicolumn{1}{c}{\cellcolor[HTML]{C0C0C0}{ 28.87}} & \cellcolor[HTML]{C0C0C0}{ 39.43} & \multicolumn{1}{c}{{ 40.45}} & \cellcolor[HTML]{C0C0C0}{ 53.42} & \multicolumn{1}{c}{-} &- \\ 
 & Refactor & \multicolumn{1}{c}{{ 25.11}} & { 38.57} & \multicolumn{1}{c}{{ 37.19}} & { 52.65} & \multicolumn{1}{c}{{ 39.13}} & { 52.17} & \multicolumn{1}{c}{{ 42.56}} & { 63.22} & \multicolumn{1}{c}{15.02} & 26.14 \\
    & Rename Operator& \multicolumn{1}{c}{{ 25.65}} & { 39.15} & \multicolumn{1}{c}{{ 38.45 }} & { 54.14} & \multicolumn{1}{c}{{ 47.13}} & { 58.31} & \multicolumn{1}{c}{{ 45.59}} & { 57.56} & \multicolumn{1}{c}{16.13} & 28.09 \\

 & Dead Operator & \multicolumn{1}{c}{{ 26.45}} & { 40.13} & \multicolumn{1}{c}{{ 37.89}} & { 53.27} & \multicolumn{1}{c}{{ 45.52}} & { 47.97} & \multicolumn{1}{c}{{ 43.27}} & { 58.65} & \multicolumn{1}{c}{15.34} & 25.57 \\

 & Inside Operator& \multicolumn{1}{c}{{ 26.89}} & { 39.98} & \multicolumn{1}{c}{{ 42.66}} & { 57.78} & \multicolumn{1}{c}{{ 48.21}} & { 59.23} & \multicolumn{1}{c}{{ 45.78}} & { 59.65} & \multicolumn{1}{c}{16.73} & 29.91 \\

 & Outside Operator & \multicolumn{1}{c}{{ 27.14}} & { 41.45} & \multicolumn{1}{c}{{ 41.24}} & { 56.24} & \multicolumn{1}{c}{{ 49.34}} & { 62.46} & \multicolumn{1}{c}{{ 46.12}} & { 55.89} & \multicolumn{1}{c}{ 17.01} & 30.12 \\
 
 & SR & \multicolumn{1}{c}{{ 26.32}} & { 43.42} & \multicolumn{1}{c}{{ 40.52}} & { 54.62} & \multicolumn{1}{c}{{ 52.17}} & { 47.83} & \multicolumn{1}{c}{{ 45.67}} & { 65.34} & \multicolumn{1}{c}{17.33} & 34.42 \\ 
 & RI & \multicolumn{1}{c}{{ 28.76}} & { 44.98} & \multicolumn{1}{c}{{ 43.45}} & { 56.88} & \multicolumn{1}{c}{{ 65.38}} & { 73.08} & \multicolumn{1}{c}{{ 42.87}} & { 55.67} & \multicolumn{1}{c}{20.52} & 28.35 \\ 
 & RS & \multicolumn{1}{c}{{ 25.02}} & { 38.83} & \multicolumn{1}{c}{{ 38.13}} & { 51.47} & \multicolumn{1}{c}{{ 60.12}} & { 70.01} & \multicolumn{1}{c}{{ 46.74}} & { 54.68} & \multicolumn{1}{c}{16.92} & 28.06 \\ 
 & RD & \multicolumn{1}{c}{\cellcolor[HTML]{C0C0C0}{ 24.96}} & { 38.78} & \multicolumn{1}{c}{{ 42.52}} & { 55.11} & \multicolumn{1}{c}{{ 38.09}} & { 42.86} & \multicolumn{1}{c}{\cellcolor[HTML]{C0C0C0}{ 39.87}} & { 56.45} & \multicolumn{1}{c}{14.87} & \cellcolor[HTML]{C0C0C0}24.26 \\ 
\multirow{-13}{*}{\begin{tabular}[c]{@{}c@{}}CodeBERT\\ (10\%)\end{tabular}} & BT & \multicolumn{1}{c}{{ 29.76}} & { 39.65} & \multicolumn{1}{c}{{ 44.73}} & { 52.12} & \multicolumn{1}{c}{{ 30.76}} & { 46.15} & \multicolumn{1}{c}{{ 48.98}} & { 64.33} & \multicolumn{1}{c}{\cellcolor[HTML]{C0C0C0}14.81} & 30.38 \\ \hline
 & No Aug (Baseline) & \multicolumn{1}{c}{{ 34.74}} & { 53.59} & \multicolumn{1}{c}{{ 45.64}} & { 64.21} & \multicolumn{1}{c}{{ 47.83}} & { 56.52} & \multicolumn{1}{c}{{ 50.87}} & { 64.33} & \multicolumn{1}{c}{17.27} & 28.76 \\ 
 & WordMixup & \multicolumn{1}{c}{{ 34.32}} & { 53.42} & \multicolumn{1}{c}{{ 43.21}} & { 58.17} & \multicolumn{1}{c}{{ 39.62}} & { 58.11} & \multicolumn{1}{c}{{ 52.55}} & { 63.22} & \multicolumn{1}{c}{-} & - \\ 
 & SenMixup & \multicolumn{1}{c}{{ 33.52}} & \cellcolor[HTML]{C0C0C0}{ 53.15} & \multicolumn{1}{c}{{ 46.32}} & { 59.63} & \multicolumn{1}{c}{\cellcolor[HTML]{C0C0C0}{ 33.78}} & { 50.53} & \multicolumn{1}{c}{\cellcolor[HTML]{C0C0C0}{ 45.33}} & \cellcolor[HTML]{C0C0C0}{ 61.45} & \multicolumn{1}{c}{-} & - \\ 
 & Refactor & \multicolumn{1}{c}{{ 34.45}} & { 53.35} & \multicolumn{1}{c}{{ 47.33}} & { 66.56} & \multicolumn{1}{c}{{ 48.83}} & { 58.82} & \multicolumn{1}{c}{{ 49.56}} & { 63.44} & \multicolumn{1}{c}{19.67} & 28.57 \\
    & Rename Operator & \multicolumn{1}{c}{{ 35.68}} & { 55.67} & \multicolumn{1}{c}{{ 49.13}} & { 65.67} & \multicolumn{1}{c}{{ 52.35}} & { 60.11} & \multicolumn{1}{c}{{ 52.11}} & { 66.65} & \multicolumn{1}{c}{ 23.56} & 34.67 \\

  & Dead Operator  & \multicolumn{1}{c}{{ 34.52}} & { 53.87} & \multicolumn{1}{c}{{ 48.56}} & { 64.35} & \multicolumn{1}{c}{{ 49.95}} & { 59.67} & \multicolumn{1}{c}{{ 50.98}} & { 63.23} & \multicolumn{1}{c}{ 22.54} &  31.18\\

  & Inside Operator  & \multicolumn{1}{c}{{ 37.23}} & { 56.34} & \multicolumn{1}{c}{{ 49.76}} & { 62.67} & \multicolumn{1}{c}{{ 51.14}} & { 59.96} & \multicolumn{1}{c}{{ 53.56}} & { 68.05} & \multicolumn{1}{c}{ 22.98} & 32.87 \\

 & Outside Operator  & \multicolumn{1}{c}{{ 38.54}} & { 55.47} & \multicolumn{1}{c}{{ 50.14}} & { 69.68} & \multicolumn{1}{c}{{ 51.68}} & { 56.87} & \multicolumn{1}{c}{{ 54.41}} & { 67.18} & \multicolumn{1}{c}{ 23.51} & 34.76 \\
 & SR & \multicolumn{1}{c}{{ 36.98}} & { 54.09} & \multicolumn{1}{c}{{ 48.52}} & { 68.32} & \multicolumn{1}{c}{{ 35.87}} & \cellcolor[HTML]{C0C0C0}{ 43.45} & \multicolumn{1}{c}{{ 48.33}} & { 63.28} & \multicolumn{1}{c}{22.74} & 43.59 \\
 & RI & \multicolumn{1}{c}{{ 39.46}} & { 57.87} & \multicolumn{1}{c}{{ 49.11}} & { 67.21} & \multicolumn{1}{c}{{ 55.33}} & { 59.67} & \multicolumn{1}{c}{{ 58.51}} & { 70.24} & \multicolumn{1}{c}{24.42} &  53.66\\ 
 & RS & \multicolumn{1}{c}{{ 33.13}} & { 53.39} & \multicolumn{1}{c}{{ 46.23}} & { 65.34} & \multicolumn{1}{c}{{ 46.56}} & { 51.41} & \multicolumn{1}{c}{{ 47.22}} & { 62.45} & \multicolumn{1}{c}{18.68} & 30.95 \\ 
 & RD & \multicolumn{1}{c}{\cellcolor[HTML]{C0C0C0}{ 32.39}} & { 53.34} & \multicolumn{1}{c}{\cellcolor[HTML]{C0C0C0}{ 42.16}} & \cellcolor[HTML]{C0C0C0}{ 56.86} & \multicolumn{1}{c}{{ 42.83}} & { 53.51} & \multicolumn{1}{c}{{ 57.45}} & { 62.57} & \multicolumn{1}{c}{\cellcolor[HTML]{C0C0C0}16.46} & \cellcolor[HTML]{C0C0C0}27.91 \\ 
\multirow{-13}{*}{\begin{tabular}[c]{@{}c@{}}CodeBERT\\ (5\%)\end{tabular}} & BT & \multicolumn{1}{c}{{ 40.87}} & { 59.76} & \multicolumn{1}{c}{{ 46.44}} & { 68.31} & \multicolumn{1}{c}{{ 50.01}} & { 52.42} & \multicolumn{1}{c}{{ 46.56}} & { 61.78} & \multicolumn{1}{c}{18.32} & 28.15 \\ \hline
 & No Aug (Baseline) & \multicolumn{1}{c}{{ 45.01}} & { 62.78} & \multicolumn{1}{c}{{ 50.33}} & { 69.52} & \multicolumn{1}{c}{{ 56.32}} & { 68.56} & \multicolumn{1}{c}{{ 56.33}} & { 75.21} & \multicolumn{1}{c}{22.45} & 48.41 \\ 
 & WordMixup & \multicolumn{1}{c}{{ 41.67}} & { 55.32} & \multicolumn{1}{c}{{ 48.45}} & { 66.83} & \multicolumn{1}{c}{{ 55.32}} & { 74.69} & \multicolumn{1}{c}{{ 55.23}} & { 74.31} & \multicolumn{1}{c}{-} & -  \\ 
 & SenMixup & \multicolumn{1}{c}{\cellcolor[HTML]{C0C0C0}{ 40.32}} & { 49.67} & \multicolumn{1}{c}{{ 47.86}} & { 65.21} & \multicolumn{1}{c}{\cellcolor[HTML]{C0C0C0}{ 53.55}} & { 67.44} & \multicolumn{1}{c}{\cellcolor[HTML]{C0C0C0}{ 51.32}} & \cellcolor[HTML]{C0C0C0}{ 72.33} & \multicolumn{1}{c}{-} & - \\
 & Refactor & \multicolumn{1}{c}{{ 41.25}} & \cellcolor[HTML]{C0C0C0}{ 45.53} & \multicolumn{1}{c}{{ 51.91}} & { 70.21} & \multicolumn{1}{c}{{ 62.56}} & { 66.43} & \multicolumn{1}{c}{{ 60.53}} & { 74.43} & \multicolumn{1}{c}{10.87} & 19.51 \\
   & Rename Operator & \multicolumn{1}{c}{{ 43.78}} & { 56.81} & \multicolumn{1}{c}{{ 53.62}} & { 72.12} & \multicolumn{1}{c}{{ 66.24}} & { 75.38} & \multicolumn{1}{c}{{ 61.34}} & { 76.92} & \multicolumn{1}{c}{16.56} & 25.12 \\
 & Dead Operator  & \multicolumn{1}{c}{{ 42.71}} & { 56.41} & \multicolumn{1}{c}{{ 50.15}} & { 71.11} & \multicolumn{1}{c}{{ 64.61}} & { 72.98} & \multicolumn{1}{c}{{ 62.78}} & { 75.32} & \multicolumn{1}{c}{ 15.13} & 22.08 \\

 & Inside Operator & \multicolumn{1}{c}{{ 45.54}} & { 58.09} & \multicolumn{1}{c}{{ 54.65}} & { 75.31} & \multicolumn{1}{c}{{ 65.51}} & { 74.31} & \multicolumn{1}{c}{{ 64.34}} & { 78.01} & \multicolumn{1}{c}{ 19.41} & 28.41 \\ 

 & Outside Operator  & \multicolumn{1}{c}{{ 47.31}} & { 58.26} & \multicolumn{1}{c}{{ 52.71}} & { 70.54} & \multicolumn{1}{c}{{ 66.67}} & { 75.45} & \multicolumn{1}{c}{{ 63.14}} & { 75.21} & \multicolumn{1}{c}{ 20.35} & 30.43\\
 & SR & \multicolumn{1}{c}{{ 46.45}} & { 68.72} & \multicolumn{1}{c}{{ 53.21}} & { 73.53} & \multicolumn{1}{c}{{ 62.87}} & { 75.44} & \multicolumn{1}{c}{{ 61.33}} & { 78.65} & \multicolumn{1}{c}{13.64} & 23.68 \\ 
 & RI & \multicolumn{1}{c}{{ 49.82}} & { 70.44} & \multicolumn{1}{c}{{ 55.32}} & { 75.32} & \multicolumn{1}{c}{{ 54.11}} & { 65.88} & \multicolumn{1}{c}{{ 52.17}} & { 74.45} & \multicolumn{1}{c}{24.76} & 38.09 \\ 
 & RS & \multicolumn{1}{c}{{ 45.88}} & { 61.45} & \multicolumn{1}{c}{{ 50.16}} & { 71.21} & \multicolumn{1}{c}{{ 55.32}} & { 66.21} & \multicolumn{1}{c}{{ 53.76}} & { 81.53} & \multicolumn{1}{c}{\cellcolor[HTML]{C0C0C0}10.63} & \cellcolor[HTML]{C0C0C0}19.05 \\ 
 & RD & \multicolumn{1}{c}{{ 44.78}} & { 61.75} & \multicolumn{1}{c}{\cellcolor[HTML]{C0C0C0}{ 45.72}} & \cellcolor[HTML]{C0C0C0}{ 64.33} & \multicolumn{1}{c}{{ 54.34}} & \cellcolor[HTML]{C0C0C0}{ 64.42} & \multicolumn{1}{c}{{ 62.42}} & { 74.53} & \multicolumn{1}{c}{22.15} & 43.91 \\ 
\multirow{-13}{*}{\begin{tabular}[c]{@{}c@{}}CodeBERT\\ (3\%)\end{tabular}} & BT & \multicolumn{1}{c}{{ 51.38}} & { 72.87} & \multicolumn{1}{c}{{ 51.35}} & { 73.26} & \multicolumn{1}{c}{{ 54.88}} & { 72.67} & \multicolumn{1}{c}{{ 51.44}} & { 73.45} & \multicolumn{1}{c}{23.87} & 35.62 \\ \hline
 & No Aug (Baseline) & \multicolumn{1}{c}{{ 80.23}} & { 82.34} & \multicolumn{1}{c}{{ 86.98}} & { 90.11} & \multicolumn{1}{c}{{ 70.44}} & { 83.65} & \multicolumn{1}{c}{{ 65.97}} & { 80.73} & \multicolumn{1}{c}{10.41} & 22.78 \\ 
 & WordMixup & \multicolumn{1}{c}{{ 76.72}} & { 79.32} & \multicolumn{1}{c}{{ 73.45}} & { 88.21} & \multicolumn{1}{c}{{ 69.11}} & { 78.31} & \multicolumn{1}{c}{{ 65.53}} & { 78.56} & \multicolumn{1}{c}{-} & - \\ 
 & SenMixup & \multicolumn{1}{c}{{ 75.26}} & { 77.32} & \multicolumn{1}{c}{\cellcolor[HTML]{C0C0C0}{ 69.77}} & \cellcolor[HTML]{C0C0C0}{ 85.44} & \multicolumn{1}{c}{{ 68.32}} & \cellcolor[HTML]{C0C0C0}{ 76.33} & \multicolumn{1}{c}{{ 64.86}} & \cellcolor[HTML]{C0C0C0}{ 76.45} & \multicolumn{1}{c}{-} & - \\ 
 & Refactor & \multicolumn{1}{c}{{ 82.59}} & { 86.63} & \multicolumn{1}{c}{{ 88.34}} & { 92.38} & \multicolumn{1}{c}{{ 72.23}} & { 79.42} & \multicolumn{1}{c}{{ 65.69}} & { 86.34} & \multicolumn{1}{c}{10.93} & 22.94 \\ 
    & Rename Operator & \multicolumn{1}{c}{{ 85.31}} & { 89.91} & \multicolumn{1}{c}{{ 90.01}} & { 94.57} & \multicolumn{1}{c}{{ 78.52}} & { 83.21} & \multicolumn{1}{c}{{ 68.38}} & { 78.42} & \multicolumn{1}{c}{ 12.89} & 34.66 \\
 & Dead Operator  & \multicolumn{1}{c}{{ 83.35}} & { 87.02} & \multicolumn{1}{c}{{ 89.34}} & { 93.36} & \multicolumn{1}{c}{{ 74.65}} & { 81.45} & \multicolumn{1}{c}{{ 67.11}} & { 77.35} & \multicolumn{1}{c}{ 11.38} & 25.34 \\

 & Inside Operator & \multicolumn{1}{c}{{ 84.67}} & { 88.14} & \multicolumn{1}{c}{{ 91.56}} & { 96.81} & \multicolumn{1}{c}{{ 76.31}} & { 85.52} & \multicolumn{1}{c}{{ 69.23}} & { 81.24} & \multicolumn{1}{c}{ 13.66} & 30.14 \\ 

 & Outside Operator  & \multicolumn{1}{c}{{ 84.21}} & { 88.76} & \multicolumn{1}{c}{{ 91.23}} & { 96.09} & \multicolumn{1}{c}{{ 77.51}} & { 88.08} & \multicolumn{1}{c}{{ 70.12}} & { 83.15} & \multicolumn{1}{c}{ 14.02} & 32.86 \\
 & SR & \multicolumn{1}{c}{{ 82.36}} & { 86.55} & \multicolumn{1}{c}{{ 91.44}} & { 93.43} & \multicolumn{1}{c}{{ 78.33}} & { 94.42} & \multicolumn{1}{c}{{ 70.94}} & { 88.45} & \multicolumn{1}{c}{\cellcolor[HTML]{C0C0C0} 9.81} & 38.42 \\ 
 & RI & \multicolumn{1}{c}{{ 84.52}} & { 87.11} & \multicolumn{1}{c}{{ 89.87}} & { 96.47} & \multicolumn{1}{c}{{ 67.25}} & { 91.31} & \multicolumn{1}{c}{{ 72.32}} & { 91.56} & \multicolumn{1}{c}{22.36} & 38.26 \\ 
 & RS & \multicolumn{1}{c}{{ 75.45}} & { 82.42} & \multicolumn{1}{c}{{ 82.56}} & { 87.43} & \multicolumn{1}{c}{{ 69.56}} & { 78.43} & \multicolumn{1}{c}{{ 66.34}} & { 79.45} & \multicolumn{1}{c}{10.14} & 36.67 \\ 
 & RD & \multicolumn{1}{c}{\cellcolor[HTML]{C0C0C0}{ 73.37}} & \cellcolor[HTML]{C0C0C0}{ 76.43} & \multicolumn{1}{c}{{ 78.32}} & { 88.87} & \multicolumn{1}{c}{{ 77.38}} & { 87.42} & \multicolumn{1}{c}{{ 65.67}} & { 80.45} & \multicolumn{1}{c}{10.33} & \cellcolor[HTML]{C0C0C0}21.85 \\ 
\multirow{-13}{*}{\begin{tabular}[c]{@{}c@{}}CodeBERT\\ (1\%)\end{tabular}} & BT & \multicolumn{1}{c}{{ 82.22}} & { 86.23} & \multicolumn{1}{c}{{ 81.34}} & { 89.22} & \multicolumn{1}{c}{\cellcolor[HTML]{C0C0C0}{ 65.32}} & { 77.42} & \multicolumn{1}{c}{\cellcolor[HTML]{C0C0C0}{ 64.56}} & { 77.33} & \multicolumn{1}{c}{21.47} & 35.89 \\ \hline
\end{tabular}}
\vspace{-3mm}
\end{table}

\begin{table}[!tb]
\caption{Effectiveness of data augmentation methods w.r.t. ASR $\downarrow$ (\%) on test data. \textbf{No Aug}: without data augmentation. The best results are highlighted in gray. The victim DNN model is GraphCodeBERT. Particularly, the training data volume drops to 10\%, 5\%, 3\%, and 1\% of the original, respectively. Tasks include \textbf{Problem Classification} (Java250, Python800), \textbf{Bug detection} (CodRep1, Refactory), and \textbf{Clone detection} (BigCloneBench)}
\label{tab:robust_GraphCodeBERT_reduce}
\resizebox{\columnwidth}{!}{
\begin{tabular}{clcccccccccc}
\hline
 & & \multicolumn{2}{c}{\textbf{Java250}} & \multicolumn{2}{c}{\textbf{Python800}} & \multicolumn{2}{c}{\textbf{Refactory}} & \multicolumn{2}{c}{\textbf{CodRep1}} & \multicolumn{2}{c}{\textbf{BigCloneBench}} \\
\multirow{-2}{*}{\textbf{Model}} & \multirow{-2}{*}{\textbf{DA method}} & \textbf{MHM} & \textbf{ALERT} & \textbf{MHM} & \textbf{ALERT} & \textbf{MHM} & \textbf{ALERT} & \textbf{MHM} & \textbf{ALERT} & \textbf{MHM} & \textbf{ALERT} \\ \hline
 & No Aug (Baseline) & \multicolumn{1}{c}{{ 23.52}} & { 35.89} & \multicolumn{1}{c}{{ 32.44}} & { 48.23} & \multicolumn{1}{c}{{ 25.67}} & { 43.64} & \multicolumn{1}{c}{{ 32.86}} & { 47.67} & \multicolumn{1}{c}{2.04} & 12.25 \\ 
 & WordMixup & \multicolumn{1}{c}{{ 22.35}} & { 35.97} & \multicolumn{1}{c}{{ 33.11}} & { 47.26} & \multicolumn{1}{c}{{ 28.45}} & { 42.03} & \multicolumn{1}{c}{{ 33.54}} & { 46.98} & \multicolumn{1}{c}{-} &  -\\ 
 & SenMixup & \multicolumn{1}{c}{\cellcolor[HTML]{C0C0C0}{ 21.45}} & \cellcolor[HTML]{C0C0C0}{ 33.54} & \multicolumn{1}{c}{{ 31.21}} & { 45.66} & \multicolumn{1}{c}{{ 24.22}} & { 41.32} & \multicolumn{1}{c}{\cellcolor[HTML]{C0C0C0}{ 27.34}} & { 43.76} & \multicolumn{1}{c}{-} & - \\ 
 & Refactor & \multicolumn{1}{c}{{ 21.89}} & { 34.42} & \multicolumn{1}{c}{{ 30.45}} & { 43.37} & \multicolumn{1}{c}{{ 28.42}} & { 45.21} & \multicolumn{1}{c}{{ 33.55}} & { 51.43} & \multicolumn{1}{c}{6.12} & 10.21 \\ 
   & Rename Operator & \multicolumn{1}{c}{{ 22.57}} & { 35.63} & \multicolumn{1}{c}{{ 32.16}} & { 47.17} & \multicolumn{1}{c}{{ 32.12}} & { 47.57} & \multicolumn{1}{c}{{ 36.98}} & { 55.41} & \multicolumn{1}{c}{ 8.13} & 14.56\\
 & Dead Operator  & \multicolumn{1}{c}{{ 22.41}} & { 33.86} & \multicolumn{1}{c}{{ 31.38}} & { 46.68} & \multicolumn{1}{c}{{ 31.78}} & { 46.19} & \multicolumn{1}{c}{{ 35.67}} & { 52.45} & \multicolumn{1}{c}{ 7.38} & 15.13 \\

 & Inside Operator & \multicolumn{1}{c}{{ 23.46}} & { 35.71} & \multicolumn{1}{c}{{ 33.57}} & { 48.91} & \multicolumn{1}{c}{{ 29.96}} & { 42.86} & \multicolumn{1}{c}{{ 38.55}} & { 57.73} & \multicolumn{1}{c}{ 7.54} & 16.81 \\ 

 & Outside Operator  & \multicolumn{1}{c}{{ 26.89}} & { 36.58} & \multicolumn{1}{c}{{ 35.49}} & { 50.14} & \multicolumn{1}{c}{{ 32.87}} & { 46.51} & \multicolumn{1}{c}{{ 34.68}} & { 57.19} & \multicolumn{1}{c}{ 8.57} & 15.35\\
 & SR & \multicolumn{1}{c}{{ 27.53}} & { 38.53} & \multicolumn{1}{c}{{ 34.21}} & { 49.54} & \multicolumn{1}{c}{{ 29.56}} & { 49.56} & \multicolumn{1}{c}{{ 35.87}} & { 53.98} & \multicolumn{1}{c}{\cellcolor[HTML]{C0C0C0}1.21} & 4.08 \\ 
 & RI & \multicolumn{1}{c}{{ 25.67}} & { 39.68} & \multicolumn{1}{c}{{ 36.33}} & { 51.67} & \multicolumn{1}{c}{{ 28.33}} & { 51.63} & \multicolumn{1}{c}{{ 39.56}} & { 56.56} & \multicolumn{1}{c}{4.08} & 12.24 \\ 
 & RS & \multicolumn{1}{c}{{ 23.35}} & { 35.56} & \multicolumn{1}{c}{{ 28.65}} & { 46.57} & \multicolumn{1}{c}{{ 24.52}} & { 48.87} & \multicolumn{1}{c}{{ 30.88}} & { 44.87} & \multicolumn{1}{c}{2.01} & 12.21 \\ 
 & RD & \multicolumn{1}{c}{{ 24.88}} & { 34.13} & \multicolumn{1}{c}{{ 31.35}} & { 42.64} & \multicolumn{1}{c}{{ 22.56}} & { 42.34} & \multicolumn{1}{c}{{ 31.65}} & \cellcolor[HTML]{C0C0C0}{ 41.49} & \multicolumn{1}{c}{6.12} & 12.18 \\ 
\multirow{-13}{*}{\begin{tabular}[c]{@{}c@{}}GraphCodeBERT\\ (10\%)\end{tabular}} & BT & \multicolumn{1}{c}{{ 23.24}} & { 36.85} & \multicolumn{1}{c}{\cellcolor[HTML]{C0C0C0}{ 27.32}} & \cellcolor[HTML]{C0C0C0}{ 41.45} & \multicolumn{1}{c}{\cellcolor[HTML]{C0C0C0}{ 21.36}} & \cellcolor[HTML]{C0C0C0}{ 40.88} & \multicolumn{1}{c}{{ 31.98}} & { 43.82} & \multicolumn{1}{c}{2.02} & \cellcolor[HTML]{C0C0C0}2.08 \\ \hline
 & No Aug (Baseline) & \multicolumn{1}{c}{{ 29.32}} & { 41.63} & \multicolumn{1}{c}{{ 37.33}} & { 51.35} & \multicolumn{1}{c}{{ 31.45}} & { 46.35} & \multicolumn{1}{c}{{ 40.56}} & { 54.45} & \multicolumn{1}{c}{11.91} & 21.42 \\
 & WordMixup & \multicolumn{1}{c}{{ 30.58}} & { 41.45} & \multicolumn{1}{c}{{ 36.64}} & { 50.42} & \multicolumn{1}{c}{{ 32.56}} & { 45.32} & \multicolumn{1}{c}{{ 37.87}} & { 52.45} & \multicolumn{1}{c}{-} & - \\
 & SenMixup & \multicolumn{1}{c}{\cellcolor[HTML]{C0C0C0}{ 26.53}} & \cellcolor[HTML]{C0C0C0}{ 37.41} & \multicolumn{1}{c}{{ 35.32}} & { 51.03} & \multicolumn{1}{c}{{ 27.24}} & { 43.67} & \multicolumn{1}{c}{\cellcolor[HTML]{C0C0C0}{ 33.56}} & \cellcolor[HTML]{C0C0C0}{ 48.56} & \multicolumn{1}{c}{-} & - \\ 
 & Refactor & \multicolumn{1}{c}{{ 27.88}} & { 38.63} & \multicolumn{1}{c}{{ 35.01}} & { 52.44} & \multicolumn{1}{c}{{ 29.55}} & { 45.22} & \multicolumn{1}{c}{{ 49.42}} & { 61.74} & \multicolumn{1}{c}{12.19} & 29.26 \\
    & Rename Operator & \multicolumn{1}{c}{{ 28.67}} & { 39.78} & \multicolumn{1}{c}{{ 38.58}} & { 55.78} & \multicolumn{1}{c}{{ 32.65}} & { 44.39} & \multicolumn{1}{c}{{ 53.46}} & { 68.21} & \multicolumn{1}{c}{ 14.46} & 26.75 \\
    
 & Dead Operator  & \multicolumn{1}{c}{{ 28.15}} & { 40.25} & \multicolumn{1}{c}{{ 37.24}} & { 54.87} & \multicolumn{1}{c}{{ 30.14}} & { 46.75} & \multicolumn{1}{c}{{ 51.45}} & { 66.12} & \multicolumn{1}{c}{ 13.27} & 23.05 \\

 & Inside Operator & \multicolumn{1}{c}{{ 31.67}} & { 45.81} & \multicolumn{1}{c}{{ 38.13}} & { 58.67} & \multicolumn{1}{c}{{ 32.56}} & { 47.16} & \multicolumn{1}{c}{{ 52.81}} & { 65.33} & \multicolumn{1}{c}{ 16.57} &  30.16\\ 

 & Outside Operator  & \multicolumn{1}{c}{{ 32.65}} & { 46.18} & \multicolumn{1}{c}{{ 39.11}} & { 57.91} & \multicolumn{1}{c}{{ 34.25}} & { 49.25} & \multicolumn{1}{c}{{ 53.09}} & { 63.29} & \multicolumn{1}{c}{ 18.24} & 28.18 \\
 
 & SR & \multicolumn{1}{c}{{ 33.02}} & { 45.22} & \multicolumn{1}{c}{{ 39.78}} & { 53.46} & \multicolumn{1}{c}{{ 33.22}} & { 48.87} & \multicolumn{1}{c}{{ 38.56}} & { 52.98} & \multicolumn{1}{c}{17.51} & 35.12 \\ 
 & RI & \multicolumn{1}{c}{{ 33.54}} & { 46.89} & \multicolumn{1}{c}{{ 36.24}} & { 50.21} & \multicolumn{1}{c}{{ 29.76}} & { 45.16} & \multicolumn{1}{c}{{ 46.31}} & { 58.57} & \multicolumn{1}{c}{20.13} & 37.51 \\ 
 & RS & \multicolumn{1}{c}{{ 31.26}} & { 43.32} & \multicolumn{1}{c}{{ 35.12}} & { 47.56} & \multicolumn{1}{c}{{ 30.75}} & { 45.53} & \multicolumn{1}{c}{{ 34.87}} & { 49.75} & \multicolumn{1}{c}{\cellcolor[HTML]{C0C0C0} 9.52} & 28.58 \\ 
 & RD & \multicolumn{1}{c}{{ 32.53}} & { 43.13} & \multicolumn{1}{c}{\cellcolor[HTML]{C0C0C0}{ 33.56}} & \cellcolor[HTML]{C0C0C0}{ 45.89} & \multicolumn{1}{c}{{ 32.45}} & { 49.61} & \multicolumn{1}{c}{{ 35.76}} & { 51.76} & \multicolumn{1}{c}{14.29} & \cellcolor[HTML]{C0C0C0}20.19 \\ 
\multirow{-13}{*}{\begin{tabular}[c]{@{}c@{}}GraphCodeBERT\\ (5\%)\end{tabular}} & BT & \multicolumn{1}{c}{{ 28.98}} & { 40.35} & \multicolumn{1}{c}{{ 36.42}} & { 48.54} & \multicolumn{1}{c}{\cellcolor[HTML]{C0C0C0}{ 25.13}} & \cellcolor[HTML]{C0C0C0}{ 40.33} & \multicolumn{1}{c}{{ 38.85}} & { 53.66} & \multicolumn{1}{c}{12.86} & 28.58 \\ \hline
 & No Aug (Baseline) & \multicolumn{1}{c}{{ 32.76}} & { 46.24} & \multicolumn{1}{c}{{ 42.21}} & { 60.12} & \multicolumn{1}{c}{{ 48.73}} & { 64.32} & \multicolumn{1}{c}{{ 51.67}} & { 64.87} & \multicolumn{1}{c}{22.51} & 53.49 \\ 
 & WordMixup & \multicolumn{1}{c}{{ 32.56}} & { 45.56} & \multicolumn{1}{c}{{ 40.45}} & { 61.53} & \multicolumn{1}{c}{{ 42.56}} & { 63.97} & \multicolumn{1}{c}{{ 49.98}} & { 62.56} & \multicolumn{1}{c}{-} & - \\ 
 & SenMixup & \multicolumn{1}{c}{{ 30.45}} & { 43.24} & \multicolumn{1}{c}{{ 39.45}} & { 58.98} & \multicolumn{1}{c}{\cellcolor[HTML]{C0C0C0}{ 39.33}} & { 63.23} & \multicolumn{1}{c}{\cellcolor[HTML]{C0C0C0}{ 48.98}} & \cellcolor[HTML]{C0C0C0}{ 60.45} & \multicolumn{1}{c}{-} & - \\ 
 & Refactor & \multicolumn{1}{c}{{ 34.67}} & { 47.53} & \multicolumn{1}{c}{{ 43.74}} & { 62.56} & \multicolumn{1}{c}{{ 53.74}} & { 62.93} & \multicolumn{1}{c}{{ 56.87}} & { 71.43} & \multicolumn{1}{c}{23.81} & 54.76 \\ 
    & Rename Operator & \multicolumn{1}{c}{{ 38.13}} & { 51.18} & \multicolumn{1}{c}{{ 46.17}} & { 67.05} & \multicolumn{1}{c}{{ 57.61}} & { 69.71} & \multicolumn{1}{c}{{ 60.14}} & { 78.35} & \multicolumn{1}{c}{ 25.16} & 55.81 \\
    
 & Dead Operator  & \multicolumn{1}{c}{{ 39.45}} & { 48.51} & \multicolumn{1}{c}{{ 44.82}} & { 65.51} & \multicolumn{1}{c}{{ 53.98}} & { 64.51} & \multicolumn{1}{c}{{ 58.26}} & { 75.31} & \multicolumn{1}{c}{ 24.24} & 55.21 \\

 & Inside Operator & \multicolumn{1}{c}{{ 40.13}} & { 53.18} & \multicolumn{1}{c}{{ 45.68}} & { 64.31} & \multicolumn{1}{c}{{ 55.61}} & { 67.71} & \multicolumn{1}{c}{{ 59.16}} & { 74.31} & \multicolumn{1}{c}{ 31.41} & 52.05 \\ 

 & Outside Operator  & \multicolumn{1}{c}{{ 39.15}} & { 50.86} & \multicolumn{1}{c}{{ 43.57}} & { 63.28} & \multicolumn{1}{c}{{ 56.35}} & { 61.25} & \multicolumn{1}{c}{{ 60.56}} & { 77.67} & \multicolumn{1}{c}{ 31.87} & 55.01 \\
 
 & SR & \multicolumn{1}{c}{{ 36.43}} & { 49.53} & \multicolumn{1}{c}{{ 46.32}} & { 64.34} & \multicolumn{1}{c}{{ 51.43}} & { 68.32} & \multicolumn{1}{c}{{ 50.32}} & { 68.32} & \multicolumn{1}{c}{29.27} & 51.21 \\ 
 & RI & \multicolumn{1}{c}{{ 38.42}} & { 50.56} & \multicolumn{1}{c}{{ 45.63}} & { 66.87} & \multicolumn{1}{c}{{ 55.21}} & { 69.33} & \multicolumn{1}{c}{{ 55.63}} & { 70.32} & \multicolumn{1}{c}{\cellcolor[HTML]{C0C0C0}17.07} & 48.78 \\ 
 & RS & \multicolumn{1}{c}{{ 31.34}} & { 43.65} & \multicolumn{1}{c}{{ 39.63}} & { 59.53} & \multicolumn{1}{c}{{ 45.32}} & { 68.43} & \multicolumn{1}{c}{{ 49.33}} & { 61.21} & \multicolumn{1}{c}{34.88} & 53.13 \\ 
 & RD & \multicolumn{1}{c}{\cellcolor[HTML]{C0C0C0}{ 28.65}} & \cellcolor[HTML]{C0C0C0}{ 42.78} & \multicolumn{1}{c}{\cellcolor[HTML]{C0C0C0}{ 37.45}} & \cellcolor[HTML]{C0C0C0}{ 57.47} & \multicolumn{1}{c}{{ 42.87}} & \cellcolor[HTML]{C0C0C0}{ 60.85} & \multicolumn{1}{c}{{ 51.43}} & { 63.21} & \multicolumn{1}{c}{27.91} & 52.81 \\ 
\multirow{-13}{*}{\begin{tabular}[c]{@{}c@{}}GraphCodeBERT\\ (3\%)\end{tabular}} & BT & \multicolumn{1}{c}{{ 31.64}} & { 44.83} & \multicolumn{1}{c}{{ 38.98}} & { 59.24} & \multicolumn{1}{c}{{ 51.33}} & { 62.11} & \multicolumn{1}{c}{{ 53.65}} & { 63.33} & \multicolumn{1}{c}{22.41} &\cellcolor[HTML]{C0C0C0}42.51 \\ \hline
 & No Aug (Baseline) & \multicolumn{1}{c}{{ 65.35}} & { 70.23} & \multicolumn{1}{c}{{ 68.11}} & { 80.66} & \multicolumn{1}{c}{{ 61.56}} & { 76.02} & \multicolumn{1}{c}{{ 64.33}} & { 75.08} & \multicolumn{1}{c}{30.15} & 47.51 \\ 
 & WordMixup & \multicolumn{1}{c}{{ 64.32}} & { 68.42} & \multicolumn{1}{c}{{ 69.32}} & { 81.23} & \multicolumn{1}{c}{{ 62.32}} & { 72.42} & \multicolumn{1}{c}{{ 62.21}} & { 72.42} & \multicolumn{1}{c}{-} & - \\ 
 & SenMixup & \multicolumn{1}{c}{\cellcolor[HTML]{C0C0C0}{ 62.11}} & \cellcolor[HTML]{C0C0C0}{ 65.56} & \multicolumn{1}{c}{\cellcolor[HTML]{C0C0C0}{ 66.34}} & { 78.54} & \multicolumn{1}{c}{{ 57.33}} & \cellcolor[HTML]{C0C0C0}{ 70.42} & \multicolumn{1}{c}{\cellcolor[HTML]{C0C0C0}{ 59.44}} & \cellcolor[HTML]{C0C0C0}{ 70.32} & \multicolumn{1}{c}{-} & - \\
 & Refactor & \multicolumn{1}{c}{{ 69.43}} & { 75.32} & \multicolumn{1}{c}{{ 72.01}} & { 82.56} & \multicolumn{1}{c}{{ 63.31}} & { 80.32} & \multicolumn{1}{c}{{ 68.56}} & { 74.98} & \multicolumn{1}{c}{26.32} & 52.63 \\ 
    & Rename Operator & \multicolumn{1}{c}{{ 71.11}} & { 76.12} & \multicolumn{1}{c}{{ 74.34}} & { 86.13} & \multicolumn{1}{c}{{ 66.37}} & { 85.53} & \multicolumn{1}{c}{{ 70.31}} & { 83.65} & \multicolumn{1}{c}{ 28.45} & 58.76 \\
    
 & Dead Operator  & \multicolumn{1}{c}{{ 70.15}} & { 73.52} & \multicolumn{1}{c}{{ 72.13}} & { 85.86} & \multicolumn{1}{c}{{ 65.14}} & { 78.16} & \multicolumn{1}{c}{{ 71.45}} & { 78.42} & \multicolumn{1}{c}{ 29.55} & 53.67 \\

 & Inside Operator & \multicolumn{1}{c}{{ 72.23}} & { 78.45} & \multicolumn{1}{c}{{ 74.48}} & { 86.15} & \multicolumn{1}{c}{{ 63.34}} & { 79.54} & \multicolumn{1}{c}{{ 75.57}} & { 84.67} & \multicolumn{1}{c}{ 31.38} & 56.72 \\ 

 & Outside Operator  & \multicolumn{1}{c}{{ 73.56}} & { 79.78} & \multicolumn{1}{c}{{ 76.13}} & { 84.45} & \multicolumn{1}{c}{{ 65.35}} & { 82.16} & \multicolumn{1}{c}{{ 74.19}} & { 82.78} & \multicolumn{1}{c}{ 32.66} & 60.08 \\
 & SR & \multicolumn{1}{c}{{ 73.21}} & { 76.98} & \multicolumn{1}{c}{{ 75.43}} & { 88.64} & \multicolumn{1}{c}{{ 77.45}} & { 83.97} & \multicolumn{1}{c}{{ 81.53}} & { 95.76} & \multicolumn{1}{c}{32.35} & 67.65 \\ 
 & RI & \multicolumn{1}{c}{{ 75.22}} & { 81.32} & \multicolumn{1}{c}{{ 73.11}} & { 86.75} & \multicolumn{1}{c}{{ 73.86}} & { 87.33} & \multicolumn{1}{c}{{ 74.54}} & { 91.35} & \multicolumn{1}{c}{58.82} & 73.52 \\ 
 & RS & \multicolumn{1}{c}{{ 65.74}} & { 69.45} & \multicolumn{1}{c}{{ 67.09}} & \cellcolor[HTML]{C0C0C0}{ 78.46} & \multicolumn{1}{c}{{ 58.41}} & { 73.76} & \multicolumn{1}{c}{{ 60.32}} & { 71.53} & \multicolumn{1}{c}{26.15} & \cellcolor[HTML]{C0C0C0}44.73 \\ 
 & RD & \multicolumn{1}{c}{{ 63.98}} & { 66.36} & \multicolumn{1}{c}{{ 69.35}} & { 83.34} & \multicolumn{1}{c}{\cellcolor[HTML]{C0C0C0}{ 53.35}} & { 72.68} & \multicolumn{1}{c}{{ 73.31}} & { 88.42} & \multicolumn{1}{c}{\cellcolor[HTML]{C0C0C0}25.64} & 48.72 \\ 
\multirow{-13}{*}{\begin{tabular}[c]{@{}c@{}}GraphCodeBERT\\ (1\%)\end{tabular}} & BT & \multicolumn{1}{c}{{ 64.44}} & { 73.21} & \multicolumn{1}{c}{{ 67.24}} & { 79.69} & \multicolumn{1}{c}{{ 70.56}} & { 79.56} & \multicolumn{1}{c}{{ 62.42}} & { 74.86} & \multicolumn{1}{c}{45.95} & 81.08 \\ \hline
\end{tabular}}
\vspace{-3mm}
\end{table}

\smallskip
\noindent
\textbf{Accuracy analysis.} Table~\ref{tab:ACC_CodeBERT_reduce}, Table ~\ref{tab:ACC_GraphCodeBERT_reduce}, and the upper part of Table~\ref{tab:GCJ_reduce} present the results of clean accuracy. For problem classification tasks (Java250 and Python800), we can see that only in one case, CodeBERT (1\%), data augmentation can improve the accuracy significantly (by up to 12.92\% accuracy improvement). A similar phenomenon also happens in the clone detection task, we can see that only in GraphCodeBERT (5\%), one data augmentation method (\emph{BT}) has a huge positive impact on the performance of the model. In contrast, for bug detection and author attribution tasks, there is a method (\emph{SenMixup}) that can always significantly improve the accuracy of models with a margin from 0.31\% to 12.92\%. In conclusion, \emph{SenMixup} is still the recommended method that has the best results in 20 (out of 36) cases. Interestingly, two noising-based methods, \emph{SR} and \emph{RI}, that perform well when using the entire training data fail to produce accurate models after reducing the size of the training data. For example, in CodeBERT (1\%), SR- and RI-produced models only have 37.06\% and 27.03\% accuracy, while the baseline method (\emph{No Aug}) has 49.62\%. 

\begin{figure}[h]
\centering
\subfigure[Java250-CodeBERT]{
\begin{minipage}[t]{0.45\linewidth}
\centering
\includegraphics[width=1.0\linewidth]{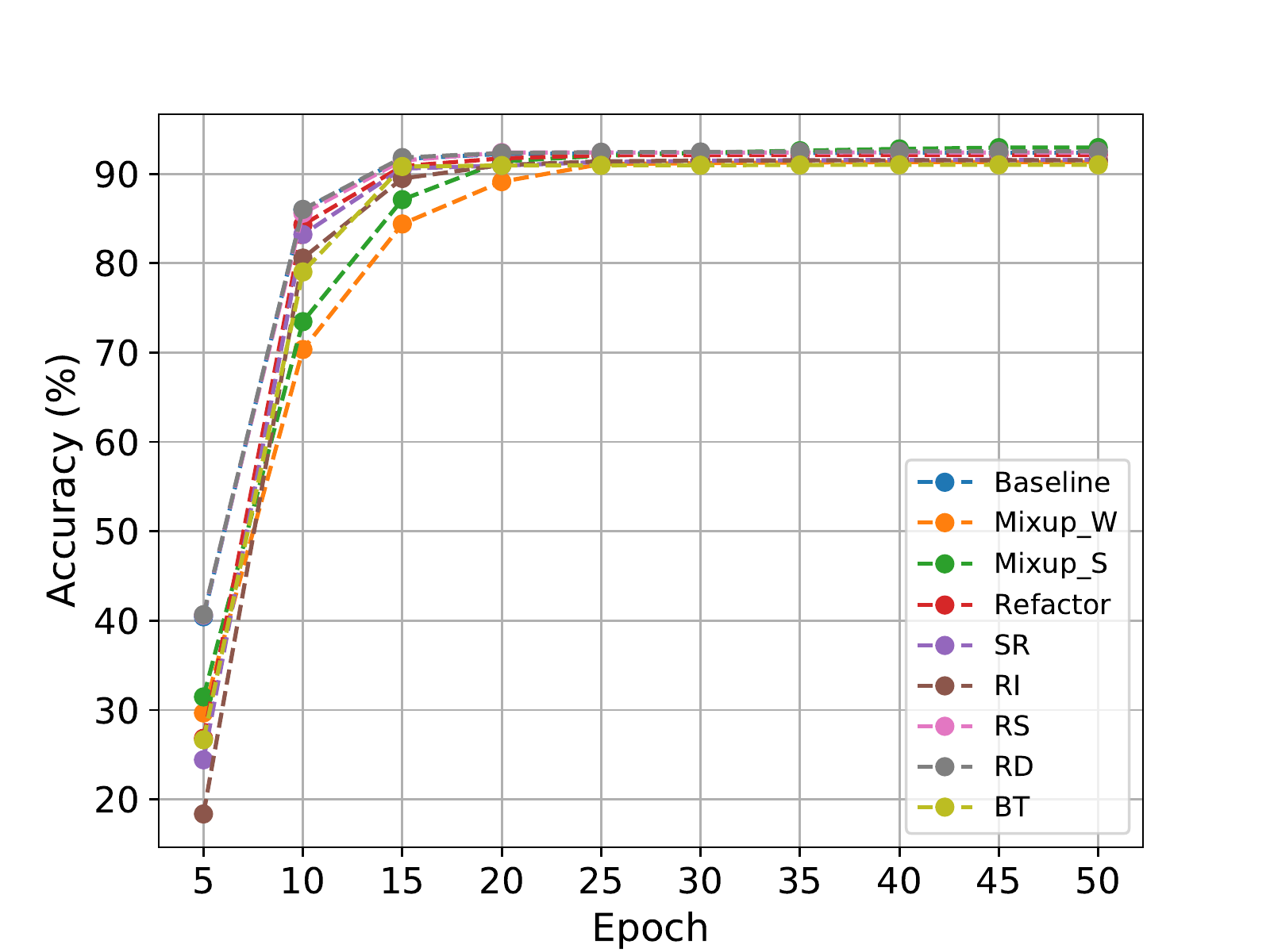}
\end{minipage}
}%
\centering
\subfigure[Java250-GraphCodeBERT ]{
\begin{minipage}[t]{0.45\linewidth}
\centering
\includegraphics[width=1.0\linewidth]{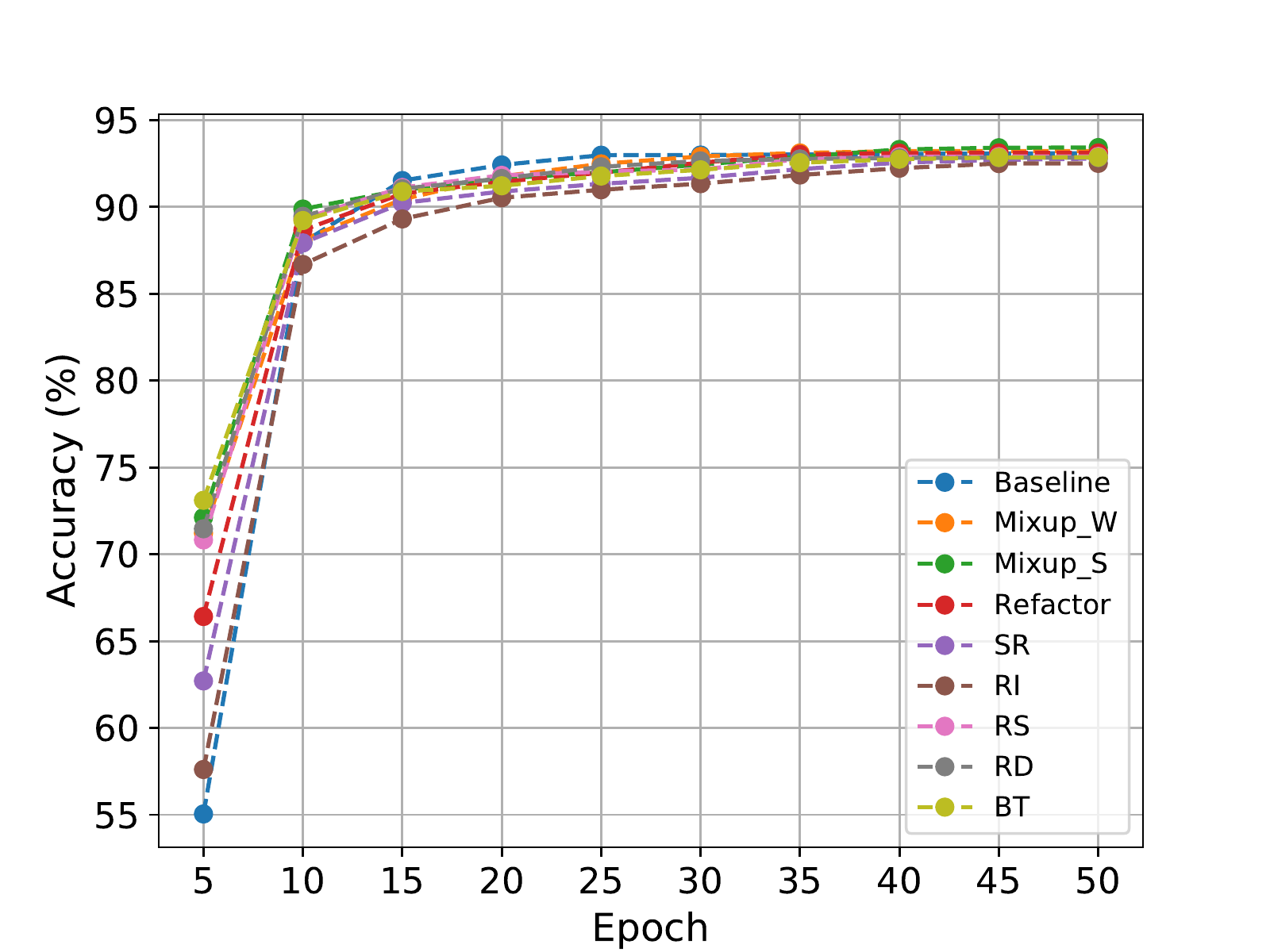}
\end{minipage}
}%

\subfigure[Python800-CodeBERT]{
\begin{minipage}[t]{0.45\linewidth}
\centering
\includegraphics[width=1.0\linewidth]{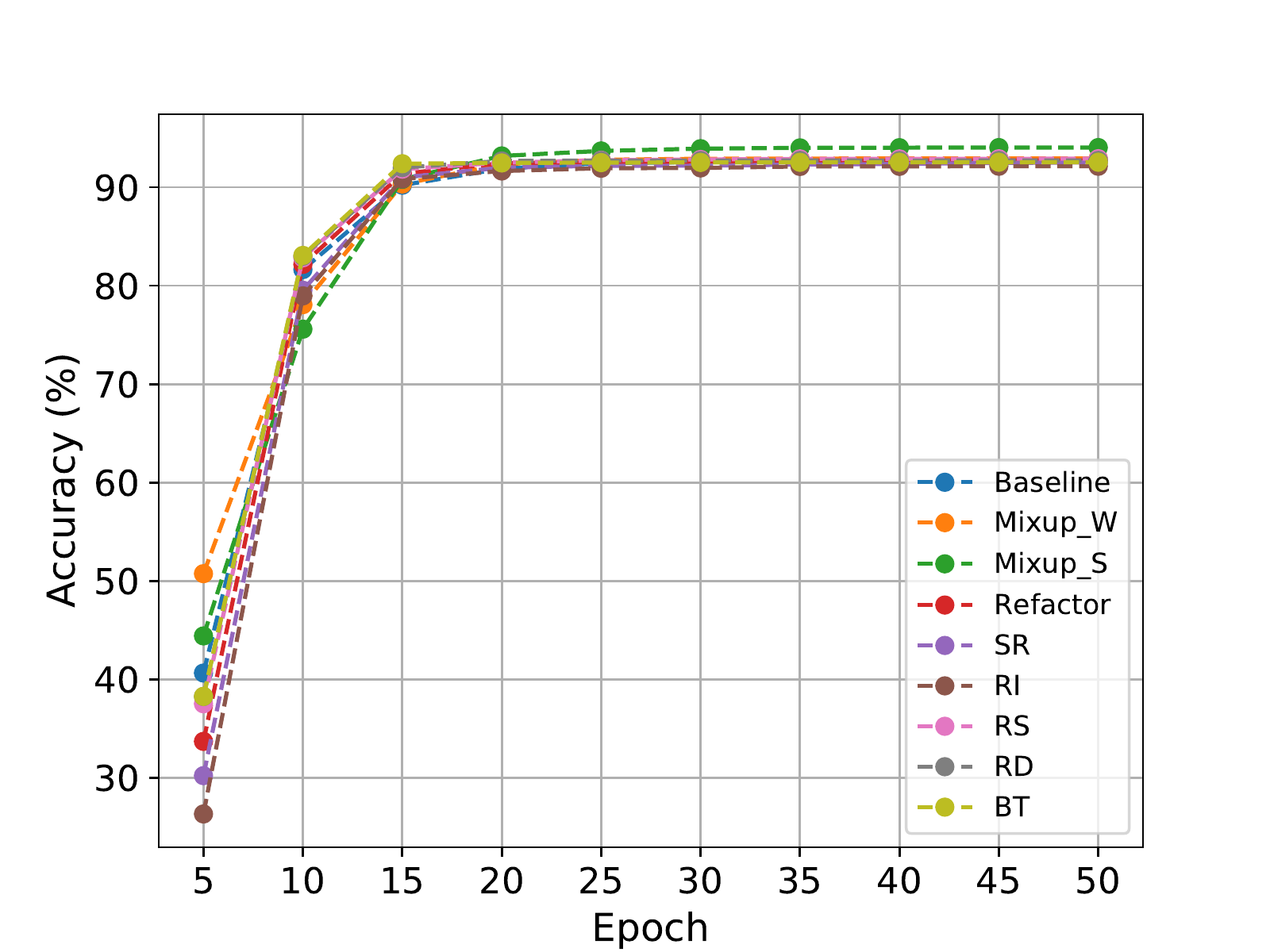}
\end{minipage}
}%
\subfigure[Python800-GraphCodeBERT ]{
\begin{minipage}[t]{0.45\linewidth}
\centering
\includegraphics[width=1.0\linewidth]{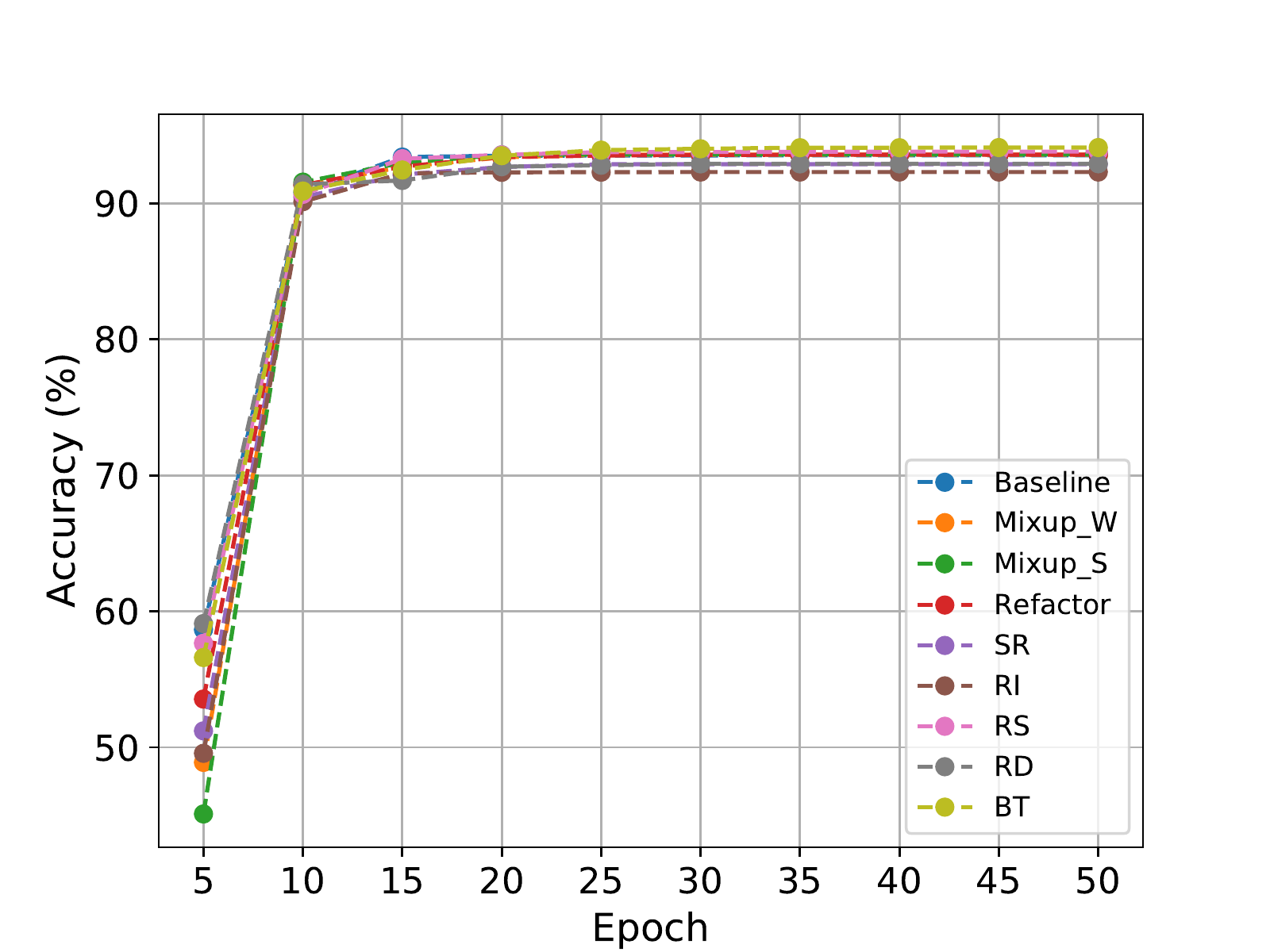}
\end{minipage}
}%

\subfigure[CodRep1-CodeBERT]{
\begin{minipage}[t]{0.45\linewidth}
\centering
\includegraphics[width=1.0\linewidth]{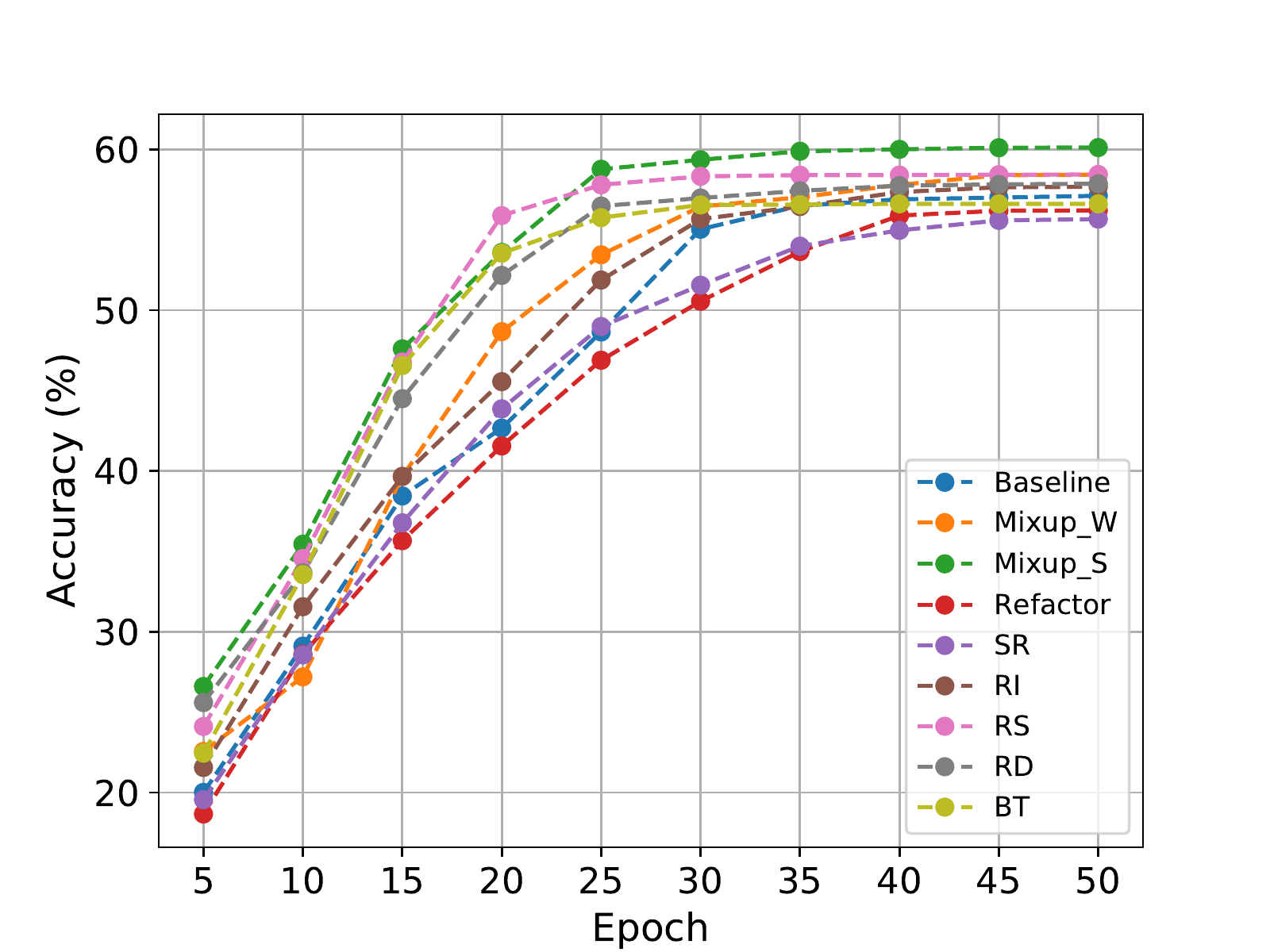}
\end{minipage}
}%
\subfigure[CodRep1-GraphCodeBERT ]{
\begin{minipage}[t]{0.45\linewidth}
\centering
\includegraphics[width=1.0\linewidth]{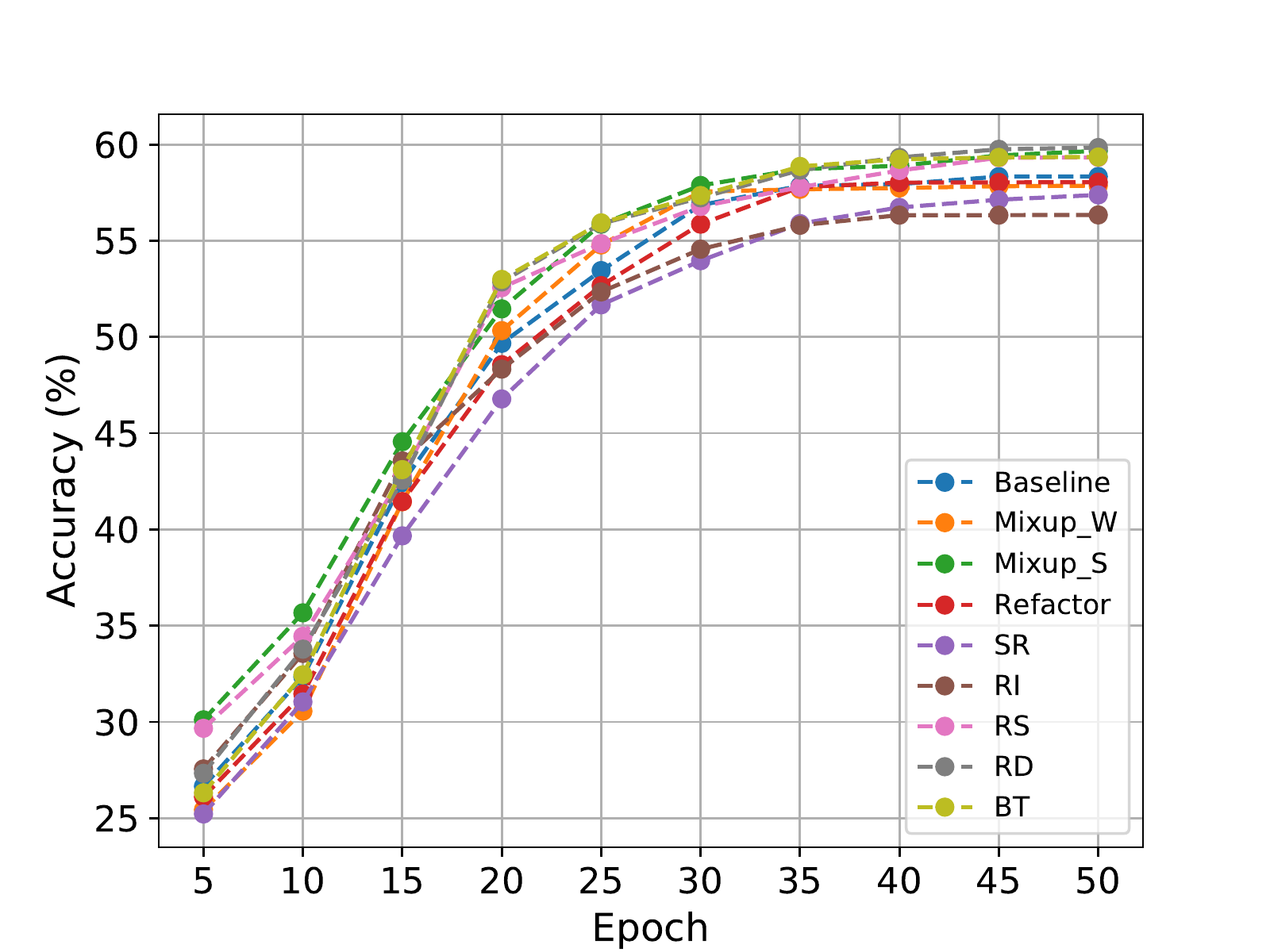}
\end{minipage}
}%
\caption{Training log of CodeBERT (10\%) and GraphCodeBERT (10\%) in Java250, Python800, and CodRep1.}
\label{fig:model_reduce}
\end{figure}

\begin{figure}[h]
\subfigure[Refactory-CodeBERT]{
\begin{minipage}[t]{0.45\linewidth}
\centering
\includegraphics[width=1.0\linewidth]{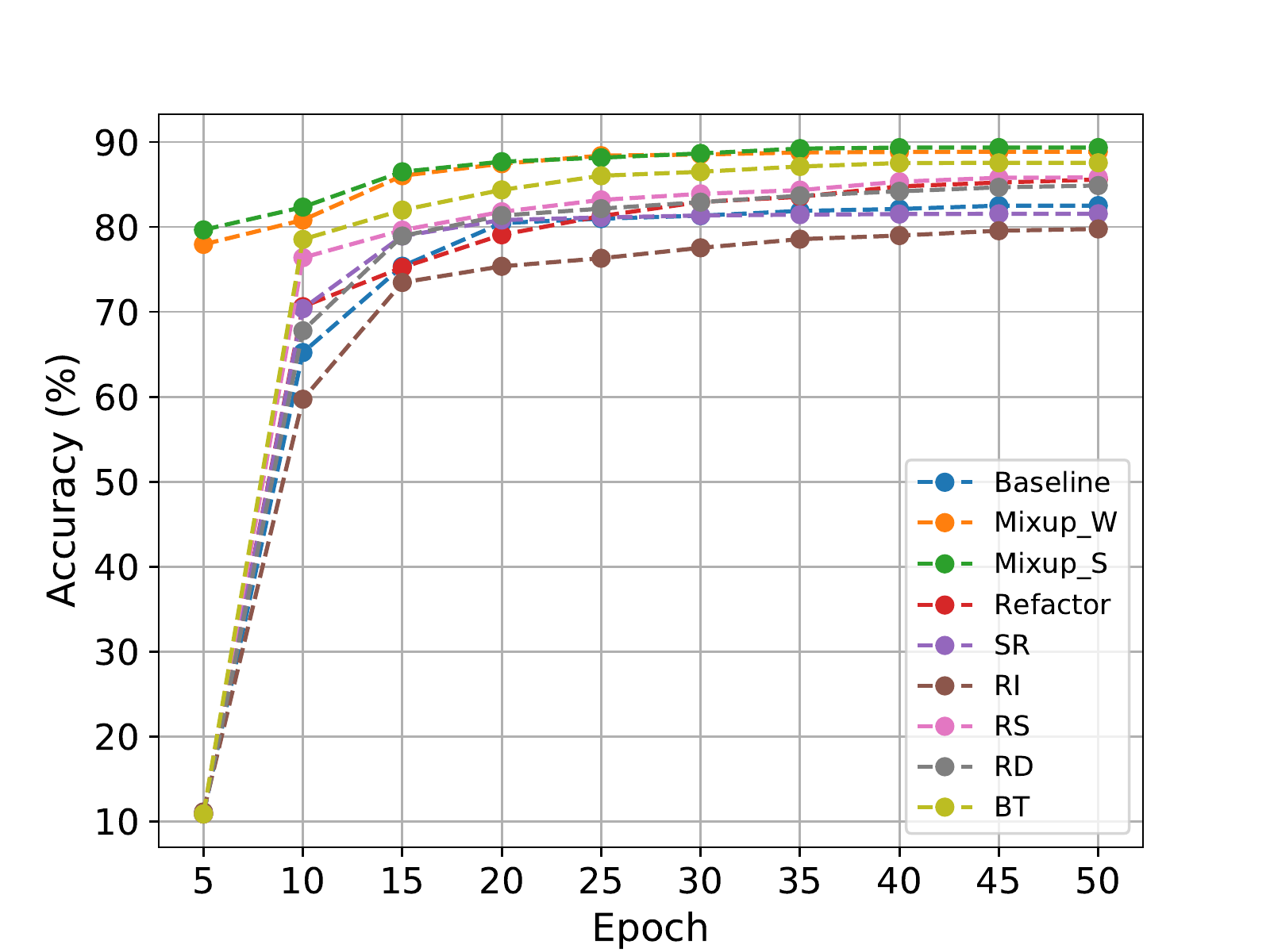}
\end{minipage}
}%
\subfigure[Refactory-GraphCodeBERT ]{
\begin{minipage}[t]{0.45\linewidth}
\centering
\includegraphics[width=1.0\linewidth]{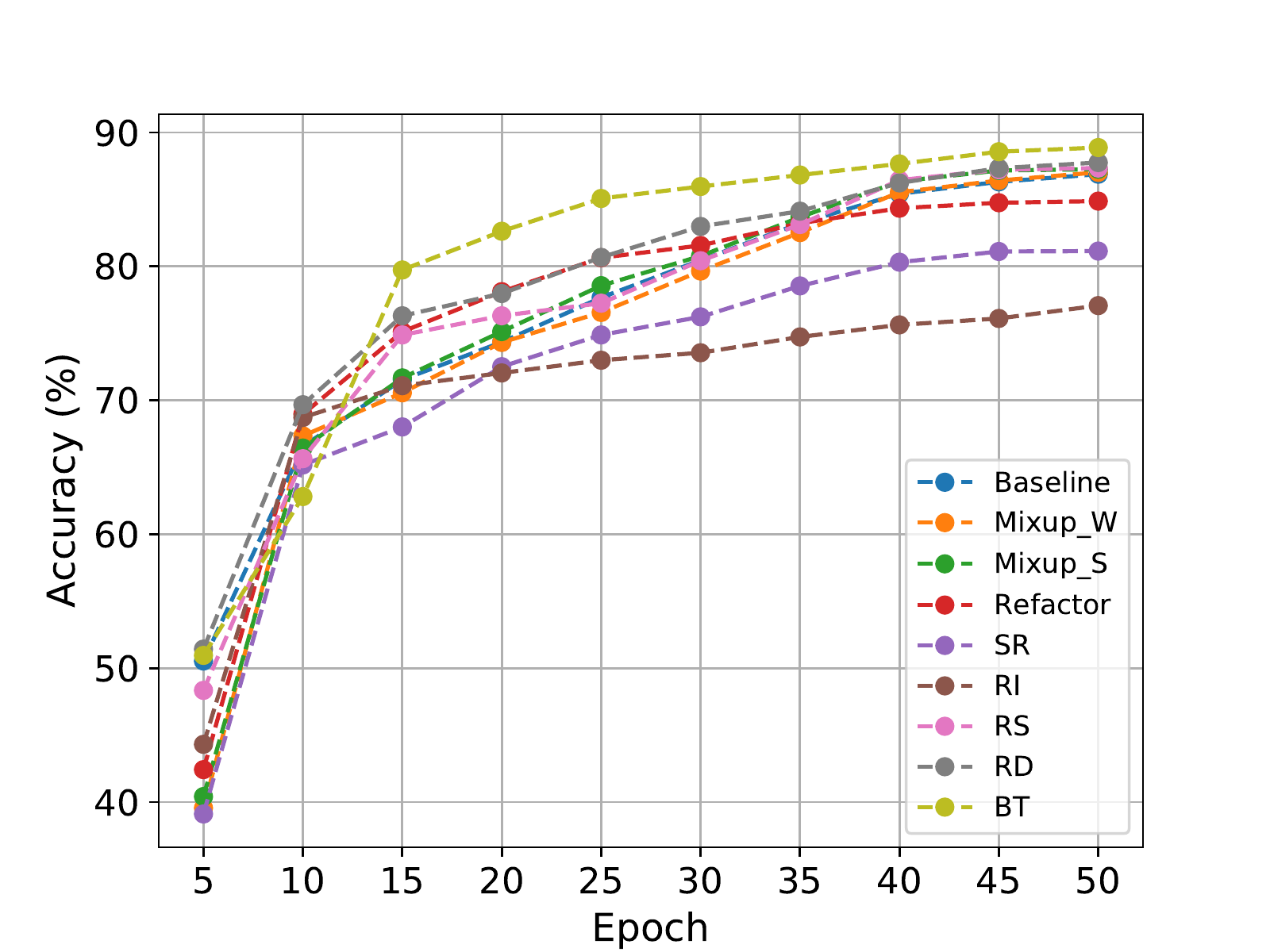}
\end{minipage}
}%

\subfigure[GCJ-CodeBERT]{
\begin{minipage}[t]{0.45\linewidth}
\centering
\includegraphics[width=1.0\linewidth]{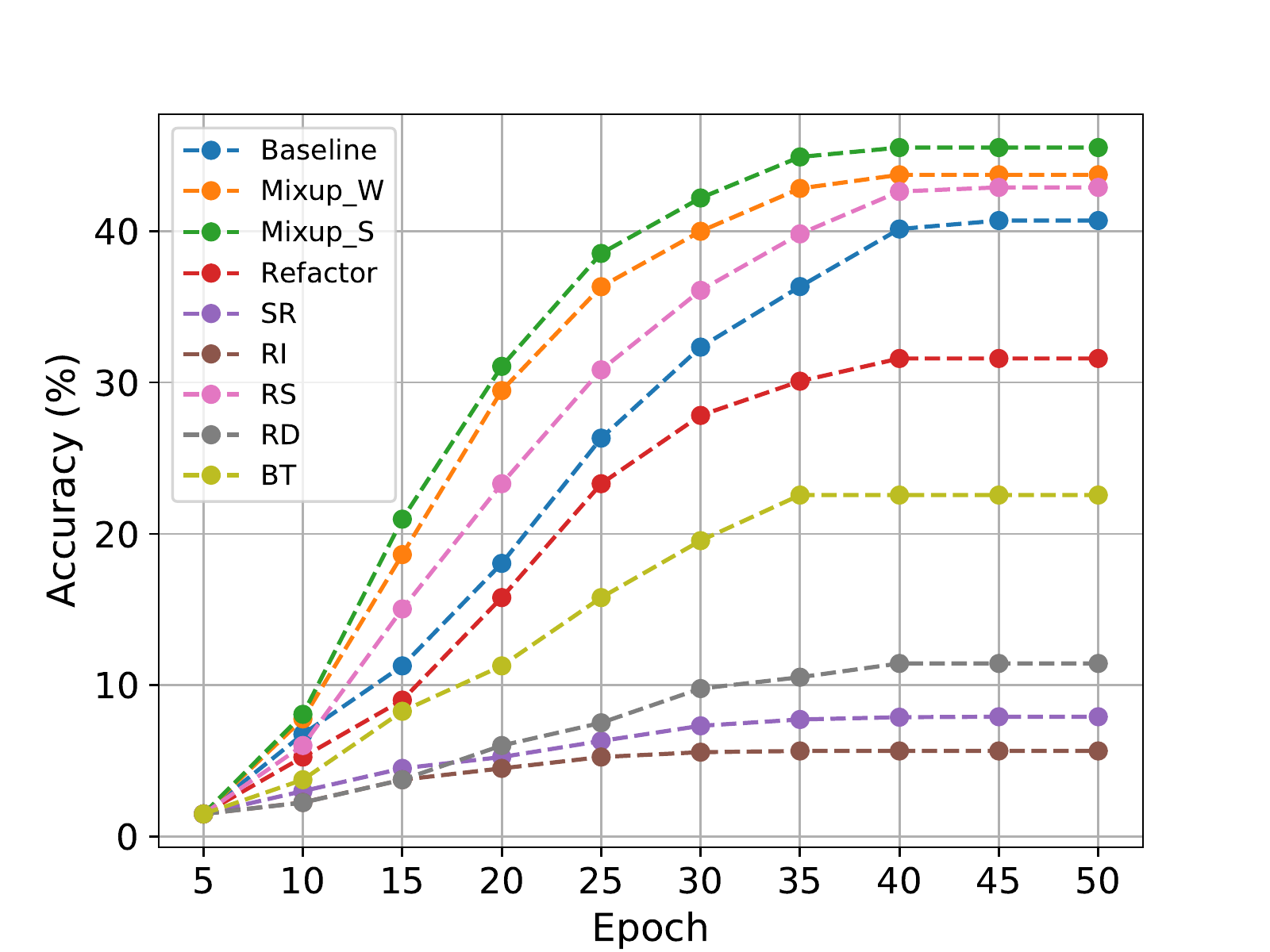}
\end{minipage}
}%
\subfigure[GCJ-GraphCodeBERT ]{
\begin{minipage}[t]{0.45\linewidth}
\centering
\includegraphics[width=1.0\linewidth]{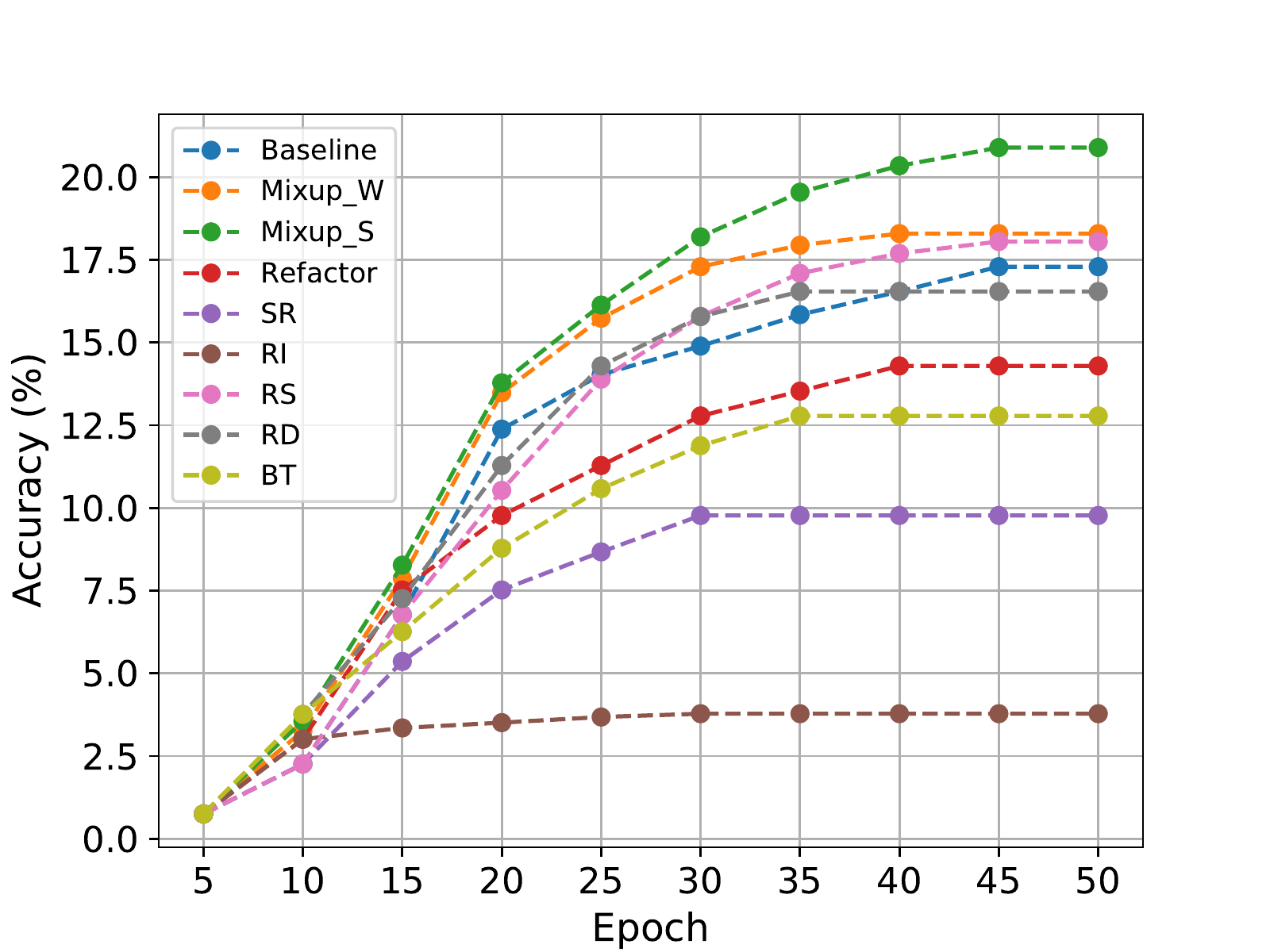}
\end{minipage}
}%

\subfigure[BigCloneBench-CodeBERT ]{
\begin{minipage}[t]{0.45\linewidth}
\centering
\includegraphics[width=1.0\linewidth]{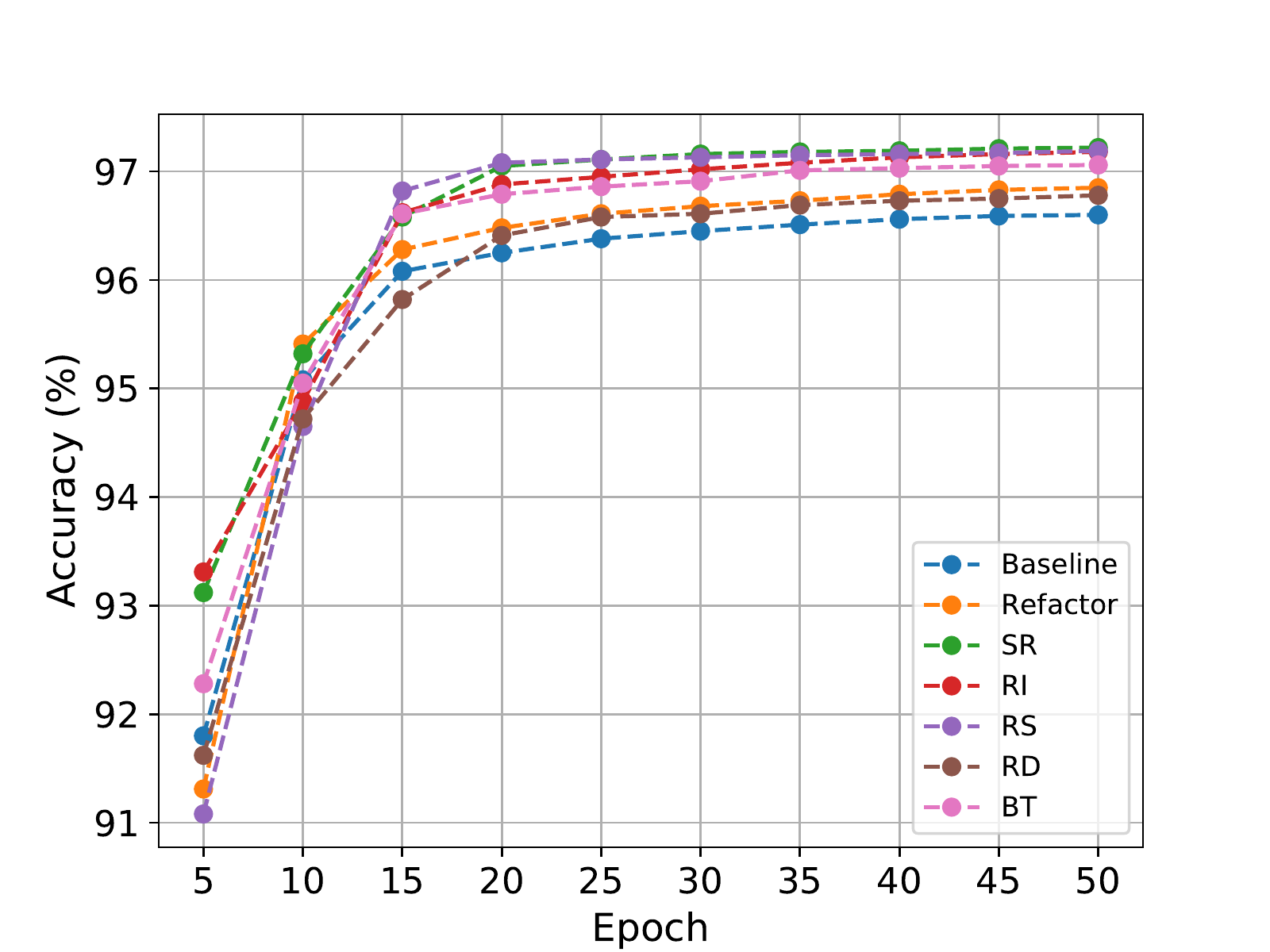}
\end{minipage}
}%
\subfigure[BigCloneBench-GraphCodeBERT ]{
\begin{minipage}[t]{0.45\linewidth}
\centering
\includegraphics[width=1.0\linewidth]{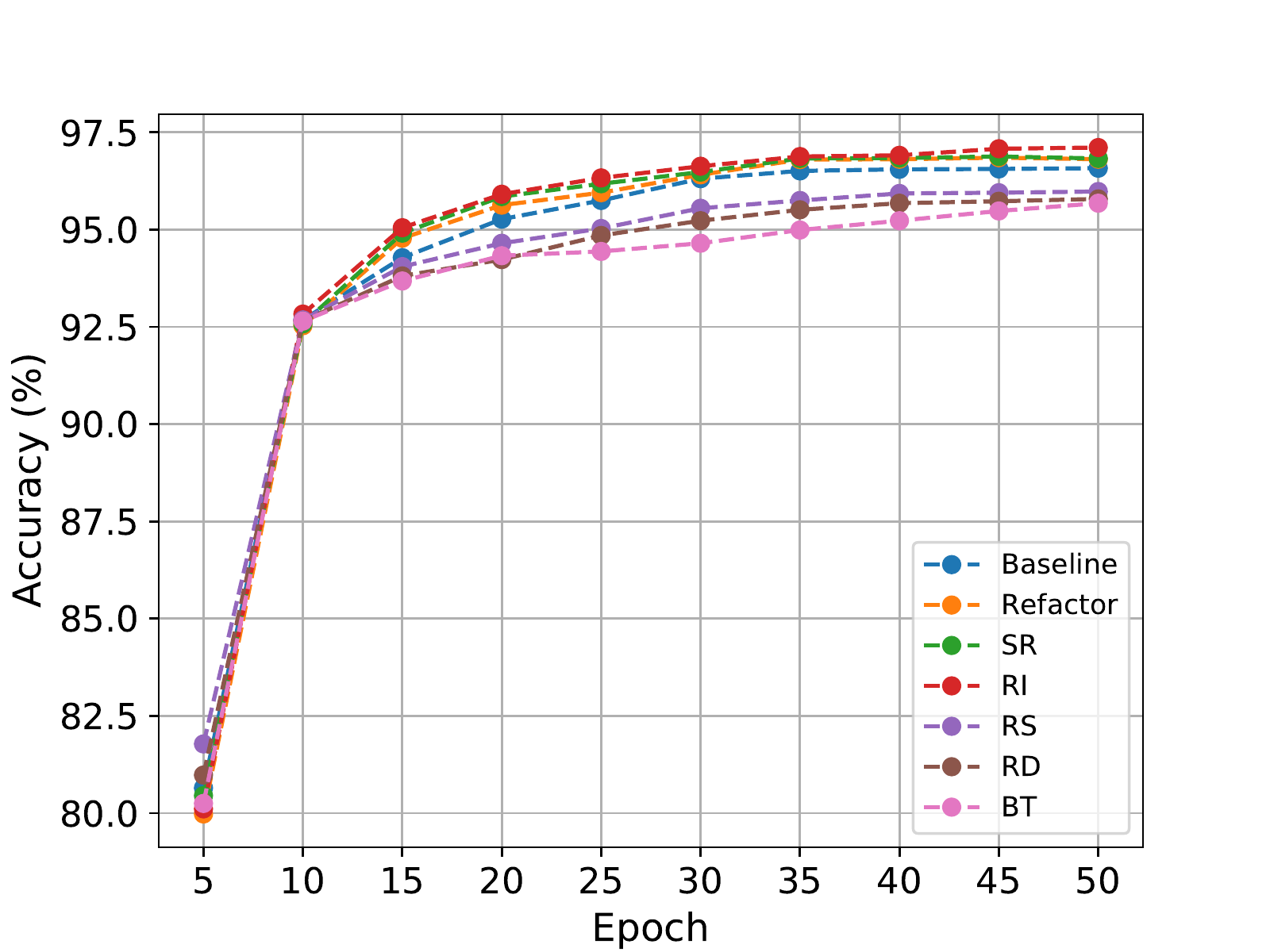}
\end{minipage}
}%
\caption{Training log of CodeBERT (10\%) and GraphCodeBERT (10\%) in Refactory, GCJ, and BigCloneBench.}
\label{fig:model_reduce_1}
\end{figure}

\smallskip
\noindent
\textbf{Convergence speed analysis.} Then, we check the convergence speed of models using different data augmentation methods. Similar to Section~\ref{sec:RQ1}, we also observe \emph{Refactor} in this visualization experiment. Fig.~\ref{fig:model_reduce}  and Fig.~\ref{fig:model_reduce_1} represent the training logs of CodeBERT (10\%) and GraphCodeBERT (10\%) models. From the results, we can see that 1) when data augmentation methods are used, the convergence speed of models is easily affected by a drop in the data scale. For instance, compared to the results of model training using the entire dataset, after 10 epochs, GCJ-CodeBERT has more significant improvement in terms of accuracy with data augmentation methods, i.e., \emph{RS}, \emph{SenMixup}, and \emph{WordMixup}. Surprisingly, for Refactory-CodeBERT, the model training using \emph{SenMixup} and \emph{WordMixup} brings almost 70.00\% significant improvement in terms of accuracy at $5^{th}$ epoch compared to other data augmentation methods as well as \emph{No Aug}. Besides, we can find all data augmentation methods effectively improve the performance of accuracy in BigCloneBench-CodeBERT compared to \emph{No Aug} after 20 epochs. This finding reflects that the smaller the training data, the more effective the data augmentation. This can be beneficial for us to select more effective data augmentation methods to reduce the time (computation budget) cost of models when the training data is in a very limited situation. 2) Compared to other data augmentation methods, models training using linear interpolation methods including \emph{SenMixup} and \emph{WordMixup} require more epochs to reach convergence, e.g., for GCJ-GraphCodeBERT, most data augmentation methods bring limited accuracy improvement after 30 epochs (except \emph{SenMixup}). This phenomenon is similar to the findings in recent studies~\cite{zhang2017mixup,yun2019cutmix}, which reveals that data augmentation methods that strongly increase the complexity of training data require more epochs to converge.

\smallskip
\noindent
\textbf{Robustness analysis.} The lower part of Table~\ref{tab:GCJ_reduce}, Table~\ref{tab:robust_CodeBERT_reduce}, and Table~\ref{tab:robust_GraphCodeBERT_reduce} depict the results of the attack success rate of MHM and ALERT attack on two pre-trained PL models. For the GCJ, we can find that although \emph{SenMixup} is the best data augmentation method that improves the accuracy of pre-trained models, it brings limited improvements in terms of robustness (e.g. 0.55\% ASR reduction in GraphCodeBERT (50\%) under ALERT attack). In contrast, the data augmentation method that slightly alters the syntactic structure, especially \emph{RI}, can effectively improve the performance of robustness. For instance, \emph{RI} reduces ASR by up to 38.47\% in CodeBERT (50\%) under MHM attack as well as 28.63\% in CodeBERT (50\%) under ALERT attack compared to \emph{No Aug}. For program classification tasks (Java250 and Python800), we observe that as the size of data decreases from 10\% to 1\%, the magnitude of robustness improvement brought by data augmentation methods becomes larger, e.g., a maximum reduction in ASR from 2.19\% (CodeBERT (10\%) ) to 17.21\% (CodeBERT (1\%)) under MHM attack. For the clone detection task, interestingly, most data augmentation methods that slightly change the code structure (except \emph{RS} and \emph{RD}) do not work on the robustness improvement at smaller data scales (CodeBERT (1\%) and GraphCodeBERT (1\%)). The bug detection task Refactory also has a similar phenomenon. In conclusion, we recommend \emph{SenMixup} as the best data augmentation method used for improving the robustness of pre-trained PL models since it has the best results in 36 (out of 72) cases. Besides, the noising-based data augmentation method \emph{BT}, which fails to produce the best robust models when using the entire training dataset, brings the greatest ASR reduction (i.e., 10.17\% in GraphCodeBERT (10\%) under ALERT attack) in the experiment using GraphCodeBERT with four different data scales.

Furthermore, we observe the same phenomenon noted in Sections~\ref{sec:RQ1} and \ref{sec:RQ2}. As the training data decreases, text-oriented data augmentation methods, especially \emph{SenMixup}, \emph{RS}, and \emph{RD}, consistently outperform the code refactoring methods in improving the accuracy and robustness of pre-trained PL models. We also find that specific code refactoring methods, such as \emph{Rename Operator} and \emph{Dead Operator}, exhibit worse robustness performance compared to the random selection method \emph{Refactor}, which aligns with findings Dong~\emph{et al.}~\cite{dong2023mixcode}.

\smallskip
\noindent
\textbf{Statistical analysis.} In conclusion, in terms of accuracy, we find that: 1) no data augmentation method consistently contributes to accuracy improvement, as shown in Table~\ref{table:statistical_testing_acc_10_codebert_graphcodebert}, Table~\ref{table:statistical_testing_acc_05_codebert_graphcodebert}, and Table~\ref{table:statistical_testing_acc_03_codebert_graphcodebert} in Appendix~\ref{sec:appendix}; and 2) however, as the dataset size decreases, linear interpolation methods such as \emph{SenMixup} demonstrate significant performance advantages compared to other data augmentation methods, particularly as illustrated in Table~\ref{table:statistical_testing_acc_01_codebert_graphcodebert} in Appendix~\ref{sec:appendix}. \emph{SenMixup} is still the recommended method that has relatively better results than others even though its advantage is not significant (i.e.,\emph{p}-values $>$ 0.05) according to our statistical analysis. 


Moving to robustness, we hold the same finding with Section~\ref{sec:RQ2} that the method of random combining different refactoring methods outperforms individual code refactoring methods, regardless of dataset size, as shown in Table~\ref{table:statistical_testing_robust_10_codebert_graphcodebert}, Table~\ref{table:statistical_testing_robust_5_codebert_graphcodebert}, Table~\ref{table:statistical_testing_robust_3_codebert_graphcodebert}, and Table~\ref{table:statistical_testing_robust_1_codebert_graphcodebert} in Appendix~\ref{sec:appendix}. Our statistical analysis shows that, in most cases, the text-oriented method \emph{RI} significantly (with \emph{p}-value $<$ 0.05) outperforms other data augmentation methods. This finding suggests that \emph{RI} is well-suited for producing robust pre-trained PL models when training data is limited. Additionally, \emph{SenMixup}  ranks second only to \emph{RI} in enhancing robustness, especially as the dataset becomes more constrained.

\begin{tcolorbox}
\textbf{Answer to RQ3}: 
Statistical analysis shows the selection of data augmentation methods becomes especially critical when training data is limited. Specifically, \emph{SenMixup} demonstrates superior performance by up to 12.92\% in terms of accuracy compared to other data augmentation methods as dataset size decreases. Moving to robustness, the text-oriented data augmentation method \emph{RI} significantly outperforms others with 38.47\% higher and is particularly well-suited for producing robust pre-trained PL models in low-data scenarios.
\end{tcolorbox}

\subsection{RQ4: Are text-oriented data augmentation methods effective for low-resource programming languages?}
\label{sec:RQ4}
Data augmentation can also be applied to another important scenario, specifically when dealing with low-resource languages. Therefore, it is necessary to explore whether text-oriented data augmentation methods can effectively improve the accuracy and robustness of pre-trained PL models in such contexts.

\smallskip
\noindent
\textbf{Accuracy analysis.} We first check the test accuracy of text-oriented data augmentation methods for low-resource programming languages. The left part of Table~\ref{tab:RQ4} presents the results of clean test accuracy. For Ruby300, we observe that the text-oriented data augmentation method \emph{SenMixup} improves accuracy by up to 0.14\% in CodeBERT and 0.22\% in GraphCodeBERT compared to the \emph{No Aug} baseline. Additionally, when comparing code refactoring methods to text transformation methods, \emph{Rename operator} achieves the best accuracy among the code refactoring methods in both CodeBERT and GraphCodeBERT. However, \emph{RD} from the text transformation methods family outperforms it by up to 0.45\%. For Go250, the text-oriented data augmentation method \emph{RD} demonstrates the best accuracy improvement for both CodeBERT and GraphCodeBERT. Specifically, \emph{RS} enhances accuracy by up to 0.83\% in CodeBERT and 1.12\% in GraphCodeBERT compared to model training without data augmentation.

\begin{table}[h]
\caption{Effectiveness of data augmentation methods w.r.t. test accuracy $\uparrow$ (average $\pm$ standard deviation, \%) on original test data and ASR $\downarrow$ (\%) on test data. \textbf{No Aug}: without data augmentation. The best results are highlighted in gray. The tested DNN models are CodeBERT and GraphCodeBERT. Tasks include \textbf{Problem Classification} (Ruby300, Go250)}
\label{tab:RQ4}
\resizebox{\columnwidth}{!}{
\begin{tabular}{clcc|cccc}
\cline{1-8}
&  & \multicolumn{2}{c|}{\textbf{Test Accuracy}} & \multicolumn{4}{c}{\textbf{Robustness}} \\
\textbf{Model} & \textbf{DA method} & \textbf{Ruby300} & \textbf{Go250} & \textbf{MHM(Ruby)} & \textbf{ALERT(Ruby)} & \textbf{MHM(Go)} & \textbf{ALERT(Go)}\\ \cline{1-8} 
 & No Aug (Baseline) & 95.37 ± 0.13 & 85.84 ± 0.34 & 41.32 & 56.56 & 49.12 & 60.42\\ 
 &WordMixup & 95.31 ± 0.24 & 85.85 ± 0.27 & 41.01 & 56.83 & 50.01 & 61.76\\
 &SenMixup & \cellcolor[HTML]{C0C0C0}95.51 ± 0.35 & 86.32 ± 0.39 & \cellcolor[HTML]{C0C0C0}37.34 & 53.69 & 47.39 & 58.01 \\
 & Refactor & 94.97 ± 0.31 & 84.52 ± 0.21 & 37.85 & 53.81 & 48.51 & 59.65\\
 & Rename Operator & 95.32 ± 0.16 & 85.72 ± 0.29 & 40.45 & 55.78 & 48.87 & 60.23\\ 

 & Dead Operator  & 94.55 ± 0.23 & 84.08 ± 0.26 & 38.67 & 53.12 & 48.98 & 59.89\\ 

 & Inside Operator  & 95.27 ± 0.26 & 85.85 ± 0.16 & 40.95 & 56.19 & 49.02 & 59.97\\ 

 & Outside Operator  & 95.08 ± 0.19 & 85.16 ± 0.32 & 40.51 & 56.08 & 48.86 & 60.12\\ 
 & SR & 95.33 ± 0.17 & 85.78 ± 0.12 & 40.98 & 56.66 & 51.22 & 59.69\\
 & RI & 95.35 ± 0.25 & 85.86 ± 0.18 & 41.56 & 53.87 & 49.43 & 59.01\\
 & RS & 95.49 ± 0.11 & \cellcolor[HTML]{C0C0C0}86.67 ± 0.11 & 39.87 & \cellcolor[HTML]{C0C0C0}52.76 & 47.05 & 56.97\\
 & RD & 95.44 ± 0.12 & 86.46 ± 0.15 & 37.44 & 52.89 &  \cellcolor[HTML]{C0C0C0}46.98 & \cellcolor[HTML]{C0C0C0}56.79\\ 
\multirow{-13}{*}{CodeBERT} & BT & 94.81 ± 0.15 & 86.55 ± 0.19 & 39.96 & 55.58 & 50.63 & 61.12\\ \cline{1-8} 

 & No Aug (Baseline) & 95.68 ± 0.09 & 87.23 ± 0.21  & 28.43 & 43.57 & 35.86 & 55.62\\
&WordMixup & 95.66 ± 0.23 & 87.41 ± 0.36 & 27.56 & 42.67 & 35.93 & 55.87\\
&SenMixup & 95.88 ± 0.17 & 88.16 ± 0.24 & 28.31 & 41.31 & 34.21 & 54.65\\ 
& Refactor  & 95.14 ± 0.21 & 85.57 ± 0.13 & 27.31 & 42.25 & 34.14 &54.31 \\
 & Rename Operator & 95.45 ± 0.08 & 86.57 ± 0.14 & 27.63 & 43.36 & 34.89 &54.63 \\ 

 & Dead Operator  & 94.97 ± 0.14 & 85.83 ± 0.11 & 27.43 & 42.89 & 34.51 & 54.46\\ 

 & Inside Operator  & 95.16 ± 0.25 & 86.51 ± 0.12 & 28.01 & 43.41 & 35.66 & 55.41\\ 

 & Outside Operator  & 95.11 ± 0.19 & 86.01 ± 0.09 & 28.12 & 43.51 & 35.81 & 54.81\\ 
& SR  & 95.53 ± 0.16 & 87.01 ± 0.21 & 27.68 & 45.06 & 36.03 &56.37 \\  
& RI& 95.56 ± 0.12 & 87.12 ± 0.14 & 28.04 & 45.31 & 35.26 & 55.82\\
& RS & 95.82 ± 0.24 & \cellcolor[HTML]{C0C0C0}88.35 ± 0.22 & \cellcolor[HTML]{C0C0C0}27.11 & 43.35 & \cellcolor[HTML]{C0C0C0}34.05 & \cellcolor[HTML]{C0C0C0}54.14\\
& RD & \cellcolor[HTML]{C0C0C0}95.90 ± 0.21 & 88.12 ± 0.15 & 27.22 & \cellcolor[HTML]{C0C0C0}41.09 & 34.68  & 55.31\\
\multirow{-13}{*}{GraphCodeBERT} & BT & 95.12 ± 0.15 & 88.16 ± 0.07 & 28.24 & 44.87  &  34.92 & 54.88\\ \cline{1-8}
\end{tabular}}
\vspace{-3mm}
\end{table}

\smallskip
\noindent
\textbf{Robustness analysis.} The right part of Table~\ref{tab:RQ4} shows the results of robustness. Overall, text-oriented data augmentation methods, particularly \emph{RD} and \emph{RS}, exhibit the most significant reduction in ASR. Specifically, \emph{RS} reduces ASR by up to 3.80\% under ALERT attack compared to \emph{No Aug} in Ruby300. \emph{RD} reduces ASR by up to 3.63\% under the ALERT attack and by 2.14\% under the MHM attack compared to \emph{No Aug}. Additionally, we make an interesting observation. Although single code refactoring methods such as \emph{Rename Operator} and \emph{Inside Operator} show better accuracy performance than \emph{Refactor}, which randomly selects one code refactoring method for each code data, they exhibit worse robustness, i.e., less ASR reduction.

\smallskip
\noindent
\textbf{Statistical analysis.} 
No data augmentation method consistently improves the accuracy of pre-trained PL models, as shown in Table~\ref{table:statistical_testing_acc_robust_RQ4_codebert_graphcodebert} in Appendix~\ref{sec:appendix}. The code refactoring method Dead Operator demonstrates the worst performance compared to others. Moving to robustness,  we recommend \emph{RS}, \emph{RD}, and \emph{SenMixup} as optimal data augmentation methods for improving the robustness of PL models in low-resource programming languages, as they demonstrate significant advantages (i.e.,  \emph{P}-value $<$ 0.05) compared to most other data augmentation methods. These results indicate that text-oriented data augmentation methods show the potential for enhancing source code learning, especially in low-resource programming languages. Additionally, exploring variations of text-oriented program transformations could provide valuable insights in this field.


\begin{tcolorbox}
\textbf{Answer to RQ4}: 
Text-oriented data augmentation methods provide limited benefits in enhancing the accuracy of pre-trained PL models for tasks involving low-resource programming language data. For instance, \emph{SenMixup} can only outperform \emph{No Aug} by up to 0.22\% in terms of accuracy for Ruby.  However,  statistical results show \emph{RS}, \emph{RD}, and \emph{SenMixup} demonstrate significant advantages in robustness improvement compared to most other data augmentation methods.
\end{tcolorbox}

\section{Discussion}
\label{sec:section7}

\subsection{Is the incorporation of data augmentation techniques essential for improving the effectiveness of ML models in source code learning?}
\label{sec:section7_1}
First, the most important question is whether it is necessary to use data augmentation when preparing code models. From our empirical study, the answer is yes. In the case of using a suitable data augmentation method, e.g., \emph{SenMixup}, the trained models have higher accuracy (by up to 12.92\%) and robustness (by up to 21.28\%) than the models without using data augmentation. 
However, the results also indicate that while data augmentation can improve the performance of pre-trained PL models, the improvement is not as significant as in cases without pre-trained embeddings. This may be because pre-training itself effectively serves as a form of data augmentation, enhancing the overall model training process. Consequently, additional data augmentation techniques provide less benefit than they do in non-pre-trained cases. An in-depth analysis of this phenomenon presents an interesting direction for future research.


\subsection{What factors contribute to the improved accuracy and robustness observed in code models when utilizing Mixup-based data augmentation methods?}
\label{sec:section7_2}

\begin{figure}[htbp]
\centering
\subfigure[No Aug]{
\begin{minipage}[t]{0.45\linewidth}
\centering
\includegraphics[width=1.0\linewidth]{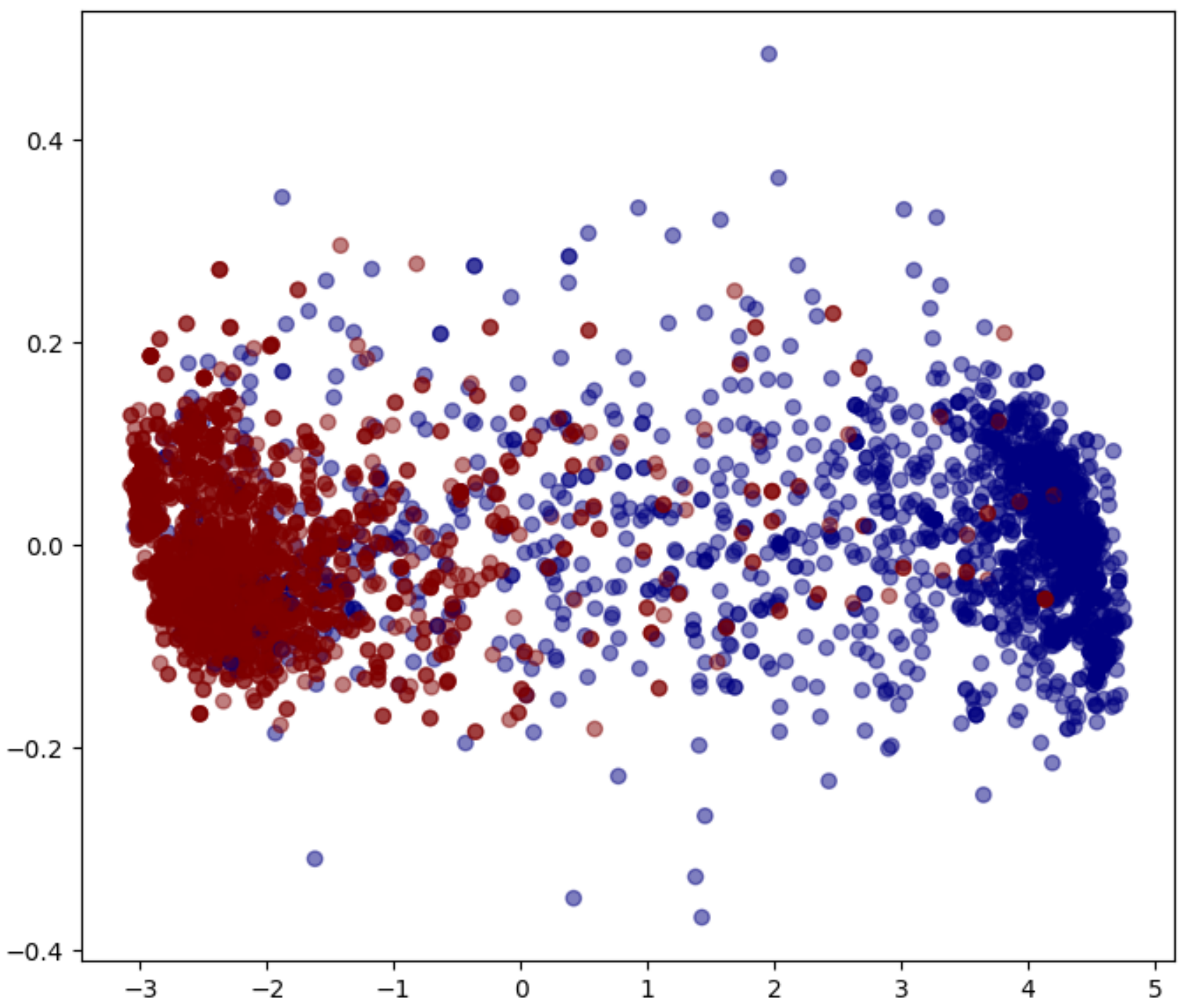}
\end{minipage}%
}%
\subfigure[Refactor]{
\begin{minipage}[t]{0.45\linewidth}
\centering
\includegraphics[width=1.0\linewidth]{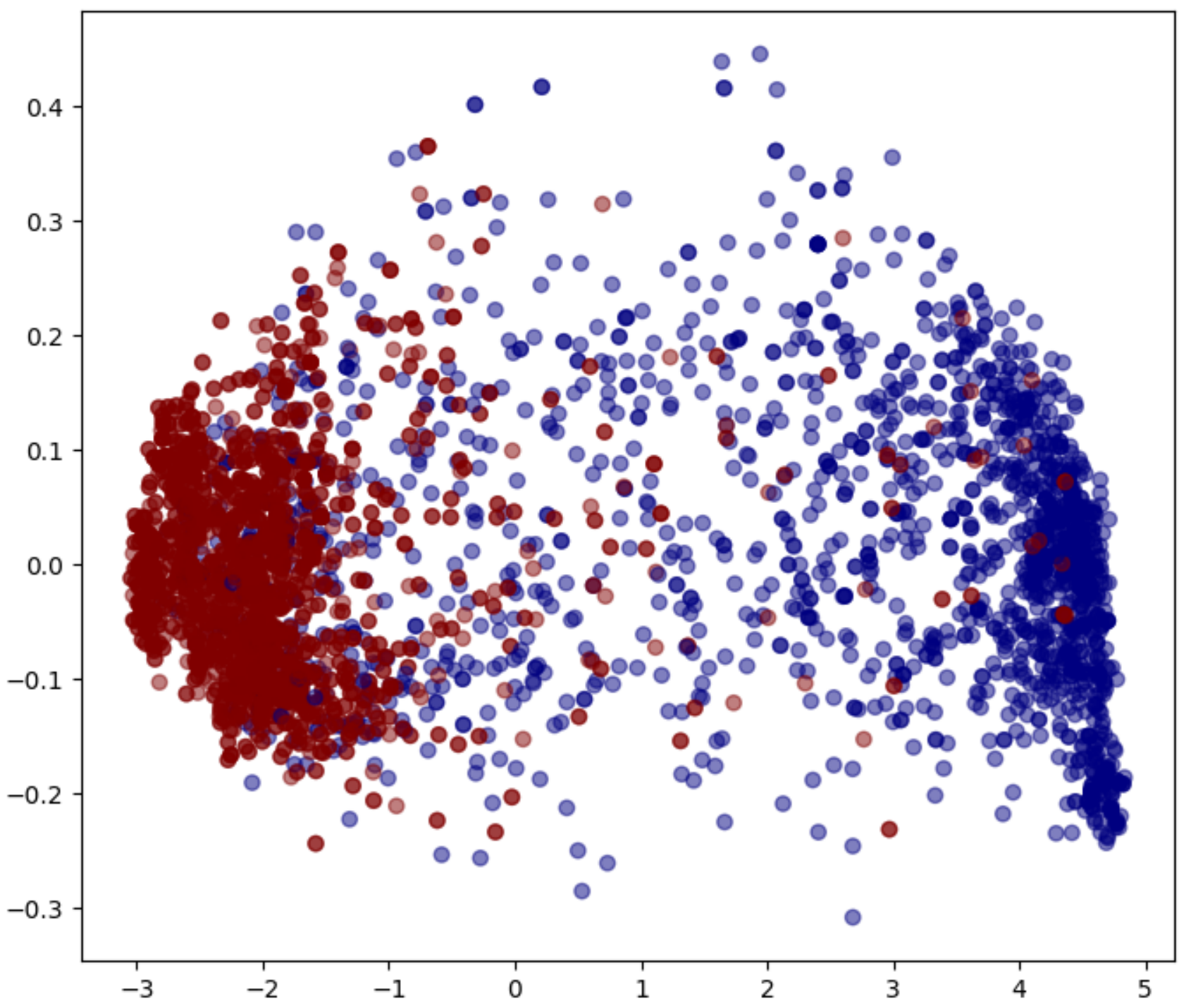}
\end{minipage}%
}%

\centering
\subfigure[SR]{
\begin{minipage}[t]{0.45\linewidth}
\centering
\includegraphics[width=1.0\linewidth]{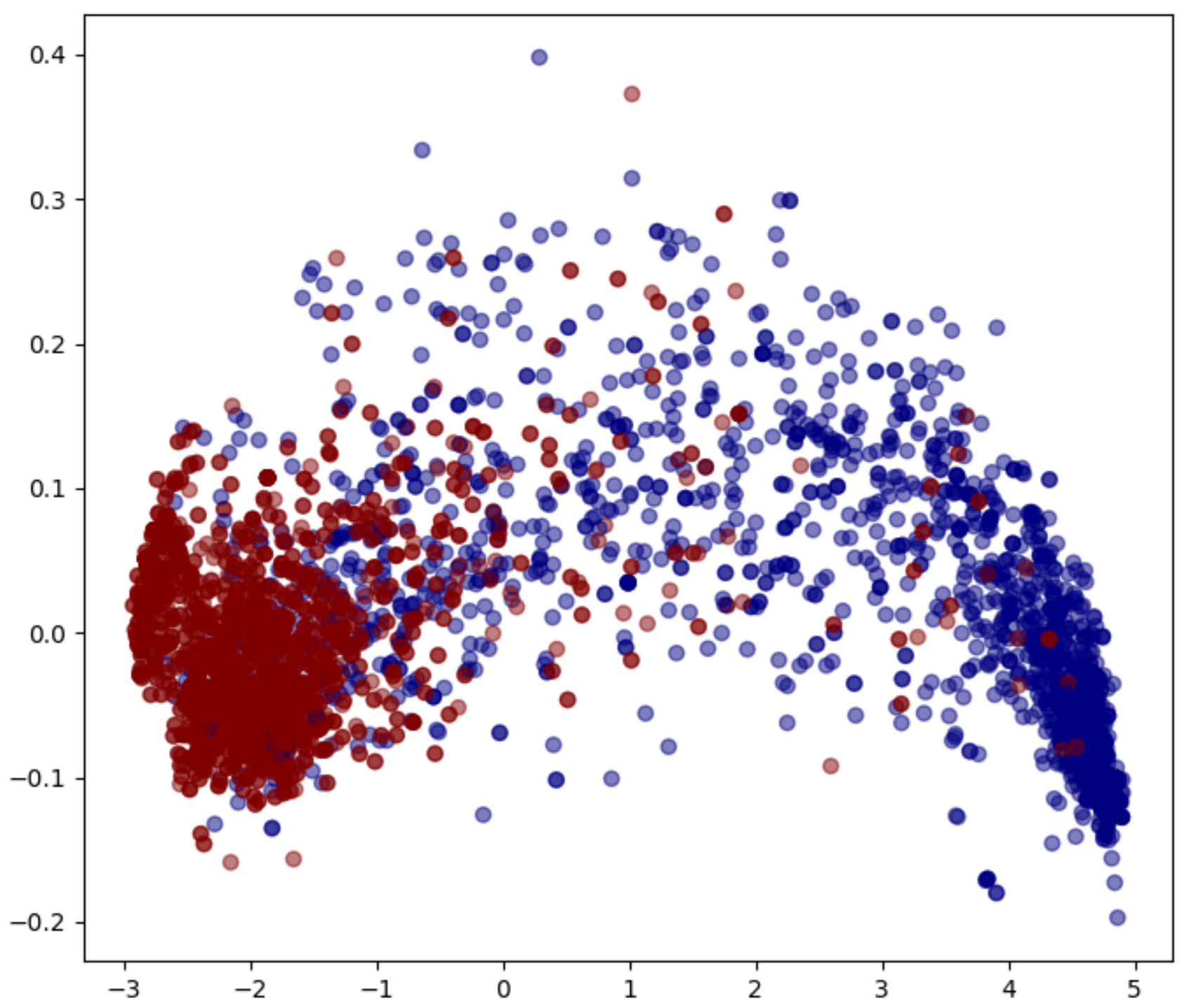}
\end{minipage}%
}%
\subfigure[RI]{
\begin{minipage}[t]{0.45\linewidth}
\centering
\includegraphics[width=1.0\linewidth]{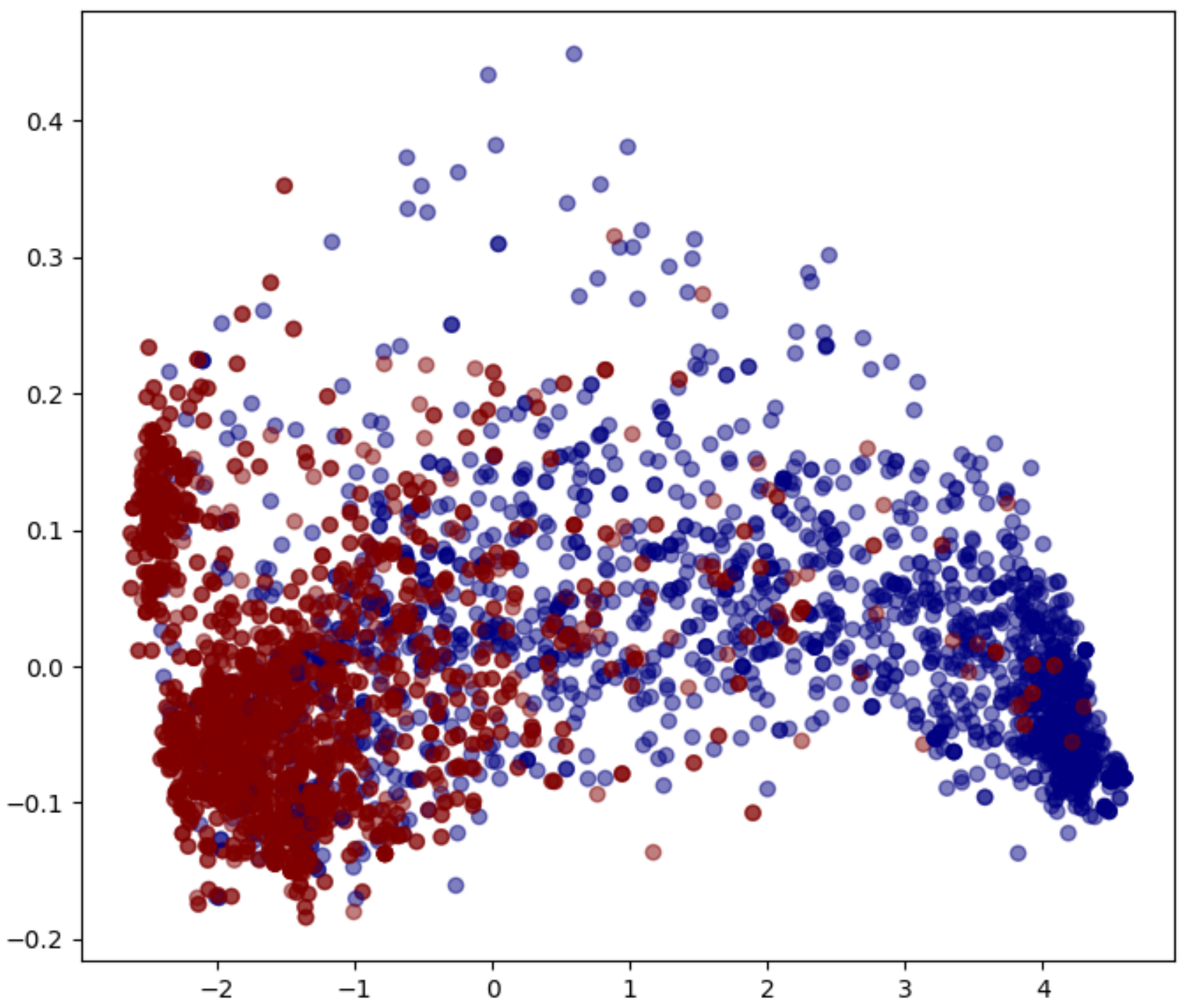}
\end{minipage}%
}%

\centering
\subfigure[RS]{
\begin{minipage}[t]{0.45\linewidth}
\centering
\includegraphics[width=1.0\linewidth]{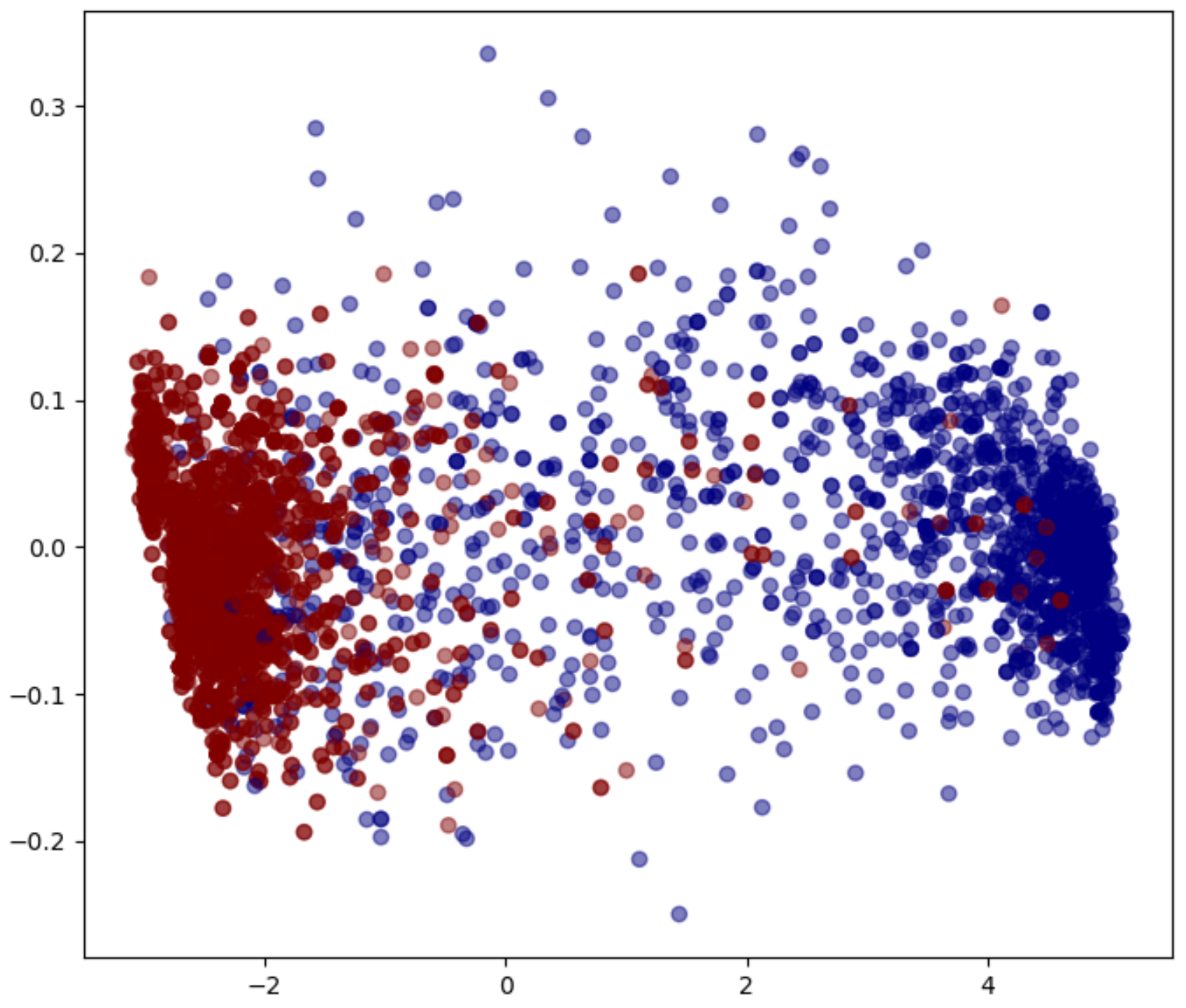}
\end{minipage}%
}%
\subfigure[RD]{
\begin{minipage}[t]{0.45\linewidth}
\centering
\includegraphics[width=1.0\linewidth]{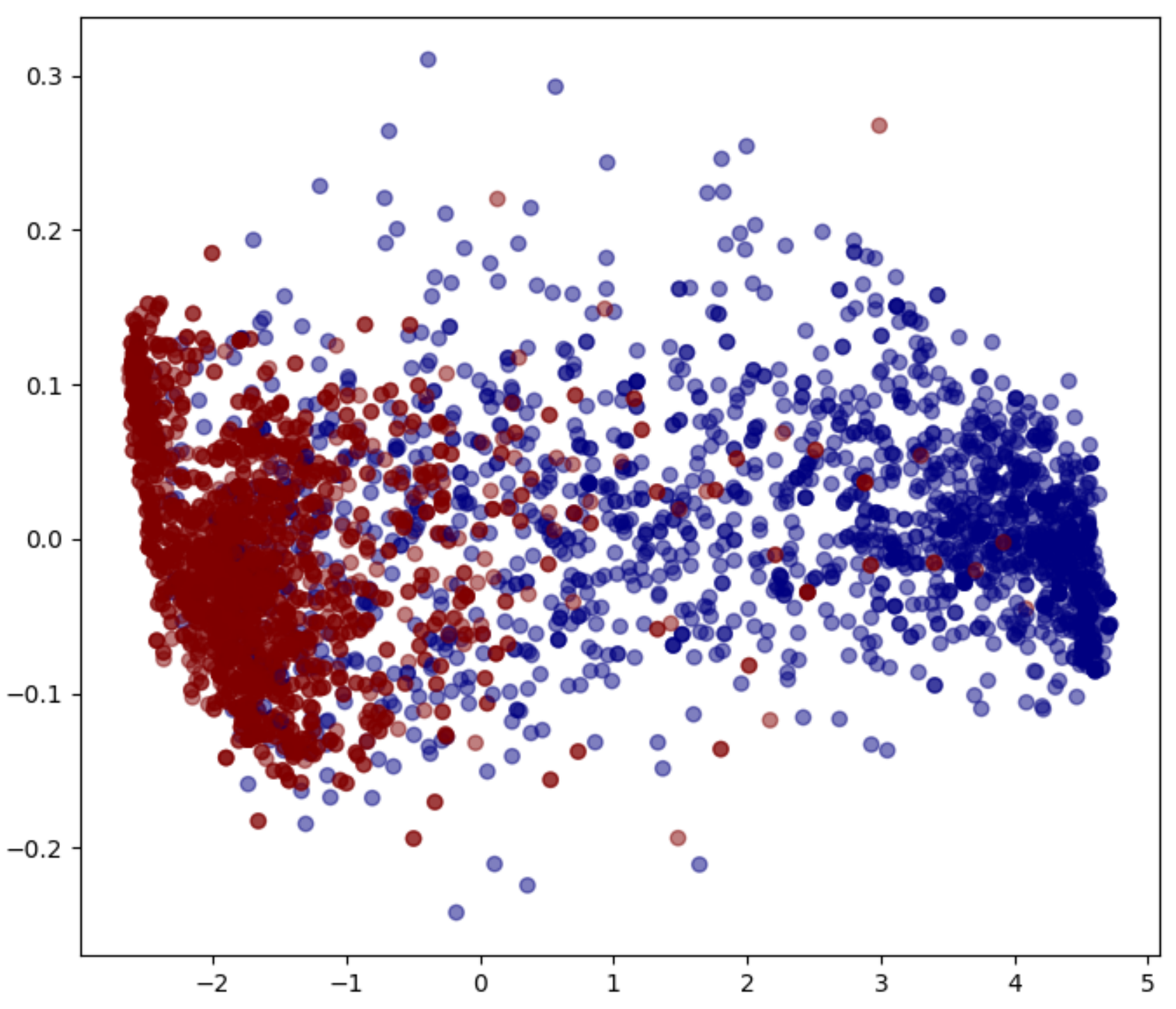}
\end{minipage}%
}%

\centering
\subfigure[BT]{
\begin{minipage}[t]{0.45\linewidth}
\centering
\includegraphics[width=1.0\linewidth]{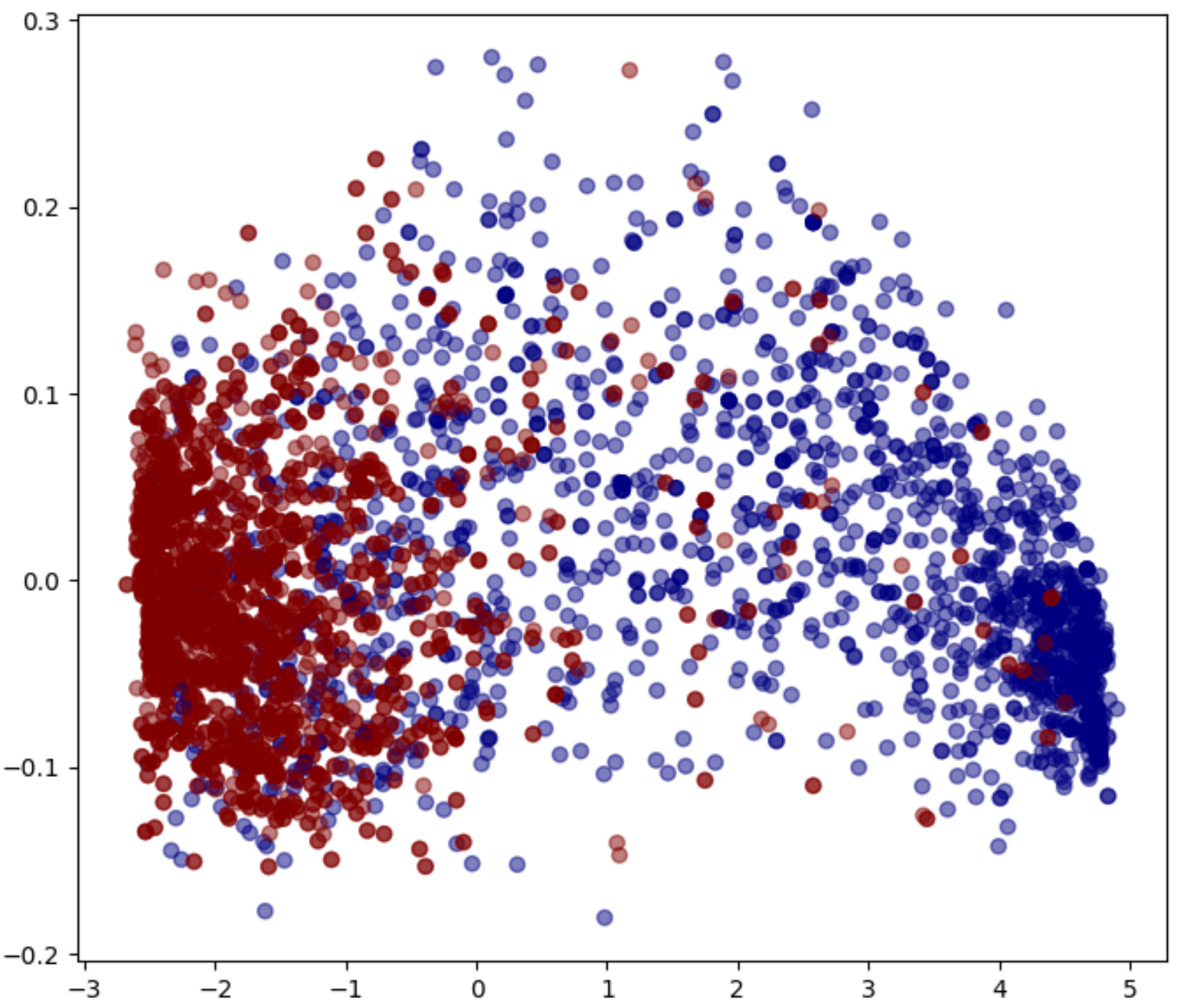}
\end{minipage}%
}%
\subfigure[SenMixup]{
\begin{minipage}[t]{0.45\linewidth}
\centering
\includegraphics[width=1.0\linewidth]{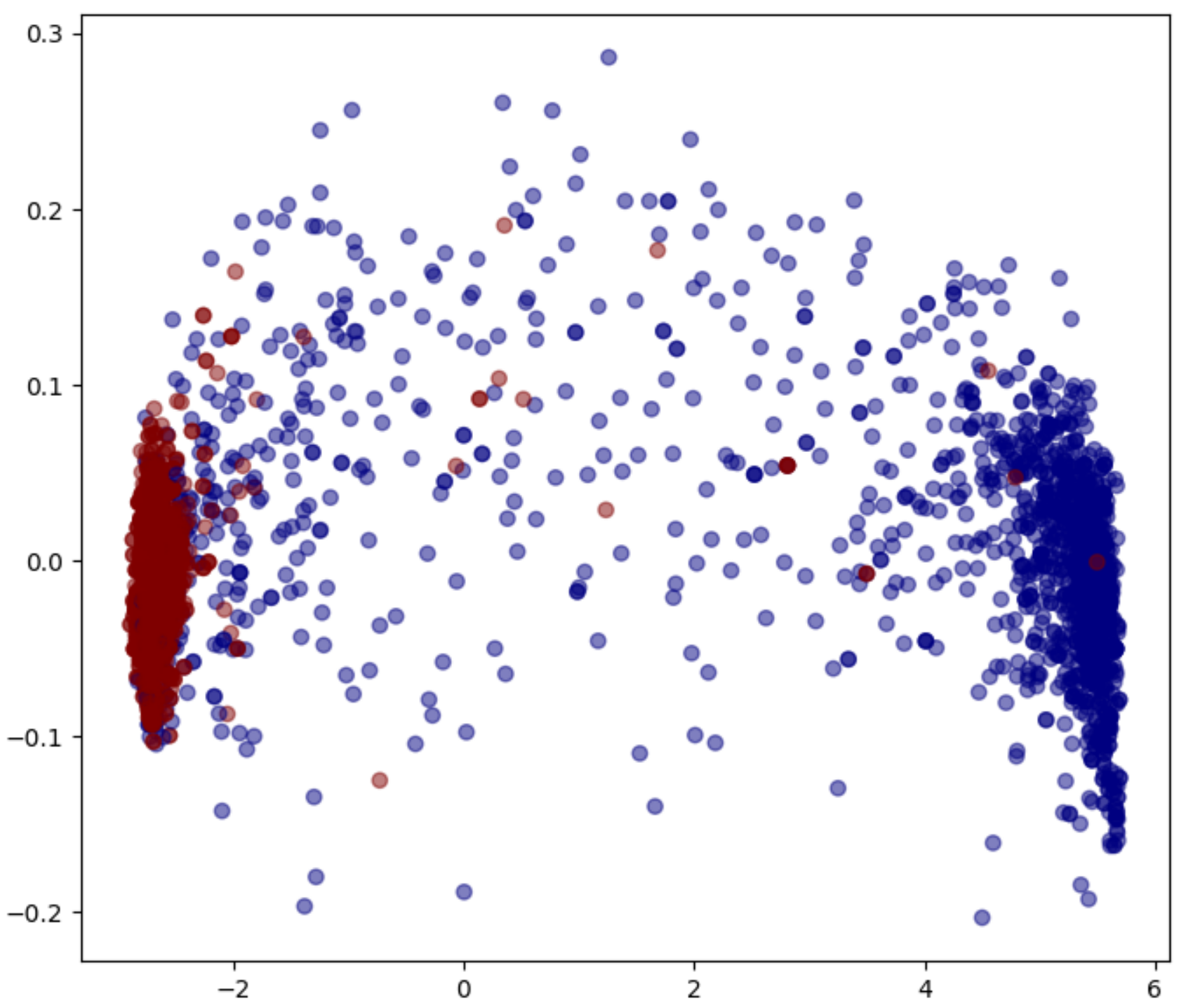}
\label{fig:pca_mixup}
\end{minipage}%
}%
\caption{Visualization of code embeddings after dimension reduction using Principal Component Analysis (PCA). Model: CodeBERT, dataset: Refactory}
\label{fig:pca}
\end{figure}
As discussed in Section~\ref{sec:section7_1}, Mixup-based data augmentation methods, such as \emph{SenMixup}, can effectively improve the performance of source code models, especially in situations with limited datasets. Therefore, we further explore the reasons why Mixup-based data augmentation methods can exhibit superior performance compared to other methods in accuracy and robustness.

\smallskip
\noindent
\textbf{Accuracy.} We discuss the accuracy via the visualization of code embeddings. Specifically, 1) we extract code embeddings produced by each trained model using all the test data as code features. 2) Since the extracted features are high-dimensional and challenging to analyze, we reduce the dimensions to two using the \emph{Principal Component Analysis~(PCA)}~\cite{mackiewicz1993principal} algorithm. 3) Subsequently, we visualize the reduced 2-dimensional vectors. Typically, features extracted by a well-trained model exhibit a clearer distribution based on the data labels.

Fig.~\ref{fig:pca} depicts the visualized features extracted from CodeBERT using the Refactory (Task: Bug detection) dataset. We can see that 1) Data augmentation methods including \emph{SR}, \emph{RI}, \emph{RS}, \emph{RD}, and~\emph{SenMixup} are beneficial in helping accuracy since they significantly improve the representation characteristics in binary classification tasks, making the data features exhibit clear classification trends compared to not using data augmentation. 2) Additionally, in this case, the code embeddings trained using Mixup-based data augmentation exhibit more evenly distributed class-specific representations compared to embeddings extracted by other data augmentation models. The distinction between buggy code and correct code is pronounced (shown in Fig.~\ref{fig:pca_mixup}).

Smoother interpolations indicate a smaller volume of hidden representations~\cite{verma2019manifold}, effectively constraining the number of directions with significant variance. Consequently, through interpolation in the hidden space between two different data points, \emph{SenMixup} assists the source code model in capturing the underlying code data structure, thereby enhancing accuracy performance. In summary, these results suggest that \emph{SenMixup} can effectively train high-quality code embeddings, thereby contributing to the development of superior code models for downstream code classification tasks.

\smallskip
\noindent
\textbf{Robustness.} We discuss the robustness based on the metric \emph{confidence}. Typically, when a DNN model exhibits higher confidence in its predictions on test data, it becomes more challenging to execute adversarial attacks on the model~\cite{ren2020adversarial}. 

Table~\ref{tab:probability} presents the results of the mean values of maximum probabilities. In most cases, it is evident that source code models trained using \emph{WordMixup} and \emph{SenMixup}, demonstrate higher confidence in their predictions on the test data compared to source code models trained using other data augmentation methods. While there are a few cases, such as GCJ and Refactory, where the mean values of maximum probabilities of using \emph{SenMixup} to enhance model training are not the absolute best, it still ranks second, indicating strong performance (e.g., compare GCJ-RS-0.0392912 to GCJ-\emph{SenMixup}-0.0392896).

As two variants of Mixup, the robustness experiments of \emph{SenMixup} and \emph{WordMixup} yield similar conclusions to the usage of Mixup.  Adversarial attacks involve making subtle alterations to input data to deceive the model's predictions. By taking two code data to generate new data as a linear combination of the original data, can effectively smooth the data distribution in the feature space. The smoothing effect on the decision boundary induced by mixing can diminish the effectiveness of these perturbations on the model's behavior. Furthermore, Zhang~\emph{et al.}~\cite{zhang2020does} has also proven that minimizing the \emph{Mixup} loss approximates the minimization of an upper bound on the adversarial loss. Therefore,  as a data-adaptive regularization, Mixup-based data augmentation methods can improve the performance of robustness. 

\begin{table*}[!tb]
\caption{Mean values of
maximum probabilities. The best and second-best results are highlighted in gray and blue respectively. Tasks include \textbf{Problem Classification} (Java250), \textbf{Bug detection} (Refactory), \textbf{Authorship Attribution} (GCJ) and \textbf{Clone detection} (BigCloneBench)}
\label{tab:probability}
\centering
\resizebox{\columnwidth}{!}{
\begin{tabular}{clcccccc}
 \cline{1-6}
\textbf{Model} & \textbf{DA method} & \textbf{GCJ}  & \textbf{Refactory} & \textbf{Java250} & \textbf{BigCloneBench} & \\ \cline{1-6} 
 & No Aug  & {0.0381842} & {0.9367431} & {0.9553056} & {0.9653085} & {} & \\ 
 &Refactor & {0.0391791} & {0.9643209} & {0.9621801} & {0.9655135} & {} & \\ 
 & SR & {0.0380601} & {0.9527591} & {0.9342893} & {0.9582997} & {} & \\ 
 & RI & {0.0371315} & {0.9189276} & {0.9328093} & \cellcolor[HTML]{C0C0C0}{0.9705281} & {} & \\  
 & RS & \cellcolor[HTML]{C0C0C0}{0.0392912}  & {0.9561052} & {0.9580978} & {0.9411782} & {} & \\  
  & RD & {0.0392801}  & \cellcolor[HTML]{C0C0C0}{0.9724103} & {0.9635246} & {0.9630691} & {} & \\  
 & BT & {0.0379615} & {0.9566809} & {0.9418421} & \cellcolor[HTML]{DBE7FC}{0.9667154} & {} & \\  
 & WordMixup & {0.0392841}& {0.9657822} & \cellcolor[HTML]{DBE7FC}{0.9639287} & {-} & {} & \\  
\multirow{-9}{*}{CodeBERT} &  SenMixup & \cellcolor[HTML]{DBE7FC}{0.0392896} & \cellcolor[HTML]{DBE7FC}{0.9681156} & \cellcolor[HTML]{C0C0C0}{0.9652706} & {-} & {} & \\    
\cline{1-6} 
\end{tabular}
}
\vspace{-3mm}
\end{table*}

\subsection{Is it necessary to keep syntax rules in data augmentation for source code learning?}
 
The previous research has shown that, for natural language, although the semantics of text data are sensitive to their syntactic change, it could remain readable to humans~\cite{wang1999reading} and valid as additional training data if the change happens within a limited range that does not break the original relations between the text data and their labels. Indeed, that is why noising-based data augmentation methods, such as \emph{RI} and \emph{RS} (see Section~\ref{sec: secction3}), are still very useful in NLP, as shown by Marivate~\emph{et al.}~\cite{marivate2020improving}.

In the context of source code learning, 15.52\% of the training samples from the augmented training datasets used in our experiments break syntax rules. Our experimental results suggest a similar conclusion, namely, that even though some data augmentation methods can produce training data that slightly breaks the syntax of the source code, these data are still useful in improving the quality of training in source code learning. Indeed, as reported by our experiments, the pre-trained PL models using the \emph{RS} method can achieve higher accuracy (by up to 17.54\%) and higher robustness (by up to 14.55\%) than the models using the baseline \emph{Refactor} method. Moreover, this finding shows the naturalness of source code and is consistent with the famous \emph{software naturalness hypothesis}~\cite{allamanis2018survey,hindle2016naturalness,buratti2020exploring}.

\subsection{How can researchers use the presented findings to improve the existing approaches in the field?}
First, our experimental results demonstrated that even though some methods cannot enhance the model's accuracy, they can significantly improve the model's robustness. Therefore, when evaluating or proposing new data augmentation methods, it is important to consider different model properties. Besides, it is not necessary to constrain the generated code during data augmentation as we found some syntax-breaking methods can better enhance code models. We observe that in most cases, no data augmentation methods including text-oriented and code refactoring techniques consistently improve the performance of models across various tasks. Thus, instead of focusing on the raw input level, trying to consider data augmentation methods from the embedding level could be a research direction.

\subsection{Threats to validity}
The internal threat to validity comes from the implementation of standard training and data augmentation methods. The code of model training, \emph{SenMixup}, and \emph{WordMixup} methods are from the official implementations of corresponding projects. The code refactoring methods for the Java language come from Pour~\emph{et al.}~\cite{pour2021search} and Wei~\emph{et al.}~\cite{wei2022cocofuzzing}, and we adapt the implementation to the Python language.  The implementation of \emph{EDA} and \emph{BT} comes from Xie~\emph{et al.}~\cite{xie2020unsupervised} and Wei~\emph{et al.}~\cite{wei-zou-2019-eda}, again we adapt them to code-related tasks.

The external threats to validity lie in the selected code-related tasks, datasets, DNNs, and data augmentation methods. In our study, we consider four different code classification tasks, including problem classification, bug detection, authorship attribution, and clone detection, and we have a total of six datasets for the above tasks. Especially we include two popular programming languages in the software engineering community (Java and Python), as well as two low-resource programming languages (Go and Ruby). We apply four types of DNN models, including two mainstream pre-trained PL models. For data augmentation methods from code, code refactoring methods cover the most common ones in the literature. Data augmentation methods of NLP come from the most classic method, which is comprehensively adopted from the number of citations of papers.

The construct threats to validity mainly come from the parameters, randomness, and evaluation measures. We follow the original recommendation of Mixup to set its parameters. The parameters of data augmentation methods from NLP also follow the original release. We repeat each experiment five times and report the average results to reduce the influence of randomness. For evaluation measures, we consider both accuracy and robustness, and the latter is used to evaluate the generalization ability of DNNs.

\section{Related work}
\label{sec:section8}
We review related work about representation for source code learning, data augmentation for source code learning, and empirical studies on source code learning.
\subsection{Code Representation for source code learning}
Code representation is a crucial technique that transforms source code into a format readable by DNNs, enabling effective source code feature learning. Existing works on code representation can be roughly divided into 1) sequential representation, 2) structural representation, and 3) hybrid representation. Sequential representation focuses on learning contextual representations of source code by either reconstructing the sequence of tokens or parsing the program into a parse tree. Alon \emph{et al.} proposed \emph{Code2Vec}~\cite{alon2019code2vec}, which represents code snippets as single fixed-length code vectors. Specifically, it decomposes a program into a collection of paths based on an \emph{abstract syntax tree (AST)}, learning the representation of each path and aggregating information from the set of paths. Meanwhile, they provided \emph{Code2Seq}~\cite{alon2018codeseq}, which uses \emph{sequence-to-sequence} models, adopted from \emph{neural machine translation}, to represent code vectors as an alternative. Feng \emph{et al.}~\cite{feng2020codebert} applied a bimodal pre-trained model to encode source code. Similarly, Kanade \emph{et al.}~\cite{kanade2020learning} employed transformer-based language models to derive the contextual embeddings of source code. Structural representation transforms the code into a graph that captures the relationship of different components in the code while preserving its syntax elements. Allamanis \emph{et al.}~\cite{allamanis2017learning} used AST augmented with data and control edges to represent the code and employed \emph{graph neural networks (GNNs)} to learn and reason over these program structures. To further improve the effectiveness of GNNs-based source code learning, Ben-Nun \emph{et al.}~\cite{ben2018neural} extracted the contextual flow graph of a program by constructing an LLVM intermediate representation. Zhou \emph{et al.}~\cite{zhou2019devign} proposed a graph representation of source code by aggregating information from AST,\emph{control flow graph}, and \emph{data flow graph}. Additionally, researchers combined sequential and structural information to represent source code. Guo \emph{et al.}~\cite{guo2020graphcodebert} introduced GraphCodeBERT, which not only captures syntactic features but also learns data flow information during the pre-training stage. Ma \emph{et al.}~\cite{ma2022graphcode2vec} conducted a synergistic combination of task-agnostic embeddings of lexical and program-dependence features.

Unlike code representation, our work focuses on code data preprocessing. Specifically, via program transformations, data augmentation can enhance feature learning by mitigating negative features of code data that hinder DNN model performance or by amplifying positive features that improve DNN model performance, all without altering the volume of data.

\subsection{Data augmentation for source code learning}
Data augmentation has achieved enormous success in the machine learning field~\cite {shorten2019survey,feng2021survey}. Inspired by its success, recently researchers devoted considerable effort to leveraging the data augmentation technique in big code tasks to improve the performance of code models in terms of accuracy and robustness. Adversarial training~\cite{goodfellow2014explaining}, which produces a set of adversarial examples to the training data, has been studied as the data augmentation method in code learning. Zhang \emph{et al.}~\cite{zhang2020training} proposed a code data augmentation method that employs the MHM algorithm~\cite{zhang2020generating} to improve the capability of deep comment generation models. Mi \emph{et al.}~\cite{mi2021effectiveness} generated the additional data from Auxiliary Classifier generative adversarial networks (GANs). Besides, as a program transformation method that is specially designed for code, code refactoring has been used as a mainstream code data augmentation method. Yu \emph{et al.}~\cite{yu2022data} designed program transformation rules for Java and evaluated the effectiveness of using these program transformations as code data augmentation in three big code-related tasks. Allamanis \emph{et al.}~\cite{allamanis2021self} used four simple code rewrite rules as code data augmentation methods for improving the generalization of the code model. 
 
Compared with the above works, our study is the first one that assesses the effectiveness of three types of data augmentation methods for code learning, namely, the methods for code data, the methods for text data, and the methods for graph data.

\subsection{Empirical studies on source code learning}

Recently, many works conducted empirical studies to explore the topic of ML4Code.
Chirkova \emph{et al.}~\cite{chirkova2021empirical} conducted a thorough empirical study to evaluate the capabilities of using Transformer~\cite{vaswani2017attention} to solve three downstream tasks related to code learning, including code completion, function naming, and bug fixing. Siow \emph{et al.}~\cite{siow2022learning} conducted an empirical study to evaluate existing program representation techniques. Zhang \emph{et al.}~\cite{zhang2020empirical} empirically analyzed current testing and debugging practices for machine learning programs. They revealed that the interaction with the platform execution environments could easily cause machine learning program failures, moreover, current debugging is insufficient to locate the fault in machine learning code well. This work is useful for programmers to improve the quality of the source code of machine learning. Jebnoun \emph{et al.}~\cite{jebnoun2020scent} performed a comparative study to explore the distribution of code smells between machine learning and traditional applications. Yan \emph{et al.}~\cite{yan2020code} conducted a comprehensive empirical study on code search using machine techniques. Their empirical evaluation results revealed that machine learning techniques are more effective for queries on reusing code. More recently, Hu \emph{et al.}~\cite{hu2022codes} empirically studied the distribution shift problem of code learning and defined five types of shift for code data. Steenhoek et al~\cite{steenhoek2022empirical} experimentally reproduced nine DNN models and two widely used vulnerability detection datasets to help understand deep learning-based models in source code learning. Mastropaolo \emph{et al.}~\cite{mastropaolo2023robustness} presented a comprehensive empirical study to evaluate the robustness of the code completion approach Github Copilot. Niu \emph{et al.}~\cite{niu2023empirical} performed a comparative study to analyze recently developed pre-trained PL models to advance the understanding of these pre-trained PL models used in source code learning.

Different from the existing empirical studies, our work investigates data augmentation on source code learning that has rarely been studied to date. 

\section{Conclusions}
\label{sec:section9}

Data augmentation has a long-term study history in the CV and NLP fields. When the model architecture is fixed, the straightforward way to enhance the code model is to prepare more high-quality training data. However, data augmentation for code learning is still in an early stage and only a few techniques have been proposed. Our study demonstrated that most of the existing methods that are widely used in the NLP field cannot work well in the code learning direction and highlighted the need to design more advanced data augmentation methods specifically for code data. Furthermore, our empirical study highlights the importance of data augmentation in code learning and offers insights into the effectiveness of various augmentation methods for different downstream tasks and DNN models. Specifically, linear interpolation methods such as \emph{SenMixup} effectively improve both the accuracy and robustness of most DNNs, with the exception of pre-trained PL models. For pre-trained PL models, slightly breaking syntax-based code data augmentation methods like Random Deletion and Random Swap, prove particularly effective in robustness improvement and could serve as a valuable future direction in this field. Notably, linear interpolation methods are also well-suited for situations with limited training data. In light of these findings, our study paves the way for further improving the effectiveness of data augmentation in code-related tasks, with the ultimate goal of enhancing program understanding and accelerating software development.

\section{Acknowledgment}
This research is supported in part by JSPS KAKENHI Grant No. JP23H03372, Japan, and by the Luxembourg National Research Fund (FNR) through the CORE project under Grant C22/IS/17426831/MeMoRIA.

Zhenya Zhang is also supported by JSPS KAKENHI Grant No. JP23K16865. Yuejun Guo is supported by the European Commission under the Horizon Europe Programme, as part of the project LAZARUS (https://lazarus-he.eu/) (Grant Agreement no. 101070303). The content of this article does not reflect the official opinion of the European Union. Responsibility for the information and views expressed therein lies entirely with the authors.

\section{Declarations}

\subsection{Conflict of Interest}
The authors declare that they have no known competing financial interests or personal relationships that could have appeared to influence the work reported in this paper.

\subsection{Data Availability Statements}
The datasets generated during and/or analyzed during the current study are available in the: \url{https://github.com/zemingd/PT4Code}.

\bibliographystyle{IEEEtran}
\bibliography{reference}

\appendix
\section{Statistical Test.}
\label{sec:appendix}

This appendix is organized as follows:
\begin{compactitem}[$\bullet$]
\item We present statistical test results for all experiments in Sections~\ref{sec:RQ1},~\ref{sec:RQ2},~\ref{sec:RQ3},~\ref{sec:RQ4}.
\end{compactitem}

\begin{sidewaystable}[h]
\centering
    \setlength{\tabcolsep}{3pt} 
    \renewcommand{\arraystretch}{1.3}
    \tiny
\caption{Results of statistical test on Accuracy. \textbf{True}: indicates that the comparison is statistically significant after the  \emph{Bonferroni correction} adjustment. Statistical test method: \emph{Wilcoxon signed-rank test}. A gray background highlights the result marked as \textbf{True}. For legibility, we simplify the terms \emph{Rename Operator}, \emph{Dead Operator}, \emph{Inside Operator}, and \emph{Outside Operator} to \emph{Rename}, \emph{Dead}, \emph{Inside}, and \emph{Outside}, respectively. Model: BagofToken \& SeqofToken.}
\label{table:statistical_testing_acc_bag_seq}
\resizebox{\textwidth}{!}{
\begin{tabular}{lcccccccccccccccc}
\hline
& & \multicolumn{14}{c}{\textbf{Accuracy}}  \\ \hline
                    & DA method & No Aug & WordMixup & SenMixup  & Refactor & Rename & Dead  & Inside & Outside  & SR & RI &  RS &  RD & BT & \\ \hline
\multirow{13}{*}{BagofToken} & No Aug                        &   -      &   \cellcolor[HTML]{C0C0C0}{\color[HTML]{000000}True }     &   \cellcolor[HTML]{C0C0C0}{\color[HTML]{000000}True }     &    -  &    -  &    -  &  -    &  -    &    -    &    -  &  \cellcolor[HTML]{C0C0C0}{\color[HTML]{000000}True }    &    -  &    -      \\
& WordMixup                         &   \cellcolor[HTML]{C0C0C0}{\color[HTML]{000000}True }       &    -     &   -     &    \cellcolor[HTML]{C0C0C0}{\color[HTML]{000000}True }   &    -  &    -  &  -    &  -    &   \cellcolor[HTML]{C0C0C0}{\color[HTML]{000000}True }    &    \cellcolor[HTML]{C0C0C0}{\color[HTML]{000000}True } &  \cellcolor[HTML]{C0C0C0}{\color[HTML]{000000}True }   &  \cellcolor[HTML]{C0C0C0}{\color[HTML]{000000}True }  &  -  \\
& SenMixup                        &   \cellcolor[HTML]{C0C0C0}{\color[HTML]{000000}True }     &    -     &   -     &    \cellcolor[HTML]{C0C0C0}{\color[HTML]{000000}True }    &    -  &    -  &  -    &  -    &    \cellcolor[HTML]{C0C0C0}{\color[HTML]{000000}True }     &    -  &  -   &  - &  \cellcolor[HTML]{C0C0C0}{\color[HTML]{000000}True } \\
& Refactor              &   -      &    \cellcolor[HTML]{C0C0C0}{\color[HTML]{000000}True }      &   \cellcolor[HTML]{C0C0C0}{\color[HTML]{000000}True }     &    -  &    \cellcolor[HTML]{C0C0C0}{\color[HTML]{000000}True }  &    -  &  \cellcolor[HTML]{C0C0C0}{\color[HTML]{000000}True }    &  \cellcolor[HTML]{C0C0C0}{\color[HTML]{000000}True }     &    -    &    \cellcolor[HTML]{C0C0C0}{\color[HTML]{000000}True } &  \cellcolor[HTML]{C0C0C0}{\color[HTML]{000000}True }    &  \cellcolor[HTML]{C0C0C0}{\color[HTML]{000000}True }  &  - \\
& Rename              &   -      &    -     &   -     &    \cellcolor[HTML]{C0C0C0}{\color[HTML]{000000}True }   &    -  &    -  &  -    &  -    &    -    &    -  &  -   &  - &  - \\
& Dead               &   -      &    -     &   -     &    -  &    -  &    -  &  \cellcolor[HTML]{C0C0C0}{\color[HTML]{000000}True }    &  -    &    -    &    -  &  -   &  - &  - \\
& Inside             &   -      &    -     &   -     &    \cellcolor[HTML]{C0C0C0}{\color[HTML]{000000}True }   &    -  &     \cellcolor[HTML]{C0C0C0}{\color[HTML]{000000}True }   &  -    &  -    &    -    &    -  &  -   &  - &  - \\
& Outside              &   -      &    -     &   -     &   \cellcolor[HTML]{C0C0C0}{\color[HTML]{000000}True }   &    -  &    -  &  -    &  -    &    -    &    -  &  -   &  - &  - \\
& SR               &   -      &    \cellcolor[HTML]{C0C0C0}{\color[HTML]{000000}True }    &   \cellcolor[HTML]{C0C0C0}{\color[HTML]{000000}True }     &    -  &    -  &    -  &  -    &  -    &    -    &    -  &  \cellcolor[HTML]{C0C0C0}{\color[HTML]{000000}True }    &  \cellcolor[HTML]{C0C0C0}{\color[HTML]{000000}True }  &  - \\
& RI               &   -      &   \cellcolor[HTML]{C0C0C0}{\color[HTML]{000000}True }     &   -     &    \cellcolor[HTML]{C0C0C0}{\color[HTML]{000000}True }  &    -  &    -  &  -    &  -    &    -    &    -  &  \cellcolor[HTML]{C0C0C0}{\color[HTML]{000000}True }   &  - &  - \\
& RS             &   \cellcolor[HTML]{C0C0C0}{\color[HTML]{000000}True }       &    \cellcolor[HTML]{C0C0C0}{\color[HTML]{000000}True }    &   -     &   \cellcolor[HTML]{C0C0C0}{\color[HTML]{000000}True }  &    -  &    -  &  -    &  -    &    \cellcolor[HTML]{C0C0C0}{\color[HTML]{000000}True }    &    \cellcolor[HTML]{C0C0C0}{\color[HTML]{000000}True }   &  -   &  - &  \cellcolor[HTML]{C0C0C0}{\color[HTML]{000000}True }  \\
& RD              &   -      &    \cellcolor[HTML]{C0C0C0}{\color[HTML]{000000}True }    &   -     &    \cellcolor[HTML]{C0C0C0}{\color[HTML]{000000}True }   &    -  &    -  &  -    &  -    &    \cellcolor[HTML]{C0C0C0}{\color[HTML]{000000}True }     &    -  &  -   &  - &  - \\
& BT             &   -      &    -     &   \cellcolor[HTML]{C0C0C0}{\color[HTML]{000000}True }     &    -  &    -  &    -  &  -    &  -    &    -    &    -  &  \cellcolor[HTML]{C0C0C0}{\color[HTML]{000000}True }    &  - &  - \\
\hline                  
\multirow{13}{*}{SeqofToken} & No Aug                        &   -      &    -     &   -     &    -  &    -  &    -  &  -    &  -    &    -    &    -  &  -   &  - &  -  \\
& WordMixup                         &   -      &    -     &   -     &    \cellcolor[HTML]{C0C0C0}{\color[HTML]{000000}True }  &    -  &    \cellcolor[HTML]{C0C0C0}{\color[HTML]{000000}True }  &  \cellcolor[HTML]{C0C0C0}{\color[HTML]{000000}True }    & \cellcolor[HTML]{C0C0C0}{\color[HTML]{000000}True }   &    -    &    -  &  -   &  - &  \cellcolor[HTML]{C0C0C0}{\color[HTML]{000000}True }  \\
& SenMixup                        &   -      &    -     &   -     &    \cellcolor[HTML]{C0C0C0}{\color[HTML]{000000}True }  &    -  &    \cellcolor[HTML]{C0C0C0}{\color[HTML]{000000}True }  &  \cellcolor[HTML]{C0C0C0}{\color[HTML]{000000}True }  &  \cellcolor[HTML]{C0C0C0}{\color[HTML]{000000}True }   &    -    &    \cellcolor[HTML]{C0C0C0}{\color[HTML]{000000}True }  &  -   &  - &  \cellcolor[HTML]{C0C0C0}{\color[HTML]{000000}True } \\
& Refactor              &   -      &    \cellcolor[HTML]{C0C0C0}{\color[HTML]{000000}True }     &   \cellcolor[HTML]{C0C0C0}{\color[HTML]{000000}True }    &    -  &    \cellcolor[HTML]{C0C0C0}{\color[HTML]{000000}True }  &    -  &  \cellcolor[HTML]{C0C0C0}{\color[HTML]{000000}True }    &  -    &    -    &    \cellcolor[HTML]{C0C0C0}{\color[HTML]{000000}True }  &  -   &  - &  - \\
& Rename              &   -      &    -     &   -     &    \cellcolor[HTML]{C0C0C0}{\color[HTML]{000000}True }  &    -  &    -  &  \cellcolor[HTML]{C0C0C0}{\color[HTML]{000000}True }    &  -    &    \cellcolor[HTML]{C0C0C0}{\color[HTML]{000000}True }    &    -  &  -   &  - &  \cellcolor[HTML]{C0C0C0}{\color[HTML]{000000}True }\\

& Dead               &   -      &    \cellcolor[HTML]{C0C0C0}{\color[HTML]{000000}True }     &   \cellcolor[HTML]{C0C0C0}{\color[HTML]{000000}True }     &    -  &    -  &    -  &  -    &  -    &    -    &    -  &  -   &  - &  \cellcolor[HTML]{C0C0C0}{\color[HTML]{000000}True } \\

& Inside             &   -      &   \cellcolor[HTML]{C0C0C0}{\color[HTML]{000000}True }    &   \cellcolor[HTML]{C0C0C0}{\color[HTML]{000000}True }    &   \cellcolor[HTML]{C0C0C0}{\color[HTML]{000000}True }  &    \cellcolor[HTML]{C0C0C0}{\color[HTML]{000000}True }  &    -  &  -    &  -    &    -    &    -  &  -   &  - &  - \\

& Outside              &   -      &    \cellcolor[HTML]{C0C0C0}{\color[HTML]{000000}True }    &   \cellcolor[HTML]{C0C0C0}{\color[HTML]{000000}True }     &    -  &    -  &    -  &  -    &  -    &    -    &    -  &  -   &  - &  - \\

& SR               &   -      &    -     &   -     &    -  &    \cellcolor[HTML]{C0C0C0}{\color[HTML]{000000}True }  &    -  &  -    &  -    &    -    &    -  & \cellcolor[HTML]{C0C0C0}{\color[HTML]{000000}True }   &  \cellcolor[HTML]{C0C0C0}{\color[HTML]{000000}True } &  - \\

& RI               &   -      &    -     &   \cellcolor[HTML]{C0C0C0}{\color[HTML]{000000}True }     &    \cellcolor[HTML]{C0C0C0}{\color[HTML]{000000}True }  &    -  &    -  &  -    &  -    &    -    &    -  &  -   &  - &  \cellcolor[HTML]{C0C0C0}{\color[HTML]{000000}True }\\

& RS             &   -      &    -     &   -     &    -  &    -  &    -  &  -    &  -    &    \cellcolor[HTML]{C0C0C0}{\color[HTML]{000000}True }    &    -  &  -   &  - &  \cellcolor[HTML]{C0C0C0}{\color[HTML]{000000}True }\\

& RD              &   -      &    -     &   -     &    -  &    -  &    -  &  -    &  -    &    \cellcolor[HTML]{C0C0C0}{\color[HTML]{000000}True }   &    -  &  -   &  - &  \cellcolor[HTML]{C0C0C0}{\color[HTML]{000000}True } \\

& BT             &   -      &   \cellcolor[HTML]{C0C0C0}{\color[HTML]{000000}True }     &   \cellcolor[HTML]{C0C0C0}{\color[HTML]{000000}True }     &    -  &    \cellcolor[HTML]{C0C0C0}{\color[HTML]{000000}True }  &    \cellcolor[HTML]{C0C0C0}{\color[HTML]{000000}True }  &  -    &  -    &    -    &    \cellcolor[HTML]{C0C0C0}{\color[HTML]{000000}True } &  \cellcolor[HTML]{C0C0C0}{\color[HTML]{000000}True }  &  \cellcolor[HTML]{C0C0C0}{\color[HTML]{000000}True } &  -\\

\hline  

\end{tabular}
}
\end{sidewaystable}

\begin{sidewaystable}[]
\centering
    \setlength{\tabcolsep}{3pt} 
    \renewcommand{\arraystretch}{1.3}
    \tiny
\caption{Results of statistical tests on Accuracy. \textbf{True}: indicates that the comparison is statistically significant after the  \emph{Bonferroni correction} adjustment. Statistical test method: \emph{Wilcoxon signed-rank test}. A gray background highlights the result marked as \textbf{True}. For legibility, we simplify the terms \emph{Rename Operator}, \emph{Dead Operator}, \emph{Inside Operator}, and \emph{Outside Operator} to \emph{Rename}, \emph{Dead}, \emph{Inside}, and \emph{Outside}, respectively. Model: CodeBERT \& GraphCodeBERT.}
\label{table:statistical_testing_acc_codebert_graphcodebert}
\centering
\resizebox{\textwidth}{!}{
\begin{tabular}{lcccccccccccccccc}
\hline
& & \multicolumn{14}{c}{\textbf{Accuracy}}  \\ \hline
                    & DA method & No Aug & WordMixup & SenMixup  & Refactor & Rename & Dead  & Inside & Outside  & SR & RI &  RS &  RD & BT & \\ \hline
\multirow{13}{*}{CodeBERT} & No Aug                         &   -      &    -     &   -     &    -  &    \cellcolor[HTML]{C0C0C0}{\color[HTML]{000000}True } &    -  &  -    &  -    &    -    &    -  &  \cellcolor[HTML]{C0C0C0}{\color[HTML]{000000}True }  &  - &  -       \\

& WordMixup                         &   -      &    -     &   \cellcolor[HTML]{C0C0C0}{\color[HTML]{000000}True }     &    -  &    -  &    -  &  -    &  -    &    -    &    -  &  -   &  - &  -   \\

& SenMixup                       &   -      &    \cellcolor[HTML]{C0C0C0}{\color[HTML]{000000}True }    &   -     &    -  &    -  &    -  &  -    &  -    &    -    &    -  &  -   &  - &  - \\

& Refactor             &   -      &    -     &   -     &    -  &    \cellcolor[HTML]{C0C0C0}{\color[HTML]{000000}True }  &   \cellcolor[HTML]{C0C0C0}{\color[HTML]{000000}True } &  -    &  -    &    -    &    -  &  \cellcolor[HTML]{C0C0C0}{\color[HTML]{000000}True }   &  - &  - \\

& Rename             &   \cellcolor[HTML]{C0C0C0}{\color[HTML]{000000}True }      &    -     &   -     &    \cellcolor[HTML]{C0C0C0}{\color[HTML]{000000}True }  &    -  &    -  & \cellcolor[HTML]{C0C0C0}{\color[HTML]{000000}True }   &  \cellcolor[HTML]{C0C0C0}{\color[HTML]{000000}True }    &    -    &    -  &  -   &  - &  \cellcolor[HTML]{C0C0C0}{\color[HTML]{000000}True }\\

& Dead              &   -      &    -     &   -     &    \cellcolor[HTML]{C0C0C0}{\color[HTML]{000000}True }  &    -  &    -  &  \cellcolor[HTML]{C0C0C0}{\color[HTML]{000000}True }    &  \cellcolor[HTML]{C0C0C0}{\color[HTML]{000000}True }   &    -    &    -  &  -   &  - &  \cellcolor[HTML]{C0C0C0}{\color[HTML]{000000}True } \\

& Inside             &   -      &    -     &   -     &    -  &    \cellcolor[HTML]{C0C0C0}{\color[HTML]{000000}True }  &    \cellcolor[HTML]{C0C0C0}{\color[HTML]{000000}True }  &  -    &  -    &    -    &    -  &  \cellcolor[HTML]{C0C0C0}{\color[HTML]{000000}True }   &  - &  - \\

& Outside             &   -      &    -     &   -     &    -  &    \cellcolor[HTML]{C0C0C0}{\color[HTML]{000000}True }  &    \cellcolor[HTML]{C0C0C0}{\color[HTML]{000000}True }  &  -    &  -    &    -    &    -  &  -   &  - &  - \\

& SR              &   -      &    -     &   -     &    -  &    -  &    -  &  -    &  -    &    -    &    -  &  \cellcolor[HTML]{C0C0C0}{\color[HTML]{000000}True }   &  - &  - \\

& RI               &   -      &    -     &   -     &    -  &    -  &    -  &  -    &  -    &    -    &    -  &  -   &  - &  -  \\

& RS             &   \cellcolor[HTML]{C0C0C0}{\color[HTML]{000000}True }      &    -     &   -     &    \cellcolor[HTML]{C0C0C0}{\color[HTML]{000000}True }  &    -  &    -  &  \cellcolor[HTML]{C0C0C0}{\color[HTML]{000000}True }   &  -    &    \cellcolor[HTML]{C0C0C0}{\color[HTML]{000000}True }   &    -  &  -   &  - &  -   \\

& RD              &   -      &    -     &   -     &    -  &    -  &    -  &  -    &  -    &    -    &    -  &  -   &  - &  - \\

& BT             &   -      &    -     &   -     &    -  &    \cellcolor[HTML]{C0C0C0}{\color[HTML]{000000}True }  &    \cellcolor[HTML]{C0C0C0}{\color[HTML]{000000}True }  &  -    &  -    &    -    &    -  &  -   &  - &  - \\

\hline                  
\multirow{13}{*}{GraphCodeBERT} & No Aug                         &   -      &    -     &   -     &    -  &    -  &    -  &  -    &  -    &    -    &    -  &  \cellcolor[HTML]{C0C0C0}{\color[HTML]{000000}True }   &  - &  -       \\

& WordMixup                         &   -      &    -     &   \cellcolor[HTML]{C0C0C0}{\color[HTML]{000000}True }     &    -  &    -  &    -  &  -    &  -    &    -    &    -  &  \cellcolor[HTML]{C0C0C0}{\color[HTML]{000000}True }   &  - &  -   \\

& SenMixup                       &   -      &    \cellcolor[HTML]{C0C0C0}{\color[HTML]{000000}True }     &   -     &    -  &    -  &    -  &  -    &  -    &    -    &    -  &  -   &  - &  - \\

& Refactor             &   -      &    -     &   -     &    -  &    \cellcolor[HTML]{C0C0C0}{\color[HTML]{000000}True }  &    -  &  -    &  -    &    -    &    -  &  -   &  - &  - \\

& Rename             &   -      &    -     &   -     &    \cellcolor[HTML]{C0C0C0}{\color[HTML]{000000}True }  &    -  &    -  &  \cellcolor[HTML]{C0C0C0}{\color[HTML]{000000}True }    &  \cellcolor[HTML]{C0C0C0}{\color[HTML]{000000}True }    &    \cellcolor[HTML]{C0C0C0}{\color[HTML]{000000}True }    &    -  &  -   &  - &  - \\

& Dead              &   -      &    -     &   -     &    -  &    -  &    -  &  -    &  -    &    -    &    -  &  -   &  - &  - \\

& Inside             &   -      &    -     &   -     &    -  &    \cellcolor[HTML]{C0C0C0}{\color[HTML]{000000}True }  &    -  &  -    &  -    &    -    &    -  &  -   &  - &  - \\

& Outside             &   -      &    -     &   -     &    -  &    \cellcolor[HTML]{C0C0C0}{\color[HTML]{000000}True }  &    -  &  -    &  -    &    -    &    -  &  -   &  - &  - \\

& SR              &   -      &    -     &   -     &    -  &    \cellcolor[HTML]{C0C0C0}{\color[HTML]{000000}True }  &    -  &  -    &  -    &    -    &    -  &  -   &  \cellcolor[HTML]{C0C0C0}{\color[HTML]{000000}True } &  - \\

& RI               &   -      &    -     &   -     &    -  &    -  &    -  &  -    &  -    &    -    &    -  &  -   &  - &  -  \\

& RS             &   \cellcolor[HTML]{C0C0C0}{\color[HTML]{000000}True }      &    \cellcolor[HTML]{C0C0C0}{\color[HTML]{000000}True }     &   -     &    -  &    -  &    -  &  -    &  -    &    -    &    -  &  -   &  - &  -   \\

& RD              &   -      &    -     &   -     &    -  &    -  &    -  &  -    &  -    &    \cellcolor[HTML]{C0C0C0}{\color[HTML]{000000}True }   &    -  &  -   &  - &  - \\

& BT             &   -      &    -     &   -     &    -  &    -  &    -  &  -    &  -    &    -    &    -  &  -   &  - &  - \\

\hline  

\end{tabular}
}
\end{sidewaystable}



\begin{sidewaystable}[]
\centering
    \setlength{\tabcolsep}{3pt} 
    \renewcommand{\arraystretch}{1.3}
    \tiny
\caption{Results of statistical tests on Robustness. \textbf{True}: indicates that the comparison is statistically significant after the  \emph{Bonferroni correction} adjustment. Statistical test method: \emph{Wilcoxon signed-rank test}. A gray background highlights the result marked as \textbf{True}. For legibility, we simplify the terms \emph{Rename Operator}, \emph{Dead Operator}, \emph{Inside Operator}, and \emph{Outside Operator} to \emph{Rename}, \emph{Dead}, \emph{Inside}, and \emph{Outside}, respectively. Model: CodeBERT \& GraphCodeBERT.}
\label{table:statistical_testing_robust_codebert_graphcodebert}
\centering
\resizebox{\textwidth}{!}{
\begin{tabular}{lcccccccccccccccc}
\hline
& & \multicolumn{14}{c}{\textbf{Accuracy}}  \\ \hline
                    & DA method & No Aug & WordMixup & SenMixup  & Refactor & Rename & Dead  & Inside & Outside  & SR & RI &  RS &  RD & BT & \\ \hline
\multirow{13}{*}{CodeBERT} & No Aug                         &   -      &    -     &   \cellcolor[HTML]{C0C0C0}{\color[HTML]{000000}True }      &    -  &    -  &    -  &  -    &  -    &    \cellcolor[HTML]{C0C0C0}{\color[HTML]{000000}True }    &    \cellcolor[HTML]{C0C0C0}{\color[HTML]{000000}True }  &  -  &  - &  \cellcolor[HTML]{C0C0C0}{\color[HTML]{000000}True }       \\

& WordMixup                         &   -      &    -     &   \cellcolor[HTML]{C0C0C0}{\color[HTML]{000000}True }    &    -  &    -  &    -  &  -    &  -    &    \cellcolor[HTML]{C0C0C0}{\color[HTML]{000000}True }    &    -  &  \cellcolor[HTML]{C0C0C0}{\color[HTML]{000000}True }  &  - &  \cellcolor[HTML]{C0C0C0}{\color[HTML]{000000}True }        \\

& SenMixup                       &   \cellcolor[HTML]{C0C0C0}{\color[HTML]{000000}True }      &    \cellcolor[HTML]{C0C0C0}{\color[HTML]{000000}True }    &   -     &    \cellcolor[HTML]{C0C0C0}{\color[HTML]{000000}True } &    \cellcolor[HTML]{C0C0C0}{\color[HTML]{000000}True }   &    \cellcolor[HTML]{C0C0C0}{\color[HTML]{000000}True }   &  \cellcolor[HTML]{C0C0C0}{\color[HTML]{000000}True }     &  \cellcolor[HTML]{C0C0C0}{\color[HTML]{000000}True }     &    \cellcolor[HTML]{C0C0C0}{\color[HTML]{000000}True }     &    \cellcolor[HTML]{C0C0C0}{\color[HTML]{000000}True }   &  -  &  - &  \cellcolor[HTML]{C0C0C0}{\color[HTML]{000000}True }       \\

& Refactor             &   -      &    -     &   \cellcolor[HTML]{C0C0C0}{\color[HTML]{000000}True }      &    -  &    \cellcolor[HTML]{C0C0C0}{\color[HTML]{000000}True }  &    \cellcolor[HTML]{C0C0C0}{\color[HTML]{000000}True }  &  \cellcolor[HTML]{C0C0C0}{\color[HTML]{000000}True }    &  \cellcolor[HTML]{C0C0C0}{\color[HTML]{000000}True }   &    \cellcolor[HTML]{C0C0C0}{\color[HTML]{000000}True }    &    -  &  -  &  - &  \cellcolor[HTML]{C0C0C0}{\color[HTML]{000000}True }       \\

& Rename            &   -      &    -     &   \cellcolor[HTML]{C0C0C0}{\color[HTML]{000000}True }      &    \cellcolor[HTML]{C0C0C0}{\color[HTML]{000000}True }  &    -  &    -  &  \cellcolor[HTML]{C0C0C0}{\color[HTML]{000000}True }    &  \cellcolor[HTML]{C0C0C0}{\color[HTML]{000000}True }    &    -    &    -  &  -  &  - &  \cellcolor[HTML]{C0C0C0}{\color[HTML]{000000}True }       \\

& Dead            &   -      &    -     &   \cellcolor[HTML]{C0C0C0}{\color[HTML]{000000}True }     &    \cellcolor[HTML]{C0C0C0}{\color[HTML]{000000}True } &    -  &    -  &  \cellcolor[HTML]{C0C0C0}{\color[HTML]{000000}True }   & \cellcolor[HTML]{C0C0C0}{\color[HTML]{000000}True }    &    -    &    -  &  -  &  \cellcolor[HTML]{C0C0C0}{\color[HTML]{000000}True } &  \cellcolor[HTML]{C0C0C0}{\color[HTML]{000000}True }       \\

& Inside            &   -      &    -     &   \cellcolor[HTML]{C0C0C0}{\color[HTML]{000000}True }     &    \cellcolor[HTML]{C0C0C0}{\color[HTML]{000000}True }  &    \cellcolor[HTML]{C0C0C0}{\color[HTML]{000000}True }  &    \cellcolor[HTML]{C0C0C0}{\color[HTML]{000000}True } &  -    &  -    &    -    &    -  &  -  &  \cellcolor[HTML]{C0C0C0}{\color[HTML]{000000}True } &  \cellcolor[HTML]{C0C0C0}{\color[HTML]{000000}True }      \\

& Outside             &   -      &    -     &  \cellcolor[HTML]{C0C0C0}{\color[HTML]{000000}True }      &    \cellcolor[HTML]{C0C0C0}{\color[HTML]{000000}True }  &   \cellcolor[HTML]{C0C0C0}{\color[HTML]{000000}True }  &    \cellcolor[HTML]{C0C0C0}{\color[HTML]{000000}True } &  -    &  -    &    -    &    -  &  -  &  \cellcolor[HTML]{C0C0C0}{\color[HTML]{000000}True }& \cellcolor[HTML]{C0C0C0}{\color[HTML]{000000}True }       \\

& SR              &   \cellcolor[HTML]{C0C0C0}{\color[HTML]{000000}True }     &    \cellcolor[HTML]{C0C0C0}{\color[HTML]{000000}True }     &   \cellcolor[HTML]{C0C0C0}{\color[HTML]{000000}True }      &    \cellcolor[HTML]{C0C0C0}{\color[HTML]{000000}True }  &    -  &    -  &  -    &  -    &    -    &    -  &  \cellcolor[HTML]{C0C0C0}{\color[HTML]{000000}True }  &  \cellcolor[HTML]{C0C0C0}{\color[HTML]{000000}True } &  \cellcolor[HTML]{C0C0C0}{\color[HTML]{000000}True }       \\

& RI               &   \cellcolor[HTML]{C0C0C0}{\color[HTML]{000000}True }      &   -    &  \cellcolor[HTML]{C0C0C0}{\color[HTML]{000000}True }      &    -  &    -  &    -  &  -    &  -    &    -    &    -  &  -  &  \cellcolor[HTML]{C0C0C0}{\color[HTML]{000000}True } &  -      \\

& RS            &   -      &    \cellcolor[HTML]{C0C0C0}{\color[HTML]{000000}True }      &   -     &    -  &    -  &    -  &  -    &  -    &   \cellcolor[HTML]{C0C0C0}{\color[HTML]{000000}True }   &    -  &  -  &  - &  \cellcolor[HTML]{C0C0C0}{\color[HTML]{000000}True }       \\

& RD              &   -      &    -     &   -     &    -  &    -  &    \cellcolor[HTML]{C0C0C0}{\color[HTML]{000000}True } &  \cellcolor[HTML]{C0C0C0}{\color[HTML]{000000}True }    &  \cellcolor[HTML]{C0C0C0}{\color[HTML]{000000}True }    &   \cellcolor[HTML]{C0C0C0}{\color[HTML]{000000}True }    &    \cellcolor[HTML]{C0C0C0}{\color[HTML]{000000}True } &  -   &  - &  \cellcolor[HTML]{C0C0C0}{\color[HTML]{000000}True } \\

& BT             &   \cellcolor[HTML]{C0C0C0}{\color[HTML]{000000}True }      &    \cellcolor[HTML]{C0C0C0}{\color[HTML]{000000}True }      &   \cellcolor[HTML]{C0C0C0}{\color[HTML]{000000}True }     &    \cellcolor[HTML]{C0C0C0}{\color[HTML]{000000}True }  &   \cellcolor[HTML]{C0C0C0}{\color[HTML]{000000}True }  &    \cellcolor[HTML]{C0C0C0}{\color[HTML]{000000}True }  & \cellcolor[HTML]{C0C0C0}{\color[HTML]{000000}True }   &  \cellcolor[HTML]{C0C0C0}{\color[HTML]{000000}True }    &    \cellcolor[HTML]{C0C0C0}{\color[HTML]{000000}True }   &    -  & \cellcolor[HTML]{C0C0C0}{\color[HTML]{000000}True }  &  \cellcolor[HTML]{C0C0C0}{\color[HTML]{000000}True } &  -       \\

\hline                  
\multirow{13}{*}{GraphCodeBERT} & No Aug                         &   -      &    \cellcolor[HTML]{C0C0C0}{\color[HTML]{000000}True }     &   \cellcolor[HTML]{C0C0C0}{\color[HTML]{000000}True }   &    -  &    -  &    -  &  -    &  -    &   \cellcolor[HTML]{C0C0C0}{\color[HTML]{000000}True }   &    -  &  \cellcolor[HTML]{C0C0C0}{\color[HTML]{000000}True }  &  - &  \cellcolor[HTML]{C0C0C0}{\color[HTML]{000000}True }      \\

& WordMixup                         &   \cellcolor[HTML]{C0C0C0}{\color[HTML]{000000}True }     &    -     &   \cellcolor[HTML]{C0C0C0}{\color[HTML]{000000}True }     &    -  &    -  &   \cellcolor[HTML]{C0C0C0}{\color[HTML]{000000}True }  &  \cellcolor[HTML]{C0C0C0}{\color[HTML]{000000}True }    &  \cellcolor[HTML]{C0C0C0}{\color[HTML]{000000}True }    &   \cellcolor[HTML]{C0C0C0}{\color[HTML]{000000}True }   &   \cellcolor[HTML]{C0C0C0}{\color[HTML]{000000}True }  &  -  &  - &  \cellcolor[HTML]{C0C0C0}{\color[HTML]{000000}True }      \\

& SenMixup                       &   \cellcolor[HTML]{C0C0C0}{\color[HTML]{000000}True }      &    \cellcolor[HTML]{C0C0C0}{\color[HTML]{000000}True }    &   -     &   \cellcolor[HTML]{C0C0C0}{\color[HTML]{000000}True }  &   \cellcolor[HTML]{C0C0C0}{\color[HTML]{000000}True } &    \cellcolor[HTML]{C0C0C0}{\color[HTML]{000000}True }  &  \cellcolor[HTML]{C0C0C0}{\color[HTML]{000000}True }   &  \cellcolor[HTML]{C0C0C0}{\color[HTML]{000000}True }   &    \cellcolor[HTML]{C0C0C0}{\color[HTML]{000000}True }    &    \cellcolor[HTML]{C0C0C0}{\color[HTML]{000000}True } &  -  &  - &  \cellcolor[HTML]{C0C0C0}{\color[HTML]{000000}True }      \\

& Refactor             &   -      &    -     &  \cellcolor[HTML]{C0C0C0}{\color[HTML]{000000}True }     &    -  &   \cellcolor[HTML]{C0C0C0}{\color[HTML]{000000}True }  &    \cellcolor[HTML]{C0C0C0}{\color[HTML]{000000}True }  &  \cellcolor[HTML]{C0C0C0}{\color[HTML]{000000}True }   &  \cellcolor[HTML]{C0C0C0}{\color[HTML]{000000}True }   &    \cellcolor[HTML]{C0C0C0}{\color[HTML]{000000}True }   &    -  &  -  &  - &  \cellcolor[HTML]{C0C0C0}{\color[HTML]{000000}True }       \\

& Rename            &   -      &    -     &   \cellcolor[HTML]{C0C0C0}{\color[HTML]{000000}True }     &    \cellcolor[HTML]{C0C0C0}{\color[HTML]{000000}True }  &    -  &   \cellcolor[HTML]{C0C0C0}{\color[HTML]{000000}True }  &  \cellcolor[HTML]{C0C0C0}{\color[HTML]{000000}True }    &  \cellcolor[HTML]{C0C0C0}{\color[HTML]{000000}True }   &   \cellcolor[HTML]{C0C0C0}{\color[HTML]{000000}True }   &    -  & \cellcolor[HTML]{C0C0C0}{\color[HTML]{000000}True }  &  - &  \cellcolor[HTML]{C0C0C0}{\color[HTML]{000000}True }      \\

& Dead            &   -      &    \cellcolor[HTML]{C0C0C0}{\color[HTML]{000000}True }     &  \cellcolor[HTML]{C0C0C0}{\color[HTML]{000000}True }     &   \cellcolor[HTML]{C0C0C0}{\color[HTML]{000000}True } &    \cellcolor[HTML]{C0C0C0}{\color[HTML]{000000}True }  &    -  &  -    &  -    &    -    &    -  &  \cellcolor[HTML]{C0C0C0}{\color[HTML]{000000}True }  &  - &  -       \\

& Inside            &   -      &    \cellcolor[HTML]{C0C0C0}{\color[HTML]{000000}True }    &  \cellcolor[HTML]{C0C0C0}{\color[HTML]{000000}True }    &    \cellcolor[HTML]{C0C0C0}{\color[HTML]{000000}True }  &   \cellcolor[HTML]{C0C0C0}{\color[HTML]{000000}True }  &    -  &  -    &  -    &    -    &    -  &  \cellcolor[HTML]{C0C0C0}{\color[HTML]{000000}True }  &  - &  -       \\

& Outside             &   -      &   \cellcolor[HTML]{C0C0C0}{\color[HTML]{000000}True }     &  \cellcolor[HTML]{C0C0C0}{\color[HTML]{000000}True }     &   \cellcolor[HTML]{C0C0C0}{\color[HTML]{000000}True }  &    \cellcolor[HTML]{C0C0C0}{\color[HTML]{000000}True }  &    -  &  -    &  -    &    -    &    -  &  \cellcolor[HTML]{C0C0C0}{\color[HTML]{000000}True }  &  - &  -       \\

& SR              &   \cellcolor[HTML]{C0C0C0}{\color[HTML]{000000}True }     &    \cellcolor[HTML]{C0C0C0}{\color[HTML]{000000}True }     &   \cellcolor[HTML]{C0C0C0}{\color[HTML]{000000}True }     &    \cellcolor[HTML]{C0C0C0}{\color[HTML]{000000}True }  &    \cellcolor[HTML]{C0C0C0}{\color[HTML]{000000}True }  &    -  &  -    &  -    &    -    &    -  &  \cellcolor[HTML]{C0C0C0}{\color[HTML]{000000}True }  &  \cellcolor[HTML]{C0C0C0}{\color[HTML]{000000}True } &  -       \\

& RI               &   -      &    \cellcolor[HTML]{C0C0C0}{\color[HTML]{000000}True }    &   \cellcolor[HTML]{C0C0C0}{\color[HTML]{000000}True }     &    -  &    -  &    -  &  -    &  -    &    -    &    -  &  -  &  - &  -       \\

& RS            &   \cellcolor[HTML]{C0C0C0}{\color[HTML]{000000}True }      &    -     &   -     &    -  &    \cellcolor[HTML]{C0C0C0}{\color[HTML]{000000}True }  &    \cellcolor[HTML]{C0C0C0}{\color[HTML]{000000}True }  &  \cellcolor[HTML]{C0C0C0}{\color[HTML]{000000}True }    &  \cellcolor[HTML]{C0C0C0}{\color[HTML]{000000}True }    &   \cellcolor[HTML]{C0C0C0}{\color[HTML]{000000}True }    &    -  &  -  &  - &  \cellcolor[HTML]{C0C0C0}{\color[HTML]{000000}True }      \\

& RD              &   -      &    -     &   -     &    -  &    -  &    -  &  -    &  -    &    \cellcolor[HTML]{C0C0C0}{\color[HTML]{000000}True }    &    -  &  -   &  - & \cellcolor[HTML]{C0C0C0}{\color[HTML]{000000}True } \\

& BT             &  \cellcolor[HTML]{C0C0C0}{\color[HTML]{000000}True }   &    \cellcolor[HTML]{C0C0C0}{\color[HTML]{000000}True }     &   \cellcolor[HTML]{C0C0C0}{\color[HTML]{000000}True }     &    \cellcolor[HTML]{C0C0C0}{\color[HTML]{000000}True }  &    \cellcolor[HTML]{C0C0C0}{\color[HTML]{000000}True }  &    -  &  -    &  -    &    -    &    -  &  \cellcolor[HTML]{C0C0C0}{\color[HTML]{000000}True }  &  \cellcolor[HTML]{C0C0C0}{\color[HTML]{000000}True } &  -       \\

\hline  

\end{tabular}
}
\end{sidewaystable}


\begin{sidewaystable}[]
\centering
    \setlength{\tabcolsep}{3pt} 
    \renewcommand{\arraystretch}{1.3}
    \tiny
\caption{Results of statistical tests on Accuracy. \textbf{True}: indicates that the comparison is statistically significant after the  \emph{Bonferroni correction} adjustment. Statistical test method: \emph{Wilcoxon signed-rank test}. A gray background highlights the result marked as \textbf{True}. For legibility, we simplify the terms \emph{Rename Operator}, \emph{Dead Operator}, \emph{Inside Operator}, and \emph{Outside Operator} to \emph{Rename}, \emph{Dead}, \emph{Inside}, and \emph{Outside}, respectively. Model: CodeBERT (10\%) \& GraphCodeBERT (10\%).}
\label{table:statistical_testing_acc_10_codebert_graphcodebert}
\centering
\resizebox{\textwidth}{!}{
\begin{tabular}{lcccccccccccccccc}
\hline
& & \multicolumn{14}{c}{\textbf{Accuracy}}  \\ \hline
                    & DA method & No Aug & WordMixup & SenMixup  & Refactor & Rename & Dead  & Inside & Outside  & SR & RI &  RS &  RD & BT & \\ \hline
\multirow{13}{*}{{\begin{tabular}[c]{@{}c@{}}CodeBERT\\ (10\%)\end{tabular}}} & No Aug                         &   -      &    -     &   -     &    -  &    - &    -  &  -    &  -    &    -    &    -  &  -  &  - &  -       \\

& WordMixup                         &   -      &    -     &   -     &    -  &    - &    -  &  -    &  -    &    -    &    -  &  -  &  - &  -       \\

& SenMixup                        &   -      &    -     &   -     &    -  &    - &    -  &  -    &  -    &    -    &    -  &  -  &  - &  -       \\

& Refactor              &   -      &    -     &   -     &    -  &    - &    -  &  -    &  -    &    -    &    -  &  -  &  - &  -       \\

& Rename             &   -      &    -     &   -     &    -  &    - &    \cellcolor[HTML]{C0C0C0}{\color[HTML]{000000}True }  &  -    &  -    &    -    &    -  &  -  &  - &  -       \\

& Dead              &   -      &    -     &   -     &    -  &    \cellcolor[HTML]{C0C0C0}{\color[HTML]{000000}True } &    -  &  -    &  -    &    -    &    -  &  -  &  - &  -       \\

& Inside              &   -      &    -     &   -     &    -  &    - &    -  &  -    &  -    &    -    &    -  &  -  &  - &  -       \\

& Outside             &   -      &    -     &   -     &    -  &    - &    -  &  -    &  -    &    -    &    -  &  -  &  - &  -       \\

& SR               &   -      &    -     &   -     &    -  &    - &    -  &  -    &  -    &    -    &    -  &  -  &  - &  -       \\

& RI            &   -      &    -     &   -     &    -  &    - &    -  &  -    &  -    &    -    &    -  &  -  &  - &  -       \\

& RS              &   -      &    -     &   -     &    -  &    - &    -  &  -    &  -    &    -    &    -  &  -  &  - &  -       \\

& RD              &   -      &    -     &   -     &    -  &    -  &    -  &  -    &  -    &    -    &    -  &  -   &  - &  - \\

& BT              &   -      &    -     &   -     &    -  &    - &    -  &  -    &  -    &    -    &    -  &  -  &  - &  -       \\

\hline                  
\multirow{13}{*}{{\begin{tabular}[c]{@{}c@{}}GraphCodeBERT\\ (10\%)\end{tabular}}} & No Aug                         &   -      &    -     &   -     &    -  &    - &    -  &  -    &  -    &    -    &    -  &  -  &  - &  -       \\

& WordMixup                         &   -      &    -     &   -     &    -  &    - &    -  &  -    &  -    &    -    &    -  &  -  &  - &  -       \\

& SenMixup                        &   -      &    -     &   -     &    -  &    - &    -  &  -    &  -    &    -    &    -  &  -  &  - &  -       \\

& Refactor              &   -      &    -     &   -     &    -  &    - &    -  &  -    &  -    &    -    &    -  &  -  &  - &  -       \\

& Rename             &   -      &    -     &   -     &    -  &    - &    -  &  -    &  -    &    -    &    -  &  -  &  - &  -       \\

& Dead              &   -      &    -     &   -     &    -  &    - &    -  &  \cellcolor[HTML]{C0C0C0}{\color[HTML]{000000}True }    &  -    &    -    &    -  &  -  &  - &  -       \\

& Inside              &   -      &    -     &   -     &    -  &    - &    \cellcolor[HTML]{C0C0C0}{\color[HTML]{000000}True }  &  -    &  -    &    -    &    -  &  -  &  - &  -       \\

& Outside             &   -      &    -     &   -     &    -  &    - &    -  &  -    &  -    &    -    &    -  &  -  &  - &  -       \\

& SR               &   -      &    -     &   -     &    -  &    - &    -  &  -    &  -    &    -    &    -  &  -  &  - &  -       \\

& RI            &   -      &    -     &   -     &    -  &    - &    -  &  -    &  -    &    -    &    -  &  -  &  - &  -       \\

& RS              &   -      &    -     &   -     &    -  &    - &    -  &  -    &  -    &    -    &    -  &  -  &  - &  -       \\

& RD              &   -      &    -     &   -     &    -  &    -  &    -  &  -    &  -    &    -    &    -  &  -   &  - &  - \\

& BT              &   -      &    -     &   -     &    -  &    - &    -  &  -    &  -    &    -    &    -  &  -  &  - &  -       \\

\hline  

\end{tabular}
}
\end{sidewaystable}

\begin{sidewaystable}[]
\centering
    \setlength{\tabcolsep}{3pt} 
    \renewcommand{\arraystretch}{1.3}
    \tiny
\caption{Results of statistical tests on Accuracy. \textbf{True}: indicates that the comparison is statistically significant after the  \emph{Bonferroni correction} adjustment. Statistical test method: \emph{Wilcoxon signed-rank test}. A gray background highlights the result marked as \textbf{True}. For legibility, we simplify the terms \emph{Rename Operator}, \emph{Dead Operator}, \emph{Inside Operator}, and \emph{Outside Operator} to \emph{Rename}, \emph{Dead}, \emph{Inside}, and \emph{Outside}, respectively. Model: CodeBERT (5\%) \& GraphCodeBERT (5\%).}
\label{table:statistical_testing_acc_05_codebert_graphcodebert}
\centering
\resizebox{\textwidth}{!}{
\begin{tabular}{lcccccccccccccccc}
\hline
& & \multicolumn{14}{c}{\textbf{Accuracy}}  \\ \hline
                    & DA method & No Aug & WordMixup & SenMixup  & Refactor & Rename & Dead  & Inside & Outside  & SR & RI &  RS &  RD & BT & \\ \hline
\multirow{13}{*}{{\begin{tabular}[c]{@{}c@{}}CodeBERT\\ (5\%)\end{tabular}}} & No Aug                         &   -      &    -     &   -     &    -  &    - &    -  &  -    &  -    &    -    &    -  &  -  &  - &  -       \\

& WordMixup                         &   -      &    -     &   -     &    -  &    - &    -  &  -    &  -    &    -    &    -  &  -  &  - &  -       \\

& SenMixup                        &   -      &    -     &   -     &    -  &    - &    -  &  -    &  -    &    -    &    -  &  -  &  - &  -       \\

& Refactor              &   -      &    -     &   -     &    -  &    - &    -  &  -    &  -    &    -    &    -  &  -  &  \cellcolor[HTML]{C0C0C0}{\color[HTML]{000000}True } &  -       \\

& Rename             &   -      &    -     &   -     &    -  &    - &    -  &  -    &  -    &    -    &    -  &  -  &  - &  -       \\

& Dead              &   -      &    -     &   -     &    -  &    - &    -  &  \cellcolor[HTML]{C0C0C0}{\color[HTML]{000000}True }  &  -    &    -    &    -  &  -  &  - &  -       \\

& Inside              &   -      &    -     &   -     &    -  &    - &    \cellcolor[HTML]{C0C0C0}{\color[HTML]{000000}True }  &  -    &  -    &    -    &    -  &  -  &  - &  -       \\

& Outside             &   -      &    -     &   -     &    -  &    - &    -  &  -    &  -    &    -    &    -  &  -  &  - &  -       \\

& SR               &   -      &    -     &   -     &    -  &    - &    -  &  -    &  -    &    -    &    -  &  \cellcolor[HTML]{C0C0C0}{\color[HTML]{000000}True }  &  - &  -       \\

& RI            &   -      &    -     &   -     &    -  &    - &    -  &  -    &  -    &    -    &    -  &  -  &  - &  -       \\

& RS              &   -      &    -     &   -     &    -  &    - &    -  &  -    &  -    &    \cellcolor[HTML]{C0C0C0}{\color[HTML]{000000}True }   &    -  &  -  &  - &  -       \\

& RD              &   -      &    -     &   -     &   \cellcolor[HTML]{C0C0C0}{\color[HTML]{000000}True } &    -  &    -  &  -    &  -    &    -    &    -  &  -   &  - &  - \\

& BT              &   -      &    -     &   -     &    -  &    - &    -  &  -    &  -    &    -    &    -  &  -  &  - &  -       \\

\hline                  
\multirow{13}{*}{{\begin{tabular}[c]{@{}c@{}}GraphCodeBERT\\ (5\%)\end{tabular}}}  & No Aug                         &   -      &    -     &   -     &    -  &    - &    -  &  -    &  -    &    -    &    -  &  -  &  - &  -       \\

& WordMixup                         &   -      &    -     &   -     &    -  &    - &    -  &  -    &  -    &    -    &    -  &  -  &  - &  -       \\

& SenMixup                        &   -      &    -     &   -     &    -  &    - &    -  &  -    &  -    &    -    &    -  &  -  &  - &  -       \\

& Refactor              &   -      &    -     &   -     &    -  &    - &    -  &  -    &  -    &    -    &    -  &  -  &  - &  \cellcolor[HTML]{C0C0C0}{\color[HTML]{000000}True }       \\

& Rename             &   -      &    -     &   -     &    -  &    - &    -  &  \cellcolor[HTML]{C0C0C0}{\color[HTML]{000000}True }    &  -    &    -    &    \cellcolor[HTML]{C0C0C0}{\color[HTML]{000000}True }  &  -  &  - &  -       \\

& Dead              &   -      &    -     &   -     &    -  &    - &    -  &  -    &  -    &    -    &    -  &  -  &  - &  -       \\

& Inside              &   -      &    -     &   -     &    -  &    \cellcolor[HTML]{C0C0C0}{\color[HTML]{000000}True } &    -  &  -    &  -    &    -    &    -  &  -  &  - &  -       \\

& Outside             &   -      &    -     &   -     &    -  &    - &    -  &  -    &  -    &    -    &    -  &  -  &  - &  -       \\

& SR               &   -      &    -     &   -     &    -  &    - &    -  &  -    &  -    &    -    &    -  &  \cellcolor[HTML]{C0C0C0}{\color[HTML]{000000}True }  &  \cellcolor[HTML]{C0C0C0}{\color[HTML]{000000}True } &  -       \\

& RI            &   -      &    -     &   -     &    -  &    \cellcolor[HTML]{C0C0C0}{\color[HTML]{000000}True } &    -  &  -    &  -    &    -    &    -  &  \cellcolor[HTML]{C0C0C0}{\color[HTML]{000000}True } &  - &  -       \\

& RS              &   -      &    -     &   -     &    -  &    - &    -  &  -    &  -    &    \cellcolor[HTML]{C0C0C0}{\color[HTML]{000000}True }   &    \cellcolor[HTML]{C0C0C0}{\color[HTML]{000000}True }  &  -  &  - &  -       \\

& RD              &   -      &    -     &   -     &    -  &    -  &    -  &  -    &  -    &    \cellcolor[HTML]{C0C0C0}{\color[HTML]{000000}True }    &    -  &  -   &  - &  - \\

& BT              &   -      &    -     &   -     &   \cellcolor[HTML]{C0C0C0}{\color[HTML]{000000}True }  &    - &    -  &  -    &  -    &    -    &    -  &  -  &  - &  -       \\

\hline  

\end{tabular}
}
\end{sidewaystable}

\begin{sidewaystable}[]
\centering
    \setlength{\tabcolsep}{3pt} 
    \renewcommand{\arraystretch}{1.3}
    \tiny
\caption{Results of statistical tests on Accuracy. \textbf{True}: indicates that the comparison is statistically significant after the  \emph{Bonferroni correction} adjustment. Statistical test method: \emph{Wilcoxon signed-rank test}. A gray background highlights the result marked as \textbf{True}. For legibility, we simplify the terms \emph{Rename Operator}, \emph{Dead Operator}, \emph{Inside Operator}, and \emph{Outside Operator} to \emph{Rename}, \emph{Dead}, \emph{Inside}, and \emph{Outside}, respectively. Model: CodeBERT (3\%) \& GraphCodeBERT (3\%).}
\label{table:statistical_testing_acc_03_codebert_graphcodebert}
\centering
\resizebox{\textwidth}{!}{
\begin{tabular}{lcccccccccccccccc}
\hline
& & \multicolumn{14}{c}{\textbf{Accuracy}}  \\ \hline
                    & DA method & No Aug & WordMixup & SenMixup  & Refactor & Rename & Dead  & Inside & Outside  & SR & RI &  RS &  RD & BT & \\ \hline
\multirow{13}{*}{{\begin{tabular}[c]{@{}c@{}}CodeBERT\\ (3\%)\end{tabular}}} & No Aug                         &   -      &    -     &   -     &    -  &    - &    -  &  -    &  -    &    -    &    -  &  -  &  - &  -       \\

& WordMixup                         &   -      &    -     &   -     &    -  &    - &    -  &  -    &  -    &    \cellcolor[HTML]{C0C0C0}{\color[HTML]{000000}True }    &    -  &  -  &  - &  -       \\

& SenMixup                        &   -      &    -     &   -     &    \cellcolor[HTML]{C0C0C0}{\color[HTML]{000000}True }  &    - &    -  &  \cellcolor[HTML]{C0C0C0}{\color[HTML]{000000}True }   &  \cellcolor[HTML]{C0C0C0}{\color[HTML]{000000}True }    &   \cellcolor[HTML]{C0C0C0}{\color[HTML]{000000}True }   &    \cellcolor[HTML]{C0C0C0}{\color[HTML]{000000}True }  &  -  &  - &  -       \\

& Refactor              &   -      &    -     &   \cellcolor[HTML]{C0C0C0}{\color[HTML]{000000}True }   &    -  &    - &    -  &  -    &  -    &    -    &    -  &  -  &  - &  -       \\

& Rename             &   -      &    -     &   -     &    -  &    - &    -  &  \cellcolor[HTML]{C0C0C0}{\color[HTML]{000000}True }  &  -    &    \cellcolor[HTML]{C0C0C0}{\color[HTML]{000000}True }   &    -  &  -  &  - &  -       \\

& Dead              &   -      &    -     &   -     &    -  &    - &    -  &  -    &  -    &    -    &    -  &  -  &  - &  -       \\

& Inside              &   -      &    -     &  \cellcolor[HTML]{C0C0C0}{\color[HTML]{000000}True }     &    -  &   \cellcolor[HTML]{C0C0C0}{\color[HTML]{000000}True } &    -  &  -    &  -    &    -    &    -  &  -  &  - &  -       \\

& Outside             &   -      &    -     &   \cellcolor[HTML]{C0C0C0}{\color[HTML]{000000}True }    &    -  &    - &    -  &  -    &  -    &    \cellcolor[HTML]{C0C0C0}{\color[HTML]{000000}True }    &    -  &  -  &  - &  -       \\

& SR               &   -      &   \cellcolor[HTML]{C0C0C0}{\color[HTML]{000000}True }    &   \cellcolor[HTML]{C0C0C0}{\color[HTML]{000000}True }    &    -  &    \cellcolor[HTML]{C0C0C0}{\color[HTML]{000000}True } &    -  &  -    & \cellcolor[HTML]{C0C0C0}{\color[HTML]{000000}True }   &    -    &    -  &  -  &  - &  -       \\

& RI            &   -      &    -     &   \cellcolor[HTML]{C0C0C0}{\color[HTML]{000000}True }     &    -  &    - &    -  &  -    &  -    &    -    &    -  &  -  &  - &  -       \\

& RS              &   -      &    -     &   -     &    -  &    - &    -  &  -    &  -    &    -    &    -  &  -  &  - &  -       \\

& RD              &   -      &    -     &   -     &    -  &    -  &    -  &  -    &  -    &    -    &    -  &  -   &  - &  - \\

& BT              &   -      &    -     &   -     &    -  &    - &    -  &  -    &  -    &    -    &    -  &  -  &  - &  -       \\

\hline                  
\multirow{13}{*}{{\begin{tabular}[c]{@{}c@{}}GraphCodeBERT\\ (3\%)\end{tabular}}} & No Aug                         &   -      &    -     &   -     &    -  &    - &    -  &  -    &  -    &    -    &     \cellcolor[HTML]{C0C0C0}{\color[HTML]{000000}True }   &  -  &   \cellcolor[HTML]{C0C0C0}{\color[HTML]{000000}True }  &  -       \\

& WordMixup                         &   -      &    -     &    \cellcolor[HTML]{C0C0C0}{\color[HTML]{000000}True }      &    -  &    - &    -  &  -    &  -    &    -    &     \cellcolor[HTML]{C0C0C0}{\color[HTML]{000000}True }   &  -  &  - &  -       \\

& SenMixup                        &   -      &     \cellcolor[HTML]{C0C0C0}{\color[HTML]{000000}True }      &   -     &    -  &    - &     \cellcolor[HTML]{C0C0C0}{\color[HTML]{000000}True }   &  -    &  -    &    -    &     \cellcolor[HTML]{C0C0C0}{\color[HTML]{000000}True }  &  -  &  - &  -       \\

& Refactor              &   -      &    -     &   -     &    -  &    \cellcolor[HTML]{C0C0C0}{\color[HTML]{000000}True }  &    -  &  -    &  -    &    -    &    -  &  -  &  - &  -       \\

& Rename             &   -      &    -     &   -     &   \cellcolor[HTML]{C0C0C0}{\color[HTML]{000000}True }   &    - &    -  &  -    &   \cellcolor[HTML]{C0C0C0}{\color[HTML]{000000}True }     &     \cellcolor[HTML]{C0C0C0}{\color[HTML]{000000}True }    &     \cellcolor[HTML]{C0C0C0}{\color[HTML]{000000}True }  &  -  &  - &  -       \\

& Dead              &   -      &    -     &   \cellcolor[HTML]{C0C0C0}{\color[HTML]{000000}True }     &    -  &    - &    -  &  -    &  -    &    -    &     \cellcolor[HTML]{C0C0C0}{\color[HTML]{000000}True }   &  -  &  - &  -       \\

& Inside              &   -      &    -     &   -     &    -  &    - &    -  &  -    &  -    &    -    &    -  &  -  &  - &  -       \\

& Outside             &   -      &    -     &   -     &    -  &     \cellcolor[HTML]{C0C0C0}{\color[HTML]{000000}True }  &    -  &  -    &  -    &    -    &    -  &  -  &  - &  -       \\

& SR               &   -      &    -     &   -     &    -  &     \cellcolor[HTML]{C0C0C0}{\color[HTML]{000000}True }  &    -  &  -    &  -    &    -    &    -  &   \cellcolor[HTML]{C0C0C0}{\color[HTML]{000000}True }   &  - &  -       \\

& RI            &    \cellcolor[HTML]{C0C0C0}{\color[HTML]{000000}True }      &     \cellcolor[HTML]{C0C0C0}{\color[HTML]{000000}True }      &    \cellcolor[HTML]{C0C0C0}{\color[HTML]{000000}True }      &    -  &     \cellcolor[HTML]{C0C0C0}{\color[HTML]{000000}True }  &     \cellcolor[HTML]{C0C0C0}{\color[HTML]{000000}True }   &  -    &  -    &    -    &    -  &   \cellcolor[HTML]{C0C0C0}{\color[HTML]{000000}True }  &   \cellcolor[HTML]{C0C0C0}{\color[HTML]{000000}True } &  -       \\

& RS              &   -      &    -     &   -     &    -  &    - &    -  &  -    &  -    &     \cellcolor[HTML]{C0C0C0}{\color[HTML]{000000}True }    &    \cellcolor[HTML]{C0C0C0}{\color[HTML]{000000}True }  &  -  &  - &  -       \\

& RD              &   \cellcolor[HTML]{C0C0C0}{\color[HTML]{000000}True }      &    -     &   -     &    -  &    -  &    -  &  -    &  -    &    -    &     \cellcolor[HTML]{C0C0C0}{\color[HTML]{000000}True }   &  -   &  - &  - \\

& BT              &   -      &    -     &   -     &    -  &    - &    -  &  -    &  -    &    -    &    -  &  -  &  - &  -       \\

\hline  

\end{tabular}
}
\end{sidewaystable}

\begin{sidewaystable}[]
\centering
    \setlength{\tabcolsep}{3pt} 
    \renewcommand{\arraystretch}{1.3}
    \tiny
\caption{Results of statistical tests on Accuracy. \textbf{True}: indicates that the comparison is statistically significant after the  \emph{Bonferroni correction} adjustment. Statistical test method: \emph{Wilcoxon signed-rank test}. A gray background highlights the result marked as \textbf{True}. For legibility, we simplify the terms \emph{Rename Operator}, \emph{Dead Operator}, \emph{Inside Operator}, and \emph{Outside Operator} to \emph{Rename}, \emph{Dead}, \emph{Inside}, and \emph{Outside}, respectively. Model: CodeBERT (1\%) \& GraphCodeBERT (1\%).}
\label{table:statistical_testing_acc_01_codebert_graphcodebert}
\centering
\resizebox{\textwidth}{!}{
\begin{tabular}{lcccccccccccccccc}
\hline
& & \multicolumn{14}{c}{\textbf{Accuracy}}  \\ \hline
                    & DA method & No Aug & WordMixup & SenMixup  & Refactor & Rename & Dead  & Inside & Outside  & SR & RI &  RS &  RD & BT & \\ \hline
\multirow{13}{*}{{\begin{tabular}[c]{@{}c@{}}CodeBERT\\ (1\%)\end{tabular}}} & No Aug                         &   -      &     \cellcolor[HTML]{C0C0C0}{\color[HTML]{000000}True }     &    \cellcolor[HTML]{C0C0C0}{\color[HTML]{000000}True }     &    -  &    - &    -  &  -    &  -    &    -    &    -  &  -  &  - &  -       \\

& WordMixup                         &    \cellcolor[HTML]{C0C0C0}{\color[HTML]{000000}True }    &    -     &   \cellcolor[HTML]{C0C0C0}{\color[HTML]{000000}True }     &     \cellcolor[HTML]{C0C0C0}{\color[HTML]{000000}True } &    - &    -  &   \cellcolor[HTML]{C0C0C0}{\color[HTML]{000000}True }   &  \cellcolor[HTML]{C0C0C0}{\color[HTML]{000000}True }    &     \cellcolor[HTML]{C0C0C0}{\color[HTML]{000000}True }   &    -  &  -  &   \cellcolor[HTML]{C0C0C0}{\color[HTML]{000000}True } &  -       \\

& SenMixup                        &    \cellcolor[HTML]{C0C0C0}{\color[HTML]{000000}True }      &     \cellcolor[HTML]{C0C0C0}{\color[HTML]{000000}True }     &   -     &     \cellcolor[HTML]{C0C0C0}{\color[HTML]{000000}True }  &    \cellcolor[HTML]{C0C0C0}{\color[HTML]{000000}True } &     \cellcolor[HTML]{C0C0C0}{\color[HTML]{000000}True } &   \cellcolor[HTML]{C0C0C0}{\color[HTML]{000000}True }    &   \cellcolor[HTML]{C0C0C0}{\color[HTML]{000000}True }    &     \cellcolor[HTML]{C0C0C0}{\color[HTML]{000000}True }    &    -  &   \cellcolor[HTML]{C0C0C0}{\color[HTML]{000000}True }  &   \cellcolor[HTML]{C0C0C0}{\color[HTML]{000000}True } &  -       \\

& Refactor              &   -      &     \cellcolor[HTML]{C0C0C0}{\color[HTML]{000000}True }     &   \cellcolor[HTML]{C0C0C0}{\color[HTML]{000000}True }     &    -  &     \cellcolor[HTML]{C0C0C0}{\color[HTML]{000000}True } &    -  &  -    &  -    &    -    &    -  &  -  &  - &  -       \\

& Rename             &   -      &    -     &    \cellcolor[HTML]{C0C0C0}{\color[HTML]{000000}True }    &    \cellcolor[HTML]{C0C0C0}{\color[HTML]{000000}True }  &    - &    -  &   \cellcolor[HTML]{C0C0C0}{\color[HTML]{000000}True }    &   \cellcolor[HTML]{C0C0C0}{\color[HTML]{000000}True }    &    -    &    -  &  -  &  - &  -       \\

& Dead              &   -      &    -     &    \cellcolor[HTML]{C0C0C0}{\color[HTML]{000000}True }     &    -  &    - &    -  &  \cellcolor[HTML]{C0C0C0}{\color[HTML]{000000}True }   &  -    &    -    &    -  &  -  &  - &  -       \\

& Inside              &   -      &    \cellcolor[HTML]{C0C0C0}{\color[HTML]{000000}True }     &    \cellcolor[HTML]{C0C0C0}{\color[HTML]{000000}True }     &    -  &     \cellcolor[HTML]{C0C0C0}{\color[HTML]{000000}True } &     \cellcolor[HTML]{C0C0C0}{\color[HTML]{000000}True } &  -    &  -    &    -    &    -  &  -  &  - &  -       \\

& Outside             &   -      &     \cellcolor[HTML]{C0C0C0}{\color[HTML]{000000}True }  &    \cellcolor[HTML]{C0C0C0}{\color[HTML]{000000}True }    &    -  &    \cellcolor[HTML]{C0C0C0}{\color[HTML]{000000}True } &    -  &  -    &  -    &    -    &    -  &  -  &  - &  -       \\

& SR               &   -      &    \cellcolor[HTML]{C0C0C0}{\color[HTML]{000000}True }    &    \cellcolor[HTML]{C0C0C0}{\color[HTML]{000000}True }    &    -  &    - &    -  &  -    &  -    &    -    &    -  &  -  &  - &  -       \\

& RI            &   -      &    -     &   -     &    -  &    - &    -  &  -    &  -    &    -    &    -  &  -  &  - &  -       \\

& RS              &   -      &    -     &    \cellcolor[HTML]{C0C0C0}{\color[HTML]{000000}True }    &    -  &    - &    -  &  -    &  -    &    -    &    -  &  -  &  - &  -       \\

& RD              &   -      &     \cellcolor[HTML]{C0C0C0}{\color[HTML]{000000}True }    &   \cellcolor[HTML]{C0C0C0}{\color[HTML]{000000}True }    &    -  &    -  &    -  &  -    &  -    &    -    &    -  &  -   &  - &  - \\

& BT              &   -      &    -     &   -     &    -  &    - &    -  &  -    &  -    &    -    &    -  &  -  &  - &  -       \\

\hline                  
\multirow{13}{*}{{\begin{tabular}[c]{@{}c@{}}GraphCodeBERT\\ (1\%)\end{tabular}}}  & No Aug                         &   -      &    -     &   -     &    -  &    - &    -  &  -    &  -    &   \cellcolor[HTML]{C0C0C0}{\color[HTML]{000000}True }   &    \cellcolor[HTML]{C0C0C0}{\color[HTML]{000000}True }  &  -  &  - &  -       \\

& WordMixup                         &   -      &    -     &   -     &     \cellcolor[HTML]{C0C0C0}{\color[HTML]{000000}True } &    \cellcolor[HTML]{C0C0C0}{\color[HTML]{000000}True } &     \cellcolor[HTML]{C0C0C0}{\color[HTML]{000000}True }  &   \cellcolor[HTML]{C0C0C0}{\color[HTML]{000000}True }    &   \cellcolor[HTML]{C0C0C0}{\color[HTML]{000000}True }   &     \cellcolor[HTML]{C0C0C0}{\color[HTML]{000000}True }   &     \cellcolor[HTML]{C0C0C0}{\color[HTML]{000000}True }  &  -  &  - &  -       \\

& SenMixup                        &   -      &    -     &   -     &     \cellcolor[HTML]{C0C0C0}{\color[HTML]{000000}True }  &     \cellcolor[HTML]{C0C0C0}{\color[HTML]{000000}True } &    \cellcolor[HTML]{C0C0C0}{\color[HTML]{000000}True }  &   \cellcolor[HTML]{C0C0C0}{\color[HTML]{000000}True }   &   \cellcolor[HTML]{C0C0C0}{\color[HTML]{000000}True }    &     \cellcolor[HTML]{C0C0C0}{\color[HTML]{000000}True }   &    \cellcolor[HTML]{C0C0C0}{\color[HTML]{000000}True }  &  -  &  - &  -       \\

& Refactor              &   -      &    \cellcolor[HTML]{C0C0C0}{\color[HTML]{000000}True }    &   \cellcolor[HTML]{C0C0C0}{\color[HTML]{000000}True }    &    -  &     \cellcolor[HTML]{C0C0C0}{\color[HTML]{000000}True } &    -  &  -    &  -    &     \cellcolor[HTML]{C0C0C0}{\color[HTML]{000000}True }    &    -  &   \cellcolor[HTML]{C0C0C0}{\color[HTML]{000000}True } &  - &  -       \\

& Rename             &   -      &     \cellcolor[HTML]{C0C0C0}{\color[HTML]{000000}True }     &    \cellcolor[HTML]{C0C0C0}{\color[HTML]{000000}True }    &    \cellcolor[HTML]{C0C0C0}{\color[HTML]{000000}True }  &    - &    -  &   \cellcolor[HTML]{C0C0C0}{\color[HTML]{000000}True }    &   \cellcolor[HTML]{C0C0C0}{\color[HTML]{000000}True }   &    \cellcolor[HTML]{C0C0C0}{\color[HTML]{000000}True }    &     \cellcolor[HTML]{C0C0C0}{\color[HTML]{000000}True }  &   \cellcolor[HTML]{C0C0C0}{\color[HTML]{000000}True }  &  - &  -       \\

& Dead              &   -      &   \cellcolor[HTML]{C0C0C0}{\color[HTML]{000000}True }     &    \cellcolor[HTML]{C0C0C0}{\color[HTML]{000000}True }    &    -  &    - &    -  &  -    &  -    &     \cellcolor[HTML]{C0C0C0}{\color[HTML]{000000}True }   &     \cellcolor[HTML]{C0C0C0}{\color[HTML]{000000}True }  &   \cellcolor[HTML]{C0C0C0}{\color[HTML]{000000}True }  &  - &  -       \\

& Inside              &   -      &     \cellcolor[HTML]{C0C0C0}{\color[HTML]{000000}True }     &    \cellcolor[HTML]{C0C0C0}{\color[HTML]{000000}True }    &    -  &     \cellcolor[HTML]{C0C0C0}{\color[HTML]{000000}True } &    -  &  -    &  -    &     \cellcolor[HTML]{C0C0C0}{\color[HTML]{000000}True }    &    \cellcolor[HTML]{C0C0C0}{\color[HTML]{000000}True }  &   \cellcolor[HTML]{C0C0C0}{\color[HTML]{000000}True } &  - &  -       \\

& Outside             &   -      &    \cellcolor[HTML]{C0C0C0}{\color[HTML]{000000}True }    &   \cellcolor[HTML]{C0C0C0}{\color[HTML]{000000}True }     &    -  &     \cellcolor[HTML]{C0C0C0}{\color[HTML]{000000}True } &    -  &  -    &  -    &     \cellcolor[HTML]{C0C0C0}{\color[HTML]{000000}True }    &     \cellcolor[HTML]{C0C0C0}{\color[HTML]{000000}True } &  \cellcolor[HTML]{C0C0C0}{\color[HTML]{000000}True } &  - &  -       \\

& SR               &   \cellcolor[HTML]{C0C0C0}{\color[HTML]{000000}True }      &    \cellcolor[HTML]{C0C0C0}{\color[HTML]{000000}True }     &    \cellcolor[HTML]{C0C0C0}{\color[HTML]{000000}True }    &     \cellcolor[HTML]{C0C0C0}{\color[HTML]{000000}True }  &     \cellcolor[HTML]{C0C0C0}{\color[HTML]{000000}True } &     \cellcolor[HTML]{C0C0C0}{\color[HTML]{000000}True } &   \cellcolor[HTML]{C0C0C0}{\color[HTML]{000000}True }   &  \cellcolor[HTML]{C0C0C0}{\color[HTML]{000000}True }    &    -    &    -  &  \cellcolor[HTML]{C0C0C0}{\color[HTML]{000000}True } &   \cellcolor[HTML]{C0C0C0}{\color[HTML]{000000}True } &   \cellcolor[HTML]{C0C0C0}{\color[HTML]{000000}True }     \\

& RI            &    \cellcolor[HTML]{C0C0C0}{\color[HTML]{000000}True }      &     \cellcolor[HTML]{C0C0C0}{\color[HTML]{000000}True }     &    \cellcolor[HTML]{C0C0C0}{\color[HTML]{000000}True }     &    -  &     \cellcolor[HTML]{C0C0C0}{\color[HTML]{000000}True } &     \cellcolor[HTML]{C0C0C0}{\color[HTML]{000000}True }  &   \cellcolor[HTML]{C0C0C0}{\color[HTML]{000000}True }    &   \cellcolor[HTML]{C0C0C0}{\color[HTML]{000000}True }    &    -    &    -  &  \cellcolor[HTML]{C0C0C0}{\color[HTML]{000000}True }  &  \cellcolor[HTML]{C0C0C0}{\color[HTML]{000000}True } &  \cellcolor[HTML]{C0C0C0}{\color[HTML]{000000}True }      \\

& RS              &   -      &    -     &   -     &    \cellcolor[HTML]{C0C0C0}{\color[HTML]{000000}True }  &    \cellcolor[HTML]{C0C0C0}{\color[HTML]{000000}True } &     \cellcolor[HTML]{C0C0C0}{\color[HTML]{000000}True } & \cellcolor[HTML]{C0C0C0}{\color[HTML]{000000}True }    &   \cellcolor[HTML]{C0C0C0}{\color[HTML]{000000}True }    &    \cellcolor[HTML]{C0C0C0}{\color[HTML]{000000}True }   &     \cellcolor[HTML]{C0C0C0}{\color[HTML]{000000}True }  &  -  &  - &  -       \\

& RD              &   -      &    -     &   -     &    -  &    -  &    -  &  -    &  -    &     \cellcolor[HTML]{C0C0C0}{\color[HTML]{000000}True }  &    \cellcolor[HTML]{C0C0C0}{\color[HTML]{000000}True } &  -   &  - &  - \\

& BT              &   -      &    -     &   -     &    -  &    - &    -  &  -    &  -    &     \cellcolor[HTML]{C0C0C0}{\color[HTML]{000000}True }    &     \cellcolor[HTML]{C0C0C0}{\color[HTML]{000000}True } &  -  &  - &  -       \\

\hline  

\end{tabular}
}
\end{sidewaystable}

\begin{sidewaystable}[]
\centering
    \setlength{\tabcolsep}{3pt} 
    \renewcommand{\arraystretch}{1.3}
    \tiny
\caption{Results of statistical tests on Robustness. \textbf{True}: indicates that the comparison is statistically significant after the  \emph{Bonferroni correction} adjustment. Statistical test method: \emph{Wilcoxon signed-rank test}. A gray background highlights the result marked as \textbf{True}. For legibility, we simplify the terms \emph{Rename Operator}, \emph{Dead Operator}, \emph{Inside Operator}, and \emph{Outside Operator} to \emph{Rename}, \emph{Dead}, \emph{Inside}, and \emph{Outside}, respectively. Model: CodeBERT (10\%) \& GraphCodeBERT (10\%).}
\label{table:statistical_testing_robust_10_codebert_graphcodebert}
\centering
\resizebox{\textwidth}{!}{
\begin{tabular}{lcccccccccccccccc}
\hline
& & \multicolumn{14}{c}{\textbf{Accuracy}}  \\ \hline
                    & DA method & No Aug & WordMixup & SenMixup  & Refactor & Rename & Dead  & Inside & Outside  & SR & RI &  RS &  RD & BT & \\ \hline
\multirow{13}{*}{{\begin{tabular}[c]{@{}c@{}}CodeBERT\\ (10\%)\end{tabular}}} & No Aug                         &   -      &    -     &   \cellcolor[HTML]{C0C0C0}{\color[HTML]{000000}True }      &    -  &    - &    -  &  \cellcolor[HTML]{C0C0C0}{\color[HTML]{000000}True }     &  \cellcolor[HTML]{C0C0C0}{\color[HTML]{000000}True }     &    \cellcolor[HTML]{C0C0C0}{\color[HTML]{000000}True }     &    \cellcolor[HTML]{C0C0C0}{\color[HTML]{000000}True }   &  -  &  - &  -       \\

& WordMixup                         &   -      &    -     &  \cellcolor[HTML]{C0C0C0}{\color[HTML]{000000}True }      &    -  &    - &    -  &  \cellcolor[HTML]{C0C0C0}{\color[HTML]{000000}True }    &  -    &    \cellcolor[HTML]{C0C0C0}{\color[HTML]{000000}True }     &    -  &  -  &  - &  -       \\

& SenMixup                        &   \cellcolor[HTML]{C0C0C0}{\color[HTML]{000000}True }       &    \cellcolor[HTML]{C0C0C0}{\color[HTML]{000000}True }     &   -     &    \cellcolor[HTML]{C0C0C0}{\color[HTML]{000000}True }  &   \cellcolor[HTML]{C0C0C0}{\color[HTML]{000000}True }  &    \cellcolor[HTML]{C0C0C0}{\color[HTML]{000000}True }   &  \cellcolor[HTML]{C0C0C0}{\color[HTML]{000000}True }    &  \cellcolor[HTML]{C0C0C0}{\color[HTML]{000000}True }    &    \cellcolor[HTML]{C0C0C0}{\color[HTML]{000000}True }     &    \cellcolor[HTML]{C0C0C0}{\color[HTML]{000000}True }   &  -  &  \cellcolor[HTML]{C0C0C0}{\color[HTML]{000000}True }  &  \cellcolor[HTML]{C0C0C0}{\color[HTML]{000000}True }        \\

& Refactor              &   -      &    -     &   \cellcolor[HTML]{C0C0C0}{\color[HTML]{000000}True }      &    -  &    \cellcolor[HTML]{C0C0C0}{\color[HTML]{000000}True }  &    -  &  \cellcolor[HTML]{C0C0C0}{\color[HTML]{000000}True }     & \cellcolor[HTML]{C0C0C0}{\color[HTML]{000000}True }     &    \cellcolor[HTML]{C0C0C0}{\color[HTML]{000000}True }     &    \cellcolor[HTML]{C0C0C0}{\color[HTML]{000000}True }  &  -  &  - &  -       \\

& Rename             &   -      &    -     &   \cellcolor[HTML]{C0C0C0}{\color[HTML]{000000}True }      &    \cellcolor[HTML]{C0C0C0}{\color[HTML]{000000}True }   &    - &    -  &  \cellcolor[HTML]{C0C0C0}{\color[HTML]{000000}True }   &  \cellcolor[HTML]{C0C0C0}{\color[HTML]{000000}True }     &    -    &   \cellcolor[HTML]{C0C0C0}{\color[HTML]{000000}True }  &  -  &  - &  -       \\

& Dead              &   -      &    -     &   \cellcolor[HTML]{C0C0C0}{\color[HTML]{000000}True }     &    -  &    -  &    -  &  \cellcolor[HTML]{C0C0C0}{\color[HTML]{000000}True }    &  \cellcolor[HTML]{C0C0C0}{\color[HTML]{000000}True }    &   \cellcolor[HTML]{C0C0C0}{\color[HTML]{000000}True }     &    \cellcolor[HTML]{C0C0C0}{\color[HTML]{000000}True }   &  -  &  - &  -       \\

& Inside              &   \cellcolor[HTML]{C0C0C0}{\color[HTML]{000000}True }      &    \cellcolor[HTML]{C0C0C0}{\color[HTML]{000000}True }      &  \cellcolor[HTML]{C0C0C0}{\color[HTML]{000000}True }      &    \cellcolor[HTML]{C0C0C0}{\color[HTML]{000000}True }   &    \cellcolor[HTML]{C0C0C0}{\color[HTML]{000000}True }  &    \cellcolor[HTML]{C0C0C0}{\color[HTML]{000000}True }  &  -    &  -    &    -    &    -  &  -  &  \cellcolor[HTML]{C0C0C0}{\color[HTML]{000000}True }  &  -       \\

& Outside             &   \cellcolor[HTML]{C0C0C0}{\color[HTML]{000000}True }       &    -     &   \cellcolor[HTML]{C0C0C0}{\color[HTML]{000000}True }     &    \cellcolor[HTML]{C0C0C0}{\color[HTML]{000000}True }   &    \cellcolor[HTML]{C0C0C0}{\color[HTML]{000000}True }  &    \cellcolor[HTML]{C0C0C0}{\color[HTML]{000000}True }   &  -    &  -    &    -    &    -  &  -  &  \cellcolor[HTML]{C0C0C0}{\color[HTML]{000000}True }  &  -       \\

& SR               &   \cellcolor[HTML]{C0C0C0}{\color[HTML]{000000}True }       &   \cellcolor[HTML]{C0C0C0}{\color[HTML]{000000}True }    &   \cellcolor[HTML]{C0C0C0}{\color[HTML]{000000}True }      &    \cellcolor[HTML]{C0C0C0}{\color[HTML]{000000}True }  &    - &   \cellcolor[HTML]{C0C0C0}{\color[HTML]{000000}True }   &  -    &  -    &    -    &    -  &  -  &  \cellcolor[HTML]{C0C0C0}{\color[HTML]{000000}True }  &  -       \\

& RI            &   -      &    -     &   \cellcolor[HTML]{C0C0C0}{\color[HTML]{000000}True }      &    \cellcolor[HTML]{C0C0C0}{\color[HTML]{000000}True }   &   \cellcolor[HTML]{C0C0C0}{\color[HTML]{000000}True }  &   \cellcolor[HTML]{C0C0C0}{\color[HTML]{000000}True }   &  -    &  -    &    -    &    -  &  \cellcolor[HTML]{C0C0C0}{\color[HTML]{000000}True }  &  \cellcolor[HTML]{C0C0C0}{\color[HTML]{000000}True }  &  -       \\

& RS              &   -      &    -     &   -     &    -  &    - &    -  &  -    &  -    &    -    &    \cellcolor[HTML]{C0C0C0}{\color[HTML]{000000}True }   &  -  &  - &  -       \\

& RD              &   -      &    -     &  \cellcolor[HTML]{C0C0C0}{\color[HTML]{000000}True }      &    -  &    -  &    -  & \cellcolor[HTML]{C0C0C0}{\color[HTML]{000000}True }     &  \cellcolor[HTML]{C0C0C0}{\color[HTML]{000000}True }    &   \cellcolor[HTML]{C0C0C0}{\color[HTML]{000000}True }     &   \cellcolor[HTML]{C0C0C0}{\color[HTML]{000000}True }   &  -   &  - &  - \\

& BT              &   -      &    -     &   \cellcolor[HTML]{C0C0C0}{\color[HTML]{000000}True }      &    -  &    - &    -  &  -    &  -    &    -    &    -  &  -  &  - &  -       \\

\hline                  
\multirow{13}{*}{{\begin{tabular}[c]{@{}c@{}}GraphCodeBERT\\ (10\%)\end{tabular}}}  & No Aug                         &   -      &    -     &   \cellcolor[HTML]{C0C0C0}{\color[HTML]{000000}True }     &    -  &    \cellcolor[HTML]{C0C0C0}{\color[HTML]{000000}True } &    -  &  \cellcolor[HTML]{C0C0C0}{\color[HTML]{000000}True }   &  \cellcolor[HTML]{C0C0C0}{\color[HTML]{000000}True }   &    -    &    \cellcolor[HTML]{C0C0C0}{\color[HTML]{000000}True }  &  -  &  - &  \cellcolor[HTML]{C0C0C0}{\color[HTML]{000000}True }       \\

& WordMixup                         &   -      &    -     &   \cellcolor[HTML]{C0C0C0}{\color[HTML]{000000}True }     &    -  &    - &    -  &  \cellcolor[HTML]{C0C0C0}{\color[HTML]{000000}True }    &  \cellcolor[HTML]{C0C0C0}{\color[HTML]{000000}True }    &    \cellcolor[HTML]{C0C0C0}{\color[HTML]{000000}True }   &    \cellcolor[HTML]{C0C0C0}{\color[HTML]{000000}True }  &  -  &  - &  \cellcolor[HTML]{C0C0C0}{\color[HTML]{000000}True }     \\

& SenMixup                        &   \cellcolor[HTML]{C0C0C0}{\color[HTML]{000000}True }      &   \cellcolor[HTML]{C0C0C0}{\color[HTML]{000000}True }     &   -     &    -  &    \cellcolor[HTML]{C0C0C0}{\color[HTML]{000000}True } &    \cellcolor[HTML]{C0C0C0}{\color[HTML]{000000}True } &  \cellcolor[HTML]{C0C0C0}{\color[HTML]{000000}True }    &  \cellcolor[HTML]{C0C0C0}{\color[HTML]{000000}True }   &    \cellcolor[HTML]{C0C0C0}{\color[HTML]{000000}True }    &    \cellcolor[HTML]{C0C0C0}{\color[HTML]{000000}True }  &  -  &  - &  -       \\

& Refactor              &   -      &    -     &   -     &    -  &  \cellcolor[HTML]{C0C0C0}{\color[HTML]{000000}True } &    \cellcolor[HTML]{C0C0C0}{\color[HTML]{000000}True }  & \cellcolor[HTML]{C0C0C0}{\color[HTML]{000000}True }    &  \cellcolor[HTML]{C0C0C0}{\color[HTML]{000000}True }    &    -    &    \cellcolor[HTML]{C0C0C0}{\color[HTML]{000000}True }  &  -  &  - &  \cellcolor[HTML]{C0C0C0}{\color[HTML]{000000}True }       \\

& Rename             &   \cellcolor[HTML]{C0C0C0}{\color[HTML]{000000}True }      &    -     &  \cellcolor[HTML]{C0C0C0}{\color[HTML]{000000}True }    &    \cellcolor[HTML]{C0C0C0}{\color[HTML]{000000}True }  &    - &    \cellcolor[HTML]{C0C0C0}{\color[HTML]{000000}True }  &  -    &  -    &    -    &    -  &  \cellcolor[HTML]{C0C0C0}{\color[HTML]{000000}True }  &  \cellcolor[HTML]{C0C0C0}{\color[HTML]{000000}True } &  \cellcolor[HTML]{C0C0C0}{\color[HTML]{000000}True }      \\

& Dead              &   -      &    -     &   \cellcolor[HTML]{C0C0C0}{\color[HTML]{000000}True }     &    \cellcolor[HTML]{C0C0C0}{\color[HTML]{000000}True }  &    \cellcolor[HTML]{C0C0C0}{\color[HTML]{000000}True } &    -  &  -   &  \cellcolor[HTML]{C0C0C0}{\color[HTML]{000000}True }    &    -    &    \cellcolor[HTML]{C0C0C0}{\color[HTML]{000000}True }  &  -  &  \cellcolor[HTML]{C0C0C0}{\color[HTML]{000000}True }& \cellcolor[HTML]{C0C0C0}{\color[HTML]{000000}True }      \\

& Inside              &   \cellcolor[HTML]{C0C0C0}{\color[HTML]{000000}True }     &   \cellcolor[HTML]{C0C0C0}{\color[HTML]{000000}True }    &   \cellcolor[HTML]{C0C0C0}{\color[HTML]{000000}True }    &   \cellcolor[HTML]{C0C0C0}{\color[HTML]{000000}True }  &    - &    - &  -    &  -    &    -    &    -  &  \cellcolor[HTML]{C0C0C0}{\color[HTML]{000000}True } &\cellcolor[HTML]{C0C0C0}{\color[HTML]{000000}True } &  \cellcolor[HTML]{C0C0C0}{\color[HTML]{000000}True }       \\

& Outside             &  \cellcolor[HTML]{C0C0C0}{\color[HTML]{000000}True }     &   \cellcolor[HTML]{C0C0C0}{\color[HTML]{000000}True }    &  \cellcolor[HTML]{C0C0C0}{\color[HTML]{000000}True }    &    \cellcolor[HTML]{C0C0C0}{\color[HTML]{000000}True }  &    - &    \cellcolor[HTML]{C0C0C0}{\color[HTML]{000000}True }  &  -    &  -    &    -    &    -  &  \cellcolor[HTML]{C0C0C0}{\color[HTML]{000000}True }  &  \cellcolor[HTML]{C0C0C0}{\color[HTML]{000000}True } &  \cellcolor[HTML]{C0C0C0}{\color[HTML]{000000}True }      \\

& SR               &   -      &   \cellcolor[HTML]{C0C0C0}{\color[HTML]{000000}True }     &   \cellcolor[HTML]{C0C0C0}{\color[HTML]{000000}True }     &    -  &    - &    -  &  -    &  -    &    -    &    \cellcolor[HTML]{C0C0C0}{\color[HTML]{000000}True } &  -  &  - &  \cellcolor[HTML]{C0C0C0}{\color[HTML]{000000}True }     \\

& RI            &   \cellcolor[HTML]{C0C0C0}{\color[HTML]{000000}True }      &    \cellcolor[HTML]{C0C0C0}{\color[HTML]{000000}True }     &   \cellcolor[HTML]{C0C0C0}{\color[HTML]{000000}True }    &    \cellcolor[HTML]{C0C0C0}{\color[HTML]{000000}True }  &    - &    \cellcolor[HTML]{C0C0C0}{\color[HTML]{000000}True } &  -    &  -    &\cellcolor[HTML]{C0C0C0}{\color[HTML]{000000}True }    &    -  &  \cellcolor[HTML]{C0C0C0}{\color[HTML]{000000}True }  &  \cellcolor[HTML]{C0C0C0}{\color[HTML]{000000}True } &  \cellcolor[HTML]{C0C0C0}{\color[HTML]{000000}True }      \\

& RS              &   -      &    -     &   -     &    -  &    \cellcolor[HTML]{C0C0C0}{\color[HTML]{000000}True } &    -  & \cellcolor[HTML]{C0C0C0}{\color[HTML]{000000}True }    &  \cellcolor[HTML]{C0C0C0}{\color[HTML]{000000}True }    &    -    &   \cellcolor[HTML]{C0C0C0}{\color[HTML]{000000}True }  &  -  &  - &  -       \\

& RD              &   -      &    -     &   -     &    -  &    \cellcolor[HTML]{C0C0C0}{\color[HTML]{000000}True }  &    \cellcolor[HTML]{C0C0C0}{\color[HTML]{000000}True }  &  \cellcolor[HTML]{C0C0C0}{\color[HTML]{000000}True }    &  \cellcolor[HTML]{C0C0C0}{\color[HTML]{000000}True }    &    -    &   \cellcolor[HTML]{C0C0C0}{\color[HTML]{000000}True }  &  -   &  - &  - \\

& BT              &  \cellcolor[HTML]{C0C0C0}{\color[HTML]{000000}True }      &    \cellcolor[HTML]{C0C0C0}{\color[HTML]{000000}True }     &   -     &    \cellcolor[HTML]{C0C0C0}{\color[HTML]{000000}True }  &    \cellcolor[HTML]{C0C0C0}{\color[HTML]{000000}True }&  \cellcolor[HTML]{C0C0C0}{\color[HTML]{000000}True }  &  \cellcolor[HTML]{C0C0C0}{\color[HTML]{000000}True }    &  \cellcolor[HTML]{C0C0C0}{\color[HTML]{000000}True }   &    \cellcolor[HTML]{C0C0C0}{\color[HTML]{000000}True }    &   \cellcolor[HTML]{C0C0C0}{\color[HTML]{000000}True }  &  -  &  - &  -       \\

\hline  

\end{tabular}
}
\end{sidewaystable}

\begin{sidewaystable}[]
\centering
    \setlength{\tabcolsep}{3pt} 
    \renewcommand{\arraystretch}{1.3}
    \tiny
\caption{Results of statistical tests on Robustness. \textbf{True}: indicates that the comparison is statistically significant after the  \emph{Bonferroni correction} adjustment. Statistical test method: \emph{Wilcoxon signed-rank test}. A gray background highlights the result marked as \textbf{True}. For legibility, we simplify the terms \emph{Rename Operator}, \emph{Dead Operator}, \emph{Inside Operator}, and \emph{Outside Operator} to \emph{Rename}, \emph{Dead}, \emph{Inside}, and \emph{Outside}, respectively. Model: CodeBERT (5\%) \& GraphCodeBERT (5\%).}
\label{table:statistical_testing_robust_5_codebert_graphcodebert}
\centering
\resizebox{\textwidth}{!}{
\begin{tabular}{lcccccccccccccccc}
\hline
& & \multicolumn{14}{c}{\textbf{Accuracy}}  \\ \hline
                    & DA method & No Aug & WordMixup & SenMixup  & Refactor & Rename & Dead  & Inside & Outside  & SR & RI &  RS &  RD & BT & \\ \hline
\multirow{13}{*}{{\begin{tabular}[c]{@{}c@{}}CodeBERT\\ (5\%)\end{tabular}}} & No Aug                         &   -      &    -     &   \cellcolor[HTML]{C0C0C0}{\color[HTML]{000000}True }     &    -  &    \cellcolor[HTML]{C0C0C0}{\color[HTML]{000000}True } &    \cellcolor[HTML]{C0C0C0}{\color[HTML]{000000}True } &  \cellcolor[HTML]{C0C0C0}{\color[HTML]{000000}True }   &  \cellcolor[HTML]{C0C0C0}{\color[HTML]{000000}True }    &    -    &   \cellcolor[HTML]{C0C0C0}{\color[HTML]{000000}True } &  -  &  - &  -       \\

& WordMixup                         &   -      &    -     &   -     &    -  &    \cellcolor[HTML]{C0C0C0}{\color[HTML]{000000}True } &    -  &  \cellcolor[HTML]{C0C0C0}{\color[HTML]{000000}True }    &  \cellcolor[HTML]{C0C0C0}{\color[HTML]{000000}True }   &    -    &    \cellcolor[HTML]{C0C0C0}{\color[HTML]{000000}True }  &  -  &  - &  -       \\

& SenMixup                        &   \cellcolor[HTML]{C0C0C0}{\color[HTML]{000000}True }     &    -     &   -     &   \cellcolor[HTML]{C0C0C0}{\color[HTML]{000000}True }  &    \cellcolor[HTML]{C0C0C0}{\color[HTML]{000000}True } &    \cellcolor[HTML]{C0C0C0}{\color[HTML]{000000}True }  &  \cellcolor[HTML]{C0C0C0}{\color[HTML]{000000}True }    &  \cellcolor[HTML]{C0C0C0}{\color[HTML]{000000}True }    &    -    &    \cellcolor[HTML]{C0C0C0}{\color[HTML]{000000}True }  &  -  &  - &  \cellcolor[HTML]{C0C0C0}{\color[HTML]{000000}True }      \\

& Refactor              &   -      &    -     &   \cellcolor[HTML]{C0C0C0}{\color[HTML]{000000}True }    &    -  &    \cellcolor[HTML]{C0C0C0}{\color[HTML]{000000}True } &    -  &  \cellcolor[HTML]{C0C0C0}{\color[HTML]{000000}True }    &  \cellcolor[HTML]{C0C0C0}{\color[HTML]{000000}True }   &    -    &    \cellcolor[HTML]{C0C0C0}{\color[HTML]{000000}True }  &  -  &  - &  -       \\

& Rename             &  \cellcolor[HTML]{C0C0C0}{\color[HTML]{000000}True }     &   \cellcolor[HTML]{C0C0C0}{\color[HTML]{000000}True }    &   \cellcolor[HTML]{C0C0C0}{\color[HTML]{000000}True }    &    \cellcolor[HTML]{C0C0C0}{\color[HTML]{000000}True }  &    - &   \cellcolor[HTML]{C0C0C0}{\color[HTML]{000000}True }  &  -    &  -    &    -    &    \cellcolor[HTML]{C0C0C0}{\color[HTML]{000000}True }  &  \cellcolor[HTML]{C0C0C0}{\color[HTML]{000000}True } &  \cellcolor[HTML]{C0C0C0}{\color[HTML]{000000}True } &  -       \\

& Dead              &   \cellcolor[HTML]{C0C0C0}{\color[HTML]{000000}True }      &    -     &   \cellcolor[HTML]{C0C0C0}{\color[HTML]{000000}True }    &    -  &    \cellcolor[HTML]{C0C0C0}{\color[HTML]{000000}True } &    -  &  \cellcolor[HTML]{C0C0C0}{\color[HTML]{000000}True }    &  \cellcolor[HTML]{C0C0C0}{\color[HTML]{000000}True }   &    -    &  \cellcolor[HTML]{C0C0C0}{\color[HTML]{000000}True }  & \cellcolor[HTML]{C0C0C0}{\color[HTML]{000000}True } & \cellcolor[HTML]{C0C0C0}{\color[HTML]{000000}True } &  -       \\

& Inside              &  \cellcolor[HTML]{C0C0C0}{\color[HTML]{000000}True }     &    \cellcolor[HTML]{C0C0C0}{\color[HTML]{000000}True }     &   \cellcolor[HTML]{C0C0C0}{\color[HTML]{000000}True }     &    \cellcolor[HTML]{C0C0C0}{\color[HTML]{000000}True }  &    - &    \cellcolor[HTML]{C0C0C0}{\color[HTML]{000000}True }  &  -    &  -    &    -    &    \cellcolor[HTML]{C0C0C0}{\color[HTML]{000000}True } &  \cellcolor[HTML]{C0C0C0}{\color[HTML]{000000}True } &  \cellcolor[HTML]{C0C0C0}{\color[HTML]{000000}True } &  -       \\

& Outside             &   \cellcolor[HTML]{C0C0C0}{\color[HTML]{000000}True }      &   \cellcolor[HTML]{C0C0C0}{\color[HTML]{000000}True }    &  \cellcolor[HTML]{C0C0C0}{\color[HTML]{000000}True }    &   \cellcolor[HTML]{C0C0C0}{\color[HTML]{000000}True }  &    - &   \cellcolor[HTML]{C0C0C0}{\color[HTML]{000000}True } &  -    &  -    &    \cellcolor[HTML]{C0C0C0}{\color[HTML]{000000}True }    &   \cellcolor[HTML]{C0C0C0}{\color[HTML]{000000}True }  &  \cellcolor[HTML]{C0C0C0}{\color[HTML]{000000}True }  &  \cellcolor[HTML]{C0C0C0}{\color[HTML]{000000}True } &  \cellcolor[HTML]{C0C0C0}{\color[HTML]{000000}True }      \\

& SR               &   -      &    -     &   -     &    -  &    - &    -  &  -    &  \cellcolor[HTML]{C0C0C0}{\color[HTML]{000000}True }    &    -    &    \cellcolor[HTML]{C0C0C0}{\color[HTML]{000000}True }  &  -  &  - &  -       \\

& RI            &  \cellcolor[HTML]{C0C0C0}{\color[HTML]{000000}True }      &    \cellcolor[HTML]{C0C0C0}{\color[HTML]{000000}True }     &  \cellcolor[HTML]{C0C0C0}{\color[HTML]{000000}True }     &    \cellcolor[HTML]{C0C0C0}{\color[HTML]{000000}True }  &    \cellcolor[HTML]{C0C0C0}{\color[HTML]{000000}True } &    \cellcolor[HTML]{C0C0C0}{\color[HTML]{000000}True }  & \cellcolor[HTML]{C0C0C0}{\color[HTML]{000000}True }    & \cellcolor[HTML]{C0C0C0}{\color[HTML]{000000}True }   &   \cellcolor[HTML]{C0C0C0}{\color[HTML]{000000}True }    &    -  &  \cellcolor[HTML]{C0C0C0}{\color[HTML]{000000}True } & \cellcolor[HTML]{C0C0C0}{\color[HTML]{000000}True } &  \cellcolor[HTML]{C0C0C0}{\color[HTML]{000000}True }       \\

& RS              &   -      &    -     &   -     &    -  &  \cellcolor[HTML]{C0C0C0}{\color[HTML]{000000}True } &    \cellcolor[HTML]{C0C0C0}{\color[HTML]{000000}True }  &  \cellcolor[HTML]{C0C0C0}{\color[HTML]{000000}True }  &  \cellcolor[HTML]{C0C0C0}{\color[HTML]{000000}True }    &    -    &   \cellcolor[HTML]{C0C0C0}{\color[HTML]{000000}True } &  -  &  - &  -       \\

& RD              &   -      &    -     &   -     &    -  &    \cellcolor[HTML]{C0C0C0}{\color[HTML]{000000}True }  &   \cellcolor[HTML]{C0C0C0}{\color[HTML]{000000}True } & \cellcolor[HTML]{C0C0C0}{\color[HTML]{000000}True }    &  \cellcolor[HTML]{C0C0C0}{\color[HTML]{000000}True }    &    -    &   \cellcolor[HTML]{C0C0C0}{\color[HTML]{000000}True } &  -   &  - &  - \\

& BT              &   -      &    -     &   \cellcolor[HTML]{C0C0C0}{\color[HTML]{000000}True }     &    -  &    - &    -  &  -    &  \cellcolor[HTML]{C0C0C0}{\color[HTML]{000000}True }    &    -    &    \cellcolor[HTML]{C0C0C0}{\color[HTML]{000000}True } &  -  &  - &  -       \\

\hline                  
\multirow{13}{*}{{\begin{tabular}[c]{@{}c@{}}GraphCodeBERT\\ (5\%)\end{tabular}}} & No Aug                         &   -      &    -     &   \cellcolor[HTML]{C0C0C0}{\color[HTML]{000000}True }    &    -  &    - &    -  &  \cellcolor[HTML]{C0C0C0}{\color[HTML]{000000}True }    &  \cellcolor[HTML]{C0C0C0}{\color[HTML]{000000}True }   &   \cellcolor[HTML]{C0C0C0}{\color[HTML]{000000}True }   &    -  &  -  &  - &  -       \\

& WordMixup                         &   -      &    -     &  \cellcolor[HTML]{C0C0C0}{\color[HTML]{000000}True }     &    -  &    - &    -  & \cellcolor[HTML]{C0C0C0}{\color[HTML]{000000}True }    &  \cellcolor[HTML]{C0C0C0}{\color[HTML]{000000}True }    &    \cellcolor[HTML]{C0C0C0}{\color[HTML]{000000}True }   &    -  &  -  &  - &  -       \\

& SenMixup                        &   \cellcolor[HTML]{C0C0C0}{\color[HTML]{000000}True }      &    \cellcolor[HTML]{C0C0C0}{\color[HTML]{000000}True }     &   -     &    \cellcolor[HTML]{C0C0C0}{\color[HTML]{000000}True }  &    \cellcolor[HTML]{C0C0C0}{\color[HTML]{000000}True }  &    \cellcolor[HTML]{C0C0C0}{\color[HTML]{000000}True }  &  \cellcolor[HTML]{C0C0C0}{\color[HTML]{000000}True }    &  \cellcolor[HTML]{C0C0C0}{\color[HTML]{000000}True }    &   \cellcolor[HTML]{C0C0C0}{\color[HTML]{000000}True }    &    \cellcolor[HTML]{C0C0C0}{\color[HTML]{000000}True }  &  -  &  - &  -       \\

& Refactor              &   -      &    -     &  \cellcolor[HTML]{C0C0C0}{\color[HTML]{000000}True }    &    -  &    \cellcolor[HTML]{C0C0C0}{\color[HTML]{000000}True } &    -  &  \cellcolor[HTML]{C0C0C0}{\color[HTML]{000000}True }    &  \cellcolor[HTML]{C0C0C0}{\color[HTML]{000000}True }    &    -    &    -  &  -  &  - &  -       \\

& Rename             &   -      &    -     &  \cellcolor[HTML]{C0C0C0}{\color[HTML]{000000}True }     &    \cellcolor[HTML]{C0C0C0}{\color[HTML]{000000}True }  &    - &    -  &  -    &  -    &    -    &    -  &  -  &  - &  \cellcolor[HTML]{C0C0C0}{\color[HTML]{000000}True }      \\

& Dead              &   -      &    -     &  \cellcolor[HTML]{C0C0C0}{\color[HTML]{000000}True }    &    -  &    - &    -  &  \cellcolor[HTML]{C0C0C0}{\color[HTML]{000000}True }    &  \cellcolor[HTML]{C0C0C0}{\color[HTML]{000000}True }    &    -    &    -  &  -  &  - &  -       \\

& Inside              &   \cellcolor[HTML]{C0C0C0}{\color[HTML]{000000}True }      &   \cellcolor[HTML]{C0C0C0}{\color[HTML]{000000}True }     &   \cellcolor[HTML]{C0C0C0}{\color[HTML]{000000}True }     &   \cellcolor[HTML]{C0C0C0}{\color[HTML]{000000}True }  &    - &   \cellcolor[HTML]{C0C0C0}{\color[HTML]{000000}True } &  -    &  -    &    -    &    -  & \cellcolor[HTML]{C0C0C0}{\color[HTML]{000000}True }  &  \cellcolor[HTML]{C0C0C0}{\color[HTML]{000000}True } &  \cellcolor[HTML]{C0C0C0}{\color[HTML]{000000}True }      \\

& Outside             &  \cellcolor[HTML]{C0C0C0}{\color[HTML]{000000}True }         &   \cellcolor[HTML]{C0C0C0}{\color[HTML]{000000}True }     &   \cellcolor[HTML]{C0C0C0}{\color[HTML]{000000}True }    &    \cellcolor[HTML]{C0C0C0}{\color[HTML]{000000}True }   &    - &    \cellcolor[HTML]{C0C0C0}{\color[HTML]{000000}True }    &  -    &  -    &    -    &    -  &  \cellcolor[HTML]{C0C0C0}{\color[HTML]{000000}True }    &  \cellcolor[HTML]{C0C0C0}{\color[HTML]{000000}True }   &  \cellcolor[HTML]{C0C0C0}{\color[HTML]{000000}True }         \\

& SR               &   \cellcolor[HTML]{C0C0C0}{\color[HTML]{000000}True }       &   \cellcolor[HTML]{C0C0C0}{\color[HTML]{000000}True }     &   \cellcolor[HTML]{C0C0C0}{\color[HTML]{000000}True }     &    -  &    - &    -  &  -    &  -    &    -    &    -  &  \cellcolor[HTML]{C0C0C0}{\color[HTML]{000000}True }   &  \cellcolor[HTML]{C0C0C0}{\color[HTML]{000000}True }   &  \cellcolor[HTML]{C0C0C0}{\color[HTML]{000000}True }         \\

& RI            &   -      &    -     &  \cellcolor[HTML]{C0C0C0}{\color[HTML]{000000}True }     &    -  &    - &    -  &  -    &  -    &    -    &    -  &  \cellcolor[HTML]{C0C0C0}{\color[HTML]{000000}True }    &  \cellcolor[HTML]{C0C0C0}{\color[HTML]{000000}True }   &  \cellcolor[HTML]{C0C0C0}{\color[HTML]{000000}True }       \\

& RS              &   -      &    -     &   -     &    -  &    - &    -  &  \cellcolor[HTML]{C0C0C0}{\color[HTML]{000000}True }     &  \cellcolor[HTML]{C0C0C0}{\color[HTML]{000000}True }     &    \cellcolor[HTML]{C0C0C0}{\color[HTML]{000000}True }     &    \cellcolor[HTML]{C0C0C0}{\color[HTML]{000000}True }    &  -  &  - &  -       \\

& RD              &   -      &    -     &   -     &    -  &    -  &    -  &  \cellcolor[HTML]{C0C0C0}{\color[HTML]{000000}True }      &  \cellcolor[HTML]{C0C0C0}{\color[HTML]{000000}True }     &   \cellcolor[HTML]{C0C0C0}{\color[HTML]{000000}True }      &   \cellcolor[HTML]{C0C0C0}{\color[HTML]{000000}True }   &  -   &  - &  - \\

& BT              &   -      &    -     &   -     &    -  &    \cellcolor[HTML]{C0C0C0}{\color[HTML]{000000}True }   &    -  &  \cellcolor[HTML]{C0C0C0}{\color[HTML]{000000}True }      &  \cellcolor[HTML]{C0C0C0}{\color[HTML]{000000}True }      &    \cellcolor[HTML]{C0C0C0}{\color[HTML]{000000}True }     &    \cellcolor[HTML]{C0C0C0}{\color[HTML]{000000}True }   & -    &  - &  -       \\

\hline  

\end{tabular}
}
\end{sidewaystable}

\begin{sidewaystable}[]
\centering
    \setlength{\tabcolsep}{3pt} 
    \renewcommand{\arraystretch}{1.3}
    \tiny
\caption{Results of statistical tests on Robustness. \textbf{True}: indicates that the comparison is statistically significant after the  \emph{Bonferroni correction} adjustment. Statistical test method: \emph{Wilcoxon signed-rank test}. A gray background highlights the result marked as \textbf{True}. For legibility, we simplify the terms \emph{Rename Operator}, \emph{Dead Operator}, \emph{Inside Operator}, and \emph{Outside Operator} to \emph{Rename}, \emph{Dead}, \emph{Inside}, and \emph{Outside}, respectively. Model: CodeBERT (3\%) \& GraphCodeBERT (3\%).}
\label{table:statistical_testing_robust_3_codebert_graphcodebert}
\centering
\resizebox{\textwidth}{!}{
\begin{tabular}{lcccccccccccccccc}
\hline
& & \multicolumn{14}{c}{\textbf{Accuracy}}  \\ \hline
                    & DA method & No Aug & WordMixup & SenMixup  & Refactor & Rename & Dead  & Inside & Outside  & SR & RI &  RS &  RD & BT & \\ \hline
\multirow{13}{*}{{\begin{tabular}[c]{@{}c@{}}CodeBERT\\ (3\%)\end{tabular}}} & No Aug                         &   -      &    -     &  \cellcolor[HTML]{C0C0C0}{\color[HTML]{000000}True }      &    -  &    - &    -  &  -    &  -    &    -    &    -  &  -  &  - &  -       \\

& WordMixup                         &   -      &    -     & \cellcolor[HTML]{C0C0C0}{\color[HTML]{000000}True }      &    -  &    \cellcolor[HTML]{C0C0C0}{\color[HTML]{000000}True }  &    -  &  \cellcolor[HTML]{C0C0C0}{\color[HTML]{000000}True }    &  \cellcolor[HTML]{C0C0C0}{\color[HTML]{000000}True }     &    \cellcolor[HTML]{C0C0C0}{\color[HTML]{000000}True }     &    -  &  -  &  - &  -       \\

& SenMixup                        &   \cellcolor[HTML]{C0C0C0}{\color[HTML]{000000}True }       &    \cellcolor[HTML]{C0C0C0}{\color[HTML]{000000}True }      &   -     &    -  &    \cellcolor[HTML]{C0C0C0}{\color[HTML]{000000}True }  &    \cellcolor[HTML]{C0C0C0}{\color[HTML]{000000}True }   &  \cellcolor[HTML]{C0C0C0}{\color[HTML]{000000}True }    &  \cellcolor[HTML]{C0C0C0}{\color[HTML]{000000}True }     &  \cellcolor[HTML]{C0C0C0}{\color[HTML]{000000}True }     &   \cellcolor[HTML]{C0C0C0}{\color[HTML]{000000}True }   &  \cellcolor[HTML]{C0C0C0}{\color[HTML]{000000}True }   &  - &  \cellcolor[HTML]{C0C0C0}{\color[HTML]{000000}True }       \\

& Refactor              &   -      &    -     &   -     &    -  &    \cellcolor[HTML]{C0C0C0}{\color[HTML]{000000}True } &    \cellcolor[HTML]{C0C0C0}{\color[HTML]{000000}True }  &  \cellcolor[HTML]{C0C0C0}{\color[HTML]{000000}True }    &  \cellcolor[HTML]{C0C0C0}{\color[HTML]{000000}True }    &    \cellcolor[HTML]{C0C0C0}{\color[HTML]{000000}True }    &    -  &  -  &  - &  -       \\

& Rename             &   -      &    \cellcolor[HTML]{C0C0C0}{\color[HTML]{000000}True }     &  \cellcolor[HTML]{C0C0C0}{\color[HTML]{000000}True }     &    \cellcolor[HTML]{C0C0C0}{\color[HTML]{000000}True }  &    - &    \cellcolor[HTML]{C0C0C0}{\color[HTML]{000000}True }  & \cellcolor[HTML]{C0C0C0}{\color[HTML]{000000}True }    &  -    &    -    &    -  &  -  &  - &  -       \\

& Dead              &   -      &    -     &   \cellcolor[HTML]{C0C0C0}{\color[HTML]{000000}True }      &   \cellcolor[HTML]{C0C0C0}{\color[HTML]{000000}True }  &    \cellcolor[HTML]{C0C0C0}{\color[HTML]{000000}True } &    -  &  \cellcolor[HTML]{C0C0C0}{\color[HTML]{000000}True }   & \cellcolor[HTML]{C0C0C0}{\color[HTML]{000000}True }   &    \cellcolor[HTML]{C0C0C0}{\color[HTML]{000000}True }    &    -  &  -  &  - &  -       \\

& Inside              &   -      &    \cellcolor[HTML]{C0C0C0}{\color[HTML]{000000}True }     &   \cellcolor[HTML]{C0C0C0}{\color[HTML]{000000}True }     &   \cellcolor[HTML]{C0C0C0}{\color[HTML]{000000}True } &    \cellcolor[HTML]{C0C0C0}{\color[HTML]{000000}True } &   \cellcolor[HTML]{C0C0C0}{\color[HTML]{000000}True }  &  -    &  -    &    -    &    -  &  \cellcolor[HTML]{C0C0C0}{\color[HTML]{000000}True }  &  - &  -       \\

& Outside             &   -      &    \cellcolor[HTML]{C0C0C0}{\color[HTML]{000000}True }     &   \cellcolor[HTML]{C0C0C0}{\color[HTML]{000000}True }      &    \cellcolor[HTML]{C0C0C0}{\color[HTML]{000000}True }  &    - &    \cellcolor[HTML]{C0C0C0}{\color[HTML]{000000}True }  &  -    &  -    &    -    &    -  &  -  &  - &  -       \\

& SR               &   -      &   \cellcolor[HTML]{C0C0C0}{\color[HTML]{000000}True }      &  \cellcolor[HTML]{C0C0C0}{\color[HTML]{000000}True }      &    \cellcolor[HTML]{C0C0C0}{\color[HTML]{000000}True }  &    - &    \cellcolor[HTML]{C0C0C0}{\color[HTML]{000000}True }  &  -    &  -    &    -    &    -  &  \cellcolor[HTML]{C0C0C0}{\color[HTML]{000000}True }  &  - &  -       \\

& RI            &   -      &    -     &   \cellcolor[HTML]{C0C0C0}{\color[HTML]{000000}True }      &    -  &    - &    -  &  -    &  -    &    -    &    -  &  -  &  - &  -       \\

& RS              &   -      &    -     &   \cellcolor[HTML]{C0C0C0}{\color[HTML]{000000}True }      &    -  &    - &    -  &  \cellcolor[HTML]{C0C0C0}{\color[HTML]{000000}True }    &  -    &    \cellcolor[HTML]{C0C0C0}{\color[HTML]{000000}True }    &    -  &  -  &  - &  -       \\

& RD              &   -      &    -     &   -     &    -  &    -  &    -  &  -    &  -    &    -    &    -  &  -   &  - &  - \\

& BT              &   -      &    -     &   \cellcolor[HTML]{C0C0C0}{\color[HTML]{000000}True }     &    -  &    - &    -  &  -    &  -    &    -    &    -  &  -  &  - &  -       \\

\hline                  
\multirow{13}{*}{{\begin{tabular}[c]{@{}c@{}}GraphCodeBERT\\ (3\%)\end{tabular}}} & No Aug                         &   -      &    -     &   \cellcolor[HTML]{C0C0C0}{\color[HTML]{000000}True }     &   \cellcolor[HTML]{C0C0C0}{\color[HTML]{000000}True } &    \cellcolor[HTML]{C0C0C0}{\color[HTML]{000000}True } &   \cellcolor[HTML]{C0C0C0}{\color[HTML]{000000}True }  &  \cellcolor[HTML]{C0C0C0}{\color[HTML]{000000}True }    &  \cellcolor[HTML]{C0C0C0}{\color[HTML]{000000}True }    &    \cellcolor[HTML]{C0C0C0}{\color[HTML]{000000}True }   &    -  &  -  &  - &  -       \\

& WordMixup                         &   -      &    -     &  \cellcolor[HTML]{C0C0C0}{\color[HTML]{000000}True }    &   \cellcolor[HTML]{C0C0C0}{\color[HTML]{000000}True }  &    \cellcolor[HTML]{C0C0C0}{\color[HTML]{000000}True } &    \cellcolor[HTML]{C0C0C0}{\color[HTML]{000000}True }  &  \cellcolor[HTML]{C0C0C0}{\color[HTML]{000000}True }   &  \cellcolor[HTML]{C0C0C0}{\color[HTML]{000000}True }   &    \cellcolor[HTML]{C0C0C0}{\color[HTML]{000000}True }   &    \cellcolor[HTML]{C0C0C0}{\color[HTML]{000000}True }  &  -  &  - &  -       \\

& SenMixup                        &   \cellcolor[HTML]{C0C0C0}{\color[HTML]{000000}True }     &    \cellcolor[HTML]{C0C0C0}{\color[HTML]{000000}True }    &   -     &    \cellcolor[HTML]{C0C0C0}{\color[HTML]{000000}True }  &    \cellcolor[HTML]{C0C0C0}{\color[HTML]{000000}True } &   \cellcolor[HTML]{C0C0C0}{\color[HTML]{000000}True }  &  \cellcolor[HTML]{C0C0C0}{\color[HTML]{000000}True }    &  \cellcolor[HTML]{C0C0C0}{\color[HTML]{000000}True }    &    \cellcolor[HTML]{C0C0C0}{\color[HTML]{000000}True }  &    \cellcolor[HTML]{C0C0C0}{\color[HTML]{000000}True }  & \cellcolor[HTML]{C0C0C0}{\color[HTML]{000000}True }  &  - &  -       \\

& Refactor              &  \cellcolor[HTML]{C0C0C0}{\color[HTML]{000000}True }      &    \cellcolor[HTML]{C0C0C0}{\color[HTML]{000000}True }     &   \cellcolor[HTML]{C0C0C0}{\color[HTML]{000000}True }     &    -  &   \cellcolor[HTML]{C0C0C0}{\color[HTML]{000000}True } &    \cellcolor[HTML]{C0C0C0}{\color[HTML]{000000}True }  &  \cellcolor[HTML]{C0C0C0}{\color[HTML]{000000}True }    &  \cellcolor[HTML]{C0C0C0}{\color[HTML]{000000}True }    &    -    &    -  &  -  &  \cellcolor[HTML]{C0C0C0}{\color[HTML]{000000}True } & \cellcolor[HTML]{C0C0C0}{\color[HTML]{000000}True }      \\

& Rename             &   \cellcolor[HTML]{C0C0C0}{\color[HTML]{000000}True }      &    \cellcolor[HTML]{C0C0C0}{\color[HTML]{000000}True }     &   \cellcolor[HTML]{C0C0C0}{\color[HTML]{000000}True }     &   \cellcolor[HTML]{C0C0C0}{\color[HTML]{000000}True }  &    - &    \cellcolor[HTML]{C0C0C0}{\color[HTML]{000000}True }  &  -    &  -    &   \cellcolor[HTML]{C0C0C0}{\color[HTML]{000000}True }    &    \cellcolor[HTML]{C0C0C0}{\color[HTML]{000000}True }  &  \cellcolor[HTML]{C0C0C0}{\color[HTML]{000000}True }  &  \cellcolor[HTML]{C0C0C0}{\color[HTML]{000000}True } &  \cellcolor[HTML]{C0C0C0}{\color[HTML]{000000}True }      \\

& Dead              &  \cellcolor[HTML]{C0C0C0}{\color[HTML]{000000}True }      &   \cellcolor[HTML]{C0C0C0}{\color[HTML]{000000}True }     &  \cellcolor[HTML]{C0C0C0}{\color[HTML]{000000}True }     &   \cellcolor[HTML]{C0C0C0}{\color[HTML]{000000}True }  &    \cellcolor[HTML]{C0C0C0}{\color[HTML]{000000}True } &    -  &  -    &  -    &    -    &    -  &  -  &  \cellcolor[HTML]{C0C0C0}{\color[HTML]{000000}True }  &  \cellcolor[HTML]{C0C0C0}{\color[HTML]{000000}True }      \\

& Inside              &   \cellcolor[HTML]{C0C0C0}{\color[HTML]{000000}True }      &    \cellcolor[HTML]{C0C0C0}{\color[HTML]{000000}True }    &  \cellcolor[HTML]{C0C0C0}{\color[HTML]{000000}True }    &    \cellcolor[HTML]{C0C0C0}{\color[HTML]{000000}True } &    - &    - &  -    &  -    &    \cellcolor[HTML]{C0C0C0}{\color[HTML]{000000}True }   &    \cellcolor[HTML]{C0C0C0}{\color[HTML]{000000}True } &  \cellcolor[HTML]{C0C0C0}{\color[HTML]{000000}True }  &  \cellcolor[HTML]{C0C0C0}{\color[HTML]{000000}True } &  \cellcolor[HTML]{C0C0C0}{\color[HTML]{000000}True }     \\

& Outside             &   \cellcolor[HTML]{C0C0C0}{\color[HTML]{000000}True }      &    \cellcolor[HTML]{C0C0C0}{\color[HTML]{000000}True }    &   \cellcolor[HTML]{C0C0C0}{\color[HTML]{000000}True }     &    \cellcolor[HTML]{C0C0C0}{\color[HTML]{000000}True }  &    - &    -  &  -    &  -    &    -    &    -  &  \cellcolor[HTML]{C0C0C0}{\color[HTML]{000000}True } &  \cellcolor[HTML]{C0C0C0}{\color[HTML]{000000}True } &  \cellcolor[HTML]{C0C0C0}{\color[HTML]{000000}True }       \\

& SR               &   \cellcolor[HTML]{C0C0C0}{\color[HTML]{000000}True }      &    \cellcolor[HTML]{C0C0C0}{\color[HTML]{000000}True }    &   \cellcolor[HTML]{C0C0C0}{\color[HTML]{000000}True }    &    -  &    \cellcolor[HTML]{C0C0C0}{\color[HTML]{000000}True } &    -  &  \cellcolor[HTML]{C0C0C0}{\color[HTML]{000000}True }    &  -    &    -    &    -  &  -  &  \cellcolor[HTML]{C0C0C0}{\color[HTML]{000000}True } &  \cellcolor[HTML]{C0C0C0}{\color[HTML]{000000}True }       \\

& RI            &   -      &    \cellcolor[HTML]{C0C0C0}{\color[HTML]{000000}True }     &   \cellcolor[HTML]{C0C0C0}{\color[HTML]{000000}True }     &    -  &    \cellcolor[HTML]{C0C0C0}{\color[HTML]{000000}True }&    -  &  \cellcolor[HTML]{C0C0C0}{\color[HTML]{000000}True }    &  -    &    -    &    -  &  -  &  \cellcolor[HTML]{C0C0C0}{\color[HTML]{000000}True }  &  \cellcolor[HTML]{C0C0C0}{\color[HTML]{000000}True }       \\

& RS              &   -      &    -     &   \cellcolor[HTML]{C0C0C0}{\color[HTML]{000000}True }     &    -  &    \cellcolor[HTML]{C0C0C0}{\color[HTML]{000000}True } &    -  &  \cellcolor[HTML]{C0C0C0}{\color[HTML]{000000}True }   &  \cellcolor[HTML]{C0C0C0}{\color[HTML]{000000}True }    &    -    &    -  &  -  &  \cellcolor[HTML]{C0C0C0}{\color[HTML]{000000}True } &  -       \\

& RD              &   -      &    -     &   -     &    \cellcolor[HTML]{C0C0C0}{\color[HTML]{000000}True }  &   \cellcolor[HTML]{C0C0C0}{\color[HTML]{000000}True }  &   \cellcolor[HTML]{C0C0C0}{\color[HTML]{000000}True }  &  \cellcolor[HTML]{C0C0C0}{\color[HTML]{000000}True }    &  \cellcolor[HTML]{C0C0C0}{\color[HTML]{000000}True }    &    \cellcolor[HTML]{C0C0C0}{\color[HTML]{000000}True }    &   \cellcolor[HTML]{C0C0C0}{\color[HTML]{000000}True }  & \cellcolor[HTML]{C0C0C0}{\color[HTML]{000000}True }   &  - &  - \\

& BT              &   -      &    -     &   -     &    \cellcolor[HTML]{C0C0C0}{\color[HTML]{000000}True }  &   \cellcolor[HTML]{C0C0C0}{\color[HTML]{000000}True } &    \cellcolor[HTML]{C0C0C0}{\color[HTML]{000000}True }  &  \cellcolor[HTML]{C0C0C0}{\color[HTML]{000000}True }   & \cellcolor[HTML]{C0C0C0}{\color[HTML]{000000}True }   &    \cellcolor[HTML]{C0C0C0}{\color[HTML]{000000}True }    &    \cellcolor[HTML]{C0C0C0}{\color[HTML]{000000}True }  &  -  &  - &  -       \\

\hline  

\end{tabular}
}
\end{sidewaystable}

\begin{sidewaystable}[]
\centering
    \setlength{\tabcolsep}{3pt} 
    \renewcommand{\arraystretch}{1.3}
    \tiny
\caption{Results of statistical tests on Robustness. \textbf{True}: indicates that the comparison is statistically significant after the  \emph{Bonferroni correction} adjustment. Statistical tests method: \emph{Wilcoxon signed-rank test}. A gray background highlights the result marked as \textbf{True}. For legibility, we simplify the terms \emph{Rename Operator}, \emph{Dead Operator}, \emph{Inside Operator}, and \emph{Outside Operator} to \emph{Rename}, \emph{Dead}, \emph{Inside}, and \emph{Outside}, respectively. Model: CodeBERT (1\%) \& GraphCodeBERT (1\%).}
\label{table:statistical_testing_robust_1_codebert_graphcodebert}
\centering
\resizebox{\textwidth}{!}{
\begin{tabular}{lcccccccccccccccc}
\hline
& & \multicolumn{14}{c}{\textbf{Accuracy}}  \\ \hline
                    & DA method & No Aug & WordMixup & SenMixup  & Refactor & Rename & Dead  & Inside & Outside  & SR & RI &  RS &  RD & BT & \\ \hline
\multirow{13}{*}{{\begin{tabular}[c]{@{}c@{}}CodeBERT\\ (1\%)\end{tabular}}} & No Aug                         &   -      &    \cellcolor[HTML]{C0C0C0}{\color[HTML]{000000}True }     &   \cellcolor[HTML]{C0C0C0}{\color[HTML]{000000}True }     &    -  &   \cellcolor[HTML]{C0C0C0}{\color[HTML]{000000}True } &    -  &  \cellcolor[HTML]{C0C0C0}{\color[HTML]{000000}True }    &  \cellcolor[HTML]{C0C0C0}{\color[HTML]{000000}True }   &    \cellcolor[HTML]{C0C0C0}{\color[HTML]{000000}True }    &    \cellcolor[HTML]{C0C0C0}{\color[HTML]{000000}True }  &  -  &  - &  -       \\

& WordMixup                         &   \cellcolor[HTML]{C0C0C0}{\color[HTML]{000000}True }      &    -     &   \cellcolor[HTML]{C0C0C0}{\color[HTML]{000000}True }     &    \cellcolor[HTML]{C0C0C0}{\color[HTML]{000000}True }  &    \cellcolor[HTML]{C0C0C0}{\color[HTML]{000000}True } &    \cellcolor[HTML]{C0C0C0}{\color[HTML]{000000}True }  &  \cellcolor[HTML]{C0C0C0}{\color[HTML]{000000}True }   &  \cellcolor[HTML]{C0C0C0}{\color[HTML]{000000}True }    &   \cellcolor[HTML]{C0C0C0}{\color[HTML]{000000}True }    &  \cellcolor[HTML]{C0C0C0}{\color[HTML]{000000}True }  &  -  &  - &  -       \\

& SenMixup                        &   \cellcolor[HTML]{C0C0C0}{\color[HTML]{000000}True }     &    \cellcolor[HTML]{C0C0C0}{\color[HTML]{000000}True }     &   -     &    \cellcolor[HTML]{C0C0C0}{\color[HTML]{000000}True }  &   \cellcolor[HTML]{C0C0C0}{\color[HTML]{000000}True } &    \cellcolor[HTML]{C0C0C0}{\color[HTML]{000000}True } &  \cellcolor[HTML]{C0C0C0}{\color[HTML]{000000}True }   &  \cellcolor[HTML]{C0C0C0}{\color[HTML]{000000}True }    &    \cellcolor[HTML]{C0C0C0}{\color[HTML]{000000}True }    &    \cellcolor[HTML]{C0C0C0}{\color[HTML]{000000}True }  &  \cellcolor[HTML]{C0C0C0}{\color[HTML]{000000}True }  & \cellcolor[HTML]{C0C0C0}{\color[HTML]{000000}True } &  -       \\

& Refactor              &   -      &    \cellcolor[HTML]{C0C0C0}{\color[HTML]{000000}True }     &   \cellcolor[HTML]{C0C0C0}{\color[HTML]{000000}True }     &    -  &    - &    -  &  \cellcolor[HTML]{C0C0C0}{\color[HTML]{000000}True }   &  \cellcolor[HTML]{C0C0C0}{\color[HTML]{000000}True }    &    \cellcolor[HTML]{C0C0C0}{\color[HTML]{000000}True }    &   \cellcolor[HTML]{C0C0C0}{\color[HTML]{000000}True }  &  -  &  - &  -       \\

& Rename             &   \cellcolor[HTML]{C0C0C0}{\color[HTML]{000000}True }      &    \cellcolor[HTML]{C0C0C0}{\color[HTML]{000000}True }     &   \cellcolor[HTML]{C0C0C0}{\color[HTML]{000000}True }     &    -  &    - &    \cellcolor[HTML]{C0C0C0}{\color[HTML]{000000}True }  &  -    &  -    &    -    &    -  &  \cellcolor[HTML]{C0C0C0}{\color[HTML]{000000}True } &  \cellcolor[HTML]{C0C0C0}{\color[HTML]{000000}True } &  -       \\

& Dead              &   -      &   \cellcolor[HTML]{C0C0C0}{\color[HTML]{000000}True }     &   \cellcolor[HTML]{C0C0C0}{\color[HTML]{000000}True }     &    -  &   \cellcolor[HTML]{C0C0C0}{\color[HTML]{000000}True } &    -  &  \cellcolor[HTML]{C0C0C0}{\color[HTML]{000000}True }    &  \cellcolor[HTML]{C0C0C0}{\color[HTML]{000000}True }   &    \cellcolor[HTML]{C0C0C0}{\color[HTML]{000000}True }   &    \cellcolor[HTML]{C0C0C0}{\color[HTML]{000000}True }  &  -  &  - &  -       \\

& Inside              &   \cellcolor[HTML]{C0C0C0}{\color[HTML]{000000}True }      &    \cellcolor[HTML]{C0C0C0}{\color[HTML]{000000}True }     &   \cellcolor[HTML]{C0C0C0}{\color[HTML]{000000}True }    &    \cellcolor[HTML]{C0C0C0}{\color[HTML]{000000}True } &    - &    \cellcolor[HTML]{C0C0C0}{\color[HTML]{000000}True }  &  -    &  -    &    -    &    -  &  \cellcolor[HTML]{C0C0C0}{\color[HTML]{000000}True }  & \cellcolor[HTML]{C0C0C0}{\color[HTML]{000000}True } &  -       \\

& Outside             &   \cellcolor[HTML]{C0C0C0}{\color[HTML]{000000}True }      &    \cellcolor[HTML]{C0C0C0}{\color[HTML]{000000}True }     &   \cellcolor[HTML]{C0C0C0}{\color[HTML]{000000}True }     &    \cellcolor[HTML]{C0C0C0}{\color[HTML]{000000}True }  &    - &   \cellcolor[HTML]{C0C0C0}{\color[HTML]{000000}True }  &  -    &  -    &    -    &    -  &  \cellcolor[HTML]{C0C0C0}{\color[HTML]{000000}True }  &  \cellcolor[HTML]{C0C0C0}{\color[HTML]{000000}True } & \cellcolor[HTML]{C0C0C0}{\color[HTML]{000000}True }       \\

& SR               &   \cellcolor[HTML]{C0C0C0}{\color[HTML]{000000}True }      &    \cellcolor[HTML]{C0C0C0}{\color[HTML]{000000}True }     &   \cellcolor[HTML]{C0C0C0}{\color[HTML]{000000}True }     &   \cellcolor[HTML]{C0C0C0}{\color[HTML]{000000}True }  &    - &    \cellcolor[HTML]{C0C0C0}{\color[HTML]{000000}True }  &  -    &  -    &    -    &    -  &  \cellcolor[HTML]{C0C0C0}{\color[HTML]{000000}True }  & \cellcolor[HTML]{C0C0C0}{\color[HTML]{000000}True } &  \cellcolor[HTML]{C0C0C0}{\color[HTML]{000000}True }       \\

& RI            &   \cellcolor[HTML]{C0C0C0}{\color[HTML]{000000}True }      &    \cellcolor[HTML]{C0C0C0}{\color[HTML]{000000}True }     &   \cellcolor[HTML]{C0C0C0}{\color[HTML]{000000}True }     &  \cellcolor[HTML]{C0C0C0}{\color[HTML]{000000}True }  &    - &    \cellcolor[HTML]{C0C0C0}{\color[HTML]{000000}True }  &  -    &  -    &    -    &    -  &  \cellcolor[HTML]{C0C0C0}{\color[HTML]{000000}True } &  \cellcolor[HTML]{C0C0C0}{\color[HTML]{000000}True } &  \cellcolor[HTML]{C0C0C0}{\color[HTML]{000000}True }       \\

& RS              &   -      &    -     &   \cellcolor[HTML]{C0C0C0}{\color[HTML]{000000}True }     &    -  &    \cellcolor[HTML]{C0C0C0}{\color[HTML]{000000}True } &    -  &  \cellcolor[HTML]{C0C0C0}{\color[HTML]{000000}True }    &  \cellcolor[HTML]{C0C0C0}{\color[HTML]{000000}True }    &    \cellcolor[HTML]{C0C0C0}{\color[HTML]{000000}True }    &    \cellcolor[HTML]{C0C0C0}{\color[HTML]{000000}True }  &  -  &  - &  -       \\

& RD              &   -      &    -     &   \cellcolor[HTML]{C0C0C0}{\color[HTML]{000000}True }     &    -  &    \cellcolor[HTML]{C0C0C0}{\color[HTML]{000000}True }  &    -  &  \cellcolor[HTML]{C0C0C0}{\color[HTML]{000000}True }    &  \cellcolor[HTML]{C0C0C0}{\color[HTML]{000000}True }    &    \cellcolor[HTML]{C0C0C0}{\color[HTML]{000000}True }   &   \cellcolor[HTML]{C0C0C0}{\color[HTML]{000000}True }  &  -   &  - &  - \\

& BT              &   -      &    -     &   -     &    -  &    - &    -  &  -    &  \cellcolor[HTML]{C0C0C0}{\color[HTML]{000000}True }    &   \cellcolor[HTML]{C0C0C0}{\color[HTML]{000000}True }    &   \cellcolor[HTML]{C0C0C0}{\color[HTML]{000000}True } &  -  &  - &  -       \\

\hline                  
\multirow{13}{*}{{\begin{tabular}[c]{@{}c@{}}GraphCodeBERT\\ (1\%)\end{tabular}}} & No Aug                         &   -      &    -     &   \cellcolor[HTML]{C0C0C0}{\color[HTML]{000000}True }      &    \cellcolor[HTML]{C0C0C0}{\color[HTML]{000000}True }   &    \cellcolor[HTML]{C0C0C0}{\color[HTML]{000000}True }  &     \cellcolor[HTML]{C0C0C0}{\color[HTML]{000000}True }   &   \cellcolor[HTML]{C0C0C0}{\color[HTML]{000000}True }     &   \cellcolor[HTML]{C0C0C0}{\color[HTML]{000000}True }     &     \cellcolor[HTML]{C0C0C0}{\color[HTML]{000000}True }    &    \cellcolor[HTML]{C0C0C0}{\color[HTML]{000000}True }   &   \cellcolor[HTML]{C0C0C0}{\color[HTML]{000000}True }  &  - &  -       \\

& WordMixup                         &   -      &    -     &   \cellcolor[HTML]{C0C0C0}{\color[HTML]{000000}True }       &     \cellcolor[HTML]{C0C0C0}{\color[HTML]{000000}True }    &     \cellcolor[HTML]{C0C0C0}{\color[HTML]{000000}True }   &     \cellcolor[HTML]{C0C0C0}{\color[HTML]{000000}True }   &   \cellcolor[HTML]{C0C0C0}{\color[HTML]{000000}True }      &   \cellcolor[HTML]{C0C0C0}{\color[HTML]{000000}True }     &     \cellcolor[HTML]{C0C0C0}{\color[HTML]{000000}True }     &     \cellcolor[HTML]{C0C0C0}{\color[HTML]{000000}True }    &  -  &  - &  -       \\

& SenMixup                        &    \cellcolor[HTML]{C0C0C0}{\color[HTML]{000000}True }      &     \cellcolor[HTML]{C0C0C0}{\color[HTML]{000000}True }       &   -     &     \cellcolor[HTML]{C0C0C0}{\color[HTML]{000000}True }   &    \cellcolor[HTML]{C0C0C0}{\color[HTML]{000000}True }   &     \cellcolor[HTML]{C0C0C0}{\color[HTML]{000000}True }    &   \cellcolor[HTML]{C0C0C0}{\color[HTML]{000000}True }      &   \cellcolor[HTML]{C0C0C0}{\color[HTML]{000000}True }      &     \cellcolor[HTML]{C0C0C0}{\color[HTML]{000000}True }      &  \cellcolor[HTML]{C0C0C0}{\color[HTML]{000000}True }    &   \cellcolor[HTML]{C0C0C0}{\color[HTML]{000000}True }   &  - &   \cellcolor[HTML]{C0C0C0}{\color[HTML]{000000}True }         \\

& Refactor              &    \cellcolor[HTML]{C0C0C0}{\color[HTML]{000000}True }       &     \cellcolor[HTML]{C0C0C0}{\color[HTML]{000000}True }       &    \cellcolor[HTML]{C0C0C0}{\color[HTML]{000000}True }      &    -  &    \cellcolor[HTML]{C0C0C0}{\color[HTML]{000000}True }  &    -  &  \cellcolor[HTML]{C0C0C0}{\color[HTML]{000000}True }     &   \cellcolor[HTML]{C0C0C0}{\color[HTML]{000000}True }   &    \cellcolor[HTML]{C0C0C0}{\color[HTML]{000000}True }     &     \cellcolor[HTML]{C0C0C0}{\color[HTML]{000000}True }  &   \cellcolor[HTML]{C0C0C0}{\color[HTML]{000000}True }  &  - &  -       \\

& Rename             &    \cellcolor[HTML]{C0C0C0}{\color[HTML]{000000}True }       &     \cellcolor[HTML]{C0C0C0}{\color[HTML]{000000}True }       &    \cellcolor[HTML]{C0C0C0}{\color[HTML]{000000}True }       &     \cellcolor[HTML]{C0C0C0}{\color[HTML]{000000}True }   &    - &     \cellcolor[HTML]{C0C0C0}{\color[HTML]{000000}True }   &  -    &  -    &     \cellcolor[HTML]{C0C0C0}{\color[HTML]{000000}True }     &     \cellcolor[HTML]{C0C0C0}{\color[HTML]{000000}True }   &   \cellcolor[HTML]{C0C0C0}{\color[HTML]{000000}True }  &   \cellcolor[HTML]{C0C0C0}{\color[HTML]{000000}True }  &  -       \\

& Dead              &   \cellcolor[HTML]{C0C0C0}{\color[HTML]{000000}True }      &     \cellcolor[HTML]{C0C0C0}{\color[HTML]{000000}True }     &    \cellcolor[HTML]{C0C0C0}{\color[HTML]{000000}True }      &    -  &     \cellcolor[HTML]{C0C0C0}{\color[HTML]{000000}True }  &    -  &   \cellcolor[HTML]{C0C0C0}{\color[HTML]{000000}True }    &   \cellcolor[HTML]{C0C0C0}{\color[HTML]{000000}True }    &     \cellcolor[HTML]{C0C0C0}{\color[HTML]{000000}True }    &     \cellcolor[HTML]{C0C0C0}{\color[HTML]{000000}True }   &   \cellcolor[HTML]{C0C0C0}{\color[HTML]{000000}True }   &  - &  -       \\

& Inside              &    \cellcolor[HTML]{C0C0C0}{\color[HTML]{000000}True }       &   \cellcolor[HTML]{C0C0C0}{\color[HTML]{000000}True }       &   \cellcolor[HTML]{C0C0C0}{\color[HTML]{000000}True }       &     \cellcolor[HTML]{C0C0C0}{\color[HTML]{000000}True }   &    - &     \cellcolor[HTML]{C0C0C0}{\color[HTML]{000000}True }  &  -    &  -    &     \cellcolor[HTML]{C0C0C0}{\color[HTML]{000000}True }     &     \cellcolor[HTML]{C0C0C0}{\color[HTML]{000000}True }  &   \cellcolor[HTML]{C0C0C0}{\color[HTML]{000000}True }  &   \cellcolor[HTML]{C0C0C0}{\color[HTML]{000000}True }  &  -       \\

& Outside             &   \cellcolor[HTML]{C0C0C0}{\color[HTML]{000000}True }       &     \cellcolor[HTML]{C0C0C0}{\color[HTML]{000000}True }    &    \cellcolor[HTML]{C0C0C0}{\color[HTML]{000000}True }     &     \cellcolor[HTML]{C0C0C0}{\color[HTML]{000000}True }   &    - &     \cellcolor[HTML]{C0C0C0}{\color[HTML]{000000}True }   &  -    &  -    &     \cellcolor[HTML]{C0C0C0}{\color[HTML]{000000}True }     &     \cellcolor[HTML]{C0C0C0}{\color[HTML]{000000}True }   &  \cellcolor[HTML]{C0C0C0}{\color[HTML]{000000}True }   &   \cellcolor[HTML]{C0C0C0}{\color[HTML]{000000}True }  &  -       \\

& SR               &   \cellcolor[HTML]{C0C0C0}{\color[HTML]{000000}True }       &    \cellcolor[HTML]{C0C0C0}{\color[HTML]{000000}True }     &    \cellcolor[HTML]{C0C0C0}{\color[HTML]{000000}True }       &     \cellcolor[HTML]{C0C0C0}{\color[HTML]{000000}True }   &  \cellcolor[HTML]{C0C0C0}{\color[HTML]{000000}True }  &   \cellcolor[HTML]{C0C0C0}{\color[HTML]{000000}True }   &   \cellcolor[HTML]{C0C0C0}{\color[HTML]{000000}True }    &   \cellcolor[HTML]{C0C0C0}{\color[HTML]{000000}True }   &    -    &    -  &  \cellcolor[HTML]{C0C0C0}{\color[HTML]{000000}True }   &   \cellcolor[HTML]{C0C0C0}{\color[HTML]{000000}True }  &  -       \\

& RI            &    \cellcolor[HTML]{C0C0C0}{\color[HTML]{000000}True }     &    \cellcolor[HTML]{C0C0C0}{\color[HTML]{000000}True }      &    \cellcolor[HTML]{C0C0C0}{\color[HTML]{000000}True }       &     \cellcolor[HTML]{C0C0C0}{\color[HTML]{000000}True }   &     \cellcolor[HTML]{C0C0C0}{\color[HTML]{000000}True }  &     \cellcolor[HTML]{C0C0C0}{\color[HTML]{000000}True }   &   \cellcolor[HTML]{C0C0C0}{\color[HTML]{000000}True }     &   \cellcolor[HTML]{C0C0C0}{\color[HTML]{000000}True }    &    -    &    -  &   \cellcolor[HTML]{C0C0C0}{\color[HTML]{000000}True }  &   \cellcolor[HTML]{C0C0C0}{\color[HTML]{000000}True }  &   \cellcolor[HTML]{C0C0C0}{\color[HTML]{000000}True }      \\

& RS              &    \cellcolor[HTML]{C0C0C0}{\color[HTML]{000000}True }      &    -     &    \cellcolor[HTML]{C0C0C0}{\color[HTML]{000000}True }      &    \cellcolor[HTML]{C0C0C0}{\color[HTML]{000000}True }   &     \cellcolor[HTML]{C0C0C0}{\color[HTML]{000000}True }  &    \cellcolor[HTML]{C0C0C0}{\color[HTML]{000000}True }   &   \cellcolor[HTML]{C0C0C0}{\color[HTML]{000000}True }     &  \cellcolor[HTML]{C0C0C0}{\color[HTML]{000000}True }   &     \cellcolor[HTML]{C0C0C0}{\color[HTML]{000000}True }    &     \cellcolor[HTML]{C0C0C0}{\color[HTML]{000000}True }  &  -  &  - &   \cellcolor[HTML]{C0C0C0}{\color[HTML]{000000}True }       \\

& RD              &   -      &    -     &    -      &    -  &     \cellcolor[HTML]{C0C0C0}{\color[HTML]{000000}True }   &    -  &   \cellcolor[HTML]{C0C0C0}{\color[HTML]{000000}True }    &   \cellcolor[HTML]{C0C0C0}{\color[HTML]{000000}True }    &     \cellcolor[HTML]{C0C0C0}{\color[HTML]{000000}True }    &     \cellcolor[HTML]{C0C0C0}{\color[HTML]{000000}True }   &  -   &  - &  - \\

& BT              &   -      &    -     &  \cellcolor[HTML]{C0C0C0}{\color[HTML]{000000}True }  -     &    -  &    - &    -  &  -    &  -    &    -    &    \cellcolor[HTML]{C0C0C0}{\color[HTML]{000000}True }  &   \cellcolor[HTML]{C0C0C0}{\color[HTML]{000000}True }  &  - &  -       \\

\hline  

\end{tabular}
}
\end{sidewaystable}


\begin{sidewaystable}[]
\centering
    \setlength{\tabcolsep}{3pt} 
    \renewcommand{\arraystretch}{1.3}
    \tiny
\caption{Results of statistical tests on Accuracy \& Robustness. \textbf{True}: indicates that the comparison is statistically significant after the  \emph{Bonferroni correction} adjustment. Statistical tests method: \emph{Wilcoxon signed-rank test}. A gray background highlights the result marked as \textbf{True}. For legibility, we simplify the terms \emph{Rename Operator}, \emph{Dead Operator}, \emph{Inside Operator}, and \emph{Outside Operator} to \emph{Rename}, \emph{Dead}, \emph{Inside}, and \emph{Outside}, respectively. Across CodeBERT \& GraphCodeBERT models.}
\label{table:statistical_testing_acc_robust_RQ4_codebert_graphcodebert}
\centering
\resizebox{\textwidth}{!}{
\begin{tabular}{lcccccccccccccccc}
\hline
                    & DA method & No Aug & WordMixup & SenMixup  & Refactor & Rename & Dead  & Inside & Outside  & SR & RI &  RS &  RD & BT & \\ \hline
\multirow{13}{*}{{\begin{tabular}[c]{@{}c@{}}Accuracy \end{tabular}}} & No Aug                         &   -      &    -     &   -     &     \cellcolor[HTML]{C0C0C0}{\color[HTML]{000000}True }  &      -  &    \cellcolor[HTML]{C0C0C0}{\color[HTML]{000000}True }  &   -   &    \cellcolor[HTML]{C0C0C0}{\color[HTML]{000000}True }   &    -    &    -  &   -   &   -  &  -       \\

& WordMixup                         &   -      &    -     &   -     &    -  &    - &    \cellcolor[HTML]{C0C0C0}{\color[HTML]{000000}True }   &  -    &  -    &    -    &    -  &  -  &  - &  -       \\

& SenMixup                        &   -      &    -     &   -     &    -  &    - &    \cellcolor[HTML]{C0C0C0}{\color[HTML]{000000}True }   &  -    &  -    &    -    &    -  &  -  &  - &  -       \\

& Refactor              &     \cellcolor[HTML]{C0C0C0}{\color[HTML]{000000}True }      &    -     &   -     &    -  &    - &    -  &  -    &  -    &    -    &    -  &  -  &  - &  -       \\

& Rename             &   -      &    -     &   -     &    -  &    - &    \cellcolor[HTML]{C0C0C0}{\color[HTML]{000000}True }  &  -    &  \cellcolor[HTML]{C0C0C0}{\color[HTML]{000000}True }    &    -    &    -  &  -  &  - &  -       \\

& Dead              &  \cellcolor[HTML]{C0C0C0}{\color[HTML]{000000}True }       &    \cellcolor[HTML]{C0C0C0}{\color[HTML]{000000}True }     &   \cellcolor[HTML]{C0C0C0}{\color[HTML]{000000}True }      &    -  &   \cellcolor[HTML]{C0C0C0}{\color[HTML]{000000}True }  &    -  &  -    &  -    &    \cellcolor[HTML]{C0C0C0}{\color[HTML]{000000}True }     &    \cellcolor[HTML]{C0C0C0}{\color[HTML]{000000}True }   &  \cellcolor[HTML]{C0C0C0}{\color[HTML]{000000}True }  & \cellcolor[HTML]{C0C0C0}{\color[HTML]{000000}True }  &  -       \\

& Inside              &   -      &    -     &   -     &    -  &    - &    -  &  -    &  -    &    -    &    -  &  -  &  - &  -       \\

& Outside             &   \cellcolor[HTML]{C0C0C0}{\color[HTML]{000000}True }       &    -     &   -     &    -  &    \cellcolor[HTML]{C0C0C0}{\color[HTML]{000000}True }  &    -  &  -    &  -    &    \cellcolor[HTML]{C0C0C0}{\color[HTML]{000000}True }    &    \cellcolor[HTML]{C0C0C0}{\color[HTML]{000000}True }   &  -  &  - &  -       \\

& SR               &   -      &    -     &   -     &    -  &    - &    \cellcolor[HTML]{C0C0C0}{\color[HTML]{000000}True }   &  -    & \cellcolor[HTML]{C0C0C0}{\color[HTML]{000000}True }    &    -    &    -  &  -  &  - &  -       \\

& RI            &   -      &    -     &   -     &    -  &    - &    \cellcolor[HTML]{C0C0C0}{\color[HTML]{000000}True }   &  -    & \cellcolor[HTML]{C0C0C0}{\color[HTML]{000000}True }    &    -    &    -  &  -  &  - &  -       \\

& RS              &   -      &    -     &   -     &    -  &    - &    \cellcolor[HTML]{C0C0C0}{\color[HTML]{000000}True }  &  -    &  -    &    -    &    -  &  -  &  - &  -       \\

& RD              &   -      &    -     &   -     &    -  &    -  &    \cellcolor[HTML]{C0C0C0}{\color[HTML]{000000}True }   &  -    &  -    &    -    &    -  &  -   &  - &  - \\

& BT              &   -      &    -     &   -     &    -  &    - &    -  &  -    &  -    &    -    &    -  &  -  &  - &  -       \\

\hline                  
\multirow{13}{*}{{\begin{tabular}[c]{@{}c@{}}Robustness \end{tabular}}} & No Aug                         &   -      &    -     &   \cellcolor[HTML]{C0C0C0}{\color[HTML]{000000}True }     &   \cellcolor[HTML]{C0C0C0}{\color[HTML]{000000}True }   &   \cellcolor[HTML]{C0C0C0}{\color[HTML]{000000}True }  &    \cellcolor[HTML]{C0C0C0}{\color[HTML]{000000}True }  & \cellcolor[HTML]{C0C0C0}{\color[HTML]{000000}True }    &  \cellcolor[HTML]{C0C0C0}{\color[HTML]{000000}True }    &    -    &    -  &  \cellcolor[HTML]{C0C0C0}{\color[HTML]{000000}True }   &  \cellcolor[HTML]{C0C0C0}{\color[HTML]{000000}True }  &  -       \\

& WordMixup                         &   -      &    -     &   \cellcolor[HTML]{C0C0C0}{\color[HTML]{000000}True }      &   \cellcolor[HTML]{C0C0C0}{\color[HTML]{000000}True }   &  \cellcolor[HTML]{C0C0C0}{\color[HTML]{000000}True }  &    \cellcolor[HTML]{C0C0C0}{\color[HTML]{000000}True }   &  -    &  -    &    -    &    -  &  \cellcolor[HTML]{C0C0C0}{\color[HTML]{000000}True }   &  \cellcolor[HTML]{C0C0C0}{\color[HTML]{000000}True } &  -       \\

& SenMixup                        &   \cellcolor[HTML]{C0C0C0}{\color[HTML]{000000}True }     &   \cellcolor[HTML]{C0C0C0}{\color[HTML]{000000}True }      &   -     &    -  &    \cellcolor[HTML]{C0C0C0}{\color[HTML]{000000}True }  &    -  &  \cellcolor[HTML]{C0C0C0}{\color[HTML]{000000}True }     &  \cellcolor[HTML]{C0C0C0}{\color[HTML]{000000}True }     &   \cellcolor[HTML]{C0C0C0}{\color[HTML]{000000}True }     &   \cellcolor[HTML]{C0C0C0}{\color[HTML]{000000}True }   &  -  &  - &  \cellcolor[HTML]{C0C0C0}{\color[HTML]{000000}True }       \\

& Refactor              &   \cellcolor[HTML]{C0C0C0}{\color[HTML]{000000}True }      &   \cellcolor[HTML]{C0C0C0}{\color[HTML]{000000}True }      &   -     &    -  &    \cellcolor[HTML]{C0C0C0}{\color[HTML]{000000}True }  &    -  &  \cellcolor[HTML]{C0C0C0}{\color[HTML]{000000}True }     &  \cellcolor[HTML]{C0C0C0}{\color[HTML]{000000}True }    &    \cellcolor[HTML]{C0C0C0}{\color[HTML]{000000}True }     &    \cellcolor[HTML]{C0C0C0}{\color[HTML]{000000}True }   &  -  &  - &  \cellcolor[HTML]{C0C0C0}{\color[HTML]{000000}True }        \\

& Rename             &   \cellcolor[HTML]{C0C0C0}{\color[HTML]{000000}True }       &    \cellcolor[HTML]{C0C0C0}{\color[HTML]{000000}True }      &   \cellcolor[HTML]{C0C0C0}{\color[HTML]{000000}True }      &    \cellcolor[HTML]{C0C0C0}{\color[HTML]{000000}True }   &    - &    \cellcolor[HTML]{C0C0C0}{\color[HTML]{000000}True }  &  \cellcolor[HTML]{C0C0C0}{\color[HTML]{000000}True }    &  -    &    \cellcolor[HTML]{C0C0C0}{\color[HTML]{000000}True }    &    -  &  \cellcolor[HTML]{C0C0C0}{\color[HTML]{000000}True }   &  \cellcolor[HTML]{C0C0C0}{\color[HTML]{000000}True }  &  -       \\

& Dead              &   \cellcolor[HTML]{C0C0C0}{\color[HTML]{000000}True }       &    \cellcolor[HTML]{C0C0C0}{\color[HTML]{000000}True }      &   -     &    -  &    \cellcolor[HTML]{C0C0C0}{\color[HTML]{000000}True }  &    -  & \cellcolor[HTML]{C0C0C0}{\color[HTML]{000000}True }    &  \cellcolor[HTML]{C0C0C0}{\color[HTML]{000000}True }     &   \cellcolor[HTML]{C0C0C0}{\color[HTML]{000000}True }   &   \cellcolor[HTML]{C0C0C0}{\color[HTML]{000000}True }  &  -  &  - &  \cellcolor[HTML]{C0C0C0}{\color[HTML]{000000}True }        \\

& Inside              &   \cellcolor[HTML]{C0C0C0}{\color[HTML]{000000}True }       &    -     &   \cellcolor[HTML]{C0C0C0}{\color[HTML]{000000}True }     &   \cellcolor[HTML]{C0C0C0}{\color[HTML]{000000}True }   &    \cellcolor[HTML]{C0C0C0}{\color[HTML]{000000}True }  &   \cellcolor[HTML]{C0C0C0}{\color[HTML]{000000}True }  &  -    &  -    &    -    &    -  &  \cellcolor[HTML]{C0C0C0}{\color[HTML]{000000}True }  & \cellcolor[HTML]{C0C0C0}{\color[HTML]{000000}True }  &  -       \\

& Outside             &   \cellcolor[HTML]{C0C0C0}{\color[HTML]{000000}True }     &    -     &   \cellcolor[HTML]{C0C0C0}{\color[HTML]{000000}True }     &    \cellcolor[HTML]{C0C0C0}{\color[HTML]{000000}True }  &    - &    \cellcolor[HTML]{C0C0C0}{\color[HTML]{000000}True }  &  -    &  -    &    -    &    -  &  \cellcolor[HTML]{C0C0C0}{\color[HTML]{000000}True }   &  \cellcolor[HTML]{C0C0C0}{\color[HTML]{000000}True }  &  -       \\

& SR               &   -      &    -     &   \cellcolor[HTML]{C0C0C0}{\color[HTML]{000000}True }      &    \cellcolor[HTML]{C0C0C0}{\color[HTML]{000000}True }   &    \cellcolor[HTML]{C0C0C0}{\color[HTML]{000000}True } &  \cellcolor[HTML]{C0C0C0}{\color[HTML]{000000}True }   &  -    &  -    &    -    &    -  & \cellcolor[HTML]{C0C0C0}{\color[HTML]{000000}True }   &  \cellcolor[HTML]{C0C0C0}{\color[HTML]{000000}True }  &  -       \\

& RI            &   -      &    -     &  \cellcolor[HTML]{C0C0C0}{\color[HTML]{000000}True }     &    \cellcolor[HTML]{C0C0C0}{\color[HTML]{000000}True }   &    - &   \cellcolor[HTML]{C0C0C0}{\color[HTML]{000000}True }  &  -    &  -    &    -    &    -  &  \cellcolor[HTML]{C0C0C0}{\color[HTML]{000000}True }  &  \cellcolor[HTML]{C0C0C0}{\color[HTML]{000000}True }  &  -       \\

& RS              &   \cellcolor[HTML]{C0C0C0}{\color[HTML]{000000}True }      &   \cellcolor[HTML]{C0C0C0}{\color[HTML]{000000}True }      &   -     &    -  &    \cellcolor[HTML]{C0C0C0}{\color[HTML]{000000}True } &    -  &  \cellcolor[HTML]{C0C0C0}{\color[HTML]{000000}True }   &  \cellcolor[HTML]{C0C0C0}{\color[HTML]{000000}True }     &   \cellcolor[HTML]{C0C0C0}{\color[HTML]{000000}True }    &    \cellcolor[HTML]{C0C0C0}{\color[HTML]{000000}True }   &  -  &  - &  \cellcolor[HTML]{C0C0C0}{\color[HTML]{000000}True }        \\

& RD              &  \cellcolor[HTML]{C0C0C0}{\color[HTML]{000000}True }       &    \cellcolor[HTML]{C0C0C0}{\color[HTML]{000000}True }      &   -     &    -  &    \cellcolor[HTML]{C0C0C0}{\color[HTML]{000000}True }  &    -  &  \cellcolor[HTML]{C0C0C0}{\color[HTML]{000000}True }    &  \cellcolor[HTML]{C0C0C0}{\color[HTML]{000000}True }     &   \cellcolor[HTML]{C0C0C0}{\color[HTML]{000000}True }     &    \cellcolor[HTML]{C0C0C0}{\color[HTML]{000000}True }   &  -   &  - & \cellcolor[HTML]{C0C0C0}{\color[HTML]{000000}True }  \\

& BT              &   -      &    -     &   \cellcolor[HTML]{C0C0C0}{\color[HTML]{000000}True }      &    \cellcolor[HTML]{C0C0C0}{\color[HTML]{000000}True }  &    - &    \cellcolor[HTML]{C0C0C0}{\color[HTML]{000000}True }  &  -    &  -    &    -    &    -  &  \cellcolor[HTML]{C0C0C0}{\color[HTML]{000000}True }  &  \cellcolor[HTML]{C0C0C0}{\color[HTML]{000000}True }  &  -       \\

\hline  

\end{tabular}
}
\end{sidewaystable}

\end{document}